\documentclass[prd,aps,showpacs,preprintnumbers,amsmath,amssymb,superscriptaddress,twocolumn,floatfix]{revtex4}
\usepackage{alltt}
\usepackage{epsfig}
\usepackage{calc,rotating,xspace}
\usepackage{pifont}
\usepackage[mediumspace,mediumqspace,Grey,squaren]{SIunits}
\usepackage{times}
\usepackage{mathptmx}
\newcommand{\met}{\mbox{${\hbox{$p$\kern-0.35em\lower-.1ex\hbox{/}}}_T\:$}}
\newcommand{\vecmet}{\mbox{${\hbox{$\vec{E}$\kern-0.6em\lower-.1ex\hbox{/}}}_T\:$}}

\begin{document}
\title{\boldmath A precise measurement of the $W$-boson mass with the Collider Detector at
Fermilab}
\affiliation{Institute of Physics, Academia Sinica, Taipei, Taiwan 11529, Republic of China}
\affiliation{Argonne National Laboratory, Argonne, Illinois 60439, USA}
\affiliation{University of Athens, 157 71 Athens, Greece}
\affiliation{Institut de Fisica d'Altes Energies, ICREA, Universitat Autonoma de Barcelona, E-08193, Bellaterra (Barcelona), Spain}
\affiliation{Baylor University, Waco, Texas 76798, USA}
\affiliation{Istituto Nazionale di Fisica Nucleare Bologna, \ensuremath{^{jj}}University of Bologna, I-40127 Bologna, Italy}
\affiliation{University of California, Davis, Davis, California 95616, USA}
\affiliation{University of California, Los Angeles, Los Angeles, California 90024, USA}
\affiliation{Instituto de Fisica de Cantabria, CSIC-University of Cantabria, 39005 Santander, Spain}
\affiliation{Carnegie Mellon University, Pittsburgh, Pennsylvania 15213, USA}
\affiliation{Enrico Fermi Institute, University of Chicago, Chicago, Illinois 60637, USA}
\affiliation{Comenius University, 842 48 Bratislava, Slovakia; Institute of Experimental Physics, 040 01 Kosice, Slovakia}
\affiliation{Joint Institute for Nuclear Research, RU-141980 Dubna, Russia}
\affiliation{Duke University, Durham, North Carolina 27708, USA}
\affiliation{Fermi National Accelerator Laboratory, Batavia, Illinois 60510, USA}
\affiliation{University of Florida, Gainesville, Florida 32611, USA}
\affiliation{Laboratori Nazionali di Frascati, Istituto Nazionale di Fisica Nucleare, I-00044 Frascati, Italy}
\affiliation{University of Geneva, CH-1211 Geneva 4, Switzerland}
\affiliation{Glasgow University, Glasgow G12 8QQ, United Kingdom}
\affiliation{Harvard University, Cambridge, Massachusetts 02138, USA}
\affiliation{Division of High Energy Physics, Department of Physics, University of Helsinki, FIN-00014, Helsinki, Finland; Helsinki Institute of Physics, FIN-00014, Helsinki, Finland}
\affiliation{University of Illinois, Urbana, Illinois 61801, USA}
\affiliation{The Johns Hopkins University, Baltimore, Maryland 21218, USA}
\affiliation{Institut für Experimentelle Kernphysik, Karlsruhe Institute of Technology, D-76131 Karlsruhe, Germany}
\affiliation{Center for High Energy Physics: Kyungpook National University, Daegu 702-701, Korea; Seoul National University, Seoul 151-742, Korea; Sungkyunkwan University, Suwon 440-746, Korea; Korea Institute of Science and Technology Information, Daejeon 305-806, Korea; Chonnam National University, Gwangju 500-757, Korea; Chonbuk National University, Jeonju 561-756, Korea; Ewha Womans University, Seoul, 120-750, Korea}
\affiliation{Ernest Orlando Lawrence Berkeley National Laboratory, Berkeley, California 94720, USA}
\affiliation{University of Liverpool, Liverpool L69 7ZE, United Kingdom}
\affiliation{University College London, London WC1E 6BT, United Kingdom}
\affiliation{Centro de Investigaciones Energeticas Medioambientales y Tecnologicas, E-28040 Madrid, Spain}
\affiliation{Massachusetts Institute of Technology, Cambridge, Massachusetts 02139, USA}
\affiliation{Institute of Particle Physics: McGill University, Montréal, Québec H3A~2T8, Canada; Simon Fraser University, Burnaby, British Columbia V5A~1S6, Canada; University of Toronto, Toronto, Ontario M5S~1A7, Canada; TRIUMF, Vancouver, British Columbia V6T~2A3, Canada}
\affiliation{University of Michigan, Ann Arbor, Michigan 48109, USA}
\affiliation{Michigan State University, East Lansing, Michigan 48824, USA}
\affiliation{Institution for Theoretical and Experimental Physics, ITEP, Moscow 117259, Russia}
\affiliation{University of New Mexico, Albuquerque, New Mexico 87131, USA}
\affiliation{The Ohio State University, Columbus, Ohio 43210, USA}
\affiliation{Okayama University, Okayama 700-8530, Japan}
\affiliation{Osaka City University, Osaka 558-8585, Japan}
\affiliation{University of Oxford, Oxford OX1 3RH, United Kingdom}
\affiliation{Istituto Nazionale di Fisica Nucleare, Sezione di Padova, \ensuremath{^{kk}}University of Padova, I-35131 Padova, Italy}
\affiliation{University of Pennsylvania, Philadelphia, Pennsylvania 19104, USA}
\affiliation{Istituto Nazionale di Fisica Nucleare Pisa, \ensuremath{^{ll}}University of Pisa, \ensuremath{^{mm}}University of Siena, \ensuremath{^{nn}}Scuola Normale Superiore, I-56127 Pisa, Italy, \ensuremath{^{oo}}INFN Pavia, I-27100 Pavia, Italy, \ensuremath{^{pp}}University of Pavia, I-27100 Pavia, Italy}
\affiliation{University of Pittsburgh, Pittsburgh, Pennsylvania 15260, USA}
\affiliation{Purdue University, West Lafayette, Indiana 47907, USA}
\affiliation{University of Rochester, Rochester, New York 14627, USA}
\affiliation{The Rockefeller University, New York, New York 10065, USA}
\affiliation{Istituto Nazionale di Fisica Nucleare, Sezione di Roma 1, \ensuremath{^{qq}}Sapienza Università di Roma, I-00185 Roma, Italy}
\affiliation{Mitchell Institute for Fundamental Physics and Astronomy, Texas A\&M University, College Station, Texas 77843, USA}
\affiliation{Istituto Nazionale di Fisica Nucleare Trieste, \ensuremath{^{rr}}Gruppo Collegato di Udine, \ensuremath{^{ss}}University of Udine, I-33100 Udine, Italy, \ensuremath{^{tt}}University of Trieste, I-34127 Trieste, Italy}
\affiliation{University of Tsukuba, Tsukuba, Ibaraki 305, Japan}
\affiliation{Tufts University, Medford, Massachusetts 02155, USA}
\affiliation{University of Virginia, Charlottesville, Virginia 22906, USA}
\affiliation{Waseda University, Tokyo 169, Japan}
\affiliation{Wayne State University, Detroit, Michigan 48201, USA}
\affiliation{University of Wisconsin, Madison, Wisconsin 53706, USA}
\affiliation{Yale University, New Haven, Connecticut 06520, USA}

\author{T.~Aaltonen}
\affiliation{Division of High Energy Physics, Department of Physics, University of Helsinki, FIN-00014, Helsinki, Finland; Helsinki Institute of Physics, FIN-00014, Helsinki, Finland}
\author{S.~Amerio\ensuremath{^{kk}}}
\affiliation{Istituto Nazionale di Fisica Nucleare, Sezione di Padova, \ensuremath{^{kk}}University of Padova, I-35131 Padova, Italy}
\author{D.~Amidei}
\affiliation{University of Michigan, Ann Arbor, Michigan 48109, USA}
\author{A.~Anastassov\ensuremath{^{w}}}
\affiliation{Fermi National Accelerator Laboratory, Batavia, Illinois 60510, USA}
\author{A.~Annovi}
\affiliation{Laboratori Nazionali di Frascati, Istituto Nazionale di Fisica Nucleare, I-00044 Frascati, Italy}
\author{J.~Antos}
\affiliation{Comenius University, 842 48 Bratislava, Slovakia; Institute of Experimental Physics, 040 01 Kosice, Slovakia}
\author{G.~Apollinari}
\affiliation{Fermi National Accelerator Laboratory, Batavia, Illinois 60510, USA}
\author{J.A.~Appel}
\affiliation{Fermi National Accelerator Laboratory, Batavia, Illinois 60510, USA}
\author{T.~Arisawa}
\affiliation{Waseda University, Tokyo 169, Japan}
\author{A.~Artikov}
\affiliation{Joint Institute for Nuclear Research, RU-141980 Dubna, Russia}
% ...

\collaboration{CDF Collaboration}
\altaffiliation[With visitors from]{
\ensuremath{^{a}}University of British Columbia, Vancouver, BC V6T 1Z1, Canada,
\ensuremath{^{b}}Istituto Nazionale di Fisica Nucleare, Sezione di Cagliari, 09042 Monserrato (Cagliari), Italy,
\ensuremath{^{c}}University of California Irvine, Irvine, CA 92697, USA,
\ensuremath{^{d}}Institute of Physics, Academy of Sciences of the Czech Republic, 182~21, Czech Republic,
\ensuremath{^{e}}CERN, CH-1211 Geneva, Switzerland,
\ensuremath{^{f}}Cornell University, Ithaca, NY 14853, USA,
\ensuremath{^{g}}University of Cyprus, Nicosia CY-1678, Cyprus,
\ensuremath{^{h}}Office of Science, U.S. Department of Energy, Washington, DC 20585, USA,
\ensuremath{^{i}}University College Dublin, Dublin 4, Ireland,
\ensuremath{^{j}}ETH, 8092 Zürich, Switzerland,
\ensuremath{^{k}}University of Fukui, Fukui City, Fukui Prefecture, Japan 910-0017,
\ensuremath{^{l}}Universidad Iberoamericana, Lomas de Santa Fe, México, C.P. 01219, Distrito Federal,
\ensuremath{^{m}}University of Iowa, Iowa City, IA 52242, USA,
\ensuremath{^{n}}Kinki University, Higashi-Osaka City, Japan 577-8502,
\ensuremath{^{o}}Kansas State University, Manhattan, KS 66506, USA,
\ensuremath{^{p}}Brookhaven National Laboratory, Upton, NY 11973, USA,
\ensuremath{^{q}}University of Manchester, Manchester M13 9PL, United Kingdom,
\ensuremath{^{r}}Queen Mary, University of London, London, E1 4NS, United Kingdom,
\ensuremath{^{s}}University of Melbourne, Victoria 3010, Australia,
\ensuremath{^{t}}Muons, Inc., Batavia, IL 60510, USA,
\ensuremath{^{u}}Nagasaki Institute of Applied Science, Nagasaki 851-0193, Japan,
\ensuremath{^{v}}National Research Nuclear University, Moscow 115409, Russia,
\ensuremath{^{w}}Northwestern University, Evanston, IL 60208, USA,
\ensuremath{^{x}}University of Notre Dame, Notre Dame, IN 46556, USA,
\ensuremath{^{y}}Universidad de Oviedo, E-33007 Oviedo, Spain,
\ensuremath{^{z}}CNRS-IN2P3, Paris, F-75205 France,
\ensuremath{^{aa}}Universidad Tecnica Federico Santa Maria, 110v Valparaiso, Chile,
\ensuremath{^{bb}}The University of Jordan, Amman 11942, Jordan,
\ensuremath{^{cc}}Universite catholique de Louvain, 1348 Louvain-La-Neuve, Belgium,
\ensuremath{^{dd}}University of Zürich, 8006 Zürich, Switzerland,
\ensuremath{^{ee}}Massachusetts General Hospital, Boston, MA 02114 USA,
\ensuremath{^{ff}}Harvard Medical School, Boston, MA 02114 USA,
\ensuremath{^{gg}}Hampton University, Hampton, VA 23668, USA,
\ensuremath{^{hh}}Los Alamos National Laboratory, Los Alamos, NM 87544, USA,
\ensuremath{^{ii}}Università degli Studi di Napoli Federico I, I-80138 Napoli, Italy
}
\noaffiliation
\begin{abstract}
We present a measurement of the $W$-boson mass, $M_W$, using data corresponding to 
2.2~fb$^{-1}$ of integrated luminosity collected in $p\bar{p}$ collisions at 
$\sqrt{s}=1.96$~TeV with the CDF II detector at the Fermilab Tevatron.  The selected 
sample of 470\,126 $W\to e\nu$ candidates and 624\,708 $W\to\mu\nu$ candidates 
yields the measurement $M_W = 80\,387 \pm 12~{\rm (stat)} \pm 15~{\rm (syst)} = 
80\,387 \pm 19$~MeV$/c^2$.  This is the most precise single measurement of the 
$W$-boson mass to date.
\end{abstract}

\maketitle


\section{Introduction}
\label{sec:introduction}
In the standard model (SM) of particle physics, all electroweak interactions are mediated by the 
$W$ boson, the $Z$ boson, and the massless photon, in a gauge theory with symmetry group 
$SU(2)_L \times U(1)_Y$~\cite{GWS}.  If this symmetry were unbroken, the $W$ and $Z$ bosons 
would be massless.  Their nonzero observed masses require a symmetry-breaking mechanism~\cite{ewsb}, 
which in the SM is the Higgs mechanism.  The mass of the resulting scalar excitation, the Higgs 
boson, is not predicted but is constrained by measurements of the weak-boson masses through loop 
corrections.  

Loops in the $W$-boson propagator contribute to the correction $\Delta r$, defined in the 
following expression for the $W$-boson mass $M_W$ in the {\it on-shell} scheme~\cite{sirlin}:

\begin{equation}
M_W^2 =\frac{\hbar^3\pi}{c}\frac{\alpha_{EM}}{\sqrt{2}G_F(1-M_W^2/M_Z^2)(1-\Delta r)},
\label{eq:mwtheory}
\end{equation}
where $\alpha_{EM}$ is the electromagnetic coupling at $Q = M_Z c^2$, $G_F$ is the Fermi 
weak coupling extracted from the muon lifetime measurement, $M_Z$ is the $Z$-boson mass, 
and $\Delta r = 3.58\%$~\cite{pdg} includes all radiative corrections.  In the SM, the 
electroweak radiative corrections are dominated by loops containing top and bottom quarks, 
but also depend logarithmically on the mass of the Higgs boson $M_H$ through loops containing 
the Higgs boson.  A global fit to SM observables yields indirect bounds on $M_H$, whose 
precision is dominated by the uncertainty on $M_W$, with smaller contributions from the 
uncertainties on the top quark mass ($m_t$) and on $\alpha_{EM}$.  A comparison of the 
indirectly-constrained $M_H$ with a direct measurement of $M_H$ is a sensitive probe for new 
particles~\cite{npconstraints}.

Following the discovery of the $W$ boson in 1983 at the UA1 and UA2 
experiments~\cite{WZdiscovery}, measurements of $M_W$ have been performed with increasing 
precision using $\sqrt{s}=1.8$~TeV $p\bar{p}$ collisions at the CDF~\cite{CDF} and 
D0~\cite{DZERO} experiments (Run I); $e^+e^-$ collisions at $\sqrt{s}=161-209$~GeV at the 
ALEPH~\cite{ALEPH}, DELPHI~\cite{DELPHI}, L3~\cite{L3}, and OPAL~\cite{OPAL} experiments 
(LEP); and $\sqrt{s}=1.96$~TeV $p\bar{p}$ collisions at the CDF~\cite{CDF2} and 
D0~\cite{DZERO2} experiments (Run II).  Combining results from Run I, LEP, and the first 
Run II measurements yields $M_W = 80399\pm 23$~MeV$/c^2$~\cite{lepewwg}.  Recent measurements 
performed with the CDF~\cite{cdf2fbprl} and D0~\cite{dzero5fbprl} experiments have improved 
the combined world measurement to $M_W = 80385\pm 15$~MeV$/c^2$ \cite{run2combo}.  The CDF 
measurement, $M_W = 80387\pm 19$~MeV$/c^2$~\cite{cdf2fbprl}, is described in this article 
and is the most precise single measurement of the $W$-boson mass to date.

This article is structured as follows.  An overview of the analysis and conventions is 
presented in Sec.~\ref{sec:overview}.  A description of the CDF II detector is presented 
in Sec.~\ref{sec:detector}.  Section~\ref{sec:model} describes the detector simulation.  
Theoretical aspects of $W$- and $Z$-boson production and decay, including constraints from 
the data, are presented in Sec.~\ref{sec:production}. The data sets are described in 
Sec.~\ref{sec:wsample}.  Sections~\ref{sec:muons} and \ref{sec:electrons} describe the 
precision calibration of muon and electron momenta, respectively.  Calibration and 
measurement of the hadronic recoil response and resolution are presented in 
Sec.~\ref{sec:recoil}, and backgrounds to the $W$-boson sample are discussed in 
Sec.~\ref{sec:background}.  The $W$-boson-mass fits to the data, and their consistency-checks 
and combinations, are presented in Sec.~\ref{sec:fits}.  Section~\ref{sec:summary} 
summarizes the measurement and provides a combination with previous measurements 
and the resulting global SM fit.


\section{Overview}
\label{sec:overview}
This section provides a brief overview of $W$-boson production and decay phenomenology 
at the Tevatron, a description of the coordinate system and conventions used in this 
analysis, and an overview of the measurement strategy. 
 
\subsection{\boldmath $W$-boson production and decay at the Tevatron}
\begin{figure}[htbp]
\begin{center}
\epsfysize = 6.cm
\includegraphics*[width=8.5cm]{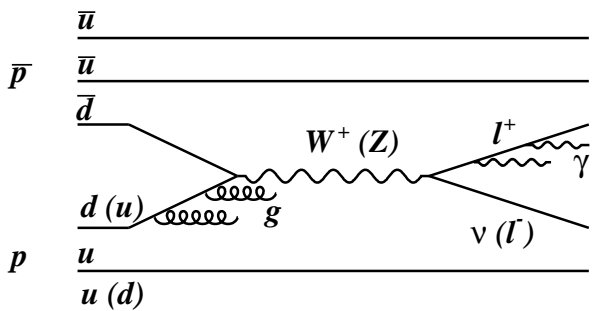}
\caption{Quark-antiquark annihilation producing a $W$ or $Z$ boson in $p\bar{p}$ 
collisions.  Higher-order processes such as initial-state gluon radiation and 
final-state photon radiation are also illustrated.}
\label{fig:wprod}
\end{center}
\end{figure}

In $p\bar{p}$ collisions at $\sqrt{s} = 1.96$~TeV, $W$ bosons are primarily produced 
via $s$-channel annihilation of valence quarks, as shown in Fig.~\ref{fig:wprod}, with 
a smaller contribution from sea-quark annihilation.  These initial-state quarks radiate 
gluons that can produce hadronic jets in the detector. The $W$ boson decays either to a 
quark-antiquark pair ($q\bar{q}'$) or to a charged lepton and neutrino ($\ell\nu$).  
The hadronic decays are overwhelmed by background at the Tevatron due to the high rate 
of quark and gluon production through quantum chromodynamics (QCD) interactions.  Decays to 
$\tau$ leptons are not included since the momentum measurement of a $\tau$ lepton is not 
as precise as that of an electron or muon.  The mass of the $W$ boson is therefore measured 
using the decays $W\to \ell\nu$ ($\ell = e,\mu$), which have about 22\% total branching 
fraction.  Samples selected with the corresponding $Z$-boson decays, $Z \to \ell\ell$, are 
used for calibration.

\subsection{Definitions}
\label{sec:definitions}
\begin{figure}
\begin{center}
\epsfysize = 4.cm
\epsffile{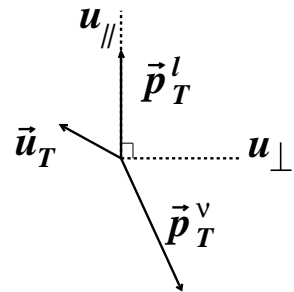}
\caption{Typical vectors associated to quantities reconstructed in a $W$-boson event, with the 
recoil hadron momentum ($\vec{u}_T$) separated into axes parallel ($u_{||}$) and perpendicular 
($u_{\perp}$) to the charged lepton. }
\label{fig:recoilw}
\end{center}
\end{figure}

The CDF experiment uses a right-handed coordinate system in which the $z$ axis is centered at the 
middle of the detector and points along a tangent to the Tevatron ring in the proton-beam direction.  
The remaining Cartesian coordinates are defined with $+x$ pointing outward and $+y$ upward 
from the Tevatron ring, respectively.  Corresponding cylindrical coordinates are defined with 
$r \equiv \sqrt{x^2+y^2}$ and azimuthal angle $\phi\equiv \tan^{-1} (y/x)$.  The rapidity 
$\zeta = -\frac{1}{2}\ln(E+p_zc)/(E-p_zc)$ is additive under boosts along the $z$ axis.  In 
the case of massless particles, $\zeta$ equals the pseudorapidity $\eta = -\ln[\tan(\theta/2)]$, 
where $\theta$ is the polar angle with respect to the $z$ axis.  Transverse quantities such as 
transverse momentum are projections onto the $x-y$ plane.  The interacting protons and 
antiprotons have negligible net transverse momentum.  Electron energy measured in the 
calorimeter is denoted as $E$ and the corresponding transverse momentum $E_T$ is derived using the 
direction of the reconstructed particle trajectory ({\it track}) and neglecting the electron mass.  
Muon transverse momentum $p_T$ is derived from its measured curvature in the magnetic field of the 
tracking system.  The recoil is defined as the negative transverse momentum of the vector boson, 
and is measured as 
\begin{equation}
\vec{u}_T = \sum_i E_i \sin(\theta_i)\hat{n}_i,
\label{eq:recoil}
\end{equation}

\noindent
where the sum is performed over calorimeter towers (Sec.~\ref{sec:calo}), with energy $E_i$, 
tower polar angle $\theta_i$, and tower transverse vector components $\hat{n}_i\equiv 
(\cos\phi_i,\sin\phi_i)$.  The tower direction is defined as the vector from the reconstructed 
collision vertex to the tower center.  The sum excludes towers that typically contain energy 
associated with the charged lepton(s).  We define the magnitude of $\vec{u}_T$ to be $u_T$, 
the component of recoil projected along the lepton direction to be $u_{||}$, and corresponding 
orthogonal component to be $u_{\perp}$ (Fig.~\ref{fig:recoilw}).  From $\vec{p}_T$ conservation, 
the transverse momentum of the neutrino in $W$-boson decay is inferred as 
$\vec{p}_T^{~\nu} \equiv  - \vec{p}_T^{~\ell} - \vec{u}_T $, where $\vec{p}_T^{~\ell}$ is 
the transverse momentum of the charged lepton.  We use units where $\hbar = c \equiv 1$ for 
the remainder of this paper.
 
\subsection{Measurement strategy}
\label{sec:strategy}
The measurement is performed by fitting for $M_W$ using three transverse quantities 
that do not depend on the unmeasured longitudinal neutrino momentum: $p_T^\ell$, $p_T^{~\nu}$, 
and the transverse mass $m_T=\sqrt{2p_T^\ell p_T^{~\nu}(1 - \cos\Delta\phi)}$~\cite{mT}, where 
$\Delta\phi$ is the angle between the charged lepton and neutrino momenta in the transverse 
plane.  Candidate events are selected with $u_T \ll p_T^\ell$, so the neutrino momentum 
can be approximated as $p_T^{~\nu} \approx p_T^\ell + u_{||}$ and the transverse mass can 
be approximated as $m_T \approx 2 p_T^\ell + u_{||}$.  These relations demonstrate the 
importance of modeling $u_{||}$ accurately relative to other recoil components.  They also 
demonstrate that the three fit variables have varying degrees of sensitivity to the modeling 
of the recoil and the $p_T$ of the $W$ boson.

High precision determination of $p_T^\ell$ is crucial to this measurement: a given fractional 
uncertainty on $p_T^\ell$ translates into an equivalent fractional uncertainty on $M_W$.  We 
calibrate the momentum scale of track measurements using large samples of $J/\psi$ and 
$\Upsilon$ meson decays to muon pairs.  These states are fully reconstructed as narrow 
peaks in the dimuon mass spectrum, with widths dominated by detector resolution.  The absolute 
scale of the calibrated track momentum is tested by measuring the $Z$-boson mass in 
$Z \rightarrow \mu\mu$ decays and comparing it to the known value.  After including the $M_Z$ 
measurement, the calibration is applied to the measurement of $M_W$ in $W\to\mu\nu$ decays and 
in the procedure used for the calibration of the electron energy scale in the calorimeter.  

The electron energy scale is calibrated using the ratio of the calorimeter energy to track 
momentum ($E/p$) in $W$ and $Z$ boson decays to electrons.  As with the track momentum 
calibration, we use a measurement of $M_Z$ to validate this energy calibration.  

During the calibration process, all $M_Z$ fit results from both $ee$ and $\mu\mu$ decay 
channels are offset by a single unknown parameter in the range $[-75,75]$~MeV.  This 
blinding offset is removed after the calibrations of momentum and energy scales are 
complete.  The $M_Z$ measurements are then included in the final calibration.

Since $W$ and $Z$ bosons are produced from a similar initial state at a similar energy 
scale, the hadronic recoil is similar in the two processes.  To model the detector response 
to this recoil, we develop a heuristic description of the contributing processes and tune 
the model parameters using fully-reconstructed $Z \rightarrow \ell\ell$ data.  The inclusive 
$p_T$ distribution of produced $W$ bosons is also tuned using $Z \rightarrow \ell\ell$ data 
by combining the measured $p_T$ distribution of $Z$ bosons with a precise calculation 
\cite{resbos} of the relative $p_T$ distributions of $W$ and $Z$ bosons.  

We employ a parametrized Monte Carlo simulation to model the line shapes of the $p_T^\ell$, 
$p_T^{~\nu}$, and $m_T$ distributions.  For each distribution, we generate templates with 
$M_W$ between 80~GeV and 81~GeV, and perform a binned likelihood fit to extract $M_W$. 
Using the statistical correlations derived from simulated experiments, we combine the $m_T$, 
$p_T^\ell$, and $p_T^{~\nu}$~fits from both $W\to e\nu$ and $W\to\mu\nu$ channels to obtain 
a final measured value of $M_W$. 

As with the fits for $M_Z$, a single blinding offset in the range $[-75,75]$~MeV is applied 
to all $M_W$ fits for the course of the analysis.  This offset differs from that applied to 
the $M_Z$ fits.  No changes are made to the analysis once the offsets to the $M_W$ fit results 
are removed.

\section{The CDF II detector}

\label{sec:detector}
\begin{figure*}
\begin{center}
\epsfysize = 11.cm
\epsffile{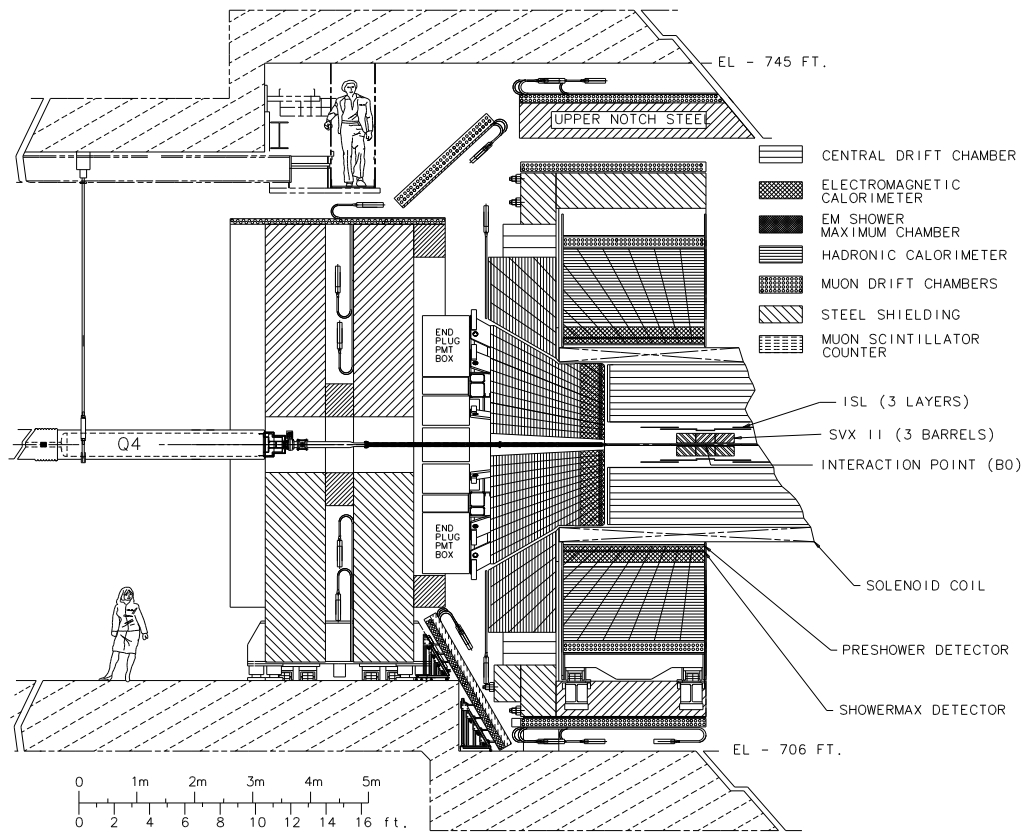}
\caption{Cut-away view of a section of the CDF II detector (the time-of-flight detector is 
not shown).  The slice is in half the $y-z$ plane at $x = 0$. }
\label{fig:detector}
\end{center}
\end{figure*}

The CDF II detector \cite{wzprd, CDF2, jpsi} is a forward-backward and cylindrically symmetric 
detector designed to study $p\bar{p}$ collisions at the Fermilab Tevatron.  The structure of 
the CDF II detector, seen in Fig.~\ref{fig:detector}, is subdivided into the following 
components, in order of increasing radius:  a charged-particle tracking system, composed of 
a silicon vertex detector \cite{SVX} and an open-cell drift chamber~\cite{COT}; a 
time-of-flight measurement detector \cite{TOF}; a system of electromagnetic 
calorimeters~\cite{CEM, cemresponse}, to contain electron and photon showers and measure their energies, 
and hadronic calorimeters~\cite{HAD}, to measure the energies of hadronic showers; and a 
muon detection system for identification of muon candidates with $p_T \gtrsim 2$ GeV.  
Events are selected online using a three-level system (trigger) designed to identify event 
topologies consistent with particular physics processes, such as $W$ and $Z$ boson 
production.  Events passing all three levels of trigger selection are recorded for offline 
analysis.  The major detector subsystems are described below.

\subsection{Tracking system}
\label{sec:tracker}
The silicon tracking detector consists of three separate subdetectors: L00, SVX II, and 
ISL~\cite{SVX}.  The L00 detector consists of a single-sided layer of silicon wafers 
mounted directly on the beampipe at a radius of 1.6~cm.  The SVX II detector consists 
of five layers of double-sided silicon wafers extending from a radius of 2.5~cm to 
10.6~cm.  Surrounding SVX II in the radial direction are port cards that transport data 
from the silicon wafers to the readout system.  The outermost layer of the silicon 
detector, the ISL, consists of one layer of double-sided silicon at a radius of 23~cm 
in the central region ($|\eta| \leq 1$), and two layers of silicon at radii of 20~cm 
and 29~cm in the forward region ($1 < |\eta| < 2$).

\begin{figure*}
\begin{center}
\epsfysize =8.cm
\epsffile{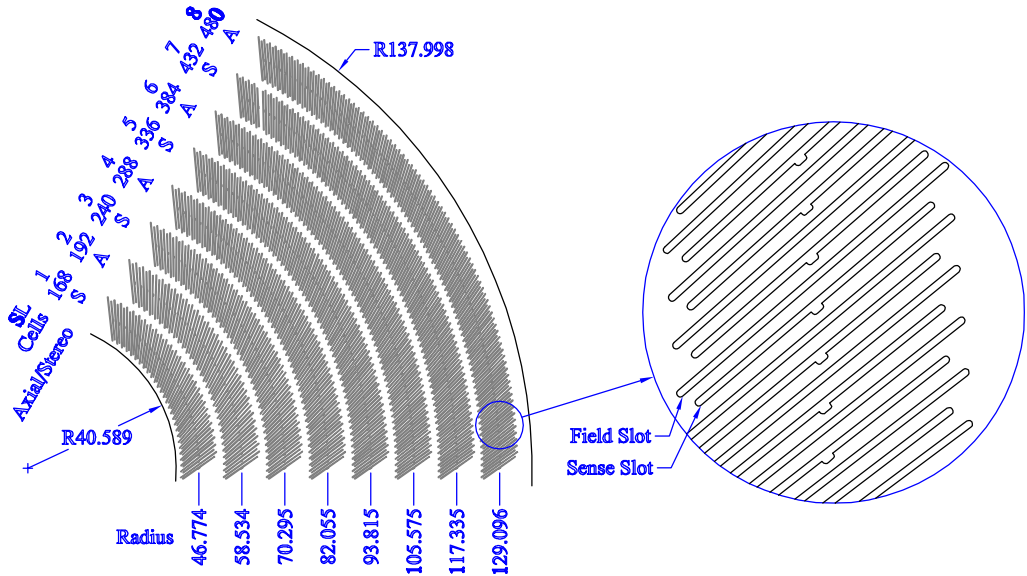}
\caption{End view of a section of a COT endplate~\cite{COT}.  
The endplates contain precision-machined slots where each cell's sense wires and 
field sheets are held under tension.  The radius at the center of each superlayer is 
shown in centimeters. }
\label{fig:COT}
\end{center}
\end{figure*}

The central outer tracking detector (COT)~\cite{COT}, an open-cell drift chamber, 
surrounds the silicon detector and covers the region $|z|<155$~cm and $40 <r<138$~cm.  
Charged particles with $p_T \gtrsim 300$ MeV and $|\eta| \lesssim 1$ traverse the 
entire radius of the COT.  The COT is segmented radially into 8 {\it superlayers} 
containing 12 sense-wire layers each.  Azimuthal segmentation consists of 12-wire 
{\it cells}, such that adjacent cells' planes are separated by $\approx 2$~cm.  The 
detector is filled with a 1:1 argon-ethane gas mixture providing an ionization drift 
velocity of 56~\micro m/ns resulting in a maximum drift time of 177 ns.  The 
superlayers alternate between stereo and axial configurations.  The axial layers 
provide $r-\phi$ measurements and consist of sense wires parallel to the $z$-axis, 
while the stereo layers contain sense wires at a $\pm 2^\circ$ angle to the $z$ axis.  
The sense wires are held under tension from an aluminum endplate at each end of the 
COT in the $z$ direction (Fig.~\ref{fig:COT}).  The wires are azimuthally sandwiched 
by field sheets that provide a 1.9~kV/cm electric field. 

The entire tracking system is immersed in a 1.4~T magnetic field generated by a 
superconducting solenoid \cite{solenoid} with a length of 5~m and a radius of 1.5~m.  
A $\chi^2$ minimization procedure is used to reconstruct the helical trajectory of a 
charged particle using COT hit positions.  The trajectory is defined in terms of five 
parameters:  the signed transverse impact parameter with respect to the nominal beam axis 
$d_0$; the azimuthal angle at closest approach to the beam $\phi_0$; the longitudinal 
position at closest approach to the beam $z_0$; the cotangent of the polar angle 
$\cot\theta$; and the curvature $c\equiv(2R)^{-1}$, where $R$ is the radius 
of curvature.  Individual COT hit positions are corrected for small nonuniformities of 
the magnetic field.  Post-reconstruction corrections to the track curvature are 
derived using $J/\psi \to \mu\mu$, $\Upsilon \to \mu\mu$, and $W \to e\nu$ data 
(Sec.~\ref{sec:muons}).  The measured track $p_T$ is a constant divided by the track 
curvature.

\subsection{Calorimeter system}
\label{sec:calo}
The central calorimeter is situated beyond the solenoid in the radial direction.  The 
calorimeter has a projective-tower geometry with 24 wedges in azimuth and a radial 
separation into electromagnetic and hadronic compartments.  Particles produced at the 
center of the detector with $|\eta| < 1.1$ have trajectories that traverse the entire 
electromagnetic compartment of the central calorimeter.  The calorimeter is split at 
$\eta = 0$ into two barrels, each of which is divided into towers of size 
$\Delta\eta\approx 0.11 \times \Delta\phi \approx 0.26$.  Two neighboring towers 
subtending $0.77 < \eta < 1.0$ and $75^\circ < \phi < 90^\circ$ are removed to allow 
a pathway for solenoid cryogenic tubes.  The forward {\it plug} region of the 
calorimeter covers $1.1 < |\eta| < 3.6$~\cite{plug}. 

The central electromagnetic calorimeter (CEM)~\cite{CEM,cemresponse} consists of 31 
layers of scintillator alternating with 30 layers of lead-aluminum plates.  There are 
$\approx 18$ radiation lengths of detector material from the collision point to the outer 
radius of the CEM.  Embedded at a depth of $R_{CES} = 184$~cm ($\approx 6 X_0$), where 
electromagnetic showers typically have their maximum energy deposition, is the central 
electromagnetic shower-maximum detector (CES).  The CES consists of multiwire proportional 
chambers whose anode wires measure the azimuthal coordinate of the energy deposition and 
whose cathodes are segmented into strips that measure its longitudinal coordinate with a 
position resolution of $\approx 2$~mm.  The position of the shower maximum is denoted as 
CES $x$ (ranging from $-24.1$~cm to 24.1~cm) in the $-R_{CES}\phi$ direction and CES $z$ 
(ranging from $\pm 6$~cm to $\pm 239$~cm) along the $z$ axis. 

The central hadronic calorimeter~\cite{HAD} is subdivided into a central region 
covering $|\eta|<0.6$ and a {\it wall} region covering $0.6<|\eta|<1.1$.  The central 
region consists of 32 alternating layers of scintillator and steel, corresponding to 
4.7 interaction lengths.  The wall region consists of 15 such layers.

\subsection{Muon detectors}
Two sets of muon detectors separately cover $|\eta|<0.6$ and $0.6 < |\eta| < 1$.  In 
the $|\eta| < 0.6$ region two four-layer planar drift chambers, the central muon detector 
(CMU)~\cite{CMU} and the central muon upgrade (CMP), sandwich 60~cm of steel and are 
situated just beyond the central hadronic calorimeter in the radial direction.  The 
 central muon extension (CMX) is an eight-layer drift chamber providing 
the remaining coverage in the forward region.

\subsection{Trigger system}
\label{sec:trigger}
The CDF data acquisition system collects and stores events at a rate of 
$\approx 100$~Hz, or about one out of every 17 000 $p\bar{p}$ crossings.
Events are selected using a three-level system consisting of two 
hardware-based triggers and one software-based trigger.
\par
The first level of triggering reconstructs charged-particle tracks, 
calorimeter energy deposits, and muon detector tracks ({\it stubs}).
Tracks are found in the COT with a trigger track processor, the 
extremely-fast-tracker (XFT) \cite{XFT}, using a lookup table of hit patterns 
in the axial superlayers.  In the CMU and CMX detectors, particle momentum 
is estimated using the timing of signals in neighboring wires.  The electron 
and muon triggers used in this analysis require either a calorimeter 
tower with electromagnetic $E_T > 8$ GeV and a matched XFT track with 
$p_T > 8$ GeV, a CMU stub with $p_T > 6$~GeV matched to a CMP stub and 
an XFT track with $p_T > 4$ GeV, or a CMX stub with $p_T > 6$~GeV matched 
to an XFT track with $p_T > 8$ GeV. 
\par
In the second trigger level, electromagnetic towers are clustered to improve 
energy resolution, allowing a higher threshold of $E_T > 16$ GeV on 
electromagnetic clusters.  The level 2 muon trigger requires both CMU and CMP 
stubs (a ``CMUP'' stub) to be matched to an XFT track with $p_T > 8$ GeV for 
the majority of the data used in this analysis.
\par
The third trigger level fully reconstructs events using an array of $\approx 300$ 
dual-processor computers.  The electron trigger applies requirements on the 
distribution of energy deposited in the calorimeter and on the relative position 
of the shower maximum and the extrapolated COT track, as well as increased energy 
($E_T > 18$ GeV) and momentum ($p_T > 9$ GeV) thresholds.  The muon triggers 
require either a CMUP stub or a CMX stub to be matched to a COT track with 
$p_T > 18$ GeV.
\par
In order to model the contribution of multiple $p\bar{p}$ collisions to the 
recoil resolution, a {\it zero bias} trigger is used.  This trigger randomly 
samples the bunch crossings without applying detector requirements.  An 
additional {\it minimum bias} trigger collects events consistent with the 
presence of an inelastic collision.  The trigger requires coincident signals 
in two small-angle gas Cherenkov luminosity counter detectors~\cite{CLC} 
arranged in three concentric layers around the beam pipe and covering 
$3.6 <|\eta|<4.6$.  These detectors are also used to determine the 
instantaneous luminosity of the $p\bar{p}$ collisions.


\section{Detector simulation}
\label{sec:model}
The measurement of $M_W$ is based on a detailed custom model of the detector response to 
muons, electrons, photons, and the hadronic recoil.  The simulation is fully tunable and 
provides a fast detector model at the required precision.  A {\sc geant}~\cite{GEANT}-based 
simulation of the CDF II detector~\cite{CDFSim} is also used in order to model the 
$Z\to \ell\ell$ background to $W\to \ell\nu$ events, where a detailed simulation of leptons 
outside the fiducial acceptance is required.

The fast simulation model of muon interactions includes the processes of ionization 
energy loss and multiple Coulomb scattering.  In addition to these processes, the 
electron simulation contains a detailed model of bremsstrahlung.  The modeled photon 
processes are $\gamma\rightarrow ee$ conversion and Compton scattering.  This section 
describes the custom simulation of the above processes, the COT response to charged 
particles, and the calorimeter response to muons and electron and photon showers.  The 
model of hadronic recoil response and resolution is discussed in Sec.~\ref{sec:recoil}.

\subsection{Charged-lepton scattering and ionization}
\label{sec:ionization}
While traversing the detector, charged leptons can undergo elastic scattering off an atomic 
nucleus or its surrounding electrons.  The ionization of atomic electrons results in 
energy loss, reducing the track momentum measured in the COT.  Scattering also affects 
the particle trajectory, thus affecting the resolution of the reconstructed track 
parameters. 

The total energy loss resulting from many individual collisions is given by the 
convolution of the collision cross section over the number of target 
electrons~\cite{bichsel}.  This convolution can be described by a Landau distribution,

\begin{equation}
L(dE) = \frac{1}{2\pi i}\int_{a-i\infty}^{a+i\infty} e^{(dE)s + s \log s} ds, 
\end{equation}

\noindent
where $a$ is a constant, $dE$ is the total energy loss, $s = dE - \langle dE \rangle$, 
and $\langle dE \rangle$ is the most probable value of the energy loss \cite{pdg}:

\begin{equation}
\langle dE \rangle = \zeta 
\left[ \ln \left(\frac{2m_e\beta^2\gamma^2 \zeta}{I^2}\right) + j - \beta^2 - 
\delta \right],
\end{equation}

\noindent
where $\zeta = (K/2)(Z/A)(x/\beta^2)$, $K = 4\pi N_A r_e^2 m_e$, $N_A$ is 
Avogadro's number, $m_e$ is the electron mass, $r_e$ is the classical electron 
radius, $Z~(A)$ is the atomic (mass) number, $x$ is the material thickness, 
$j=0.2$, $\beta$ is the particle velocity, $I$ is the mean excitation energy, 
and $\delta$ is the material-dependent density effect as a function of $\beta$.  
We use silicon for the material in the calculation of $\delta$.

We calculate the total energy loss of electrons and muons in the material 
upstream of the COT by sampling the Landau distribution after each of 32 radial 
steps using a fine-grained lookup table of $\langle Z/A \rangle$ and $I$ of the 
detector.  Within the COT we calculate the energy loss along the trajectory up 
to the radius of each sense wire.  To obtain a measured $J/\psi$ mass that is 
independent of the $\langle p_T^{-1} \rangle$ of the final-state muons, we 
multiply the energy loss upstream of the COT by a correction factor of 1.043, 
as described in Section~\ref{sec:jpsi}.

The effect of Coulomb scattering on each particle's trajectory is modeled by a 
Gaussian distribution of the scattering angle through each radial step of the 
detector.  For 98\% of the scatters \cite{ms}, the core Gaussian resolution is 
\begin{equation}
\sigma_{\vartheta} = \frac{13.6~{\rm MeV}}{\beta p}\sqrt{x/X_0},
\end{equation}

\noindent
where $x$ is the thickness of the layer and $X_0$ is the layer's radiation 
length~\cite{CDF2, tsai}.  The remaining 2\% of the scatters are modeled 
by a Gaussian with resolution $3.8 \sigma_{\vartheta}$, based on results of 
low-energy muon scattering data \cite{mstail}.

\subsection{Electron bremsstrahlung}
Bremsstrahlung radiation is modeled using the Bethe-Heitler spectrum~\cite{CDF2, tsai}
\begin{equation}
\frac{d\sigma}{dy} = \frac{A}{N_A X_0 \rho} \left[\left(\frac{4}{3} + C\right)
\left(\frac{1}{y} - 1\right) + y \right],
\end{equation}

\noindent
where $\rho$ is the material density, $y$ is the fraction of the electron momentum 
carried by the photon, and $C$ is a material-dependent constant (taken to be 0.02721, 
the value appropriate for copper).  The spectrum receives corrections~\cite{riddick} 
for the suppression of photons radiated with very low or high $y$.  For $y \gtrsim 0.8$, 
the nuclear electromagnetic field is not completely screened by the atomic 
electrons~\cite{tsai}, reducing the bremsstrahlung cross section.  For $y \lesssim 0.05$, 
interference effects from multiple Coulomb~\cite{migdal} or Compton~\cite{dielectric} 
scattering reduce the rate of photon radiation.  The Landau-Pomeranchuk-Migdal (LPM) 
suppression due to multiple Coulomb scattering is given in terms of the Bethe-Heitler 
cross section as
\begin{equation}
\label{eq:migdal}
S_{LPM} \equiv \frac{d\sigma_{LPM}/dy}{d\sigma_{BH}/dy} =
\sqrt{\frac{E_{LPM}}{E_e}\frac{y}{(1-y)}},
\end{equation}

\noindent
where $E_{LPM}$ depends on the material traversed by the electron.  Dense materials have 
low $E_{LPM}$ and more significant suppression.  

We model the material dependence of the LPM effect based on the material composition of the 
upstream detector, whose components were determined at the time of construction to a relative 
accuracy of 10\%.  To simplify the model, the low-density and high-density components are 
each modeled as a single element or mixture in each layer.  The relative fractions of the 
low-density and high-density components are determined by the ionization energy-loss constant 
and the radiation lengths of the layer.  In increasing radius, the upstream material is 
modeled as follows: a beryllium beampipe; silicon sensors mounted on hybrid readout structures 
consisting of a low-density mixture of equal parts beryllium oxide and glass, combined with 
gold; portcards consisting of a low-density mixture of 37\% beryllium oxide and 63\% kapton 
combined with a high-density mixture of 19\% gold and 81\% copper; the carbon inner COT wall; 
and the COT active volume consisting of kapton combined with a high-density mixture of 
35\% gold and 65\% copper.  The main feature in the longitudinal direction is the silicon 
beryllium bulkhead located at $z=\pm 15$~cm and $\pm 45$~cm.  The model of simplified 
components is designed to reproduce the measured components to a relative accuracy of 10\%.

In each traversed layer we calculate the number of photons radiated using the integrated 
Bethe-Heitler spectrum.  For each photon we draw a value of $y$ from this spectrum and apply 
the appropriate radiation suppression~\cite{CDF2} if $y$ is outside the range 0.05 to 0.8.

\subsection{Photon conversion and scattering}
Photons radiated in the production and the decay of the $W$ boson, or in the 
traversal of electrons through the detector, contribute to the measured 
electron energy if the photon shower is in the vicinity of the electron 
shower.  This contribution depends on photon conversions and on photon 
scattering in the material upstream of the calorimeter.  We model these 
interactions explicitly at radii less than that of the outer COT wall.

The probability for a photon to convert depends on the photon energy and 
on the number of radiation lengths traversed.  At high photon energy the 
probability is determined by integrating the screened Bethe-Heitler 
equation~\cite{tsai, CDF2} over the fraction $y$ of photon energy 
carried by the conversion electron.  The dependence of this probability on 
photon energy has been tabulated in detail \cite{hubble}; we parametrize 
this dependence to determine the conversion probability of a given photon 
in the detector \cite{CDF2}.

To account for the internal-conversion process, where an incoming electron 
produces three electrons via an internal photon, we add an effective number 
of radiation lengths due to the photon-conversion coupling, 
$\Delta (x/X_0) = (\alpha_\mathrm{EM}/\pi) \log(E_{\gamma}/m_e)$~\cite{photonsplit}.  
The radius of conversion is chosen using the radial distribution of radiation 
lengths.  The energy fraction $y$ is taken from the Bethe-Heitler spectrum.

Compton scattering reduces the photon energy and is relevant at low photon 
momentum.  We parametrize the cross section using the tables in Ref.~\cite{hubble} 
and apply a fractional energy loss ($y$) distribution according to 
$d\sigma/dy \propto 1/y+y$, using a lower bound of $y=0.001$~\cite{CDF2}.

\subsection{COT simulation and reconstruction}
\label{sec:cotsim}
The track simulation produces individual measurement points ({\it hits}) in the COT based 
on the trajectory of each charged lepton in the generated event~\cite{CDF2}.  The hit 
spatial resolution is determined for each superlayer using the reconstructed muon tracks
 in $Z\rightarrow \mu\mu$ data, with global multiplicative factors chosen to best match 
the mass distributions of the calibration resonances in data.  These factors deviate from 
one by $\lesssim 5\%$.  The resolution improves from $\approx 180$~ \micro m in the inner 
superlayer to $\approx 140$~\micro m in the outer superlayer.  Efficiencies for detecting 
hits are tuned to approximate the hit multiplicity distribution of the leptons in each 
sample~\cite{CDF2}.  A small correlated hit inefficiency in the inner superlayers accounts 
for the effects of high occupancy.  For prompt lepton candidates the transverse beam 
position is added as a constraint in the track fit, with the $42 \pm 1_{\rm stat}~$\micro m 
beam size chosen to minimize the $\chi^2$ of the reconstructed $Z \rightarrow \mu\mu$ mass 
distribution.

\subsection{Calorimeter response}
\label{sec:ecal} 
Between the outer COT wall and the outer radius of the electromagnetic 
calorimeter there are $\approx 19$ radiation lengths of material.  Using 
a detailed {\sc geant} model of this material, we parametrize the calorimeter 
response to electrons and photons as a function of energy and traversed  
radiation lengths \cite{calgeantnim}.  The parametrization models the 
longitudinal leakage of the shower into the hadronic calorimeter, the 
fraction of energy deposited in the scintillators (including fluctuations), 
and the energy dependence of the response due to the material upstream of 
the scintillators and to the lead absorbers.  

The measured transverse energy $E_T^{\rm meas}$ is parametrized as 
\begin{equation}
\label{eq:ecal}
E_T^{\rm meas} = S_E \left(1 + \xi \log \frac{E_T^{\rm inc}}{39~{\rm GeV}}\right) 
E_T^{\rm inc}, 
\end{equation}

\noindent
where $E_T^{\rm inc}$ is the incident transverse energy, the empirical correction 
$\xi$ accounts for the depth dependence of the calorimeter response due to aging or 
attenuation in the light guides, and $S_E$ is the energy scale determined using the 
same data (see Sec.~\ref{sec:electrons}).  The measured energy receives corrections 
dependent on the measured CES position of the electron shower~\cite{CDF2}. The 
correction $\xi = (5.25 \pm 0.70_{\rm stat}) \times 10^{-3}$ is determined using 
the observed energy dependence of the electron response in $W$ and $Z$ boson 
data.  This correction is adjusted for photons, which produce an electromagnetic 
shower deeper in the calorimeter, by simulating conversion of the photon at an 
average depth and applying the appropriate correction to each conversion 
electron.

Electrons and photons in the same tower, and those in the closest tower 
in $\eta$, are combined to produce a calorimeter electron cluster.  A 
Gaussian smearing is applied to the energy of this cluster with fractional 
resolution $\sigma_E/E = \sqrt{0.126^2/E_T + \kappa^2}$, where $E_T$ is in 
GeV and $\kappa = [0.58 \pm 0.05$(stat)$]\%$ is determined by minimizing the 
$\chi^2$ of the $E/p$ distribution of electrons from the $W$-boson data sample.  
To model the resolution of the $Z\rightarrow ee$ mass peak for electrons 
radiating a high-momentum photon, {\it i.e.}, those electrons with $E/p > 1.11$, 
we apply an additional constant term of $\kappa_{\gamma} = [7.4 \pm 1.8]\%$ 
to all radiated photons and electrons within the simulated electron cluster.  
Electron showers in the two towers nearest $|\eta|=0$ can leak into the gap 
between the central calorimeters.  The resulting loss in energy degrades the 
measurement resolution of the cluster.  To account for this degradation, an 
additional constant term of $\kappa_0 = 0.96\%$ is added in quadrature to $\kappa$ 
for these two towers.  

To improve the modeling of the low tail of the $E/p$ distribution for electrons, 
which is typically populated by electron showers with high leakage out of the 
electromagnetic calorimeter, we multiply the nominal radiation lengths of the 
calorimeter by a pseudorapidity-dependent value between 1 and 1.027.  We improve 
the modeling of the high tail of the $E/p$ distribution for electrons, typically 
populated by electrons with significant photon radiation in the material upstream 
of the COT, by multiplying the nominal radiation lengths of this material by 1.026.

The energy deposited by muons in the calorimeter is simulated using a distribution 
from identified cosmic rays with no additional tracks in the event~\cite{CDF2}.  
The underlying event contribution is modeled from the observed distribution in 
$W$-boson data and scaled to account for its dependence on $u_{||}$, $u_{\perp}$, 
and tower $\eta$.  The distribution is determined using towers at a wide angle 
relative to the lepton in the event and is thus sensitive to the lower threshold 
on tower energy of 60~MeV, which is easily exceeded in a tower traversed by a 
high-momentum lepton.  To correct for this threshold bias we add 25~MeV to the 
underlying event energy in the lepton calorimeter towers, where 25~MeV is the mean 
energy of the extrapolated observed distribution below 60 MeV.


\section{Production and decay models}
\label{sec:production}
The $W$-boson mass is extracted from fits to kinematic distributions, requiring a 
comprehensive theoretical description of boson production and decay.  We describe the 
production of $W$ and $Z$ bosons using CTEQ6.6 parton distribution functions (PDFs) 
\cite{CTEQ} and the {\sc resbos} generator~\cite{resbos}, which combines perturbative 
QCD with a parametrization of nonperturbative QCD effects.  The parameters are 
determined {\it in situ} with fits to $Z$-boson data.  The boson polarization is 
accounted for perturbatively in QCD when modeling the boson decay.  Radiation of photons 
from the final-state charged lepton is simulated using the {\sc photos}~\cite{photos} 
generator and calibrated to the {\sc horace}~\cite{horace} generator for the $M_W$ and 
$M_Z$ mass measurements.

\subsection{Parton distribution functions}
At the Tevatron the longitudinal momentum of a given $W$ or $Z$ boson is unknown, 
but its distribution is well constrained by the parton distribution functions 
(PDFs) describing the fraction $x_i$ of a hadron's momentum carried by a given 
interacting parton.  We consider two independent PDF parametrizations performed 
by the CTEQ~\cite{CTEQ} and MSTW~\cite{MSTW} collaborations.

The mass measurement is performed using the next-to-leading-order CTEQ6.6 parton 
distribution functions to model the parton momentum fraction in $p\bar{p}$ 
collisions.  Variations in the PDFs affect the lepton acceptance as a function 
of the lepton's decay angle with respect to the beam axis.  Since the $W$-boson 
mass is measured using transverse quantities, this change in acceptance impacts the 
measurement.  The CTEQ and MSTW collaborations independently determine a set of 
eigenvectors to form an orthonormal basis, from which uncertainties due to PDF 
variations can be calculated.  The sets calculated by the CTEQ collaboration 
correspond to 90\% C.L. uncertainty, while the sets calculated by the MSTW 
collaboration correspond to both 90\% C.L. and 68\% C.L. uncertainties.  We 
calculate the total PDF uncertainty on $M_W$ from a quadrature sum of all 
eigenvector contributions in a given set of eigenvectors, 
$\delta M_W^\mathrm{PDF}=\frac{1}{2}\sqrt{\sum_i (M_W^{i+} - M_W^{i-})^2},$
where $M_W^{i\pm}$ represents the fitted mass obtained using the $\pm n\sigma$ 
shifts in the $i$th eigenvector.  In the cases where the signs of $M_W^{i+}$ and 
$M_W^{i-}$ are the same, we use half the maximum deviation between the nominal $M_W$ 
and $M_W^{i+}$ or $M_W^{i-}$.  Using events generated with {\sc horace}~\cite{horace}, 
we find $\delta M_W$ to be consistent between the CTEQ6.6 and MSTW2008 PDF 90\% C.L. 
sets.  We calculate the systematic uncertainty due to PDFs using the 68\% C.L. 
eigenvectors for the MSTW2008 PDF sets and obtain $\delta M_W$ of 10, 9, and 11 MeV 
for the $m_T$, $p_T^\ell$, and $p_T^\nu$ fits, respectively~\cite{zeng,beecher}.  As 
a consistency check we find that fits using the nominal CTEQ6.6 and MSTW2008 PDF sets 
yield $M_W$ values that differ by 6~MeV.

\subsection{$W$ and $Z$ boson $p_T$}
\begin{figure}[!tp]
\begin{center}
\epsfysize = 6.cm
\includegraphics*[width=8.8cm]{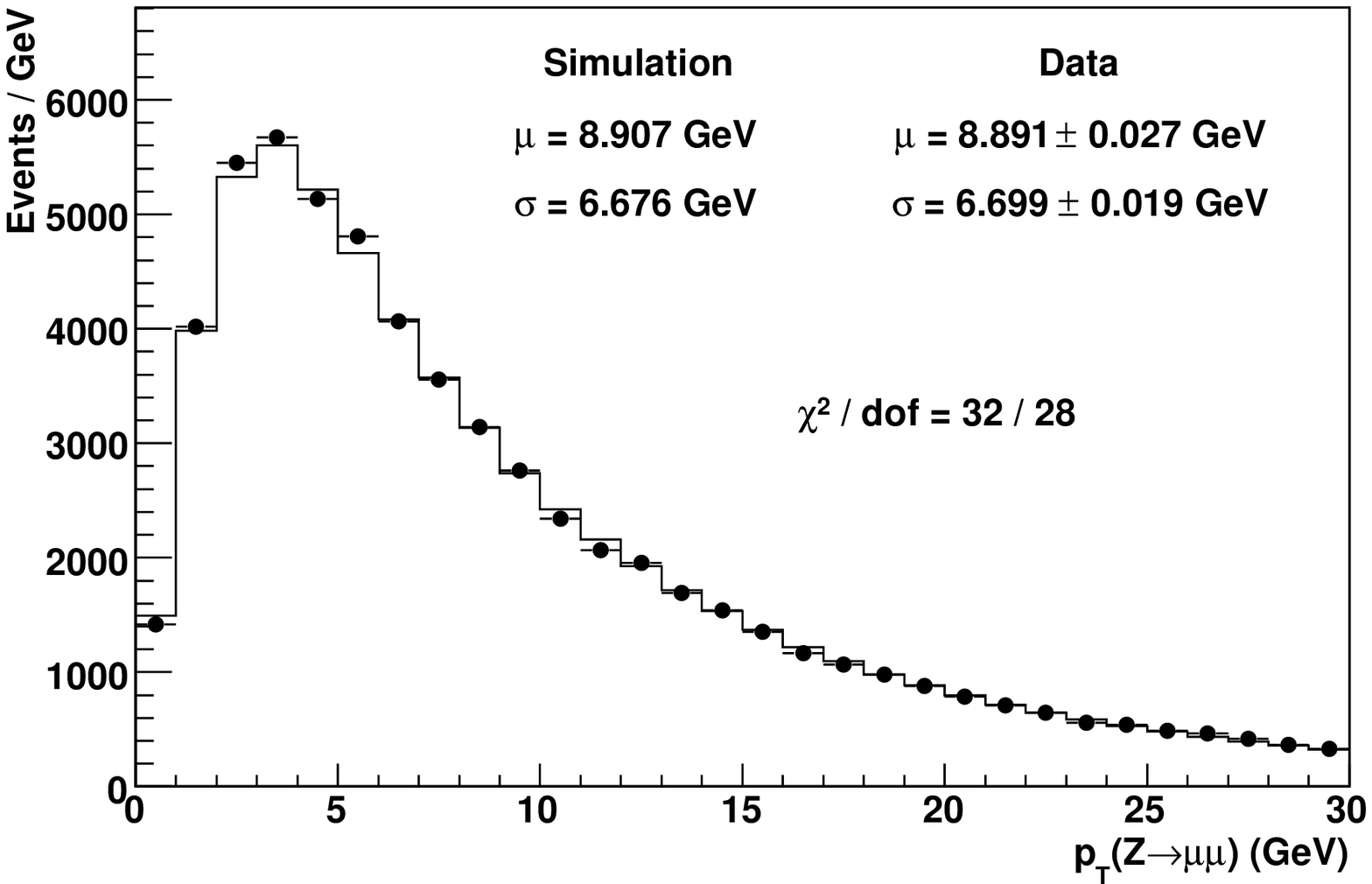}
\includegraphics*[width=8.8cm]{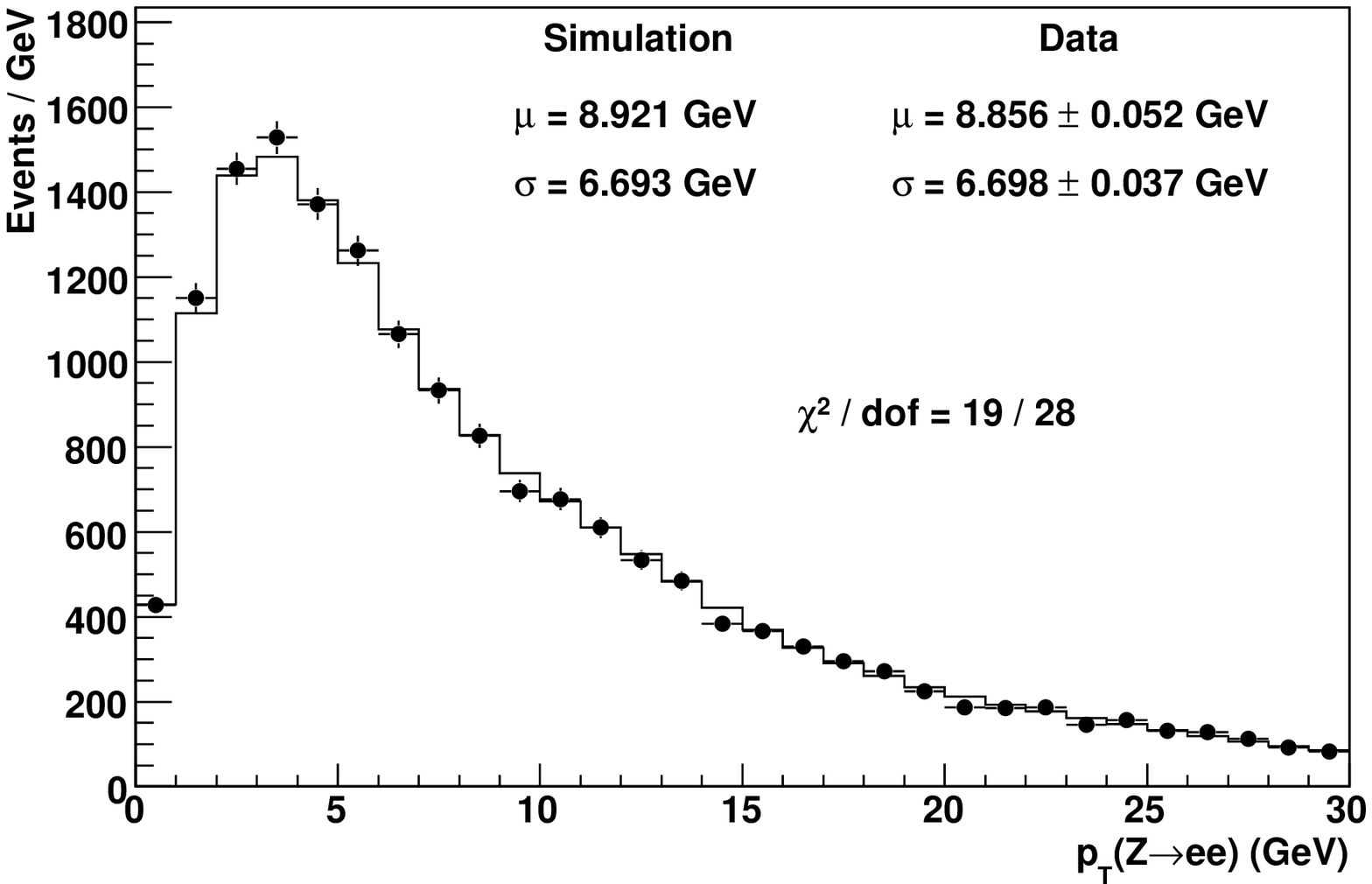}
\caption{Distributions of $p_T^Z$ from simulation (histogram) and data (circles) for $Z$-boson 
 decays to $\mu\mu$ (top), and to $ee$ (bottom).  The distributions are used to fit for 
 the nonperturbative parameter $g_2$ and for $\alpha_s$. }
\label{fig:zpt}
\end{center}
\end{figure}

The $p_T$ of the $W$ boson affects the kinematic distributions used to fit for 
$M_W$, particularly the distribution of charged lepton $p_T$.  We model the $p_T$ 
of the vector boson $V$ using the {\sc resbos} generator, which merges a fixed-order 
perturbative QCD calculation at large boson $p_T$ with a resummed perturbative QCD 
calculation at intermediate $p_T$ and a nonperturbative form factor at low $p_T$.  
{\sc resbos} uses the Collins-Soper-Sterman~\cite{css} resummation formalism to 
describe the cross section for vector-boson production as 
\begin{equation}
\begin{split}
\frac{d\sigma(jk\rightarrow V + X)}{d\hat{s} d^2\vec{p}_T^{~V} dy}  &\propto  
\int d^2\vec{b} e^{i\vec{p}_T^{~V}\cdot\vec{b}} \\
& \times \tilde{W}_{jk}(\vec{b},\hat{s},x_j,x_k)\times e^{-S} + 
Y_{jk}(p_T^V,\hat{s},x_j,x_k),
\label{wxsec}
\end{split}
\end{equation}

\noindent
where $\sqrt{\hat{s}}$ is the partonic center-of-mass energy, $y$ is the boson rapidity, 
$x_j$ and $x_k$ are the momentum fractions of partons $j$ and $k$, respectively, and 
$\vec{b}$ is the relative impact parameter of the partons in the collision.  The 
functions $\tilde{W}_{jk}$ and $Y_{jk}$ are perturbative terms, while $S$ parametrizes 
the nonperturbative part of the transition amplitude.  {\sc resbos} uses the 
Brock-Landry-Nadolsky-Yuan form to characterize the nonperturbative function as~\cite{resbos}
\begin{equation}
S = \Bigg[g_1-g_2\log \Bigg( \frac{\sqrt{\hat{s}}}{2 Q_0} \Bigg)-g_1 g_3\log (100 x_j 
x_k) \Bigg] b^2,
\end{equation}

\noindent
where $Q_0$ is the cutoff parameter of 1.6~GeV and $g_1$, $g_2$, and $g_3$ are parameters 
to be determined experimentally. At fixed beam energy and $\hat{s}$, the $g_i$ parameters 
are completely correlated~\cite{beecher}.  The parameter $g_2$ is particularly sensitive to 
the position of the peak of the boson $p_T$ spectrum.  We fit for $g_2$ using the dilepton 
$p_T$ spectra from $Z\to ee$ and $Z\to \mu\mu$ candidate events (Fig.~\ref{fig:zpt}), 
obtaining a statistical uncertainty on $g_2$ of 0.013~GeV$^2$~\cite{g2value}.  We vary 
$g_3$ by $\pm 0.3$ (the uncertainty obtained in a global fit~\cite{resbos}) and find that 
this variation is equivalent to a $g_2$ variation of $\pm 0.007$~GeV$^2$.  Thus, the 
combined effective uncertainty on $g_2$ is $\pm 0.015$~GeV$^2$, which translates to 
uncertainties on $M_W$ of 1, 3, and 2~MeV for the $m_T$, $p_T^\ell$, and $p_T^\nu$ 
fits, respectively. 

The boson $p_T$ spectrum is sensitive to the value of the strong-interaction coupling 
constant $\alpha_s$, particularly at high boson $p_T$ ($\gtrsim 5$ GeV).  We parametrize 
the variation of the boson $p_T$ spectrum with $\alpha_s$ variations in {\sc resbos} and 
use this parametrization to propagate the constraint from the dilepton $p_T$ spectra to 
an uncertainty on $M_W$.  The resulting uncertainties on $M_W$ are 3, 8, and 4~MeV for 
the $m_T$, $p_T^\ell$, and $p_T^\nu$ fits, respectively. 

We perform a simultaneous fit of the data to $g_2$ and $\alpha_s$ and determine their 
correlation coefficient to be $-0.71$~\cite{beecher}.  Including this anticorrelation, 
the uncertainties on $M_W$ due to the modeling of the $p_T^{W}$ distribution are 
3, 9, and 4~MeV for the $m_T$, $p_T^\ell$, and $p_T^\nu$ fits, respectively.

\subsection{Boson decay}
The polarization of a vector boson produced in proton-antiproton collisions is 
affected by the initial-state QCD radiation associated with the boson production.  
This polarization, together with the $V-A$ coupling of the weak interactions, 
determines the angular distributions of the final-state leptons in the 
vector-boson rest frame.  {\sc resbos} models the boson polarization to 
next-to-next-to-leading-order (NNLO) in $\alpha_s$.

We validate the {\sc resbos} prediction by comparing the angular distribution 
of the charged lepton to that predicted by the NLO $W+\geq 1$-jet generator 
{\sc dyrad}~\cite{dyrad}.  Using the Collins-Soper frame~\cite{cs}, defined as 
the rest frame of the $W$ boson with the $x$ axis along the direction of 
$p_T^W$, the angular distribution of the charged lepton is expressed as 

\begin{equation}
\begin{split}
\frac{d\sigma}{d\Omega} \propto & (1 + \cos^2\theta) + \frac{1}{2}A_0(1-3\cos^2\theta) \\
 & + A_1\sin2\theta \cos\phi + \frac{1}{2}A_2\sin^2\theta \cos2\phi  \\
 & + A_3 \sin\theta \cos\phi + A_4 \cos\theta + A_5 \sin^2\theta\sin2\phi  \\
 & + A_6 \sin2\theta \sin\phi + A_7 \sin\theta\sin\phi,
\end{split}
\label{eq:wdecayang}
\end{equation}

\noindent
where the coefficients $A_i$ are calculated to NNLO in $\alpha_s$ as functions 
of $p_T^W$.  We compare each $A_i(p_T^W)$ value obtained from {\sc resbos} with 
that from {\sc dyrad} and find the generators to give consistent coefficients for 
$p_T^W > 50$~GeV.  At lower $p_T^W$ the coefficients from {\sc resbos} evolve 
continuously to the expected behavior for $p_T^W \to 0$, since {\sc resbos} 
includes the QCD resummation calculation at low $p_T^W$, while {\sc dyrad } is 
a fixed-order calculation whose result does not asymptotically approach the 
expected behavior at $p_T^W = 0$.
 
To check the effect of the difference between the fixed-order and resummed 
calculation on a measurement of $M_W$, we reweight the {\sc resbos} events 
such that the $A_i$ values from {\sc resbos} match the values from {\sc dyrad} 
at $p_T^W = 25$~GeV.  Fitting the reweighted events with the default {\sc resbos} 
templates results in a change in the fitted $W$-boson mass of 3 MeV.  Since the 
{\sc resbos} model includes the resummation calculation while {\sc dyrad} does 
not, the uncertainty in the {\sc resbos} model of the decay angular distribution 
is considered to be negligible.

\subsection{QED radiation}
Final-state photon radiation (FSR) from the charged lepton produced in the $W$-boson 
decay reduces the lepton's momentum, biasing the measurement of $M_W$ in the absence 
of an FSR simulation.  For small-angle radiation ($\Delta R \lesssim 0.1$, where 
$\Delta R \equiv \sqrt{(\Delta\phi)^2 + (\Delta\eta)^2}$), the photon energy is 
recovered by the reconstruction of the electron energy in the calorimeter; for 
radiation at wide angles, or from muons, the photon energy is not included in the 
measured lepton energy.

To simulate FSR, we use the {\sc photos}~\cite{photos} generator with an energy 
cutoff of $E_\gamma > 0.4$~MeV.  {\sc photos} uses a leading-log calculation to 
produce $n$ final-state photons, with a reweighting factor applied to each photon 
such that the complete NLO QED calculation is reproduced for $n=1$.  Ordering the 
photons in $p_T$, we include $n \leq 4$ in the event generation.  Raising the 
$E_\gamma$ threshold to $4$~MeV shifts the value of $M_W$ fitted in pseudoexperiments 
by 2 MeV, which is taken as a systematic uncertainty on the choice of $E_{\gamma}$ 
threshold.

The simulation of QED radiation is improved with a calibration to the {\sc horace} 
generator~\cite{horace}.  {\sc horace} performs a similar leading-log reweighting 
scheme to {\sc photos}, but matches single-photon radiation to the NLO electroweak
calculation~\cite{wgrad} and includes initial-state radiation (ISR) and interference 
between ISR and FSR.  Fitting for $M_W$ in simulated {\sc horace} events yields a 
shift of $-3\pm 4_{\textrm{MC stat}}$ MeV in the electron channel and 
$4 \pm 4_{\textrm{MC stat}}$ MeV in the muon channel.  We apply these corrections 
in the data $M_W$ fit.  Residual uncertainties on the {\sc horace} simulation of 
radiated photons are estimated to be 1~MeV on the $M_W$ measurement.  

A higher-order process contributing to QED energy loss is the radiation of an 
electron-positron pair.  To model this process, we use the effective radiator 
approximation~\cite{photonsplit} to simulate the conversion of radiated photons 
with a probability $(\alpha_\mathrm{EM}/\pi) \log(E_{\gamma}/m_e)$; we estimate 
the remaining uncertainty on $M_W$ to be 1 MeV.  The combined uncertainty on $M_W$ 
due to QED radiation is 4 MeV and is correlated between the channels and the fit 
distributions.


\section{$W$ and $Z$ boson event selection}
\label{sec:wsample}
$W$ and $Z$ boson candidate events are selected by triggers that require a 
muon (electron) with $p_T>18$ ($E_T > 18$)~GeV (see Sec.~\ref{sec:trigger}).  
Events in the $W$-boson sample contain one identified lepton and the 
following kinematic selection: $u_T< 15$~GeV, $30<p_T^\ell<55$~GeV, 
$30<p_T^\nu<55$~GeV, and $60<m_T<100$~GeV.  Candidate $Z$-boson events 
have two oppositely charged same-flavor identified leptons with invariant 
mass $m_{\ell\ell}$ in the range $66 < m_{\ell\ell} < 116$~GeV, and with 
$p_T(Z\to\ell\ell) < 30$~GeV.  Common lepton identification criteria are 
used in the $W$- and $Z$-boson selection.  To suppress the contribution of 
$Z$-boson decays to the $W$-boson sample, loosened lepton identification 
criteria are used to reject events with additional leptons in this sample.  
The number of candidate events in each sample is shown in Table~\ref{tab:sample}.  
In the following we describe the selection criteria and efficiencies for 
electron and muon candidates.

\begin{table}
\begin{ruledtabular}
\caption{Number of events passing all selection criteria for $W$ and $Z$ 
boson candidates in the data.}
\begin{tabular}{lc}
Sample & Candidate events\\
\hline
$W\to\mu\nu$ & 624\,708\\
$W\to e\nu$ & 470\,126\\
$Z\to\mu\mu$ & 59\,738\\
$Z\to ee$ & 16\,134\\
\end{tabular}
\label{tab:sample}
\end{ruledtabular}
\end{table}

\subsection{Muon selection}
\label{sec:muselection}
Muon reconstruction is based on high-momentum tracks reconstructed in the COT, with 
muon-chamber track stubs required when necessary for consistency with trigger selection.  
The selection ensures high-resolution COT tracks, with high purity achieved via tracking 
and calorimeter quality requirements.

A large number of position measurements in multiple COT superlayers leads to high 
precision of the measured track parameters.  We require at least five hits in three or 
more axial superlayers, and a total of 25 hits or more in all axial superlayers.  These 
requirements are also applied to the hits in stereo superlayers.  To suppress the 
potentially large background from the decays of long-lived hadrons such as $K$ or $\pi$ 
mesons to muons, or {\it decays-in-flight} (DIF), we impose requirements on the transverse 
impact parameter ($|d_0|<0.1$~cm) and the quality of the track fit ($\chi^2/$dof$<3$).  
In addition, we identify hit patterns characteristic of a kink in the apparent trajectory, 
due to a particle decay.  A kinked trajectory typically leads to significant numbers of 
consecutive hits deviating from the helical fit in the same direction, since the trajectory 
is a combination of two helices.  We require the number of transitions of hit deviations from 
one side of the track to the other to be larger than $30 + 2 \chi^2/$dof, where dof is the 
number of degrees of freedom.  

Tracks associated with muon candidates are required to originate from the luminous region 
($|z_0|<60$~cm) and to have $p_T > 30$ GeV, measured including a constraint to the transverse 
position of the beam.  The tracks are geometrically extrapolated to the calorimeter and muon 
detectors.  The total energy $E_{\rm EM}$ measured in the electromagnetic towers traversed by 
the extrapolated track is required to be less than 2~GeV; the peak from minimum ionization is 
about 350 MeV.  Similarly, the total energy $E_{\rm had}$ in traversed hadronic towers is 
required to be less than 6~GeV, where the typical energy from minimum-ionizing particles is 
about 2 GeV.  Candidate COT tracks are matched to muon track stubs if the $r-\phi$ distance 
between the extrapolated track and the stub is less than 3~cm in the CMX detector, or 5~cm 
and 6~cm in the CMU and CMP detectors, respectively.

To reduce the background of $Z/\gamma^*\to\mu\mu$ events in the $W$-boson candidate 
sample, we reject events with a second muon candidate satisfying either the above criteria 
or the following criteria, which are independent of the presence of a muon-chamber stub:  
$p_T>20$~GeV, track $\chi^2/$dof $<3$, $|d_0|<0.1$~cm, $\geq 2$ axial and $\geq 2$ 
stereo superlayers with $\geq 5$ hits each, $E_{\rm EM} < 2$~GeV, $E_{\rm had}<6$~GeV, 
and $z_0$ within 5 cm of the candidate muon from the $W$-boson decay.  Cosmic-ray 
background is highly suppressed by fitting for a single track crossing the entire diameter 
of the COT, with sets of azimuthally opposed hits~\cite{cosmic}.

\begin{figure}
\begin{center}
\epsfysize = 6.cm
\epsffile{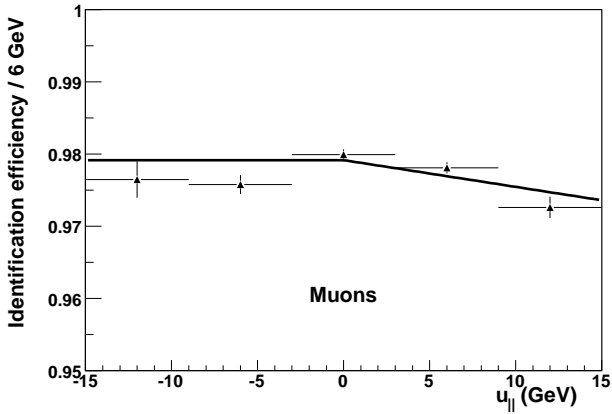}
\caption{Muon identification efficiency as a function of the recoil component in the 
direction of the muon ($u_{||}$).  }
\label{fig:uparmu}
\end{center}
\end{figure}

The muon identification efficiency depends on the projection of the recoil along the 
direction of the muon ($u_{||}$).  Large $u_{||}$ is typically associated with significant 
hadronic activity in the vicinity of the muon, affecting muon identification.  We model this 
dependence through an explicit model of $E_{\rm EM}$, as described in Sec.~\ref{sec:ecal}, 
and a $u_{||}$-dependent efficiency measurement in data for the remaining identification 
requirements.  This measurement uses $Z\to\mu\mu$ events with low recoil $(u_T<15$ GeV) and 
one muon passing the candidate criteria and a second {\it probe} muon identified as a track 
with $p_T>30$~GeV and $E_{\rm EM} < 2$ GeV.  The two muons are required to have opposite 
charge and an invariant mass in the range $81 < m_{\ell\ell} < 101$~GeV.  The small 
background is subtracted using same-charge muon pairs.  The fraction of probe muons 
passing the full $W$-boson muon-candidate criteria as a function of $u_{||}$ is shown in 
Fig.~\ref{fig:uparmu}.  We characterize the observed dependence on $u_{||}$ using the 
parametrization

\begin{equation}
\epsilon_u = a[1+b(u_{||} + |u_{||}|)],
\label{eq:epsupar}
\end{equation}

\noindent 
where $a$ is a normalization parameter and $b$ is a slope parameter for $u_{||}>0$.  
Based on this measurement we simulate a muon identification efficiency with 
$b = [-0.17 \pm 0.07_{\rm stat}]\times 10^{-3}$.  The value of $a$ does not impact the 
$M_W$ measurement.  The statistical uncertainty on $b$ results in an uncertainty 
$\delta M_W$ of 1~MeV and 2~MeV for the $p_T^\ell$ and $p_T^\nu$ fits, respectively.  
The uncertainty on the $m_T$ fit is negligible.

\subsection{Electron selection}
\label{sec:wesample}
Electron candidates are reconstructed from the energy deposited in a pair of 
EM calorimeter towers neighboring in $\eta$ and matched to a COT track 
extrapolated to the position of the shower maximum.  Electromagnetic showers 
are required to be loosely consistent with that of an electron and to be 
fully within the fiducial volume of the EM calorimeter, based on the electron 
track trajectory.  

Measurements of CES deposits are used to determine the energy-weighted $\phi-z$ 
position of the electron shower maximum.  The cluster position must be separated 
from the edges of towers: $|\textrm{CES } x| < 18$~cm, CES~$z$ more than 1.58~cm 
from each tower edge, and CES~$z$ more than 11~cm from the central 
division between east and west calorimeters.  Requiring the shower to be fully 
within the fiducial volume of the EM calorimeter removes additional electron 
candidates in regions near $|\eta| = 0$ and beyond $|\eta| = 0.9$.  We require 
electron $E_T > 30$ GeV, where the energy is measured using the electromagnetic 
calorimeter and the direction is determined using the associated track.

Tracks matched to electromagnetic clusters must be fully within the fiducial 
volume of the COT and must pass the same hit requirements as imposed on muon 
tracks (see Sec.~\ref{sec:muselection}).  The difference in $z$ between the 
extrapolated track and the cluster is required to be less than 5~cm.  The track 
must have $p_T > 18$ GeV and the ratio of calorimeter energy to track momentum, 
$E/p$, is required to be less than 1.6; this requirement significantly reduces 
the misidentified hadron background.

Misidentified hadrons are further suppressed with loose lateral and longitudinal 
shower shape requirements.  The ratio of energy in the hadronic calorimeter to that 
in the electromagnetic calorimeter, $E_{\rm had}/E_{\rm EM}$, must be less than 0.1.  
A lateral shower discriminator quantifying the difference between the observed and 
expected energies in the two electron towers is defined as~\cite{wzrun1}
\begin{equation}
L_\mathrm{shr} = 0.14 \sum_i \frac{E_i^\mathrm{adj} - E_i^\mathrm{exp}}
{\sqrt{0.14^2E_i^\mathrm{adj} + (\Delta E_i^\mathrm{exp})^2}},
\end{equation}

\noindent 
where $E_i^\mathrm{adj}$ is the energy in a neighboring tower, $E_i^\mathrm{exp}$ is 
the expected energy contribution to that tower, $\Delta E_i^\mathrm{exp}$ is the rms 
spread of the expected energy, and the sum is over the two towers.  All energies are 
measured in GeV.  We require $L_\mathrm{shr} < 0.3$.

Candidate events for the $W\to e\nu$ sample are required to have one electron satisfying the 
above criteria.  The $Z/\gamma^*\to ee$ process is highly suppressed by the $u_T<15$~GeV 
requirement.  Further suppression is achieved by rejecting events that have an 
additional high-$p_T$ track extrapolating to a crack between electromagnetic towers 
($|\textrm{CES }x|>21$~cm, $|\textrm{CES }z|<6$~cm, or $|\textrm{CES }z|>235$~cm).  The track 
must also have $p_T>20$~GeV, $|d_0|< 0.3$~cm, and track isolation fraction less than 0.1, in 
order for the event to be rejected.  The track isolation fraction is defined as the sum of 
track $p_T$ contained in a cone $\sqrt{(\Delta\eta)^2+(\Delta\phi)^2} = 0.4$ surrounding 
(and not including) the candidate track, divided by the candidate track $p_T$.

\begin{figure}
\begin{center}
\epsfysize = 6.cm
\epsffile{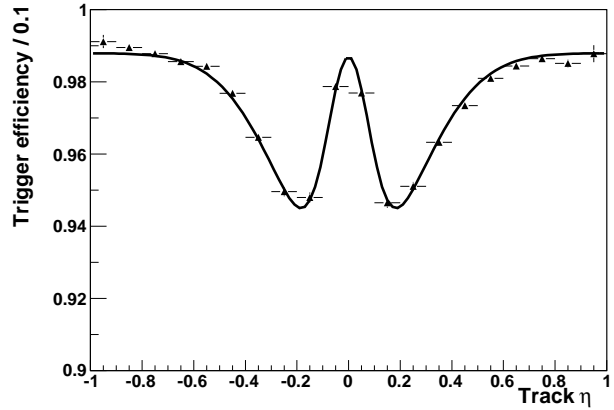}
\caption{Trigger efficiency as a function of track $\eta$ for electrons identified in the 
calorimeter. }
\label{fig:cemetaphieff}
\end{center}
\end{figure}

The efficiency for reconstructing electrons is dependent on $\eta$ due to the track trigger 
requirements.  The efficiency is measured using $W$-boson events collected with a trigger 
with no track requirement, and modeled using the sum of two Gaussian distributions 
(Fig.~\ref{fig:cemetaphieff}).  The drop in efficiency as $|\eta|$ decreases is due to the 
presence of structural supports for the COT wires near $z=0$.  The peak at $|\eta| = 0$ 
arises because the gap between calorimeters overlaps with these supports, so measured 
electrons at $|\eta| = 0$ do not traverse the supports.

\begin{figure}
\begin{center}
\epsfysize = 6.cm
\epsffile{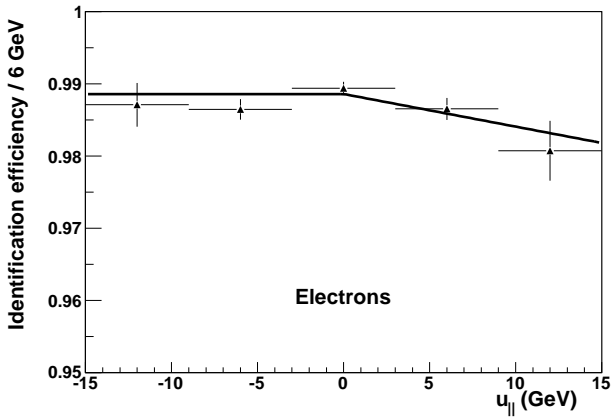}
\caption{Electron identification efficiency as a function of the recoil component in the 
direction of the electron ($u_{||}$).  }
\label{fig:uparel}
\end{center}
\end{figure}

As with the muon identification, the electron identification has a $u_{||}$-dependent 
efficiency.  We measure this efficiency using a sample of $Z\to ee$ events with $u_T<15$~GeV 
where one electron passes the $W$-boson candidate criteria and the other {\it probe} electron 
has an EM energy cluster with $E_T>30$ GeV, an associated track with $p_T>18$~GeV, and 
$E/p<1.6$.  The two electrons must have opposite charge and an invariant mass in the range 
$81 < m_{\ell\ell} < 101$~GeV; background is subtracted using same-charge electrons.  The 
fraction of probe electrons passing the full $W$-boson electron-candidate criteria as a 
function of $u_{||}$ is shown in Fig.~\ref{fig:uparel}.  We characterize the observed 
dependence on $u_{||}$ using the parametrization in Eq.~(\ref{eq:epsupar}) and apply it in 
the simulation with $b=(-0.20\pm 0.10) \times 10^{-3}$.  The statistical uncertainty on $b$ 
results in uncertainties $\delta M_W$ of 3~MeV and 2~MeV for the $p_T^\ell$ and $p_T^\nu$ 
fits, respectively.  The uncertainty on the $m_T$ fit is negligible.

\section{Muon momentum measurement}
\label{sec:muons}
The momentum of a muon produced in a $p\bar{p}$ collision is measured using a 
helical track fit to the hits in the COT, with a constraint to the transverse 
position of the beam for promptly produced muons \cite{CDF2}.  The initial 
momentum calibration has an uncertainty determined by the precision on the 
average radius of the COT and on the average magnetic field.  To maximize 
precision, we perform an additional momentum calibration with data samples of 
$J/\psi$ and $\Upsilon(1S)$ meson decays, and $Z$-boson decays to muons.  Uniformity of the 
calibration is significantly enhanced by an alignment of the COT wire positions 
using cosmic-ray data.

\subsection{COT alignment}
\label{sec:alignment}
The nominal positions of the COT wires are based on measurements of cell 
positions during construction, a finite element analysis of endplate 
distortions due to the load of the wires, and the expected wire 
deflection between endplates due to gravitational and electrostatic 
effects~\cite{COT}.  An alignment procedure~\cite{CDF2} using 
cosmic-ray data taken during Tevatron proton-antiproton bunch crossings 
improves the accuracy of the relative positions of the wires.  The 
procedure determines relative cell positions at the endplates using the 
differences between measured and expected hit positions using a 
single-helix track fit through the entire COT for each cosmic-ray 
muon~\cite{cosmic}.  The deflection of the wires from endplate to 
endplate is determined by comparing parameters of separate helix fits 
on opposite sides of the beam axis for each muon.  

\begin{figure}[!tp]
\begin{center}
\epsfysize = 6.25cm
\epsffile{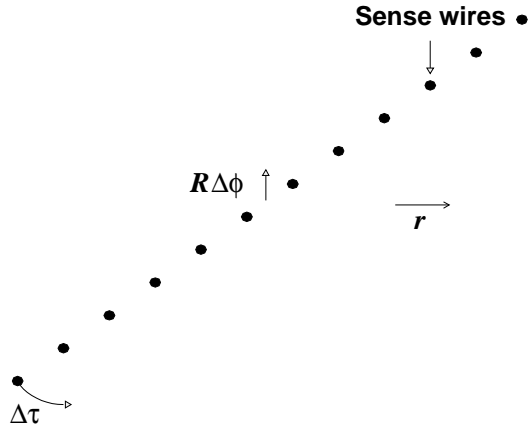}
\caption{Definitions of the tilt ($\Delta \tau$) and shift 
($R\Delta\phi$) corrections derived for each twelve-wire COT cell 
using cosmic-ray data. }
\label{fig:alignment}
\end{center}
\end{figure}

The cosmic-ray sample is selected by requiring no more than two tracks 
from the standard reconstruction.  A single-helix track fit is then 
performed, and fit-quality and kinematic criteria are applied.  The 
sample used for the alignment consists of 136\,074 cosmic-ray muons, 
weighted such that muons with positive and negative charge have equal 
weight.  Using differences between the expected and measured hit 
positions, the tilt and shift of every twelve-wire cell is determined 
for each endplate (see Fig.~\ref{fig:alignment}).  Constraints are 
applied to prevent a global rotation of the endplates and a relative 
twist between endplates.

To reduce biases in track parameters as a function of $z_0$, a correction 
is applied to the nominal amplitude of the electrostatic deflection of the 
wires from endplate to endplate.  The correction is a quadratic function of 
detector radius, with separate coefficients for axial and stereo superlayers.

The cosmic-ray-based alignment is used in the track reconstruction and 
validated with tracks from electrons and positrons from $W$-boson decays.  
Global misalignments to which the cosmic rays are insensitive are 
corrected at the track level using the difference in $\langle E/p \rangle$ 
between electrons and positrons, where $E/p$ is in the range 0.9--1.1.  
Additive corrections are applied to $q/p_T$, a quantity proportional to the 
track's curvature, where $q$ is the particle charge.  The corrections take the form
\begin{equation}
q\Delta p_T^{-1} = f(\theta) + g(\phi) + h(\theta,\phi),
\end{equation}

\noindent
with 
\begin{eqnarray} 
f(\theta) &=& A + B \cot \theta + C \cot^2 \theta, 
\label{eq:cot}\\
g(\phi) &=& a_i \sin(\phi - \alpha_i) + b \sin(3\phi - \beta), 
\label{eq:phi}\\
h(\theta,\phi) &=&  d \sin(\phi - \delta)\cot\theta +  e \sin(3\phi - \epsilon)\cot^2\theta. 
\label{eq:thetaphi}
\end{eqnarray}

\begin{figure*}[!htb]
\begin{center}
\begin{minipage}{0.495\textwidth}
\epsfysize=2.1in
\epsffile{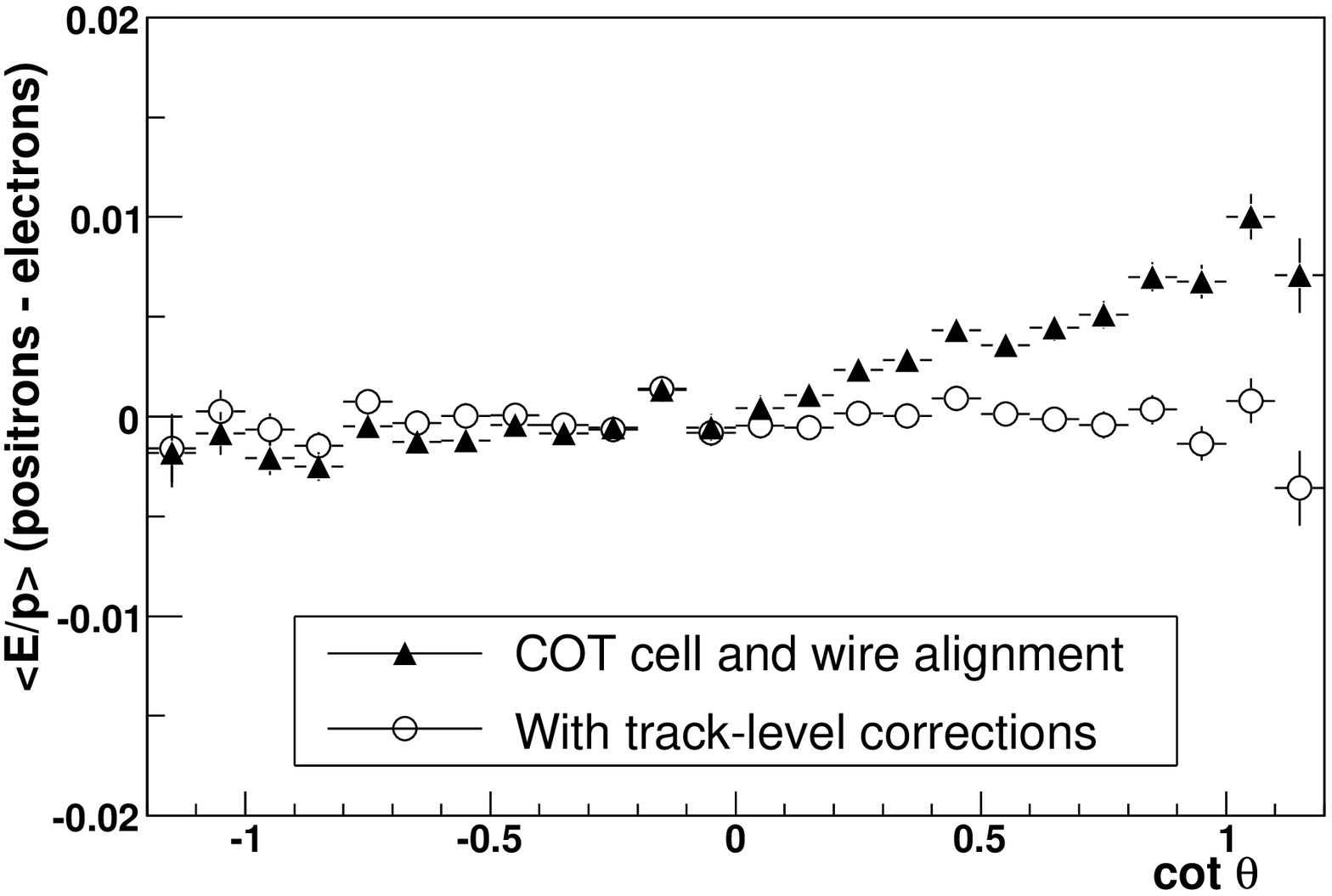}
\end{minipage}
\begin{minipage}{0.495\textwidth}
\epsfysize=2.1in
\epsffile{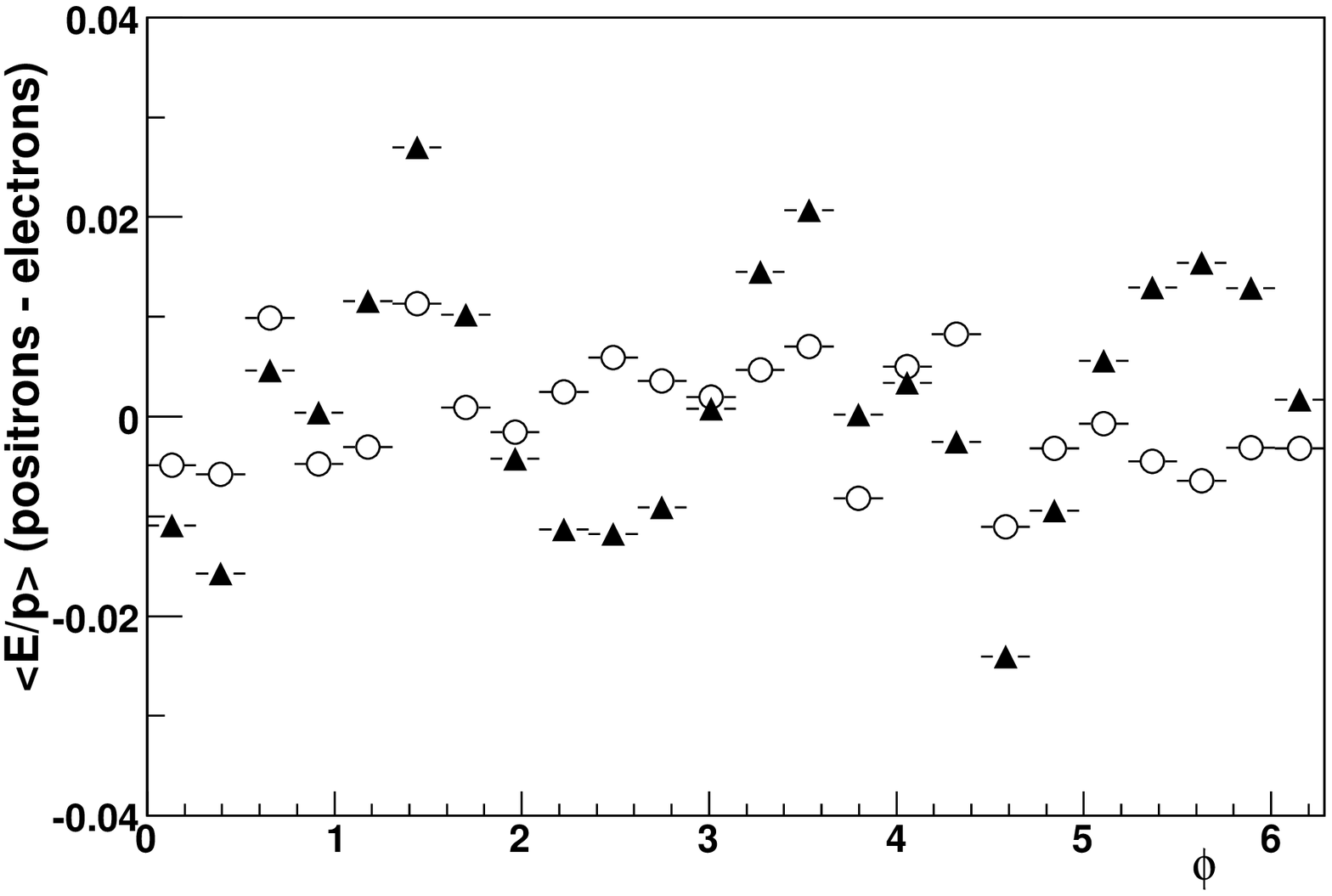}
\end{minipage}
\end{center}
\caption{Difference in $\langle E/p \rangle$ between positrons and 
electrons as a function of $\cot\theta$ (left) and $\phi$ (right).  
The closed triangles correspond to measurements after the cosmic-ray 
alignment, and the open circles correspond to measurements after 
curvature corrections based on the $\langle E/p \rangle$ difference. }
\label{fig:eopcotphi}
\end{figure*}

\noindent
The measured values of the parameters in Eqs. (\ref{eq:cot})--(\ref{eq:thetaphi})
are shown in Table~\ref{tbl:eoverpcor}, with the coefficient and 
phase of the sinusoidal term separated approximately into the first 
($a_1$, $\alpha_1$) and second ($a_2$, $\alpha_2$) halves of the collected 
sample. None of the other parameters show significant variation 
between the two halves of the data sample.  The quoted uncertainties on the 
corrections are given by the statistical uncertainties on the data.  The 
differences in $\langle E/p \rangle$ between positrons and electrons as 
functions of $\phi$ and $\cot\theta$ are shown in Fig.~\ref{fig:eopcotphi}, 
before and after corrections.  The coefficients for the correlated terms 
are determined using the $\langle E/p \rangle$ difference as a function of 
$\cot\theta$ in four equal ranges of $\phi$ centered on 0, $\pi/2$, $\pi$, 
and $3\pi/2$. 

\begin{table}[!tb]
\caption{Values of the correction parameters measured using the difference 
in $\langle E/p \rangle$ between positrons and electrons as functions of 
$\cot\theta$ and $\phi$.  The uncertainties on the phase parameters ($\alpha_i, 
\beta, \delta, \epsilon$), which are quoted in radians, have a negligible impact on the overall uncertainty 
due to misalignment. }
\label{tbl:eoverpcor}
\begin{center}
\begin{ruledtabular}
\begin{tabular}{cc}
Parameter & Value \\
\hline
$A$ & $-(12 \pm 8) \times 10^{-6}$ GeV$^{-1}$ \\
$B$ & $-(91 \pm 5) \times 10^{-6}$ GeV$^{-1}$ \\
$C$ & $-(57 \pm 13) \times 10^{-6}$ GeV$^{-1}$\\\hline
$a_1$ & $(70 \pm 70) \times 10^{-6}$ GeV$^{-1}$\\
$\alpha_1$ & 1.3 \\
$a_2$ & $-(43 \pm 43) \times 10^{-6}$ GeV$^{-1}$ \\
$\alpha_2$ & $-0.2$ \\
$b$ & $(28 \pm 3) \times 10^{-5}$ GeV$^{-1}$ \\
$\beta$ & $-0.5$ \\\hline
$d$ & $-(14 \pm 2) \times 10^{-5}$ GeV$^{-1}$ \\
$\delta$ & 1.5 \\
$e$ & $(16 \pm 3) \times 10^{-5}$ GeV$^{-1}$\\
$\epsilon$ & 0.9 \\
\end{tabular}
\end{ruledtabular}
\end{center}
\end{table}

\subsection{$J/\psi\rightarrow \mu\mu$ calibration}
\label{sec:jpsi}
The large $J/\psi \rightarrow \mu \mu$ production rate allows studies 
of the differential muon momentum scale to test and improve the uniformity 
of its calibration.  Because the $J/\psi$ has a precisely known mass, 
$M_{J/\psi}=3096.916\pm 0.011$ MeV, and narrow width, 
$\Gamma_{J/\psi}=0.0929 \pm 0.0028$~MeV~\cite{pdg}, the main limitation of 
a $J/\psi$-based momentum calibration is the small systematic uncertainty 
on the modeling of the $J/\psi$ mass lineshape~\cite{zeng}.

\subsubsection{Data selection}
\label{sec:jpsiselection}
Online, $J/\psi$ candidates are collected with a level 1 trigger requiring
two XFT tracks matched to two CMU stubs or one CMU and one CMX stub.  The 
$p_T$ threshold on XFT tracks matched to CMU stubs is 1.5 GeV for the early 
data-taking period and 2 GeV for the remainder; for tracks matched to CMX 
stubs the threshold is 2 GeV.  For the later data-taking period the level 2 
trigger requires the tracks to have opposite-sign curvature, $\Delta\phi < 
2\pi/3$, and $m_T < 20$ GeV, where $m_T$ is the two-track transverse mass.  
The level 3 requirements on the corresponding pair of COT tracks are 
opposite-sign curvature, $z$ vertex positions less than 5~cm apart, and an 
invariant mass in the range 2.7--4~GeV.  An additional requirement of 
$\Delta\phi < 2.25$ is imposed when a $\Delta\phi$ requirement is applied 
at level 2.

Offline requirements on COT tracks are $p_T>2.2$~GeV, $|d_0|<0.3$~cm, 
and seven or more hits in each COT superlayer.  The $p_T$ requirements are 
tightened from those required online to avoid trigger bias.  Additionally, 
the two muons are required to be separated by less than 3~cm in $z$ at the 
beamline.  Since approximately 20\% of the selected $J/\psi$ mesons result 
from decays of long-lived $B$ hadrons, we do not constrain the COT tracks 
to the measured beam position.  The resulting sample has approximately 
6 million $J/\psi$ candidates.

\subsubsection{Monte Carlo generation}
\label{sec:jpsimc}
We use {\sc pythia}~\cite{pythia} to generate muon four-momenta from
$J/\psi\to\mu\mu$ decays.  The generator does not model QED final-state 
radiation, so we simulate it using a Sudakov form factor~\cite{CDF2, sudakov} 
with the factorization scale set to the mass of the $J/\psi$ meson.  The 
curvature of the simulated muon track is increased according to the energy 
fraction taken by the radiation.

\begin{figure}
\begin{center}
\includegraphics[width=8.5cm]{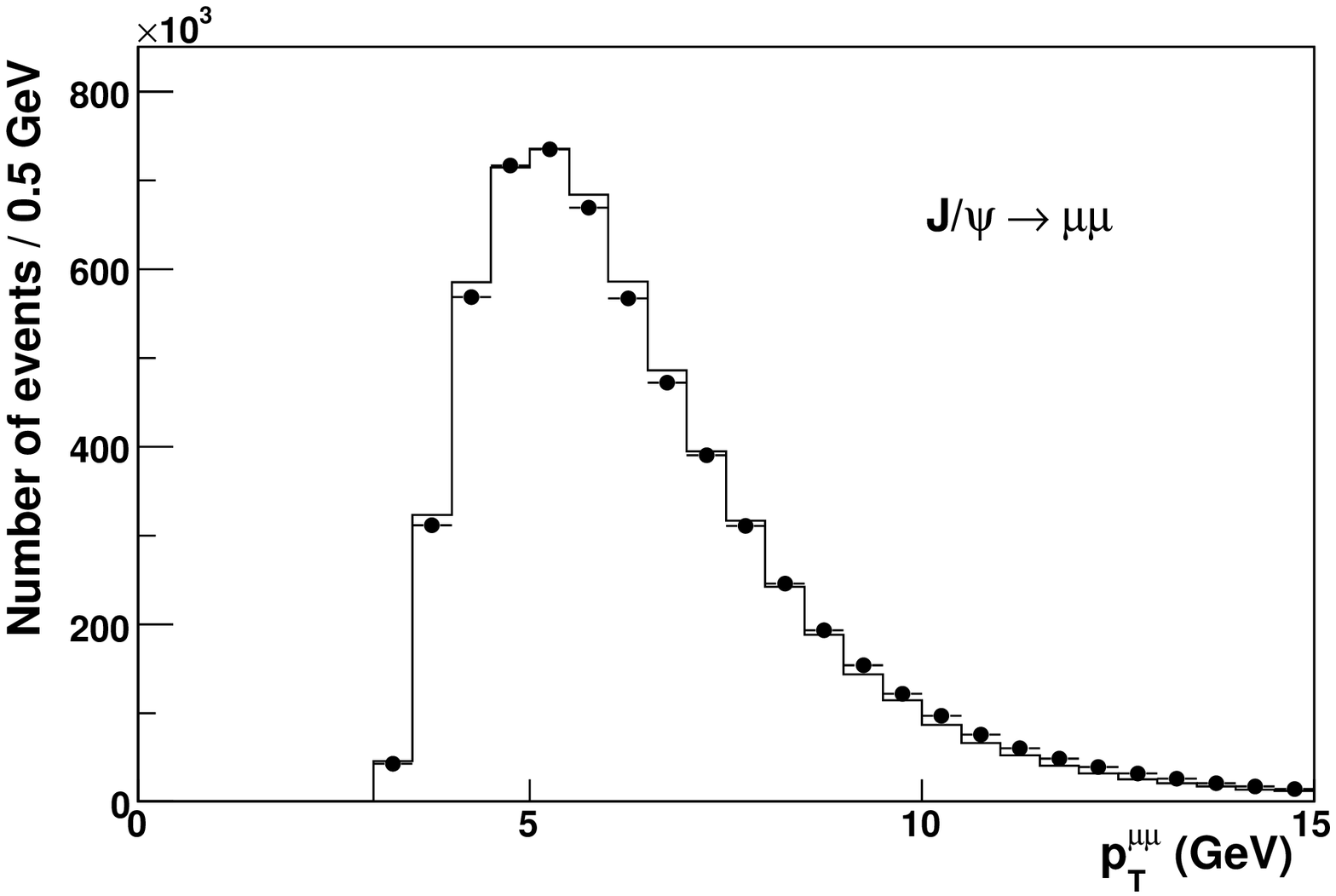}
\includegraphics[width=8.5cm]{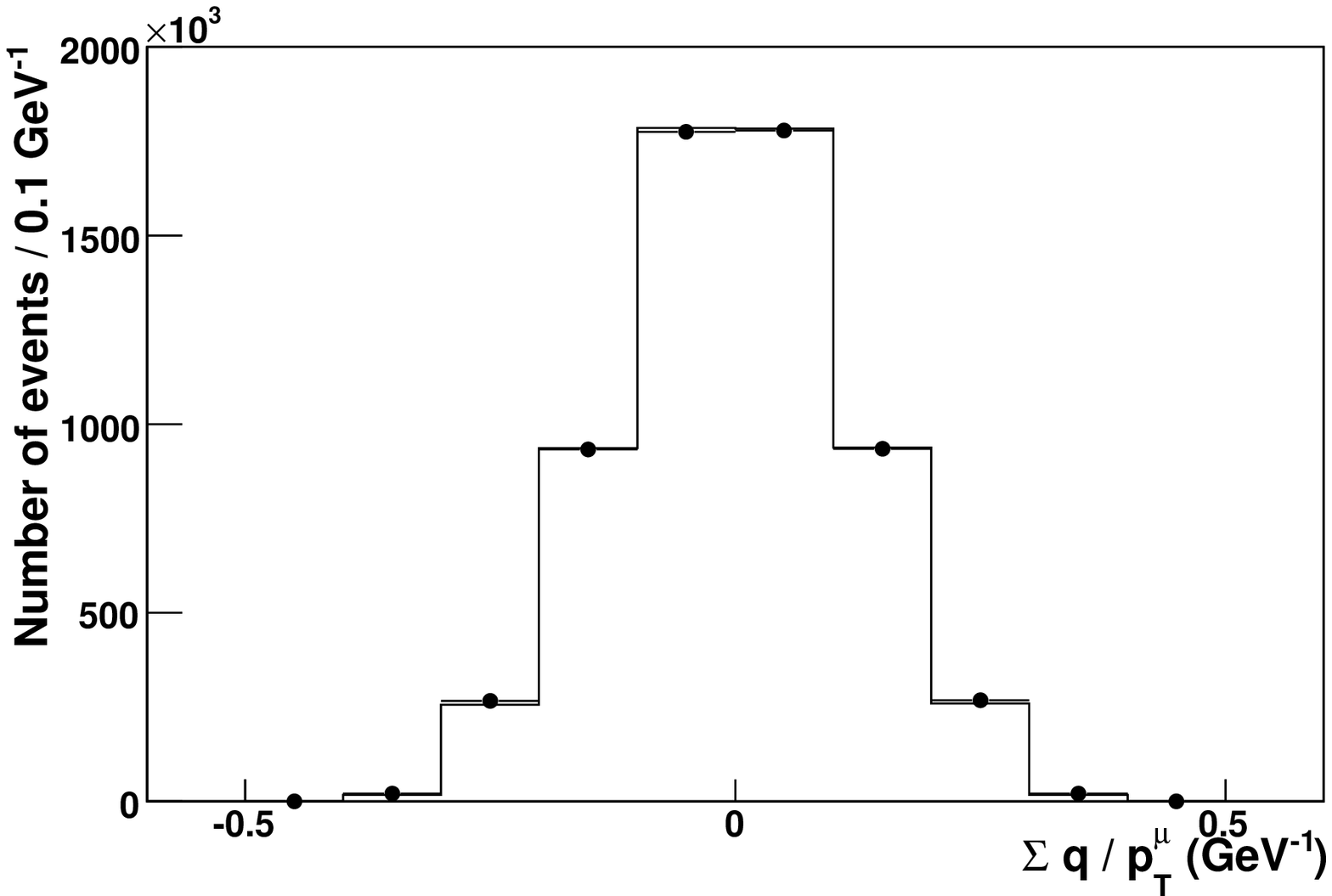}
\caption{Distributions of $p_T^{\mu\mu}$ (top) and $\sum q/p_T^\mu$ (bottom) 
in the data (circles) and the tuned simulation of $J/\psi$ decays (histogram).  
The data distributions are background-subtracted using events in the mass 
range 3.17--3.31 GeV.  }
\label{fig:jpsitune}
\end{center}
\end{figure}

The {\sc pythia} sample is generated with only prompt $J/\psi$ production, 
for which the $p_T^{\mu\mu}$ spectrum peaks at a lower value than in 
$B\rightarrow J/\psi X$ production.  Since $p_T^{\mu\mu}$ affects the mass
resolution, and thus the shape of the observed $J/\psi$ meson lineshape, we tune the 
simulation of this distribution by scaling the rapidity of the $J/\psi$ 
meson along its direction of motion by a factor of 1.2 for half of the 
mesons and 1.5 for the other half.  The resulting tuned $p_T^{\mu\mu}$ 
distribution agrees well with those of the data in the mass range 
3.01--3.15 GeV (Fig.~\ref{fig:jpsitune}).  

The fractional muon momentum resolution degrades linearly with transverse 
momentum, so the mass resolution tends to be dominated by the higher-$p_T$ 
muon.  The $p_T$ asymmetry of the two muons is thus an important quantity to 
model, and is affected by the decay angle $\theta^*$ between the
$\mu^+$ momentum vector and the $J/\psi$ momentum vector,  as computed
in the latter's rest frame.  We multiply $\cot\theta^*$ by a factor of 1.3 to improve the 
modeling of the distribution of the sum of track curvatures of the two muons, 
which is a measure of their $p_T$ asymmetry.  The result of the tuning is 
shown in Fig.~\ref{fig:jpsitune}.

\subsubsection{Momentum scale measurement}
\label{sec:jpsicalib}
The large size of the $J/\psi\to\mu\mu$ data sample allows for detailed corrections 
of nonuniformities in the magnetic field and alignment, and of mismodeling of the 
material in the silicon tracking detector.  These corrections are determined by 
fitting for $\Delta p/p$, the relative momentum correction to each simulated muon, 
as a function of the mean $\cot\theta$ of the muons, the $\cot\theta$ difference 
between muons, or the mean inverse $p_T$ of the muons, respectively.

Nonuniformities in the magnetic field were determined prior to the tracking system 
installation and their effects are included in the trajectory reconstruction.  
Global COT misalignments can lead to additional nonuniformities, in particular 
at the longitudinal ends of the tracking detector.  We measure the corresponding 
effect on the momentum scale using the mean $\cot\theta$ dependence of $\Delta p/p$ 
for the $J/\psi$ sub-sample with small longitudinal opening angle
between the final-state muons, 
$|\Delta\cot\theta| < 0.1$.  Based on this dependence we apply the following 
correction to the measured track $p_T$ in data:

\begin{equation}
p_T^{cor} = (1 - 0.00019\cdot\cot\theta + 0.00034 \cdot{\cot^2\theta}) p_T.
\end{equation}

\begin{figure}
\begin{center}
\includegraphics[width=9.cm]{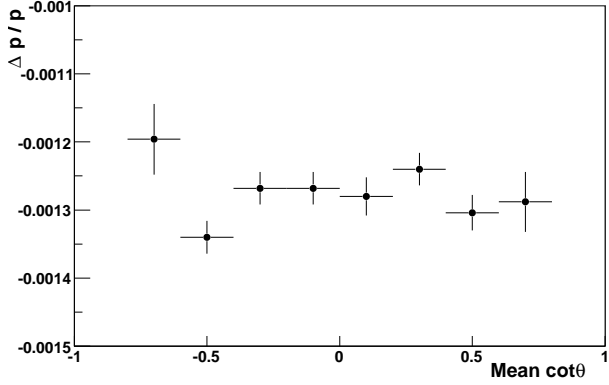}
\caption{Measured $\Delta p/p$ as a function of the mean $\cot\theta$ of the 
muon pair from $J/\psi$ decay, after requiring $|\Delta\cot\theta|<0.1$ and 
including corrections. }
\label{fig:psi_bfield_corr}
\end{center}
\end{figure}

\noindent
After applying this correction, the fitted $\Delta p/p$ shows no significant 
dependence on $\cot\theta$ (Fig.~\ref{fig:psi_bfield_corr}).

We study COT misalignments by measuring $\Delta p/p$ as a function of the 
difference in $\cot\theta$ between the muons tracks from a $J/\psi$ decay.  A 
$z$-scale factor different from unity, equivalent to a scale factor on $\cot\theta$, 
can be caused by a small deviation of the stereo angles from their nominal values; 
this would lead to a quadratic variation of $\Delta p/p$ with $\Delta\cot\theta$.  
A relative rotation of the east and west endplates of the COT would lead to a 
linear dependence of $\Delta p/p$ on $\Delta\cot\theta$.  These effects are 
reduced with respective corrections on the track $\cot\theta$ and curvature $c$ 
of the form 

\begin{equation}
\begin{split}
 \cot\theta &\rightarrow s_z \cot\theta;  \\
 c &\rightarrow c - t \cot\theta .
\end{split}
\label{eq:aligncorrjpsi}
\end{equation}

\noindent
For muons from $J/\psi$ decay the dependence on $\Delta\cot\theta$ is removed 
with a $z$-scale correction $s_z = 1.001640 \pm 0.000018$ and a twist correction 
$t = (1.320\pm 0.092)\times 10^{-7}$~cm$^{-1}$.

\begin{figure}
\begin{center}
\includegraphics[width=9.cm]{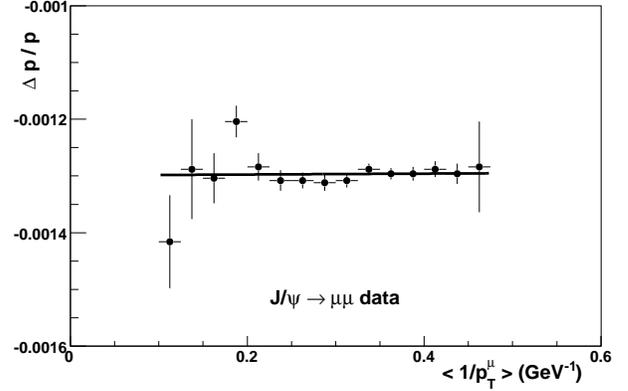}
\includegraphics[width=9.cm]{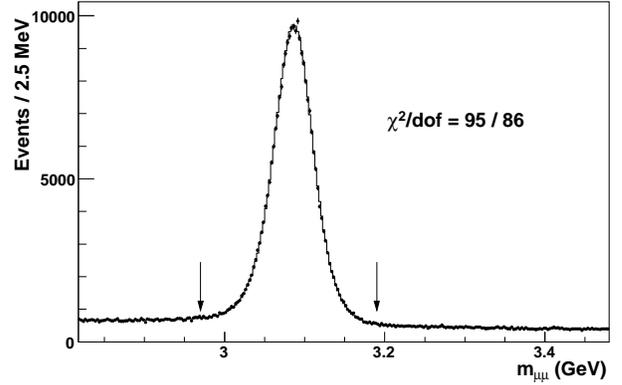}
\caption{Top: Fractional momentum correction $\Delta p/p$ as a function of 
the mean inverse transverse momentum of the muons from $J/\psi$ decay.  Bottom:  
Representative $m_{\mu\mu}$ fit (histogram) to data (circles), here in the range 
$\langle 1/p_T^\mu\rangle = (0.2,0.225)$~GeV$^{-1}$.  The fit region is indicated 
by arrows.  }
\label{fig:psi_pt}
\end{center}
\end{figure}

The modeling of energy loss of muons traversing the silicon tracking detector is 
probed by measuring $\Delta p/p$ as a function of $\langle 1/p_T^{\mu} \rangle$, the 
mean unsigned curvature of the two muons.  A bias in the modeling of ionization energy 
loss appears as a linear dependence of this measurement \cite{CDF2}.  After 
applying a scale factor of 1.043 to the simulated amount of ionizing material in the 
tracking detectors, a linear fit in the range $\langle 1/p_T^{\mu} \rangle = 
(0.1,0.475)$~GeV$^{-1}$ gives a slope consistent with zero (Fig.~\ref{fig:psi_pt}, 
top).  Using the fit to extrapolate to zero mean curvature, we find $\Delta p /p = 
(-1.311 \pm 0.004_{\rm stat} \pm 0.022_{\rm slope / material}) \times 10^{-3}$.

\subsubsection{Systematic uncertainties}
\label{sec:jpsisyst}
Systematic uncertainties on the momentum-scale correction extracted from 
$J/\psi\to\mu\mu$ decays are listed in Table~\ref{tab:pscalesys}.  The 
dominant uncertainty arises from the modeling of the rising portion of 
the $m_{\mu\mu}$ lineshape.  Since we model final-state QED radiation with 
a leading-log Sudakov factor~\cite{CDF2, sudakov}, the modeling of this region is imperfect.  
We estimate the corresponding uncertainty by varying the factorization 
scale $Q$ in the Sudakov form factor to minimize the sum-$\chi^2$ of the 
$\langle 1/p_T^{\mu} \rangle$-binned $J/\psi$ mass fits (one of these 
fits is shown in the bottom of Fig.~\ref{fig:psi_pt}).  The change in the 
fitted $\Delta p / p$ for this $Q$ value, compared to the nominal value 
of $Q = m_{J/\psi}$, is $0.080 \times 10^{-3}$.  

We determine the impact of the nonuniformity of the magnetic field by 
applying the magnetic field correction obtained from $J/\psi\to\mu\mu$ 
data to $W\to\mu\nu$ data.  The resulting shift 
in $M_W$ is in the same direction as the shift in the $J/\psi$ momentum 
scale, resulting in a partial cancellation of the corresponding 
uncertainty.  The residual shift in $M_W$ corresponds to a momentum 
correction shift of $0.064 \times 10^{-3}$.  The uncertainty on the 
magnetic field correction is estimated to be 50\%, resulting in an 
uncertainty of $0.032 \times 10^{-3}$ on $\Delta p / p$ for the $M_W$ 
fit.

Fixing the slope in the fit to $\Delta p / p$ as a function of 
$\langle 1/p_T^{\mu} \rangle$ gives a statistical uncertainty of 
$0.004 \times 10^{-3}$ on the $\Delta p / p$ correction at zero curvature.  
Including the slope variation, the uncertainty is $0.022 \times 10^{-3}$, 
which is the effective uncertainty due to the ionizing material correction.  

We quantify the uncertainty due to COT hit-resolution modeling by varying 
the resolution scale factor (see Sec.~\ref{sec:cotsim}) determined using 
the sum-$\chi^2$ of the highest momentum bins in the 
$\langle 1/p_T^{\mu} \rangle$-binned $J/\psi$ mass fits.  Fitting for 
this factor in individual $\langle 1/p_T^{\mu} \rangle$ bins, we observe a 
maximum spread of 3\%.  Assuming a uniform distribution gives a $1\sigma$ 
variation of 1.7\%, which corresponds to an uncertainty on $\Delta p/p$ of 
$0.020 \times 10^{-3}$.

The background in each $J/\psi\to\mu\mu$ mass distribution is described by 
a linear fit to the regions on either side of the peak.  Varying the slope 
and intercept by their uncertainties in the inclusive $J/\psi$ sample leads 
to a shift in $\Delta p/p$ of $0.011 \times 10^{-3}$, which is taken as the 
uncertainty due to background modeling. 

The alignment corrections in Eq.~(\ref{eq:aligncorrjpsi}) are varied by 
their uncertainties to obtain an uncertainty on $\Delta p/p$ of 
$0.009 \times 10^{-3}$.  To study the impact of unmodeled effects (such 
as trigger efficiencies) near the muon $p_T$ threshold, we increase this 
threshold by 200~MeV.  The shift affects $\Delta p/p$ by 
$0.004 \times 10^{-3}$, which is taken as an associated uncertainty.

The sensitivity of $\Delta p/p$ to the modeling of resolution tails is 
studied by changing the fit range by $\pm 20\%$.  The $0.004 \times 10^{-3}$ 
change in $\Delta p/p$ is taken as an uncertainty.  Templates are simulated 
in $0.004 \times 10^{-3}$ steps of $\Delta p/p$; we take half the step size 
as a systematic uncertainty due to the resolution of the $\Delta p/p$ fit.  
Finally, the uncertainty on the world-average $J/\psi$ mass contributes 
$0.004 \times 10^{-3}$ to the uncertainty on $\Delta p/p$.

Including all systematic uncertainties, the momentum scale correction 
estimated using $J/\psi\to\mu\mu$ data is
\begin{equation}
[\Delta p /p]_{J/\psi} = (-1.311 \pm 0.092) \times 10^{-3}.
\label{eq:jpsiscale}
\end{equation}

\begin{table*}
\begin{ruledtabular}
\caption{Fractional uncertainties on the muon momentum scale determined from 
$J/\psi$ and $\Upsilon$ mass measurements without a beam constraint on the 
muon tracks.  The last column shows the uncertainty for each source that is 
common to the $J/\psi$ and $\Upsilon$ results. }
\begin{tabular}{lccc}
Source & $J/\psi$ ($\times 10^{-3}$) & $\Upsilon$ ($\times 10^{-3}$) & Common ($\times 10^{-3}$) \\
\hline
QED and energy-loss model       &  0.080       &  0.045    &  0.045  \\
Magnetic field nonuniformities 	&  0.032       &  0.034    &  0.032  \\
Ionizing material correction    &  0.022       &  0.014    &  0.014  \\
Resolution model                &  0.020       &  0.005    &  0.005  \\
Background model                &  0.011       &  0.005    &  0.005  \\
COT alignment corrections       &  0.009       &  0.018    &  0.009  \\
Trigger efficiency		&  0.004       &  0.005    &  0.004  \\
Fit range		        &  0.004       &  0.005    &  0.004  \\
$\Delta p/p$ step size		&  0.002       &  0.003    &  0      \\
World-average mass value        &  0.004       &  0.027    &  0      \\
\hline
 Total systematic               &  0.092       &  0.068    &  0.058 \\
\hline
 Statistical                    &  0.004       &  0.025    &  0 \\
\hline
 Total                          &  0.092       &  0.072    &  0.058 \\
\end{tabular}
\label{tab:pscalesys}
\end{ruledtabular}
\end{table*}

\subsection{$\Upsilon\to\mu\mu$ calibration}
\label{sec:upsilon}
With a mass of $M_\Upsilon = 9460.30\pm 0.26$~MeV~\cite{pdg}, the $\Upsilon(1S)$ 
resonance provides an intermediate-mass calibration reference between the $J/\psi$ 
meson and the $Z$ boson.  Unlike $J/\psi$ mesons, all $\Upsilon$ mesons are produced 
promptly, so the reconstructed muon tracks from their decays can be constrained to 
the transverse beam position to improve momentum resolution.  This allows a test 
for beam-constraint bias in a larger calibration sample than the $Z$-boson data 
sample~\cite{zeng}.

The online selection for $\Upsilon$ candidates is the same at level 1 as for 
selecting $J/\psi$ candidates (see Sec.~\ref{sec:jpsiselection}).  At level 2 
at least one CMUP muon with $p_T > 3$~GeV is required.  The level 3 selection 
increases this threshold to 4~GeV and the $p_T$ threshold of the other muon 
to 3~GeV.  The muons must have opposite charge and a pair invariant mass between 
8 and 12 GeV.  In the offline selection the $p_T$ thresholds are increased by 
200~MeV and the muons are required to have $|d_0| < 0.3$~cm and a small $z_0$ 
difference ($|\Delta z_0| < 3$~cm).  The COT hit requirements are the same as those 
applied to tracks from $W$- and $Z$-boson decays (see Sec.~\ref{sec:muselection}).  

\begin{figure}
\begin{center}
\includegraphics[width=8.5cm]{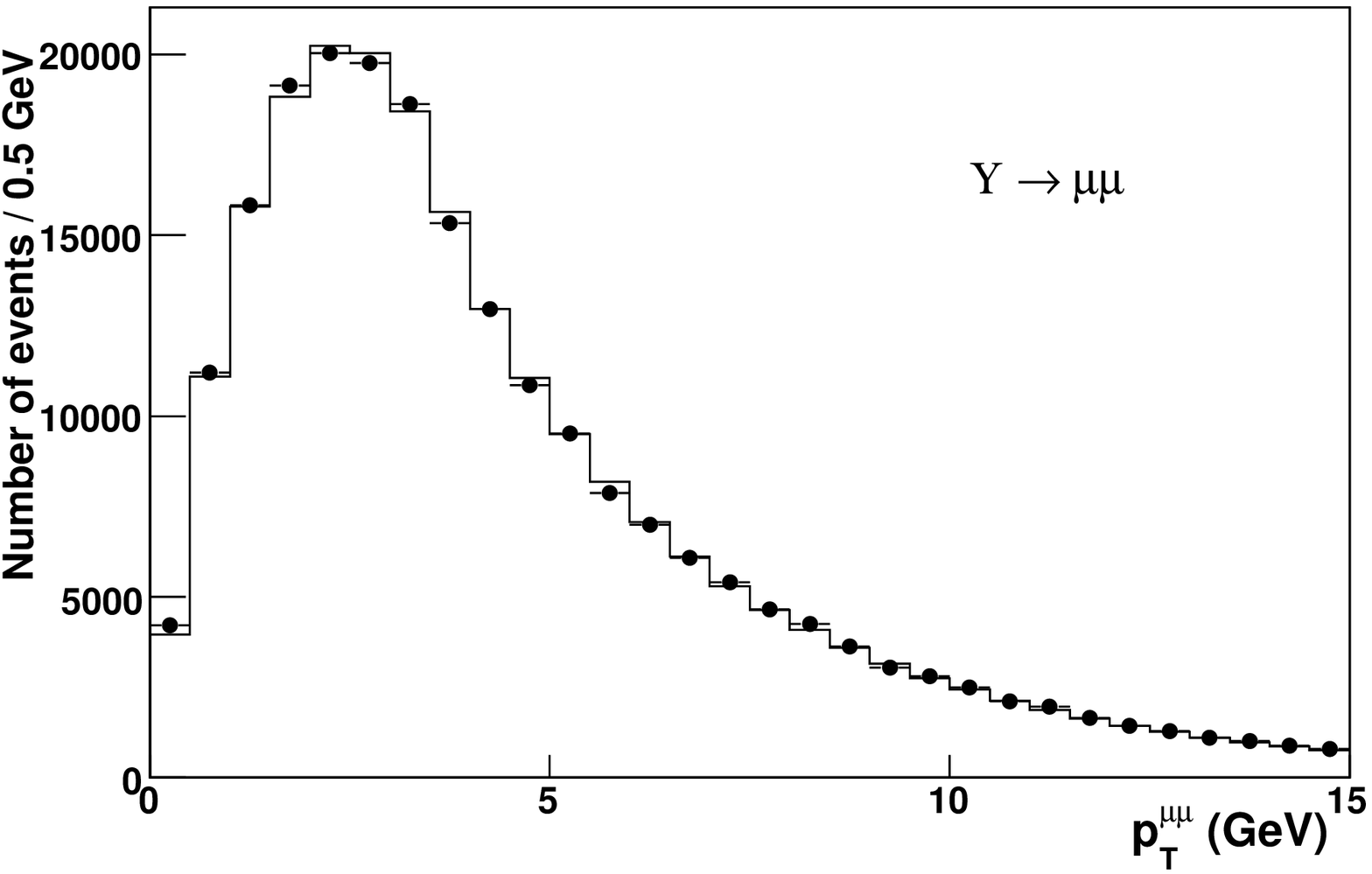}
\includegraphics[width=8.5cm]{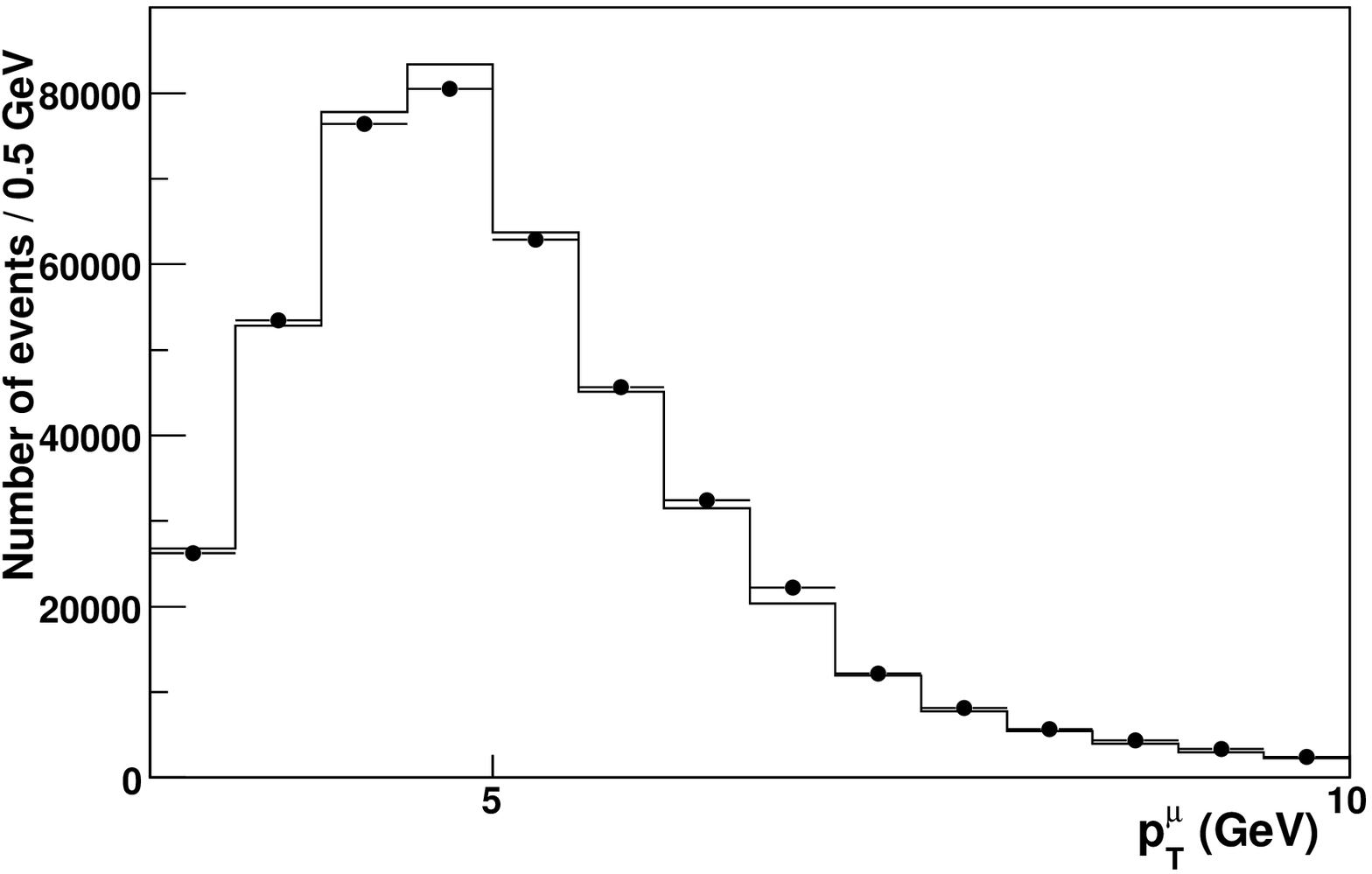}
\includegraphics[width=8.5cm]{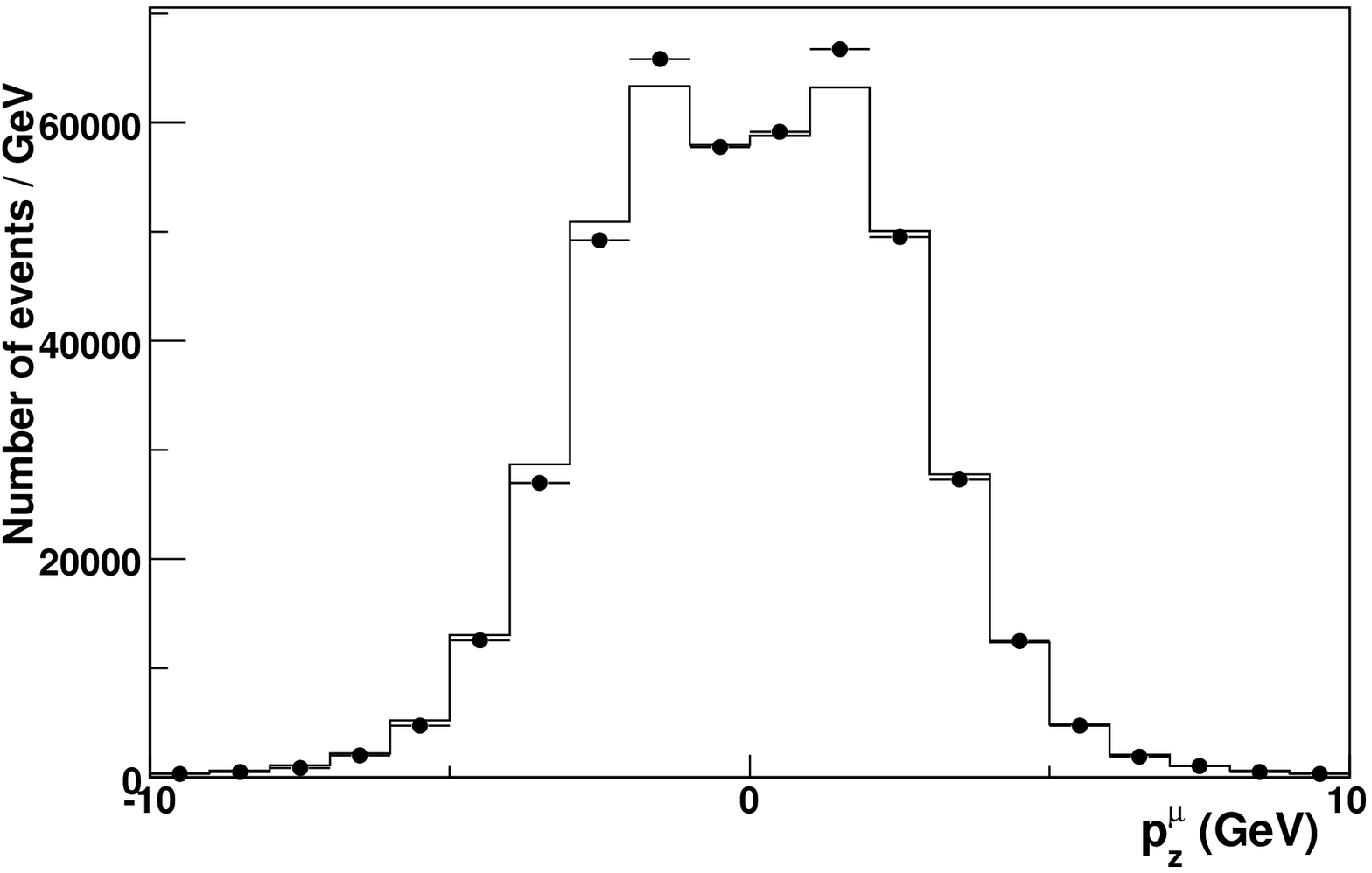}
\caption{Distributions of $p_T^{\mu\mu}$ (top), $p_T^{\mu}$ (middle), and 
$p_z^{\mu}$ (bottom) in the data (circles) and the tuned simulation of 
$\Upsilon$ decays (histogram).  The data distributions correspond to the 
mass range $9.30-9.56$ GeV and are background-subtracted using events in 
the mass ranges $9.17-9.3$ GeV and $9.56-9.69$ GeV. }
\label{fig:upstune}
\end{center}
\end{figure}

As with the $J/\psi\to\mu\mu$-based calibration, we use {\sc pythia}~\cite{pythia} 
to generate muon four-momenta from $\Upsilon(1S)\to\mu\mu$ decays.  We tune the 
simulation by increasing the rapidity of the $\Upsilon$ by 
$\Delta \zeta_{\upsilon} = k y_{\Upsilon}$, where $k = 0.1$ for half of the 
mesons and $k = 0.6$ for the other half.  With this tuning, the
kinematic properties  of 
the $\Upsilon$ and the final-state muons are well described, as shown in 
Fig.~\ref{fig:upstune}.

The correction for magnetic field nonuniformity measured in $J/\psi$ data 
(see Sec.~\ref{sec:jpsicalib}) is applied to the $\Upsilon$ data.  By fitting 
for $\Delta p/p$ as a function of $\langle 1/p_T \rangle$, we find that the 
material scale value of 1.043 determined with $J/\psi$ data 
 removes any dependence 
on $\langle 1/p_T \rangle$.  

The intermediate momentum range of the muons from $\Upsilon$-meson decays can 
lead to different sensitivity to misalignments than muons from $J/\psi$-meson  
or $W$- or $Z$-boson decays.  We measure the $z$-scale and twist corrections 
of Eq.~(\ref{eq:aligncorrjpsi}) separately in $\Upsilon$ data, finding 
$s_z = 1.00160 \pm 0.00025$ ($1.00148 \pm 0.00019$) and 
$t = 0.50 \pm 0.36~(2.10 \pm 0.28) \times 10^{-7}$~cm$^{-1}$ for muon tracks without 
(with) a beam constraint.

\begin{figure}
\begin{center}
\includegraphics[width=9.cm]{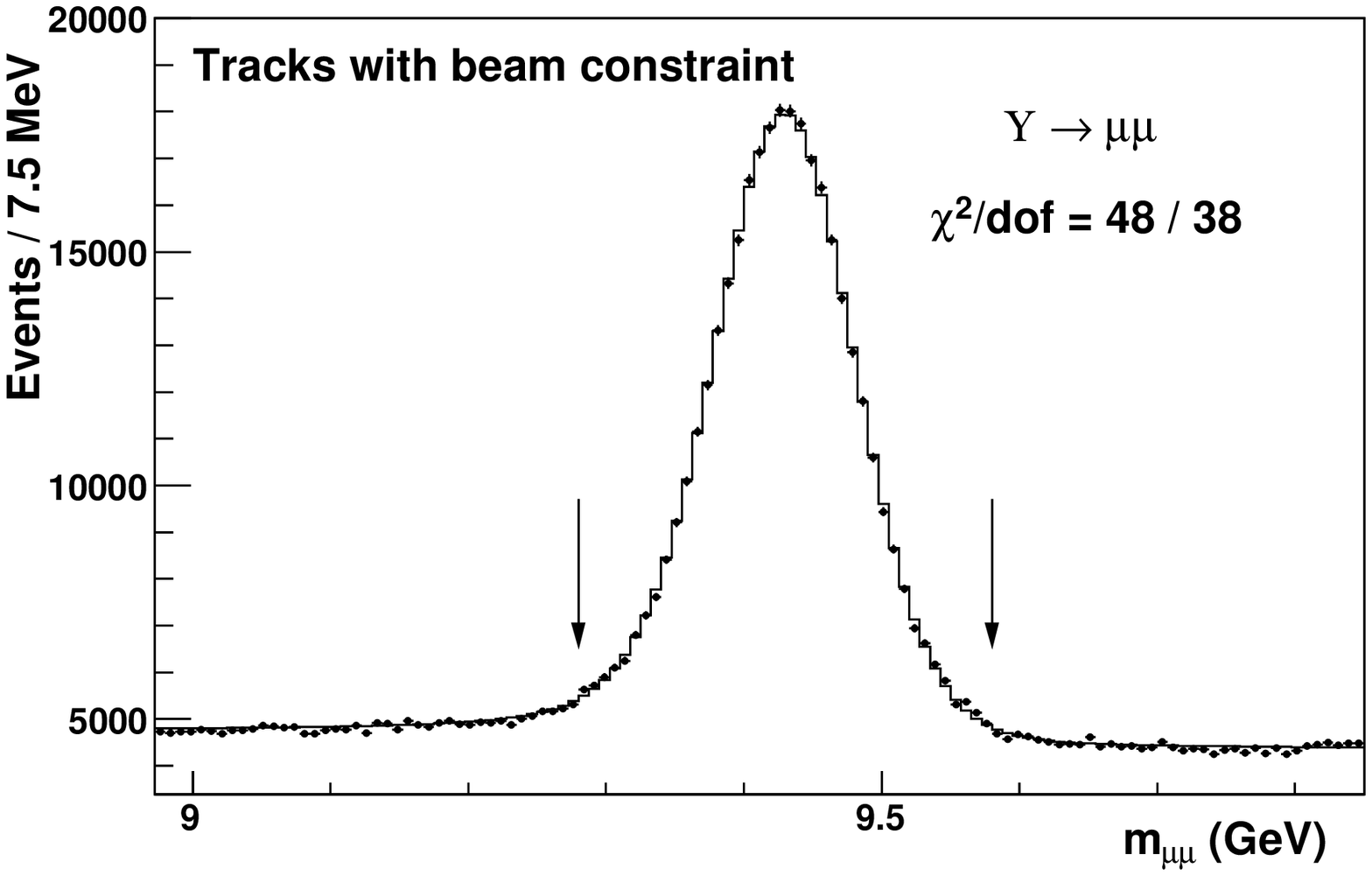}
\includegraphics[width=9.cm]{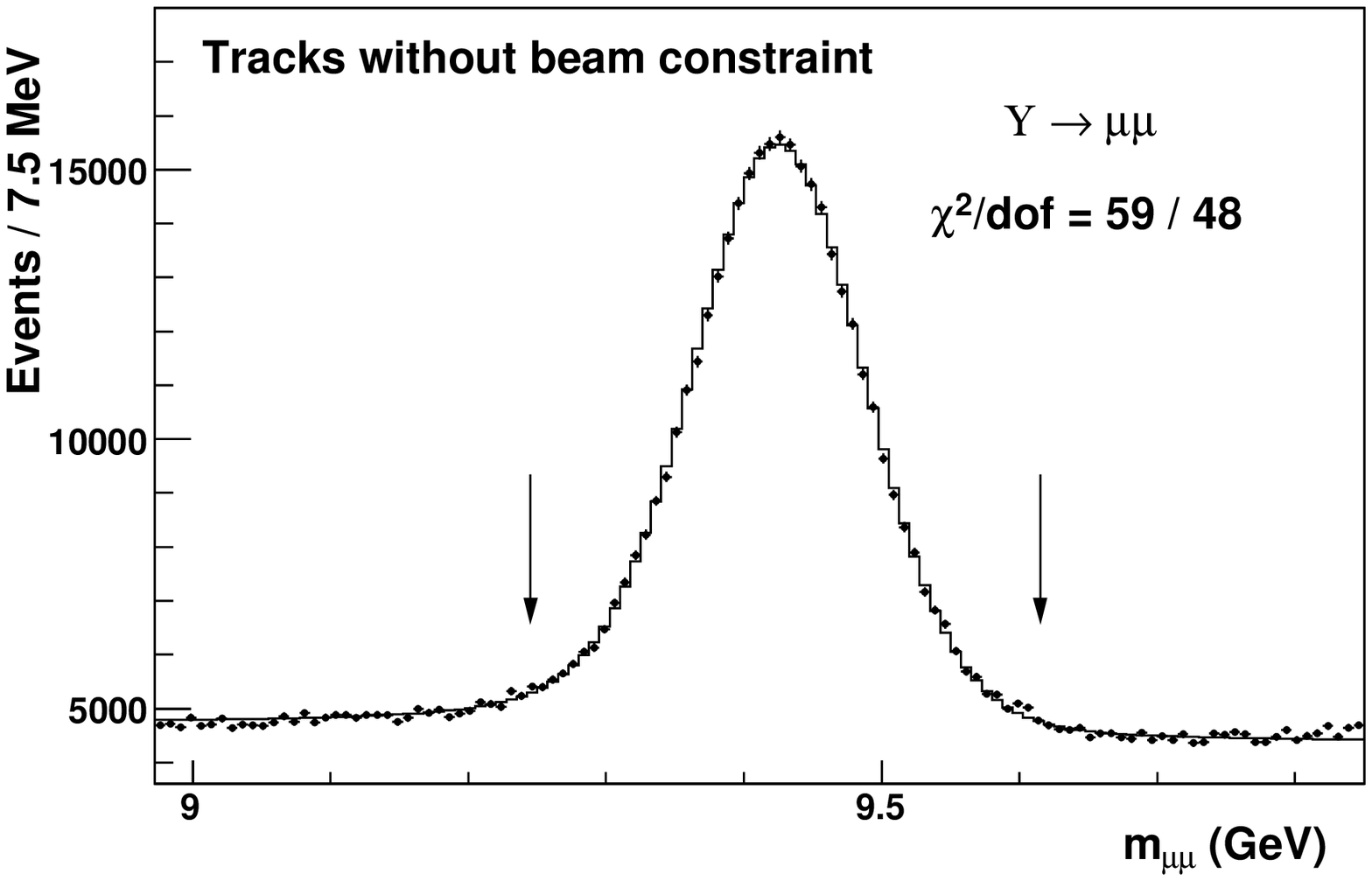}
\caption{Distribution of $m_{\mu\mu}$ for the best-fit templates (histograms) and 
the data (circles) in the $\Upsilon\to\mu\mu$ sample used to calibrate the muon 
momentum scale.  The muon tracks are reconstructed with (top) or without (bottom) 
a constraint to the beam position in the transverse plane.  The arrows enclose the 
fit region. }
\label{fig:upsilonmass}
\end{center}
\end{figure}

In order to test for a beam-constraint bias, we fit for $\Delta p/p$ with and 
without incorporating the beam constraint.  The fits are performed in the mass 
ranges $9.28< m_{\mu\mu} < 9.58$~GeV and $9.245< m_{\mu\mu} < 9.615$ GeV for 
the constrained and unconstrained tracks, respectively, and are shown in 
Fig.~\ref{fig:upsilonmass}.  The measurement with unconstrained tracks yields 
$\Delta p/p = (-1.335\pm 0.025_{\rm stat}\pm 0.068_{\rm syst}) \times 10^{-3}$, 
where the systematic uncertainties are evaluated in a similar manner to the 
$J/\psi$-based calibration and are shown in Table~\ref{tab:pscalesys}.  Using 
constrained tracks, the measurement yields 
$\Delta p/p = (-1.185\pm 0.020_{\rm stat}\pm 0.068_{\rm syst}) \times 10^{-3}$.  
We correct the $\Upsilon$-based calibration with unconstrained tracks by half 
the difference between measurements obtained with unconstrained and constrained 
tracks, and take the correction ($\Delta p/p = 0.075\times 10^{-3}$) as a 
systematic uncertainty on the calibration.  The momentum scale correction 
estimated using $\Upsilon\to\mu\mu$ data is therefore

\begin{equation}
[\Delta p /p]_{\Upsilon} = (-1.260 \pm 0.103) \times 10^{-3}.
\label{eq:upsilonscale}
\end{equation}

\subsection{Combination of $J/\psi$ and $\Upsilon$ calibrations}
\label{sec:psiups}
Table~\ref{tab:jpsiupsilon} summarizes the measured momentum scales from reconstructed 
samples of $J/\psi$ mesons, $\Upsilon$ mesons without a beam-constraint (NBC), and 
$\Upsilon$ mesons with a beam-constraint (BC).  Since the $J/\psi$-based measurement 
is performed using tracks without a beam-constraint, we combine the results from $J/\psi$ 
and NBC $\Upsilon$ meson fits.  Using the Best Linear Unbiased Estimator (BLUE) 
algorithm~\cite{BLUE} and accounting for the correlations listed in Table~\ref{tab:pscalesys}, 
we obtain

\begin{equation}
[\Delta p/p]_{J/\psi+NBC~\Upsilon} =  (-1.329\pm 0.068) \times 10^{-3}.
 \label{eq:nbccombination}
\end{equation}
 
\noindent
As with the scale determination based on $\Upsilon$ meson decays only, we correct this result by half the 
difference with respect to the BC $\Upsilon$ meson result, and take the full correction 
as a systematic uncertainty.  The final combined momentum scale based on measurements 
of $J/\psi$ and $\Upsilon$ mesons is

\begin{equation}
[\Delta p /p]_{J/\psi+\Upsilon} = (-1.257 \pm 0.101) \times 10^{-3}.
\label{eq:pscale}
\end{equation}

\begin{table}
\begin{ruledtabular}
\caption{Summary of momentum scale determinations using $J/\psi$-meson data and 
$\Upsilon$-meson data with (BC) and without (NBC) beam-constrained tracks.  
The systematic uncertainties do not include the uncertainty stemming from the 
difference between the BC and NBC $\Upsilon$-meson results.  The systematic 
uncertainties for the $\Upsilon$ samples are obtained using BC $\Upsilon$ data 
and assumed to be the same for NBC $\Upsilon$ data, since the sources are 
completely correlated. }
\begin{tabular}{lc}
Sample & $\Delta p/p (\times 10^{-3})$\\
\hline
$J/\psi\to\mu\mu$ & $-1.311\pm 0.004_{\rm stat}\pm 0.092_{\rm syst}$ \\
$\Upsilon\to\mu\mu$ (NBC) & $-1.335\pm 0.025_{\rm stat}\pm 0.068_{\rm syst}$ \\
$\Upsilon\to\mu\mu$ (BC) & $-1.185\pm 0.020_{\rm stat}\pm 0.068_{\rm syst}$ \\
\end{tabular}
\label{tab:jpsiupsilon}
\end{ruledtabular}
\end{table}

\subsection{$Z\to\mu\mu$ mass measurement and calibration}
\label{sec:zmm}
Using the precise momentum scale calibration obtained from $J/\psi$ and $\Upsilon(1S)$ 
decays, we perform a measurement of the $Z$-boson mass in $Z\to\mu\mu$ decays.  The 
measurement result was hidden during the calibration process, following the procedure 
described in Sec.~\ref{sec:strategy}.  After unblinding and testing the consistency of 
the measured $M_Z$ with the known value of $M_Z = 91187.6\pm 2.1$~MeV~\cite{pdg}, we 
use the latter to further constrain $\Delta p/p$.  The resulting calibration is then 
applied to the $W$-boson data for the $M_W$ measurement.

\begin{figure}
\begin{center}
\epsfysize = 6.cm
\includegraphics*[width=8.5cm]{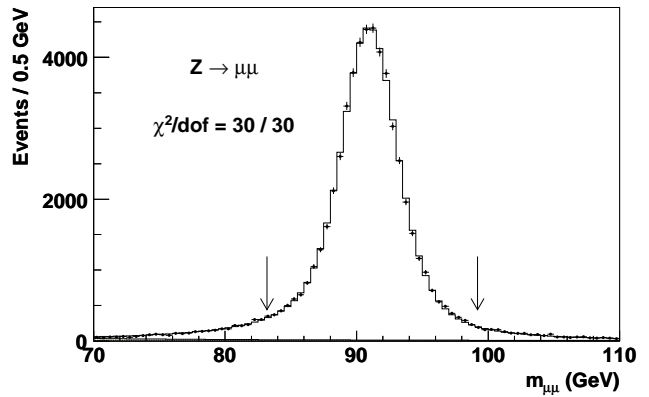}
\caption{Distribution of $m_{\mu\mu}$ for the best-fit template (histogram) 
and data (circles) in the $Z\rightarrow\mu\mu$ candidate sample.  The filled histogram 
shows the $\gamma^*\to\mu\mu$ contribution.  The fit region is enclosed by the arrows.}
\label{fig:zmm}
\end{center}
\end{figure}

The $Z\to\mu\mu$ sample of 59\,738 events is selected as described in 
Sec.~\ref{sec:muselection} and includes the momentum scale calibration given in 
Eq.~(\ref{eq:pscale}).  We form templates for the $Z\to\mu\mu$ invariant mass 
lineshapes using the {\sc resbos} generator, with final-state photon emission 
simulated using the {\sc photos} generator and calibrated to the {\sc horace} 
generator (Sec.~\ref{sec:production}).  We measure $M_Z$ using a binned likelihood 
template fit to the data in the range $83190 < m_{\mu\mu} < 99190$~MeV 
(Fig.~\ref{fig:zmm}).  Systematic uncertainties on $M_Z$ are due to uncertainties 
on the COT momentum scale (9~MeV), alignment corrections (2~MeV), and QED radiative 
corrections (5~MeV).  The alignment uncertainty is dominated by the uncertainty on 
the $z$-scale parameter $t$ of Eq.~(\ref{eq:aligncorrjpsi}), as determined using 
BC $\Upsilon\rightarrow \mu\mu$ data.

The measurement of the $Z$-boson mass in the muon decay channel is 

\begin{equation}
M_Z = 91180\pm 12_{\rm stat}\pm 10_{\rm syst}~\textrm{MeV}.
\label{eq:zmm}
\end{equation}

\noindent
This result is the most precise determination of $M_Z$ at a hadron collider and is 
in excellent agreement with the world-average value of $M_Z$, providing a sensitive 
consistency check of the momentum scale calibration.  Combining this measurement with 
the calibration of Eq.~(\ref{eq:pscale}) from $J/\psi$ and $\Upsilon$ data, and taking 
the alignment and QED uncertainties to be fully correlated, we obtain

\begin{equation}
[\Delta p /p]_{J/\psi+\Upsilon+Z} = (-1.29 \pm 0.09) \times 10^{-3}.
\label{eq:pscalewithz}
\end{equation}

\section{Electron Momentum Measurement}
\label{sec:electrons}
The mean fraction of traversed radiation lengths for an electron in the CDF 
tracking volume is approximately 19\%~\cite{CDF2}. Hence, electron track momentum measurements do not 
provide as high precision as calorimeter measurements.  For the high-energy 
electrons used in this analysis, the bremsstrahlung photons are absorbed by 
the same calorimeter tower as the primary electron.  We perform a precise 
calibration of the calorimeter response using the measured ratio of 
calorimeter energy to track momentum ($E/p$).  We validate the calibration 
by measuring the mass of the $Z$ boson in $Z\to ee$ events and then combine 
the $E/p$ and $Z$-mass calibrations to obtain the calorimeter calibration 
used for the $M_W$ measurement.

\subsection{$E/p$ calibration}
\label{sec:eop}
The precise track momentum calibration is applied to calorimeter-based measurements
through the ratio $E/p$.  The calibration includes several corrections:
the data are corrected for response variations near tower edges in $\phi$
and $z$ and the simulation is corrected for limitations in the knowledge 
of the number of radiation lengths in the tracking detector and the 
calorimeter, and for the observed energy dependence of the calorimeter 
response.  Including in the model the energy resolution determined from 
the $E/p$ peak region and from $Z\to ee$ data (see Sec.~\ref{sec:ecal}), 
the calorimeter energy scale $S_E$ is extracted using a likelihood fit to 
the $E/p$ peak.

\begin{figure}
\begin{center}
\includegraphics[width=8.5cm]{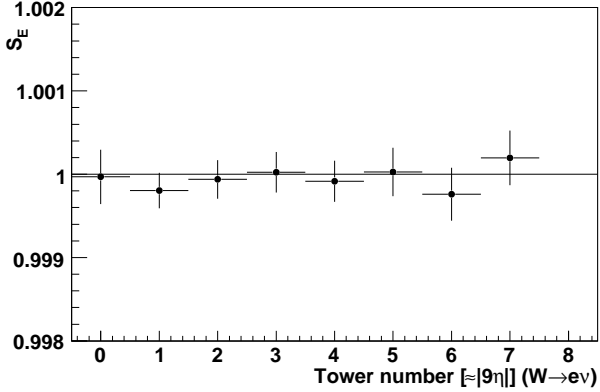}
\end{center}
\caption{Measured calorimeter energy scale in bins of electron tower in 
$W\to e\nu$ data after corrections are applied.  The towers are numbered 
in order of increasing $|\eta|$ and each tower subtends 
$\Delta \eta \approx 0.11$. }
\label{fig:eop_eta}
\end{figure}

The dominant spatial nonuniformities in the CEM response are corrected in 
the event reconstruction~\cite{cemresponse}.  Residual nonuniformities near 
gaps between towers are at the 1--2\% level, as determined using the mean 
$E/p$ in the range 0.9--1.1.  After correcting for these nonuniformities, 
the likelihood fits for the calorimeter energy scale are independent of 
electron $|\eta|$ (Fig.~\ref{fig:eop_eta}).

\begin{figure}
\begin{center}
\includegraphics*[width=8.5cm]{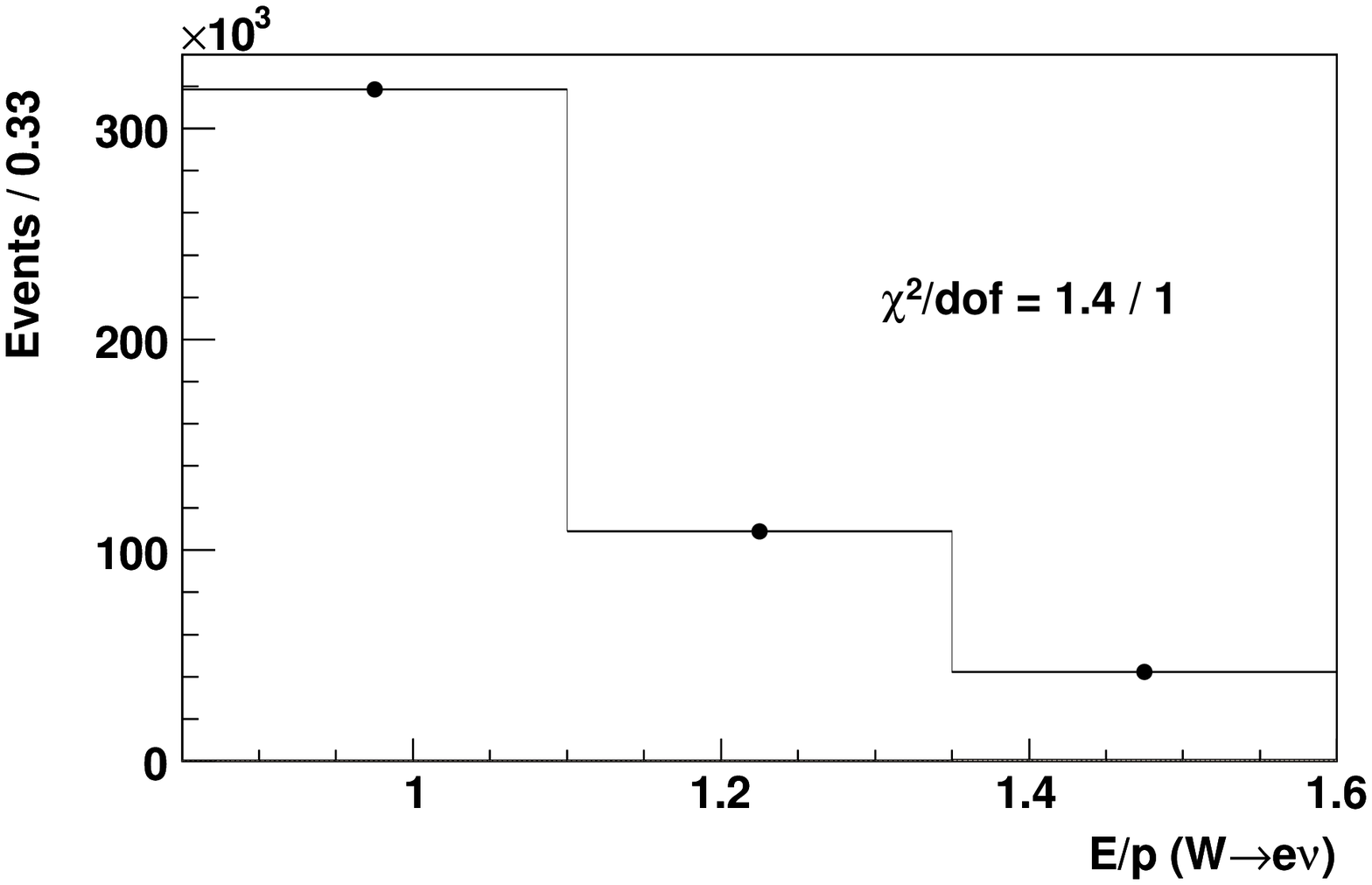}
\includegraphics*[width=8.5cm]{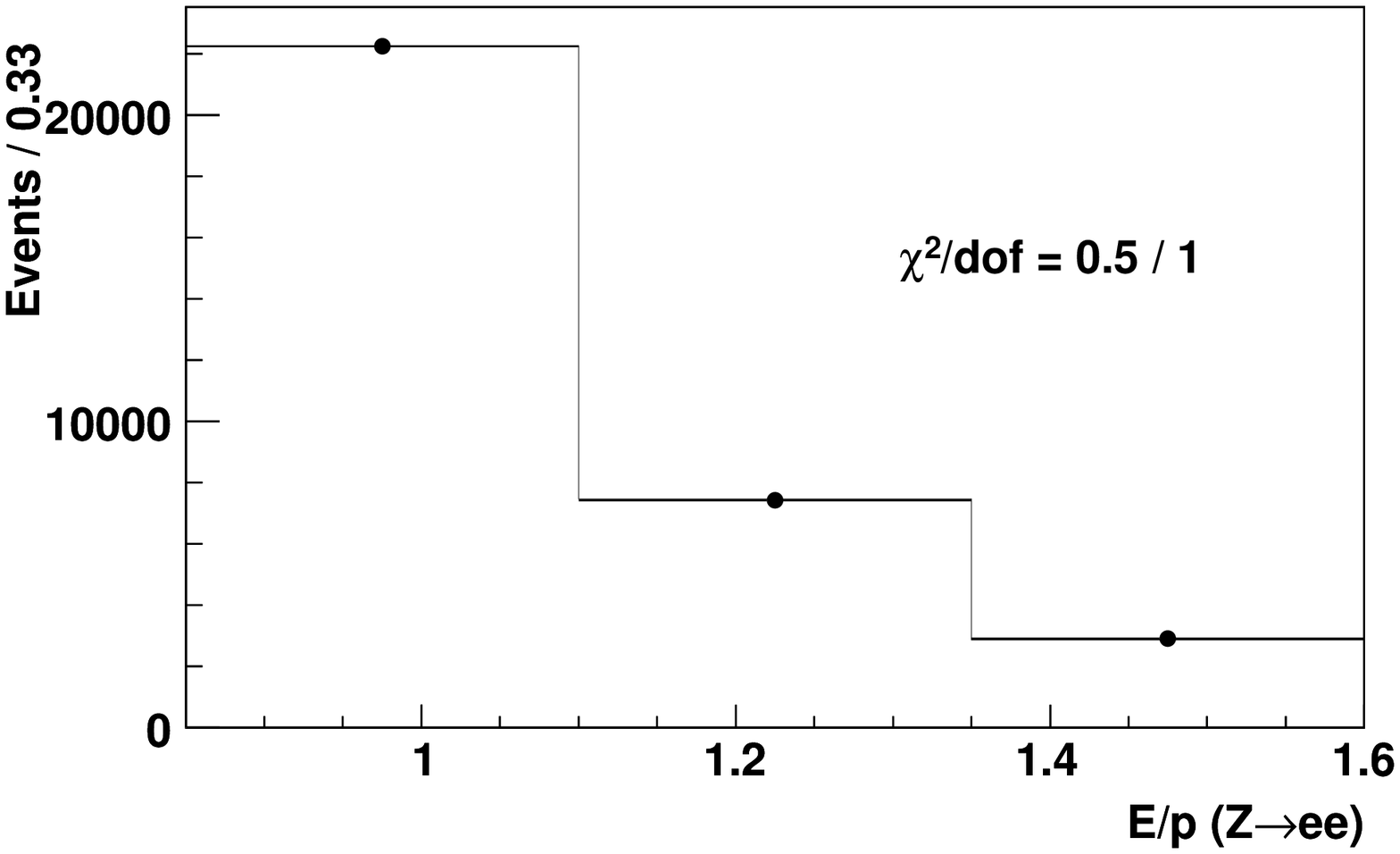}
\caption{Distributions of $E/p$ in data (circles) and simulation with the 
best-fit value of $S^{W,Z}_{\rm mat}$ (histograms) in $W\rightarrow e\nu$ 
(top) and $Z\to ee$ (bottom) events. }
\label{fig:eopmat}
\end{center}
\end{figure}

The radiative detector material is mapped into a three-dimensional lookup table, 
as described in Sec.~\ref{sec:model}.  We fine-tune this material model with 
a likelihood fit to two ranges in the tail of the $E/p$ distribution 
($1.1 < E/p < 1.6$), which is sensitive to the total number of radiation 
lengths traversed.  The region $0.85 < E/p < 1.1$ effectively normalizes the 
simulation.  From a maximum likelihood fit to electrons in $W\to e\nu$ 
($Z\to ee$) data, we obtain a multiplicative factor 
$S^W_{\rm mat} = 1.027 \pm 0.004$ ($S^Z_{\rm mat} = 1.001 \pm 0.011$) to the 
number of radiation lengths in the simulation.  The results from $W$ and $Z$ 
data are statistically consistent within $2.2\sigma$ and are combined to 
give the correction $S^{W,Z}_{\rm mat} = 1.024 \pm 0.003$ applied to the 
simulation for mass measurements.  Figure~\ref{fig:eopmat} shows the 
three-bin $E/p$ distributions for both $W\to e\nu$ and $Z\to ee$ data after 
the correction factor is applied. 

Electron candidates with low $E/p$ are predominantly electron showers that 
are not fully contained in the EM calorimeter.  Accurate simulation of 
these showers relies on a knowledge of the amount and composition of the CEM 
material.  We tune the {\it a priori} estimate of this material using the 
relative fraction of electron candidates with low $E/p$ ($0.85 < E/p < 0.93$) 
to those at low $E/p$ or in the peak ($0.85 < E/p < 1.09$).  From a 
comparison of data to simulation of this ratio as a function of the amount of 
tower material, we find that the data are accurately reproduced by adding a 
thin layer to each simulated calorimeter tower.  The thickness of the 
additional tower increases linearly from zero for the central towers 
($|\eta| \approx 0$) to $0.51 X_0$ for the most forward tower 
($|\eta| \approx 1$).  The estimated uncertainty on the forward tower 
correction is $0.07 X_0$. 

\begin{figure}
\begin{center}
\includegraphics*[width=8.5cm]{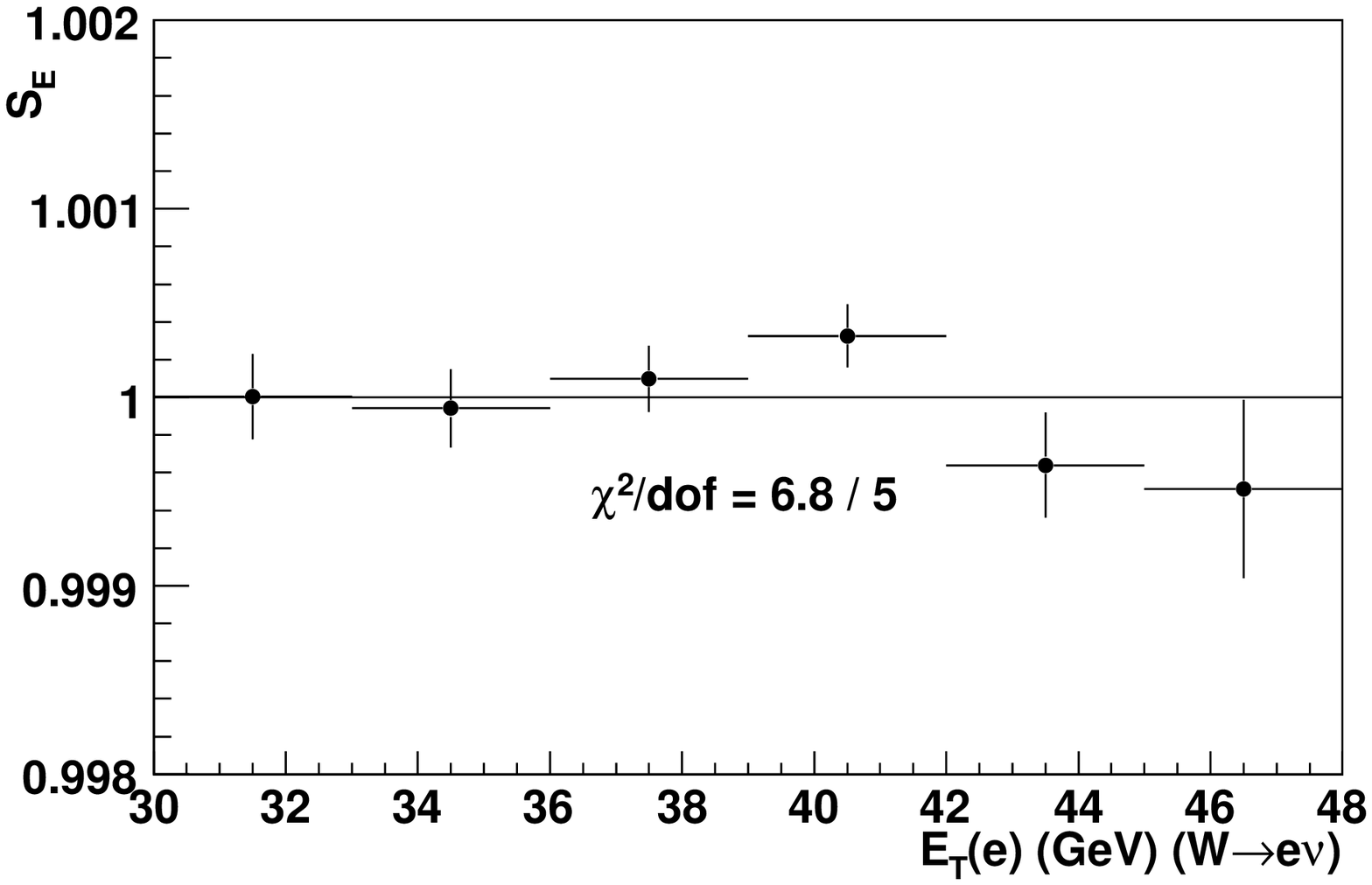}
\includegraphics*[width=8.5cm]{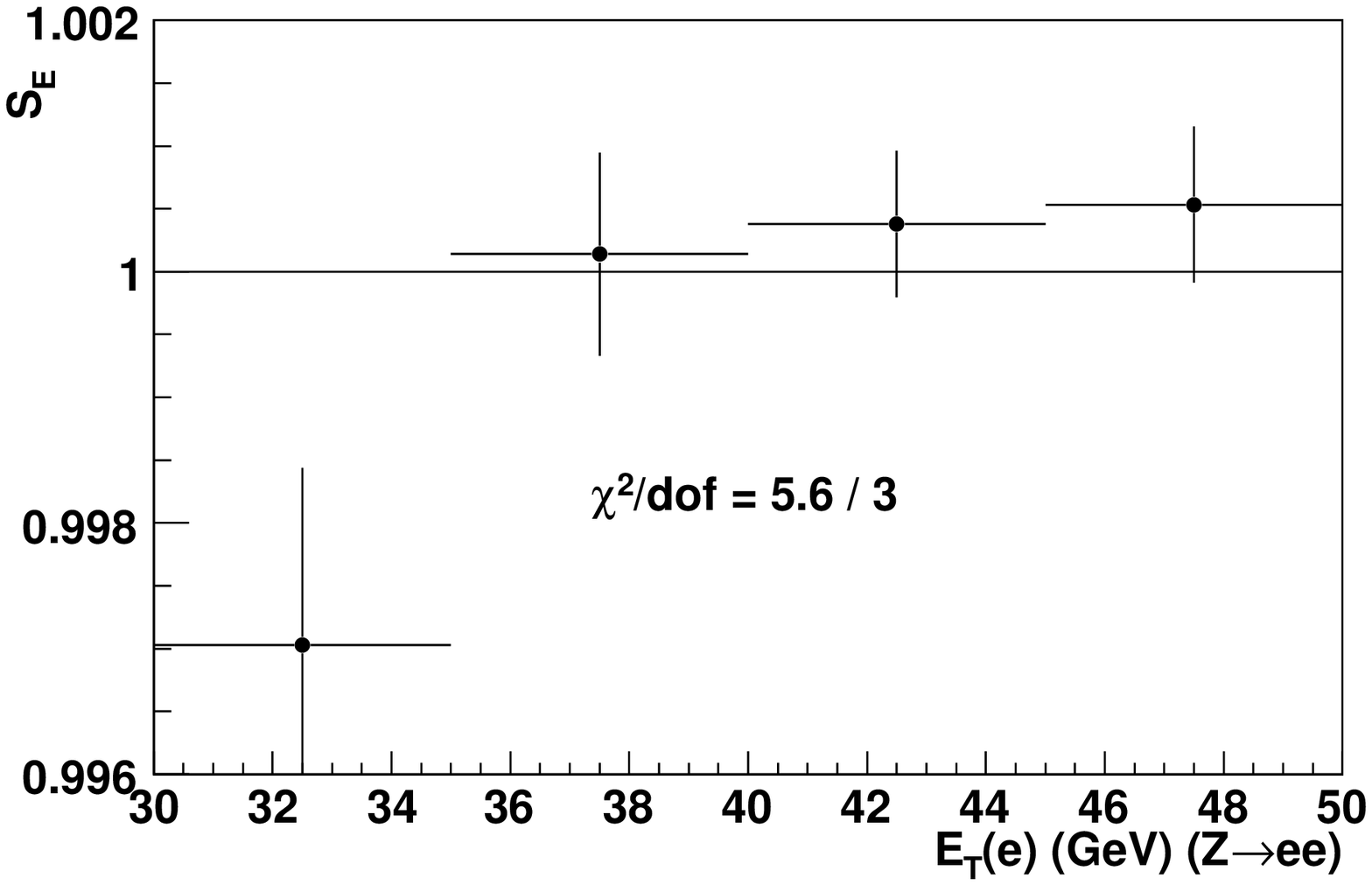}
\caption{Measured energy scale as a function of electron $E_T$ for 
$W \to e\nu$ (top) and $Z \to ee$ (bottom) data.  The simulation is 
corrected with the best-fit value of $\xi = (5.25 \pm 0.70) \times 10^{-3}$. }
\label{fig:eopet}
\end{center}
\end{figure}

We correct the energy dependence of the detector response by applying a 
per-particle scale in the simulation (Sec.~\ref{sec:ecal}).  We measure this 
correction, $\xi$ in Eq.~(\ref{eq:ecal}), using the fit energy scale as a 
function of measured calorimeter $E_T$ in $W\to e\nu$ and $Z\to ee$ data. 
Figure~\ref{fig:eopet} shows the results of these fits after including the 
correction from the combined data, $\xi = (5.25\pm 0.70) \times 10^{-3}$.

\begin{figure}
\begin{center}
\includegraphics*[width=8.5cm]{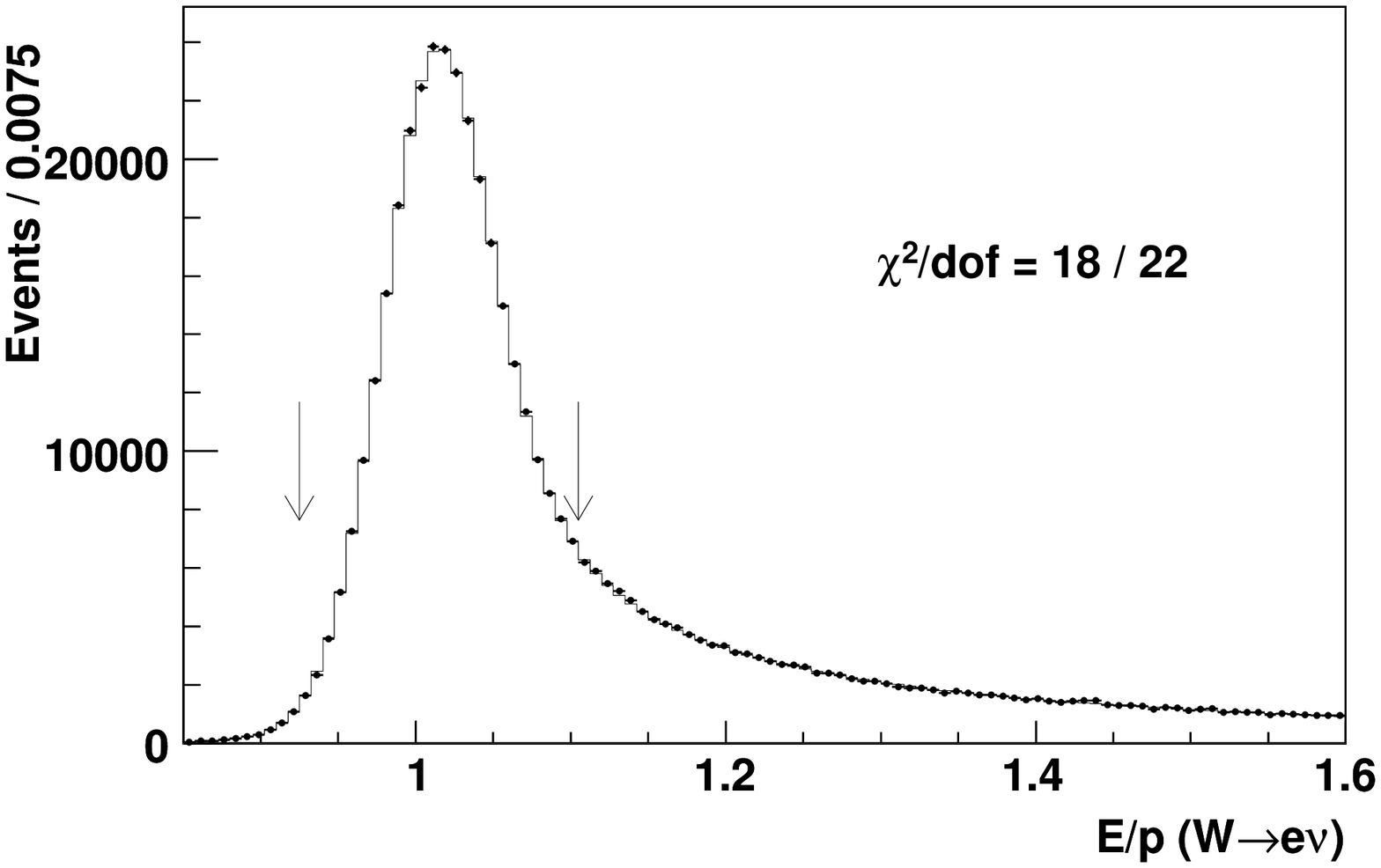}
\includegraphics*[width=8.5cm]{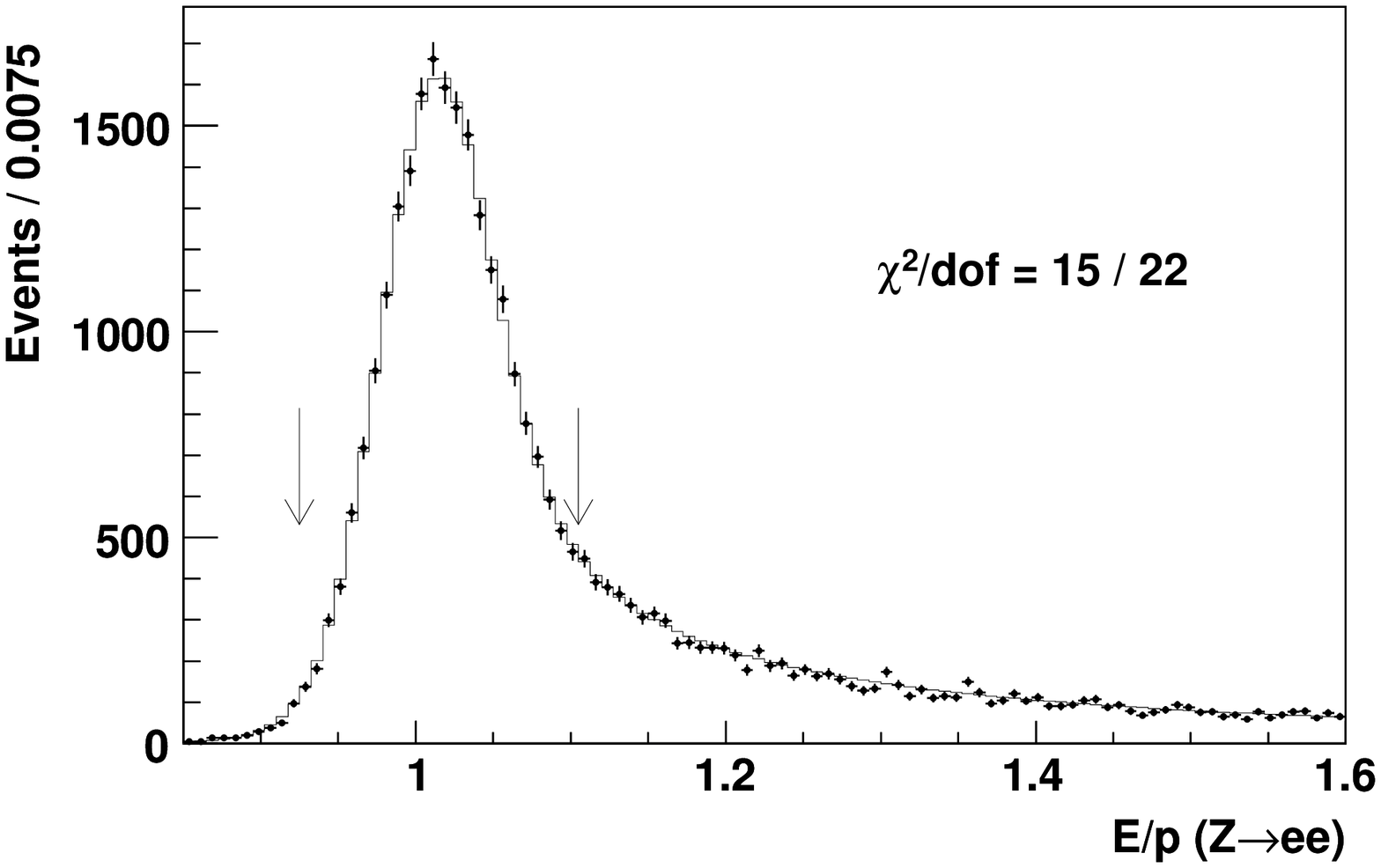}
\caption{Distribution of $E/p$ for $W\to e\nu$ (top) and $Z\to ee$ (bottom) data 
(circles) after the full energy-scale calibration; the best-fit templates 
(histograms) are overlaid.  The fit region is enclosed by arrows.}
\label{fig:eop}
\end{center}
\end{figure}

After applying the complete set of corrections described above, we fit the peak 
region ($0.93 < E/p < 1.11$) of the $E/p$ distribution for $S_E$ in both 
$W\to e\nu$ and $Z\to ee$ data.  The fits results are statistically consistent, 
differing by $(0.019 \pm 0.030)\%$ between the two data sets; their combination 
has a statistical uncertainty of 0.008\%.  After applying the calibrated energy 
scale, the simulated $E/p$ distribution shows good agreement with the data for 
both $W\to e\nu$ and $Z\to ee$ events (Fig.~\ref{fig:eop}).  

By varying the simulation parameters we determine the correlations between the 
uncertainties on the energy scale estimated using $E/p$ and on $M_W$ obtained 
from the mass-fit distributions.  The $E/p$-based calibration uncertainties on 
$M_W$ using the $m_T$ fit are due to $S_{\rm mat}$ (4 MeV), the tracker material 
model (3 MeV), calorimeter material (2 MeV), calorimeter nonlinearity (4 MeV), 
track momentum scale (7 MeV), and resolution (4 MeV).  Including the statistical 
uncertainty gives a total $E/p$-based calibration uncertainty on $M_W$ of 12 MeV.

\subsection{$Z\to ee$ mass measurement and calibration}
\label{sec:zee}
As with the meson-based calibration of track momentum, the $E/p$-based calorimeter 
energy calibration is validated with a measurement of the $Z$-boson mass.  After 
comparing the mass measured in $Z\to ee$ decays to the known value of $M_Z$, we 
incorporate the result into the electron energy calibration used for the $M_W$ 
measurement. 

\begin{figure}
\begin{center}
\epsfysize = 6.cm
\includegraphics*[width=8.5cm]{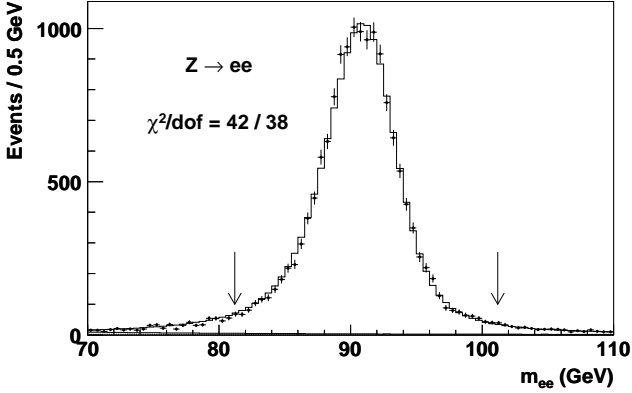}
\caption{Best-fit $M_Z$ template (histogram) compared to data (circles) in $Z\to ee$ 
decays.  The fit region is enclosed by arrows. }
\label{fig:zee}
\end{center}
\end{figure}

The $Z\to ee$ candidate sample contains 16\,134 events.  We use the same simulation 
and fit procedure as for the mass measurement using $Z\to \mu\mu$ decays, but with 
a broader fit range of $81\,190 < m_{ee} < 101\,190$~MeV (Fig.~\ref{fig:zee}).  We 
measure $M_Z = 91\,230 \pm 30_{\rm stat}$~MeV.

Systematic uncertainties on $M_Z$ are due to the $E/p$ calibration (10~MeV), the COT 
momentum-scale calibration (8~MeV), alignment corrections (2~MeV), and the QED radiative 
corrections (5~MeV).  Including these uncertainties, the $Z$ boson mass determined using 
electron decays is

\begin{equation}
M_Z = 91\,230\pm 30_{\rm stat}\pm 14_{\rm syst}~\textrm{MeV},
\label{eq:zee}
\end{equation}
which is consistent with the known value of $M_Z$ at the level of $1.3\sigma$.  The 
measurement is converted into an energy-scale calibration and combined with the 
$E/p$-based calibration to define the energy scale for the $M_W$ measurement.  Taking 
into account correlations between uncertainties on the energy scale and on the fits 
for $M_W$, the uncertainty on $M_W$ due to the combined energy-scale calibration is 
10 MeV.

\begin{figure}
\begin{center}
\epsfysize = 6.cm
\includegraphics*[width=8.5cm]{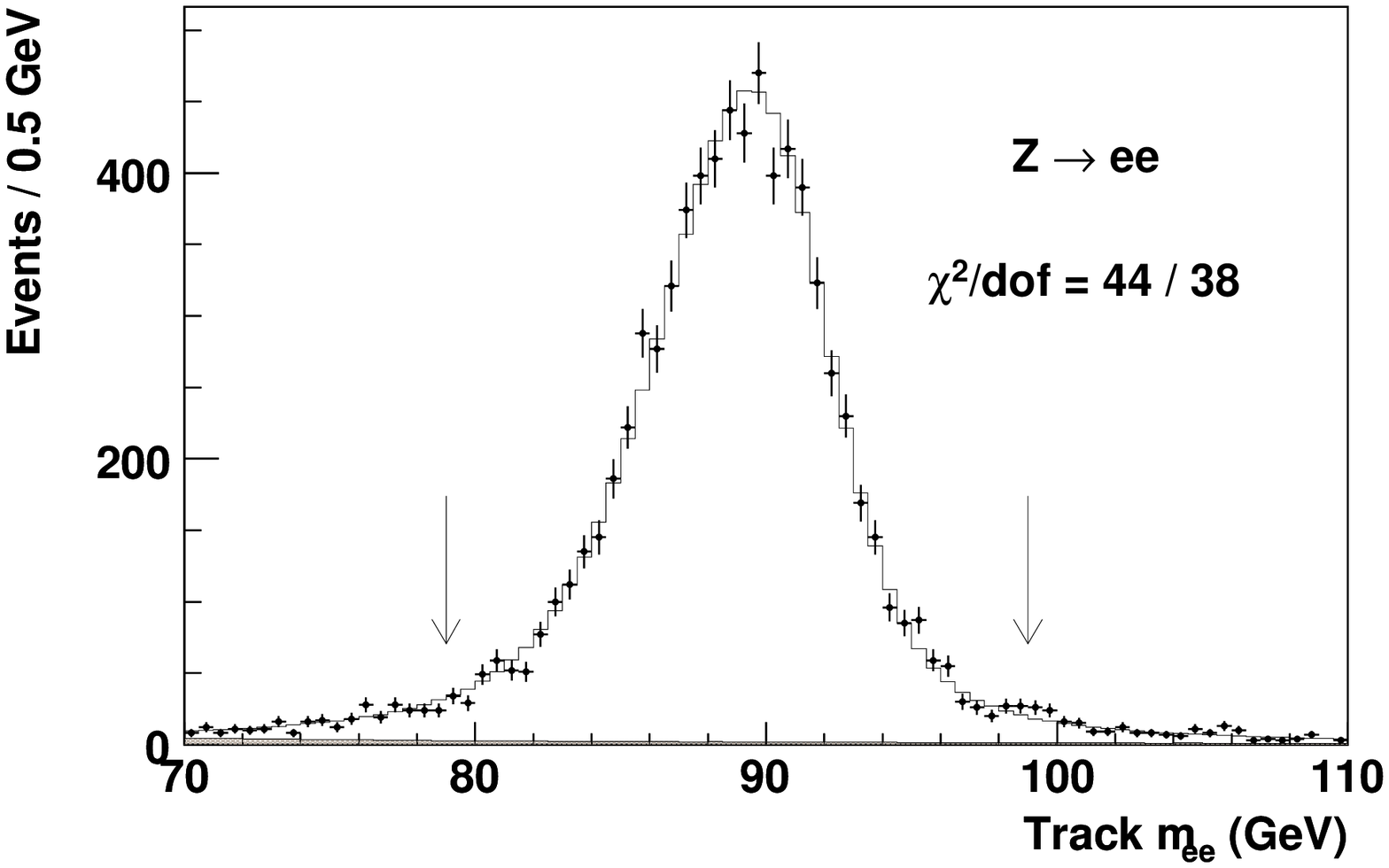}
\includegraphics*[width=8.5cm]{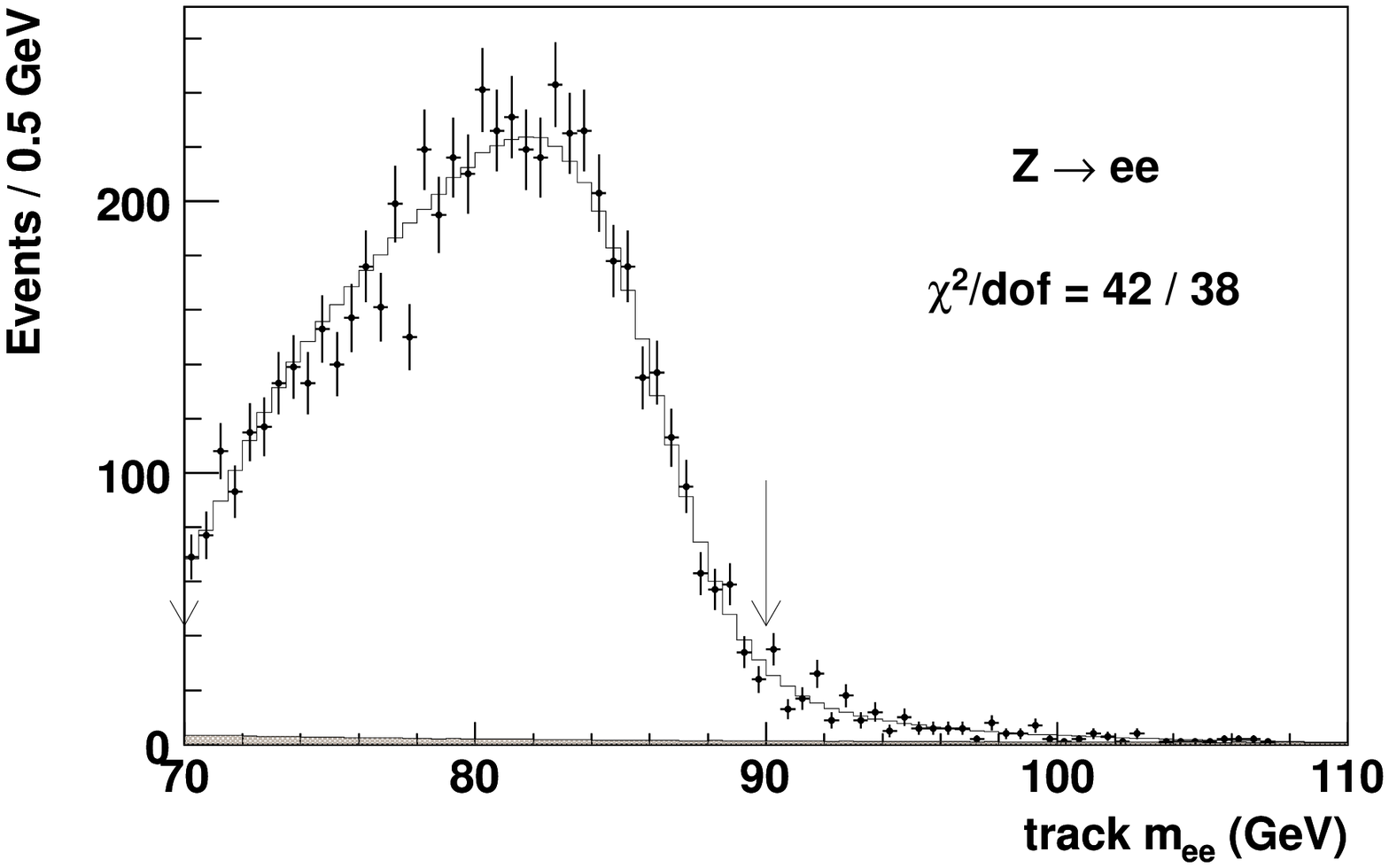}
\includegraphics*[width=8.5cm]{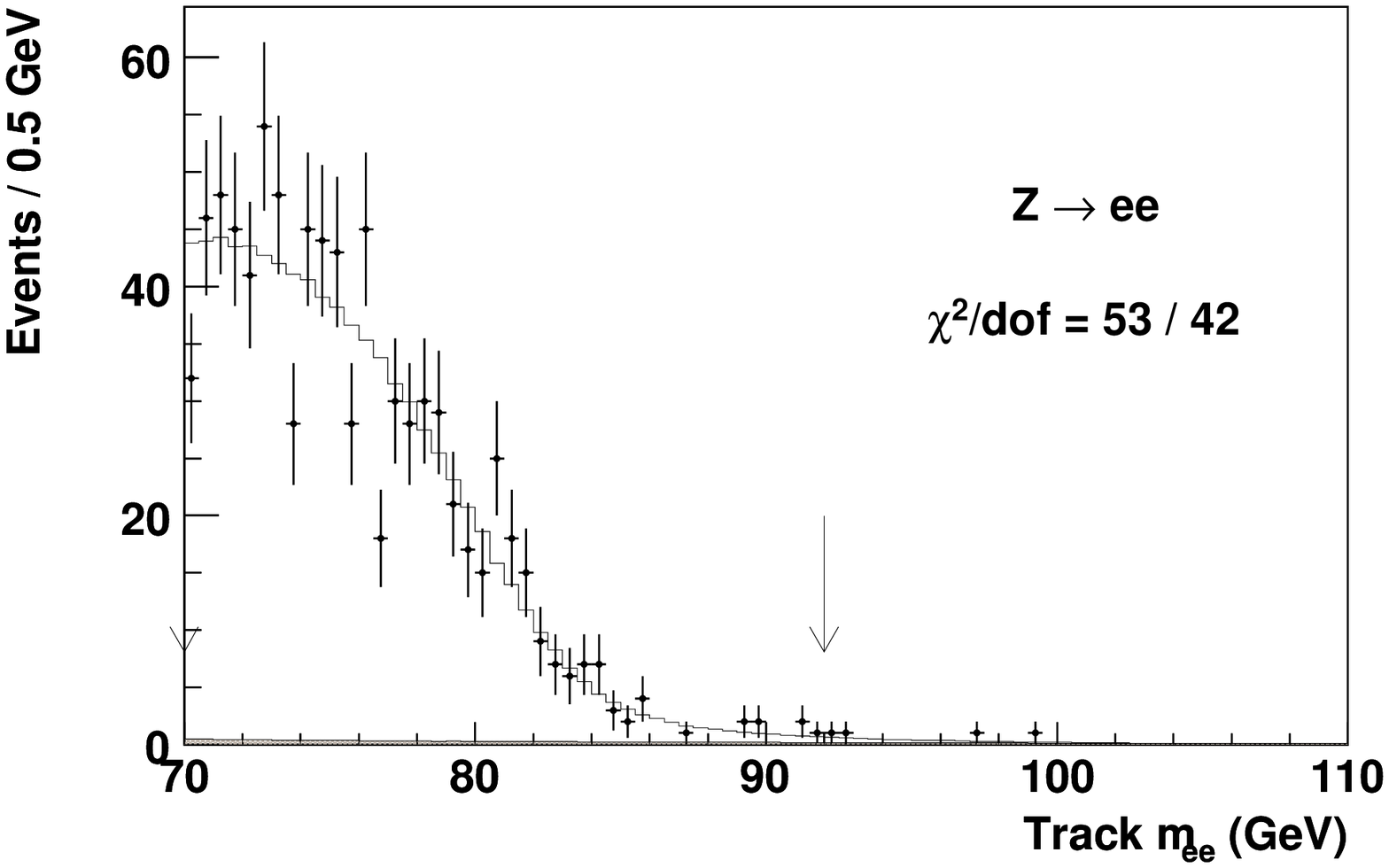}
\caption{Best-fit $M_Z$ templates (histogram) compared to data (circles) in $Z\to ee$ decays 
using only reconstructed track information in events with two nonradiative electrons (top), 
one radiative electron (middle), or two radiative electrons (bottom).  The nonradiative fit 
region is enclosed by arrows; the other fit regions are to the left of the arrows.}
\label{fig:zee_track_subs}
\end{center}
\end{figure}

\begin{figure}
\begin{center}
\epsfysize = 6.cm
\includegraphics*[width=8.5cm]{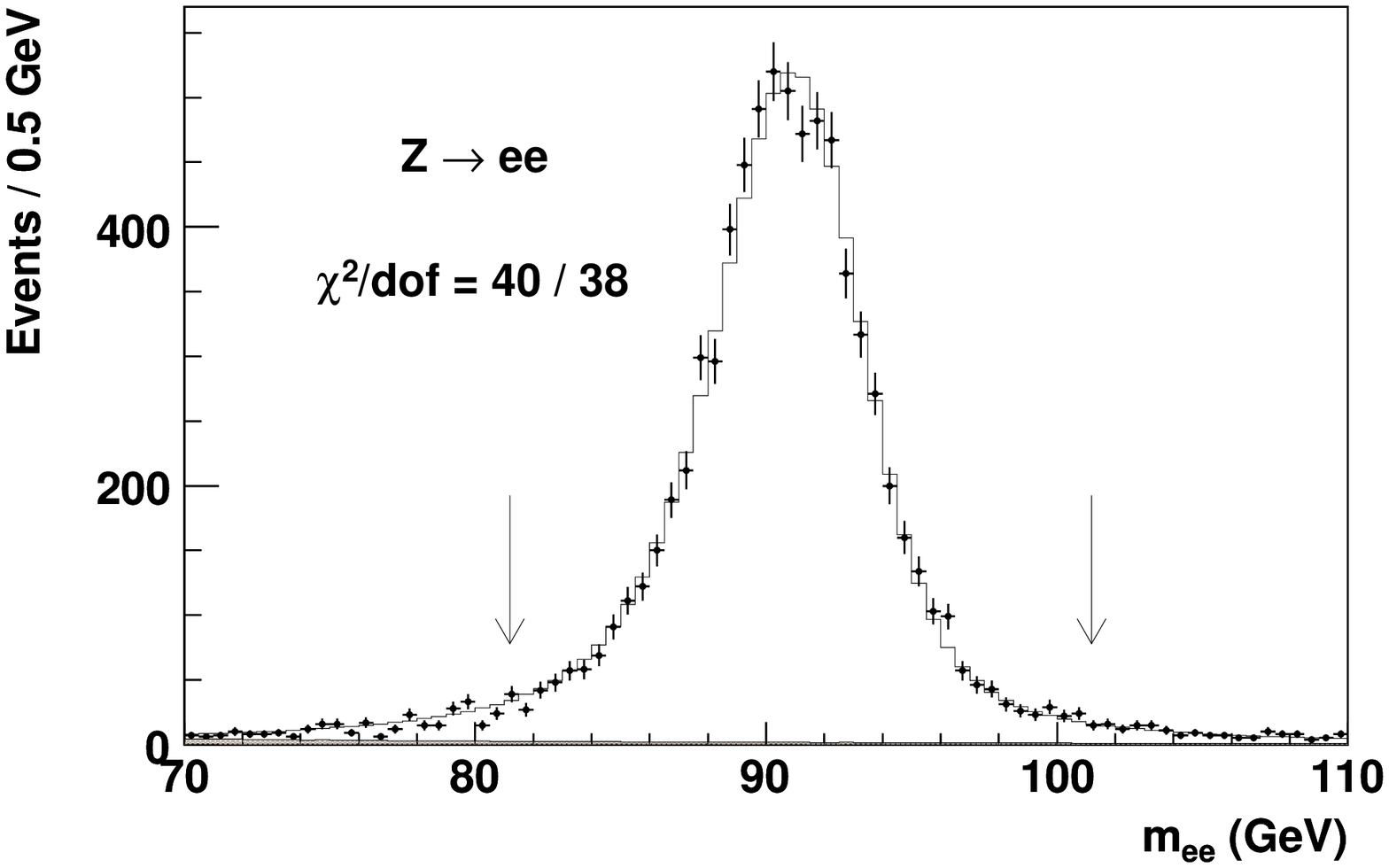}
\includegraphics*[width=8.5cm]{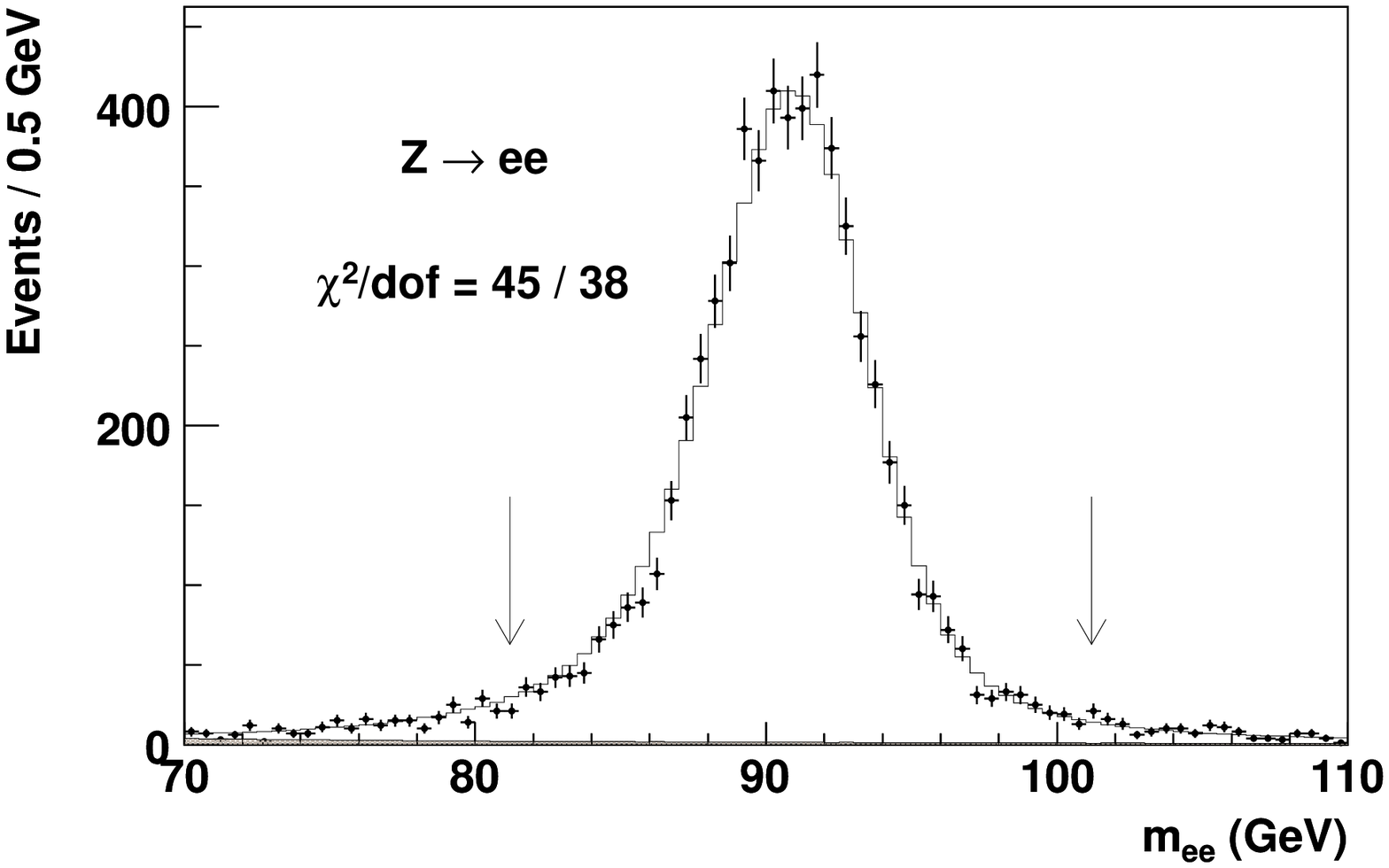}
\includegraphics*[width=8.5cm]{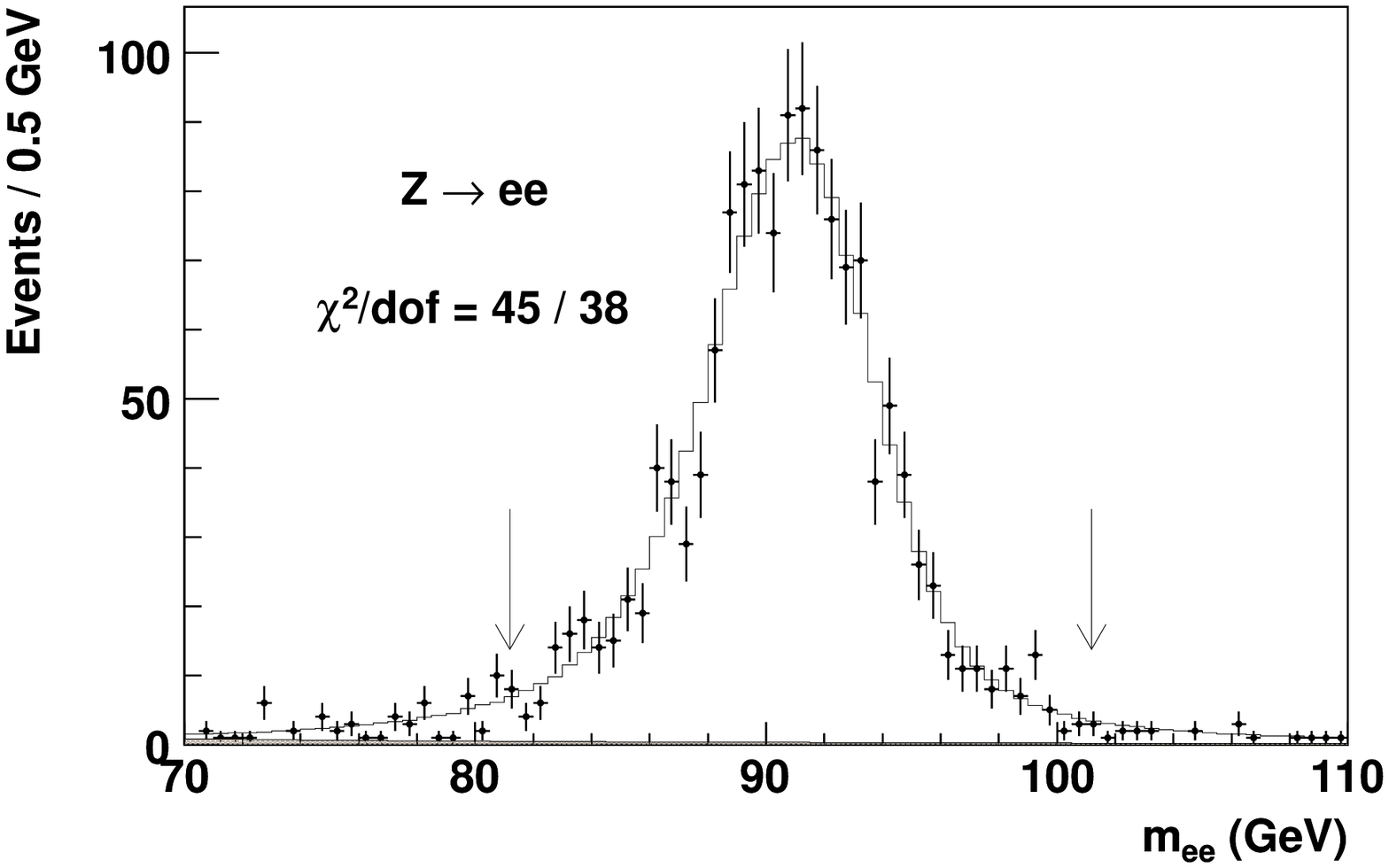}
\caption{Best-fit $M_Z$ templates (histogram) compared to data (circles) in $Z\to ee$ decays 
in events with two nonradiative electrons (top), one radiative electron (middle), and two 
radiative electrons (bottom).  Fit regions are enclosed by arrows.}
\label{fig:zee_clus_subs}
\end{center}
\end{figure}

The application of the momentum-scale calibration to a calorimeter 
energy calibration via $E/p$ relies on an accurate simulation of the electron radiation 
and the track reconstruction.  We test the simulation by measuring $M_Z$ using electron track 
momenta only.  The measurement is performed for three configurations: neither electron 
radiative ({\it i.e.}, both with $E/p < 1.1$), one electron radiative ($E/p > 1.1$), and both 
electrons radiative.  The results of the fits are shown in Table~\ref{tab:zee_subs} and 
Fig.~\ref{fig:zee_track_subs}.  Combining the measurements of events with at least one 
radiative electron gives $M_Z = 91\,240 \pm 38_{\rm stat}$ MeV, in good agreement with the 
known $M_Z$ and with the measurement determined using calorimeter energy.  As an 
additional check, we split the calorimeter-based measurement into the same categories of 
radiative and nonradiative electrons, and obtain consistent results 
(Table~\ref{tab:zee_subs} and Fig.~\ref{fig:zee_clus_subs}). 

\begin{table}
\begin{ruledtabular}
\caption{Summary of $M_Z$ measurements obtained using subsamples of data containing events 
with nonradiative electrons ($E/p < 1.1$), one radiative electron ($E/p>1.1$), or two 
radiative electrons.  Calorimeter-based and track-based measurements are shown for each 
category; uncertainties are statistical only.}
\begin{tabular}{lcc}
Electrons & Calorimeter $M_Z$ (MeV) & Track $M_Z$ (MeV) \\
\hline
$E/p < 1.1$ only & $91\,208\pm 39$ & $91\,231 \pm 41$ \\
$E/p > 1.1$ and $E/p < 1.1$ & $91\,234\pm 51$ & $91\,294\pm 98$ \\
$E/p > 1.1$ only & $91\,370\pm 127$ & $91\,176\pm 407$ \\
\end{tabular}
\label{tab:zee_subs}
\end{ruledtabular}
\end{table}

\section{Recoil measurement} 
\label{sec:recoil}
The neutrino transverse momentum is determined using a measurement of the recoil 
$\vec{u}_T$, defined in Eq.~(\ref{eq:recoil}).  To minimize bias in the recoil measurement, 
we correct the data to improve the uniformity of the calorimeter response.  The 
recoil simulation models the removal of underlying event energy in the vicinity 
of each lepton, the response to QCD and QED initial-state radiation through a 
parametrization, the response to final-state photons using the same detailed 
accounting as for the lepton momentum calibration, and the response to the 
energy from the underlying event and additional $p\bar{p}$ collisions.  The 
parameters are determined using events containing $Z$-boson decays to electrons 
or muons, since the dilepton transverse momentum is measured to high precision.

\subsection{Data corrections}
The modeling of the recoil projected along the lepton direction directly impacts 
the $M_W$ measurement, as described in Sec.~\ref{sec:strategy}.  To simplify the 
modeling of the recoil direction, we apply corrections to the data to reduce 
nonuniformities in recoil response.  

The uncorrected recoil has a sinusoidal distribution as a function of $\phi$, due in 
part to the offset of the collision point from the origin (in the radial direction).  
Calorimeter towers in the direction of the offset subtend a larger angle than those 
in the opposite direction, resulting in a higher energy measurement on average.  A 
relative misalignment between the calorimeter and the beam has a similar effect, 
with an additional bias due to the mismeasured azimuthal angle of the tower.  The 
azimuthal dependence increases with $|\eta|$~\cite{CDF2}, so the plug calorimeter 
towers have the largest dependence.  For simplicity we remove the recoil variation 
by adjusting the origin of each plug calorimeter in our simulation.  We use the 
minimum-bias data to parametrize these effective shifts in three time periods to 
correct for the sinking of the detector into the earth.  Uniformity is improved by 
increasing the transverse energy threshold to 5~GeV for the two most forward towers 
in each plug detector, corresponding to the region $|\eta| > 2.6$.

In addition to the azimuthal uniformity correction, we improve the recoil 
measurement resolution by applying a relative energy scale between the central and 
plug calorimeters~\cite{CDF2}.

\subsection{Lepton tower removal}
The measured recoil $\vec{u}_T$ in the data is determined by summing over the transverse 
momenta of all calorimeter towers with $|\eta|<2.6$, excluding towers with lepton energy 
deposits.  The excluded towers are chosen by studying the average energy deposition in 
towers in the vicinity of the lepton.  In the simulation we subtract an estimated 
underlying event energy in each event to model the lepton tower removal, with corrections 
for its dependence on $u_{||}$, $|u_{\perp}|$, and $|\eta|$.

\begin{figure}
\begin{center}
\includegraphics*[width=8.5cm]{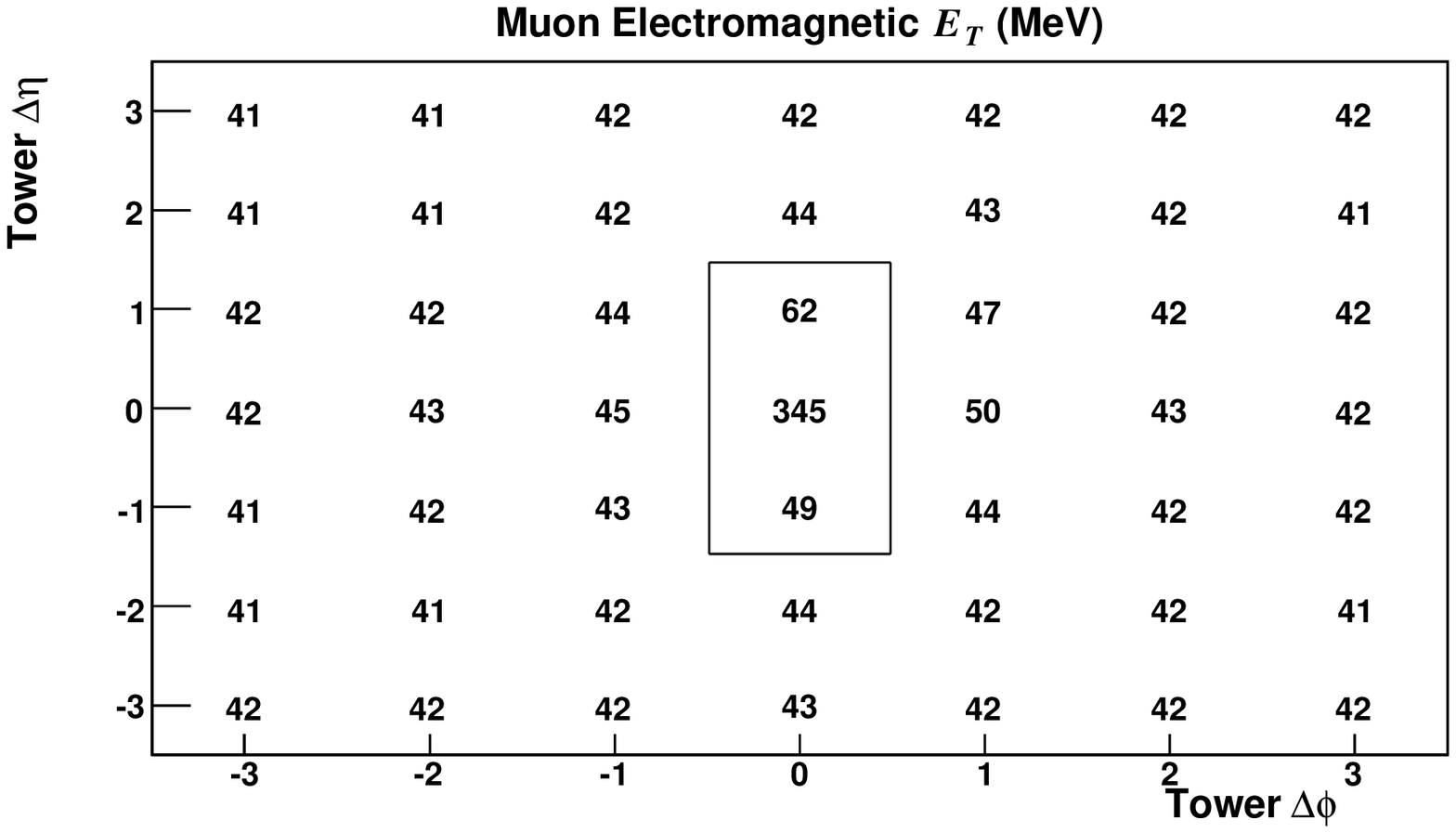}
\includegraphics*[width=8.5cm]{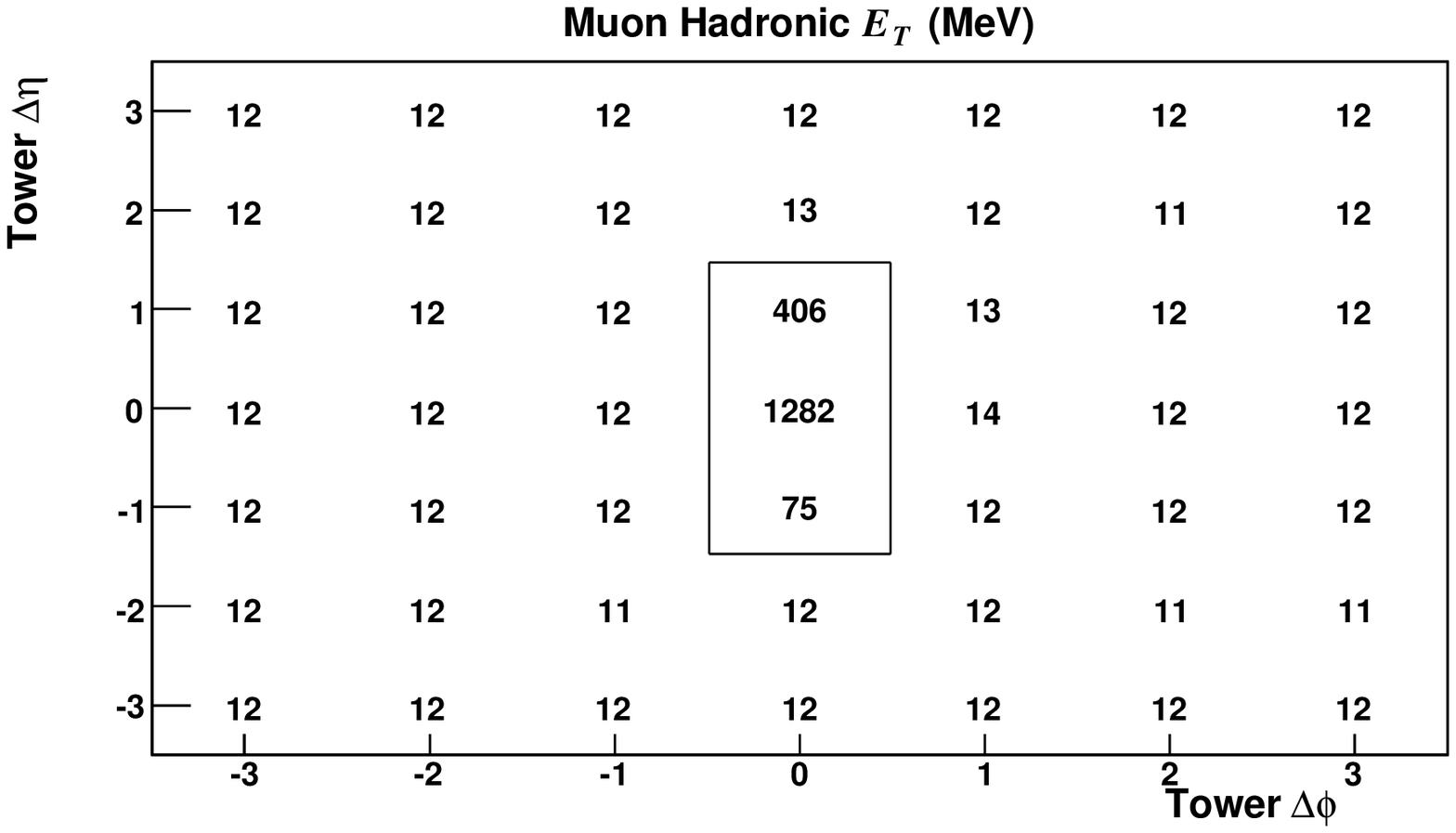}
\caption{Average measured energy (in MeV) in the electromagnetic (top) and hadronic (bottom) 
calorimeters in the vicinity of the muon in $W$-boson decays.  The differences 
$\Delta \phi$ and $\Delta \eta$ are signed such that positive differences correspond 
to towers closest to the muon position at the CES detector.  The three towers inside 
the box are removed from the recoil measurement.  }
\label{fig:muremoval}
\end{center}
\end{figure}

\begin{figure}
\begin{center}
\includegraphics*[width=8.5cm]{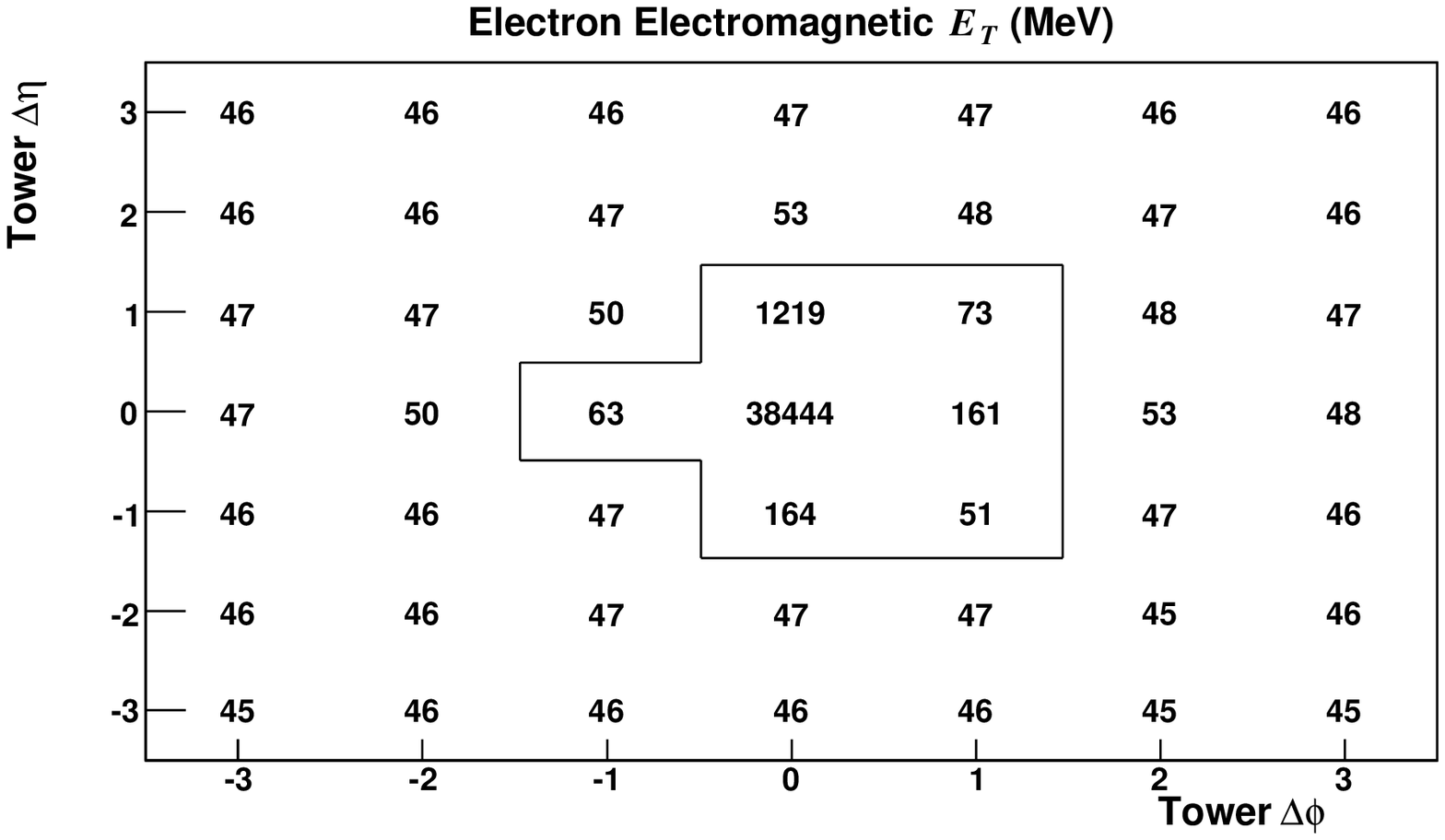}
\includegraphics*[width=8.5cm]{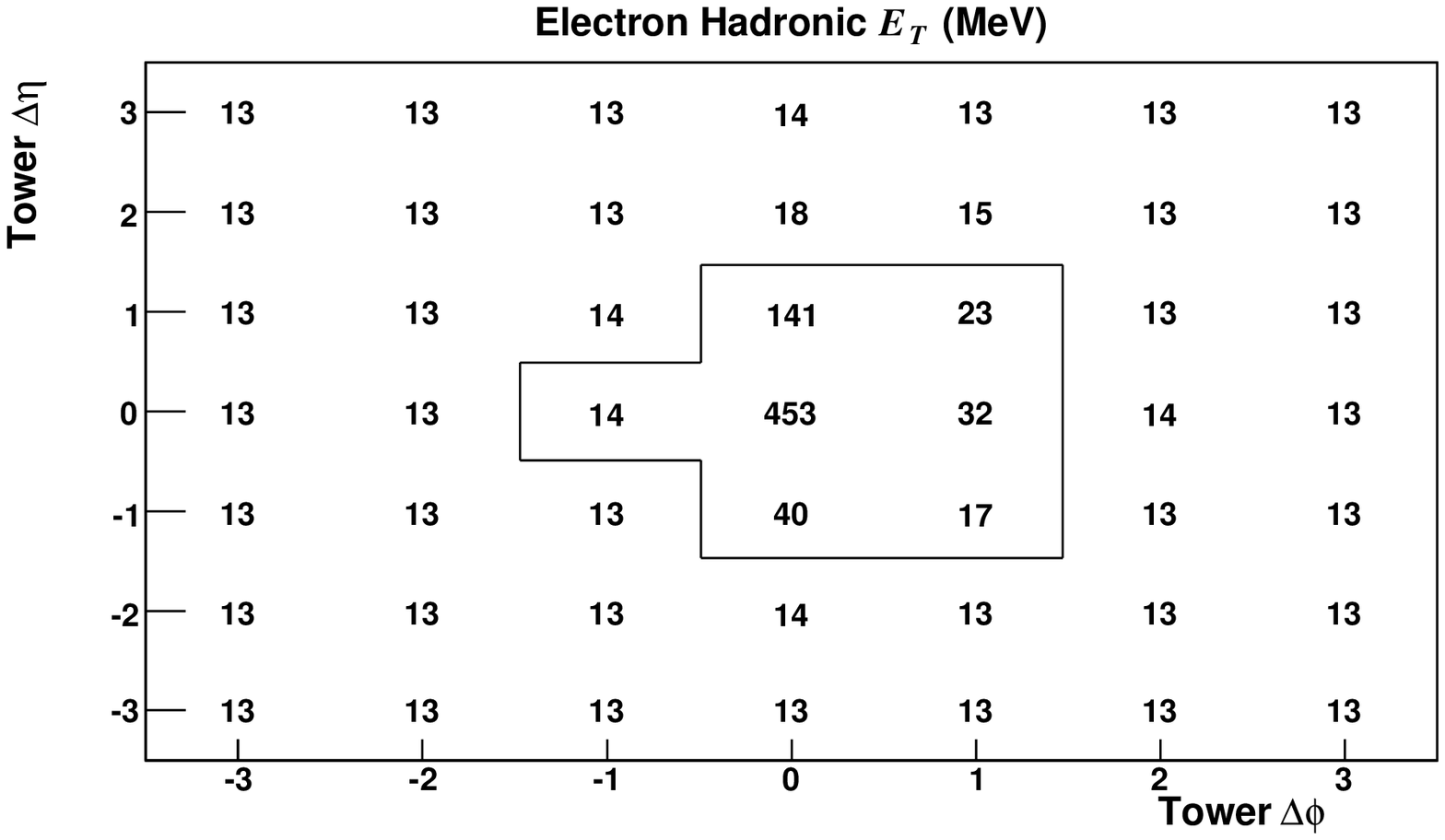}
\caption{Average measured energy (in MeV) in the electromagnetic (top) and hadronic (bottom) 
calorimeters in the vicinity of the electron shower in $W$-boson decays.  The 
differences $\Delta \phi$ and $\Delta \eta$ are signed such that positive differences 
correspond to towers closest to the electron shower position at the CES detector.  
The seven towers inside the box are removed from the recoil measurement.  }
\label{fig:eleremoval}
\end{center}
\end{figure}

We define the set of excluded calorimeter towers based on the presence of an average 
excess of energy over the uniform underlying event energy distribution.  The ionization 
energy deposited by muons is highly localized, but spans neighboring towers in $\eta$ 
when a muon originates from a vertex with large $|z_0|$.  We therefore remove the central 
tower, defined by the CES position of the muon, and both neighboring towers in $\eta$.  
The average energy in these and surrounding towers is shown in Fig.~\ref{fig:muremoval}.  
The additional observed energy in the nearest tower in $\phi$ is due to final-state QED 
radiation, which is modeled by the simulation and is accurately described in this tower.
Electrons shower across towers in both $\eta$ and $\phi$, and produce more QED 
final-state radiation.  The number of removed electron towers is therefore larger, as 
shown in Fig.~\ref{fig:eleremoval}.   

To model the underlying event energy removed from the excluded towers, we use the energy 
distribution of equivalent towers separated by $90^\circ$ in $\phi$ from the lepton.  The 
$90^\circ$ rotation is chosen to minimize bias from QED radiation and from kinematic 
selection, which depends primarily on $u_{||}$.  Muon identification is emulated by 
requiring the local hadronic energy to be less than 4.2 GeV (the muon deposits 1.8 GeV of 
ionization energy on average in the hadronic calorimeter).  The electron 
$E_{\rm had}/E_{\rm EM}$ and $L_\mathrm{shr}$ identification requirements are emulated by 
respectively requiring the local $E_{\rm had}$ to be less than 10\% of the measured 
electron energy, and the neighboring tower in $\eta$ to have less than 5\% of the 
electron energy.  

\begin{figure}
\begin{center}
\includegraphics*[width=8.5cm]{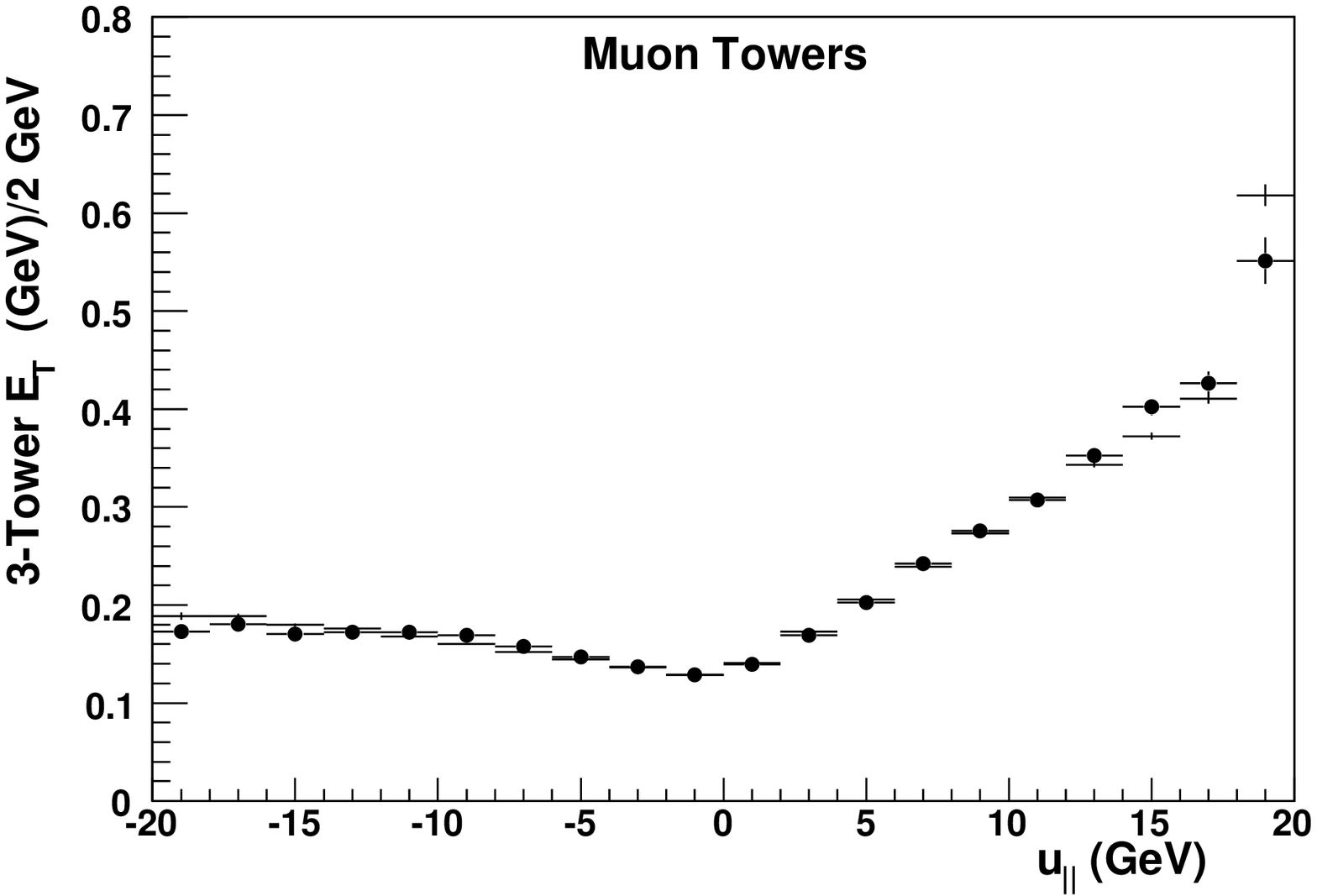}
\includegraphics*[width=8.5cm]{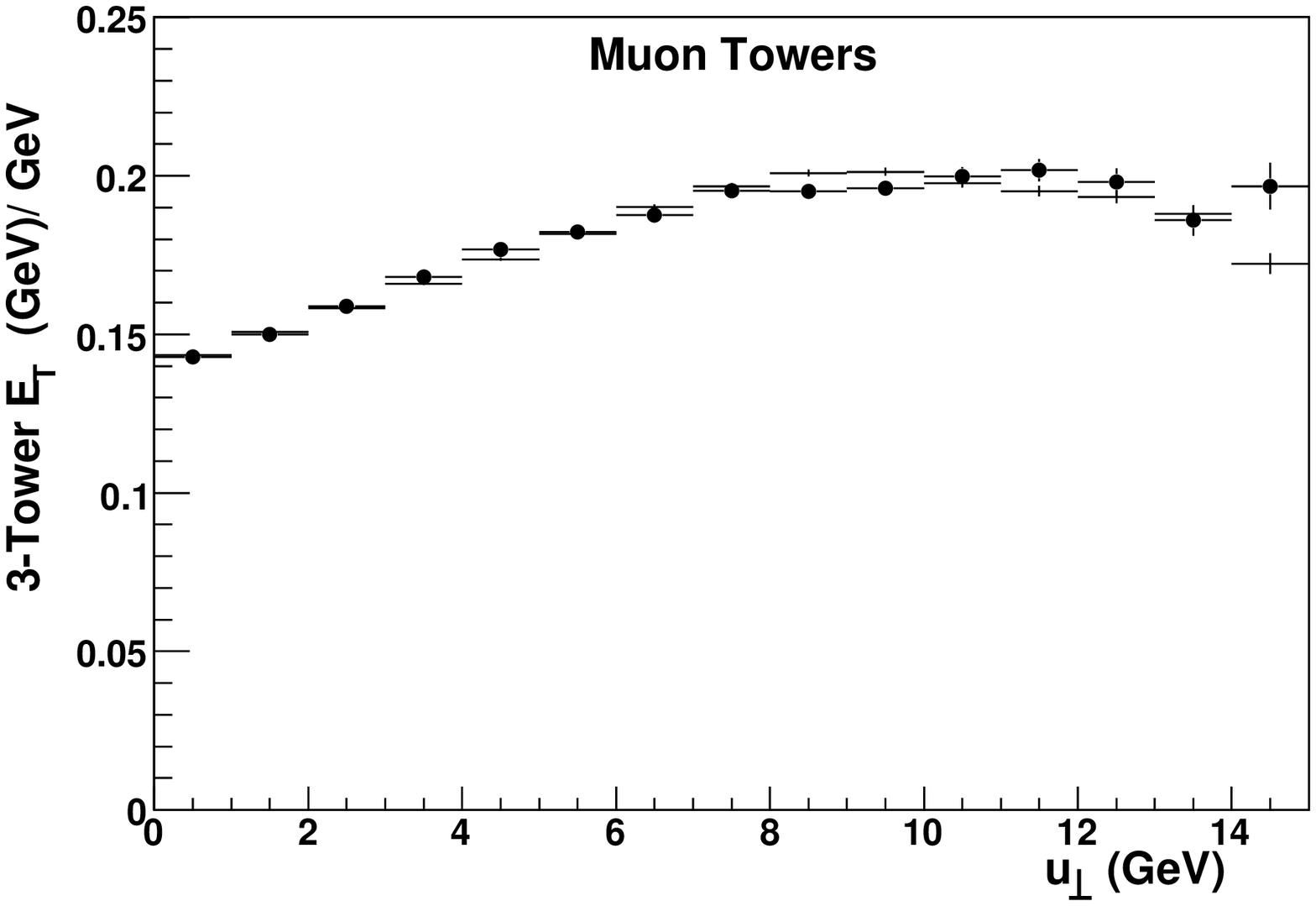}
\includegraphics*[width=8.5cm]{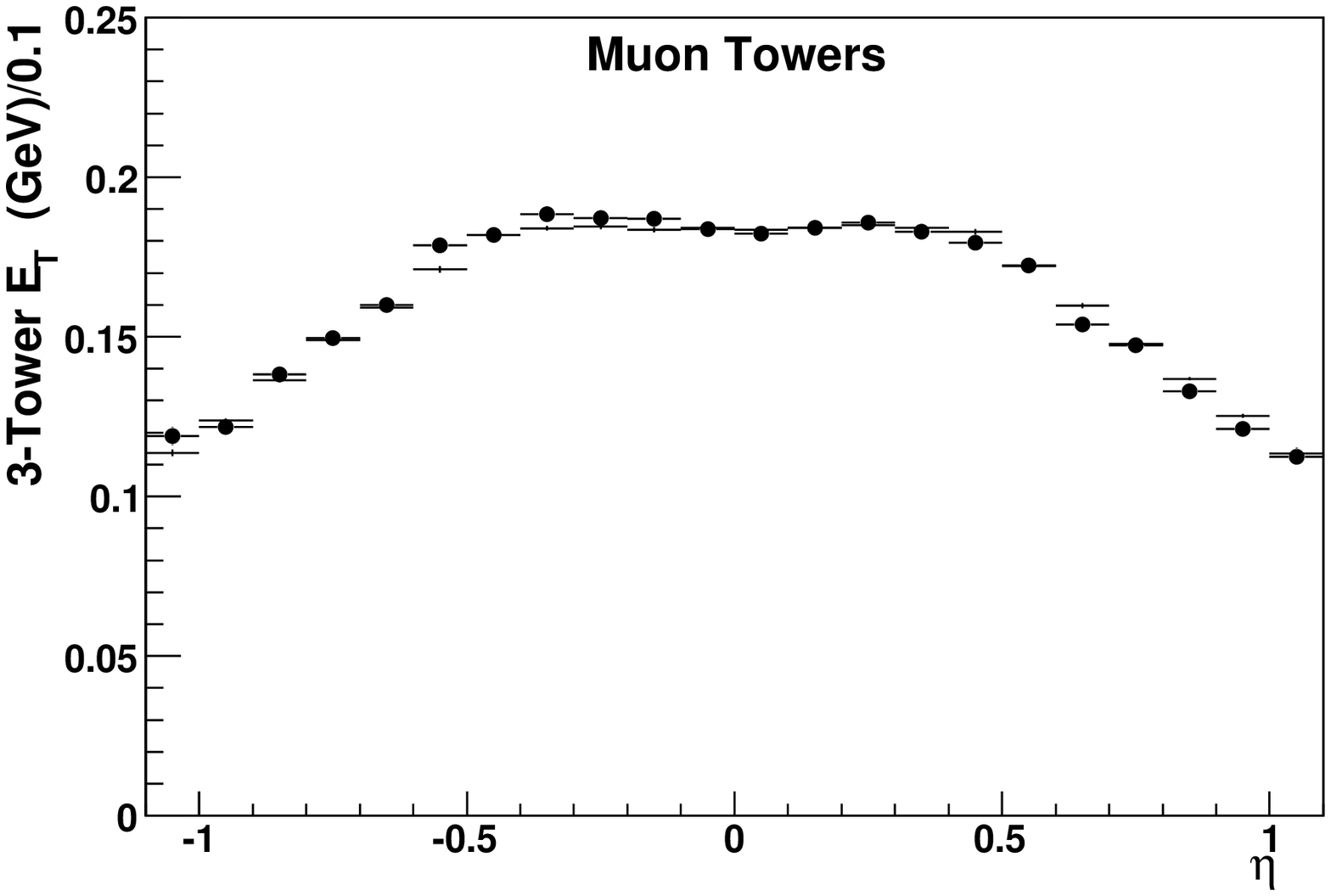}
\caption{Variation of underlying event $E_T$ in the three-tower region rotated by 
$90^\circ$ in $\phi$ from the muon as a function of $u_{||}$ (top), $|u_{\perp}|$ 
(middle), and $\eta$ (bottom) for $W\to \mu\nu$ data (circles) and simulation 
(points). }
\label{fig:muscaling}
\end{center}
\end{figure}

\begin{figure}
\begin{center}
\includegraphics*[width=8.5cm]{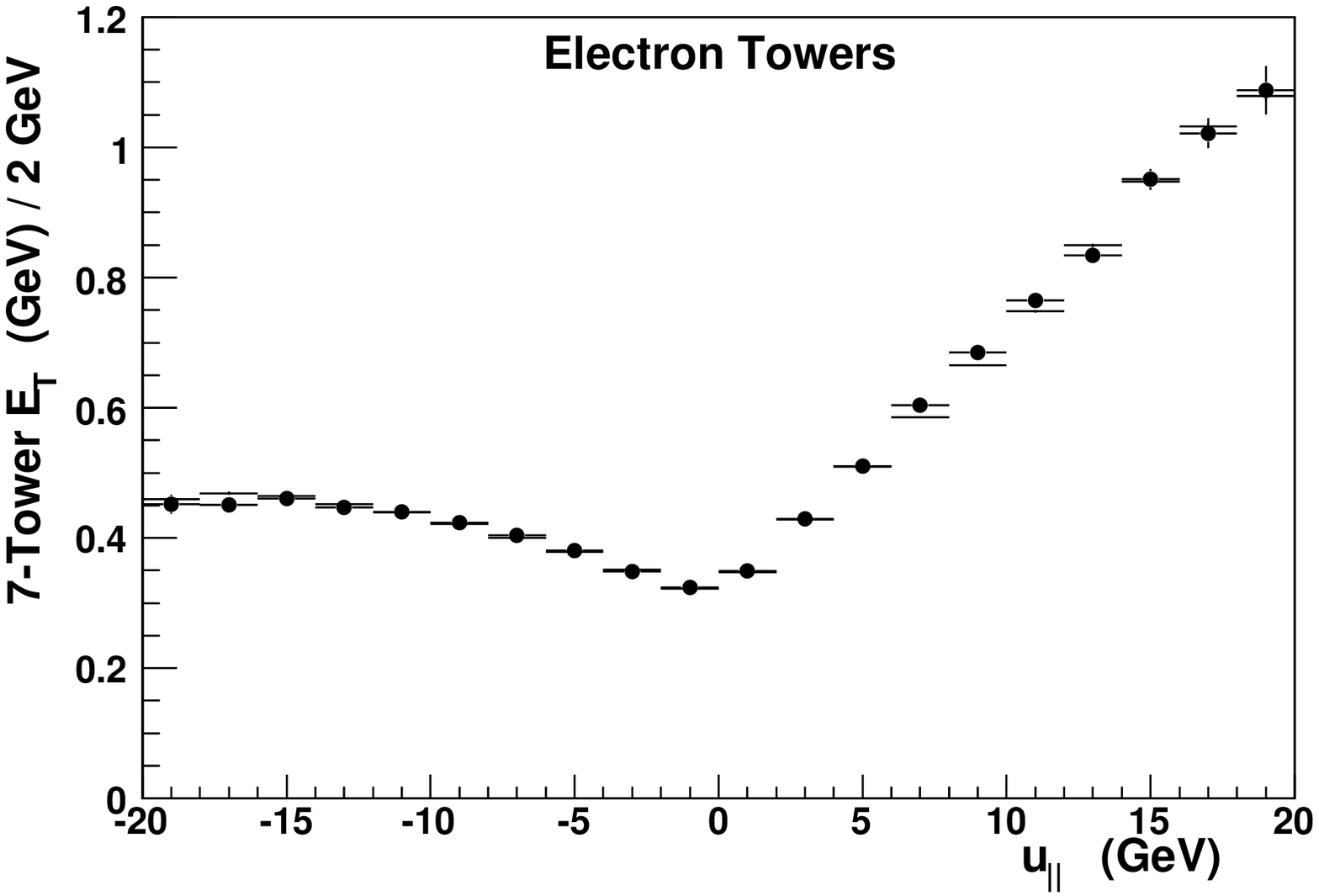}
\includegraphics*[width=8.5cm]{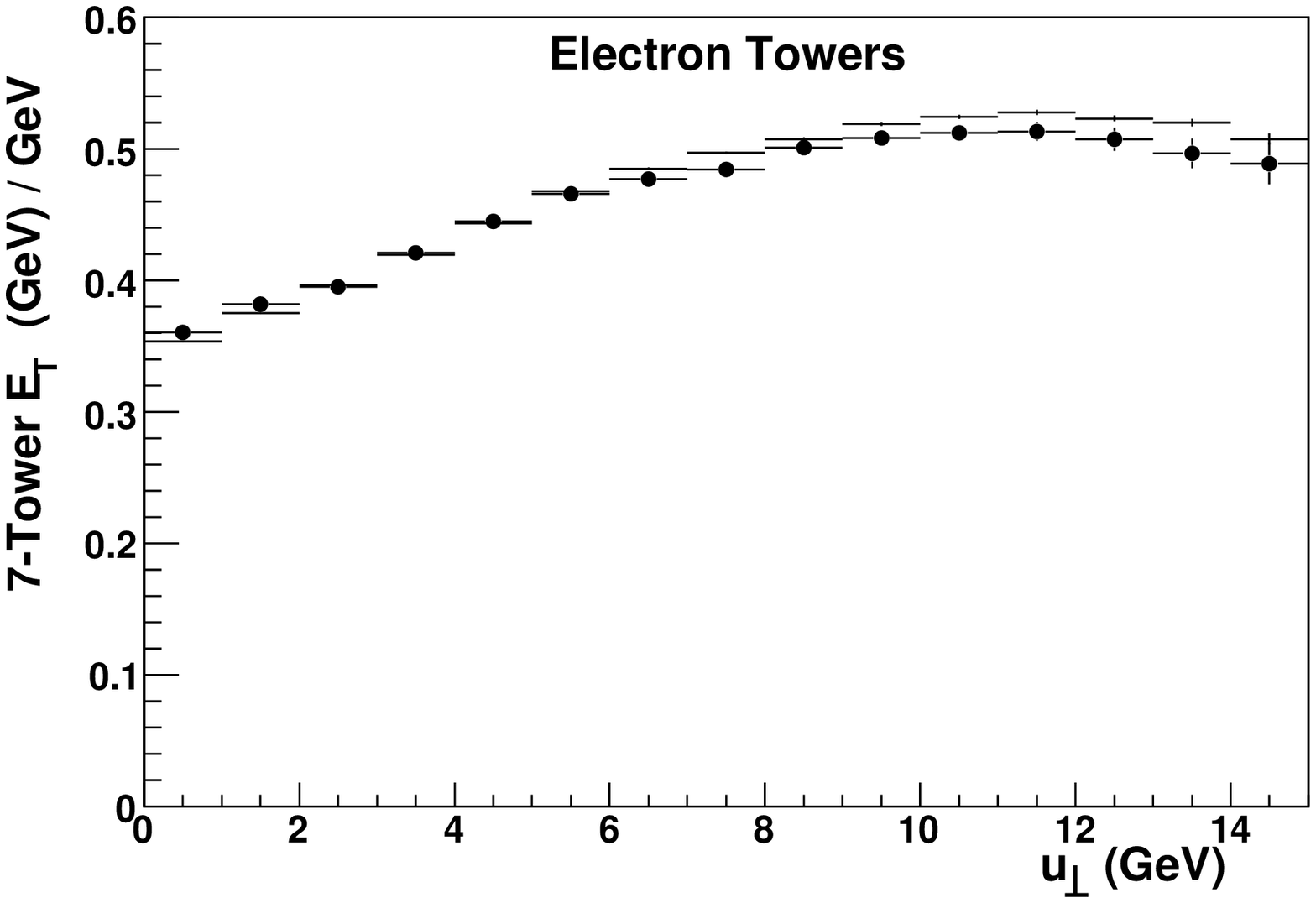}
\includegraphics*[width=8.5cm]{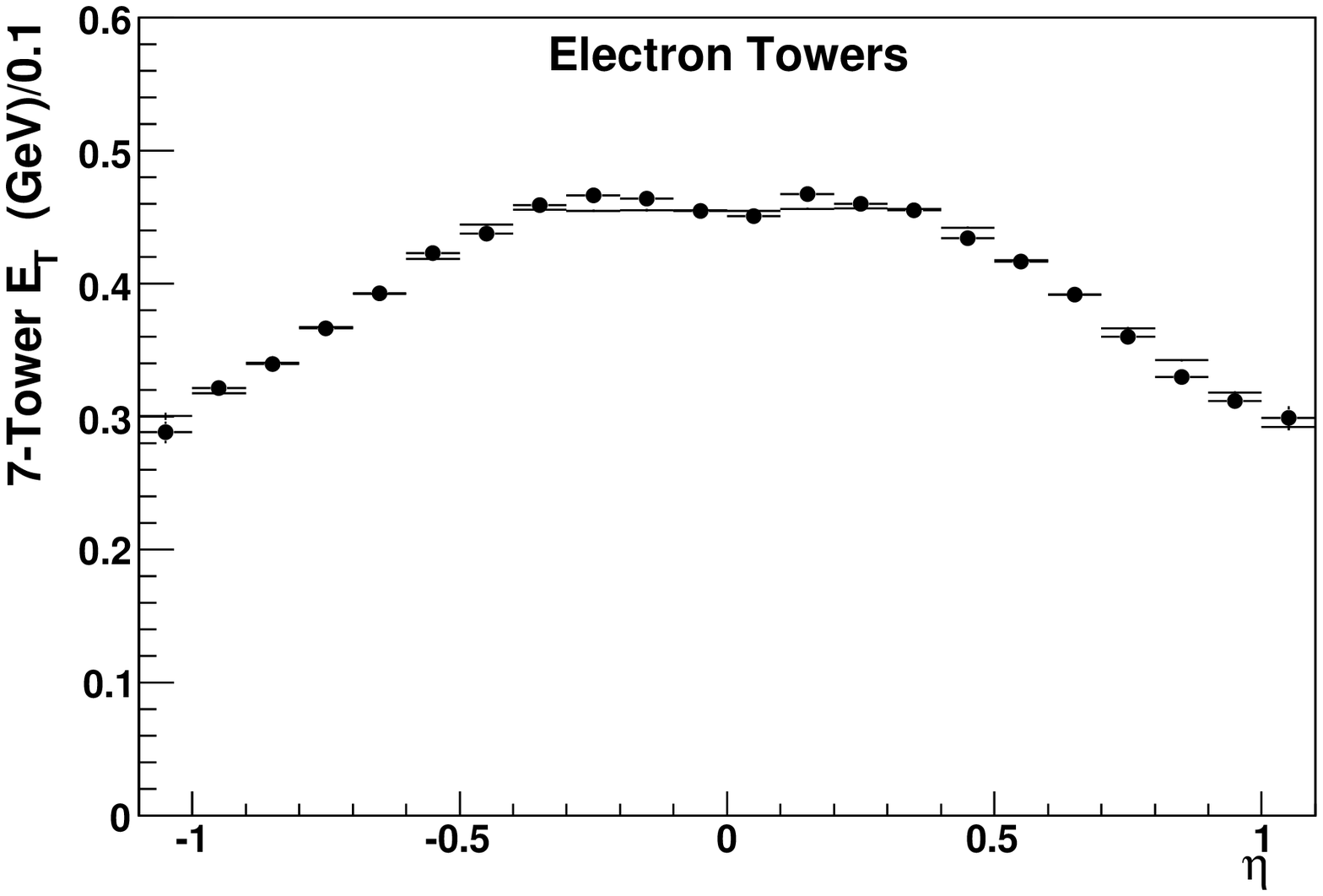}
\caption{Variation of underlying event $E_T$ in the seven-tower region rotated by 
$90^\circ$ in $\phi$ from the electron as a function of $u_{||}$ (top), $|u_{\perp}|$ 
(middle), and $\eta$ (bottom) for $W\to e\nu$ data (circles) and simulation (points).}
\label{fig:elescaling}
\end{center}
\end{figure}

In the simulation, we sample from the underlying event distribution measured in the 
rotated towers of all $W$-boson candidate events in the appropriate decay channel.  We 
scale the extracted energy to account for the measured dependence on $u_{||}$, $|u_\perp|$, 
and $|\eta|$ (Figs.~\ref{fig:muscaling} and \ref{fig:elescaling}).  The procedure is 
validated by applying the removal to a window rotated by $180^\circ$ from the lepton 
and comparing to data.  The mean underlying event energy in this region is modeled to 
an accuracy of 1 MeV (2 MeV) in the muon (electron) channel.  We take this as an 
estimate of the systematic uncertainty on the choice of rotation angle, and combine it 
with a parametrization uncertainty of 2 MeV for the electron channel and a selection 
bias uncertainty of 1 MeV for the muon channel.  The total systematic uncertainty on 
$M_W$ due to lepton-removal modeling in the muon (electron) channel is 2 MeV (3 MeV), 
2 MeV (3 MeV), and 4 MeV (6 MeV) for the $m_T$, $p_T^\ell$, $p_T^\nu$ fits, 
respectively.

\subsection{Model parametrization}
\begin{figure}[!tp]
\begin{center}
\epsfysize = 3.cm
\epsffile{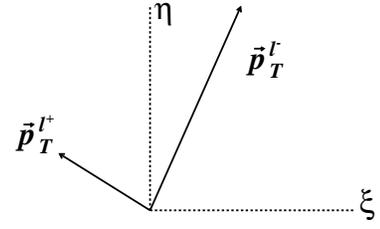}
\caption{Illustration of the $\eta$ and $\xi$ axes in $Z$ boson events. }
\label{fig:axes}
\end{center}
\end{figure}

The recoil response model consists of a parametrization of three major sources:  QCD 
and QED radiation in the parton-parton interaction producing a $W$ or $Z$ boson, 
radiation from the underlying event, and any additional $p\bar{p}$ collisions in the 
bunch crossing.  The parameters are tuned using $Z \to \ell\ell$ events, since the 
lepton-pair $p_T$ is accurately measured and the balance between $p_T(Z \to \ell \ell)$ 
and $u_T$ probes the detector response and resolution.  We define the axis parallel to 
$p_T(Z \to \ell\ell)$ as the ``$\eta$'' axis (Fig.~\ref{fig:axes})~\cite{UA2etaxi}, 
and the orthogonal axis in the transverse plane as the ``$\xi$'' axis.

\subsubsection{Recoil response}
\begin{figure}[!tp]
\begin{center}
\epsfysize = 6.cm
\includegraphics*[width=8.5cm]{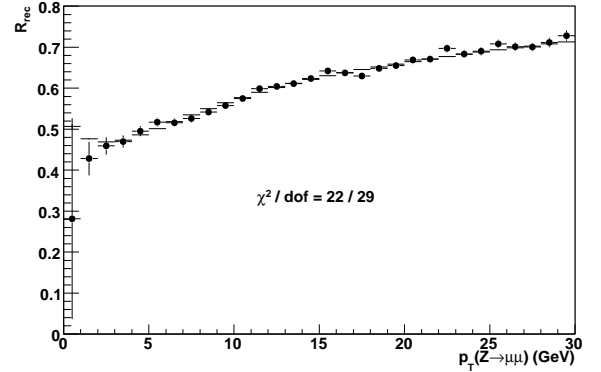}
\caption{Mean value of $R_{\textrm{rec}} \equiv -\vec{u}_T\hat{p}_T^{~\mu\mu}/p_T^{\mu\mu}$, 
which approximates the recoil response $R$, as a function of dimuon $p_T$.  The distribution 
motivates the logarithmic parametrization of the response.  The simulation (lines) models the 
data (circles) accurately. }
\label{fig:recoilscale}
\end{center}
\end{figure}

The response of the calorimeter to the radiation produced in the recoil of a $W$ or $Z$ boson 
is defined as the ratio of measured recoil to true recoil, projected along the direction of 
the true recoil ($R\equiv \vec{u}_T\cdot\hat{u}^\mathrm{true}/u_T^\mathrm{true}$, where 
$\vec{u}_T^\mathrm{true} = -\vec{p_T}^{W,Z}$ is the net $\vec{p}_T$ of the initial-state 
radiation).  The simulation parametrizes the response function as
\begin{equation}	
R = a \log (c u_T^\mathrm{true} + b)/\log(c \cdot 15~{\rm GeV} + b),
\label{eq:recoilresponse}
\end{equation}

\noindent
where $u_T^\mathrm{true}$ is expressed in units of GeV, and $a$, $b$, and $c$ are positive 
constants determined from $Z$-boson data.  This functional form is empirically motivated by an approximation 
to $R$ measured in $Z$-boson data, $-\vec{u}_T\cdot\hat{p}_T^{\ell\ell}/p_T^{\ell\ell}$ 
(Fig.~\ref{fig:recoilscale}).

\begin{figure}[!tp]
\begin{center}
\epsfysize = 6.cm
\includegraphics*[width=8.5cm]{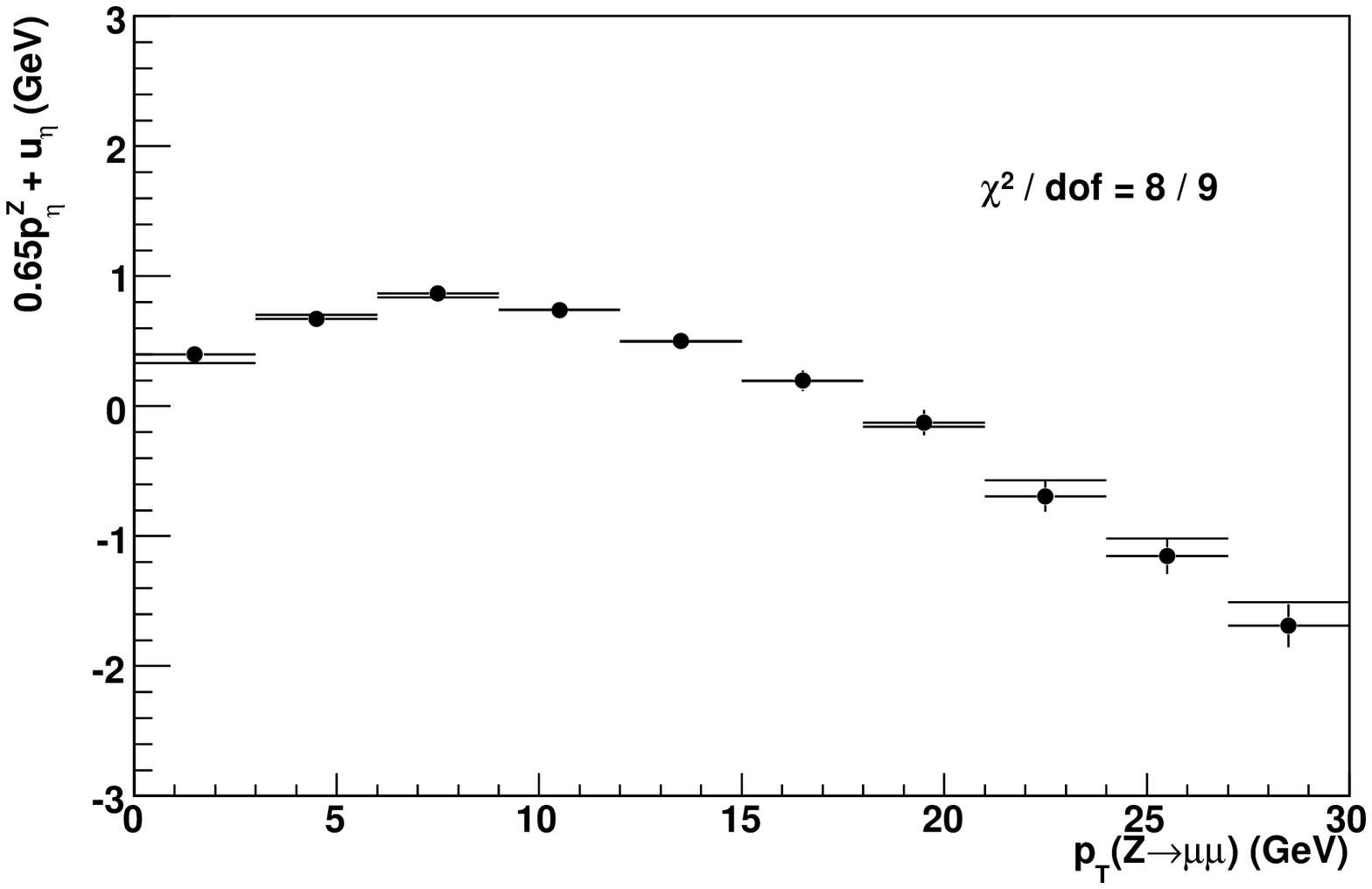}
\includegraphics*[width=8.5cm]{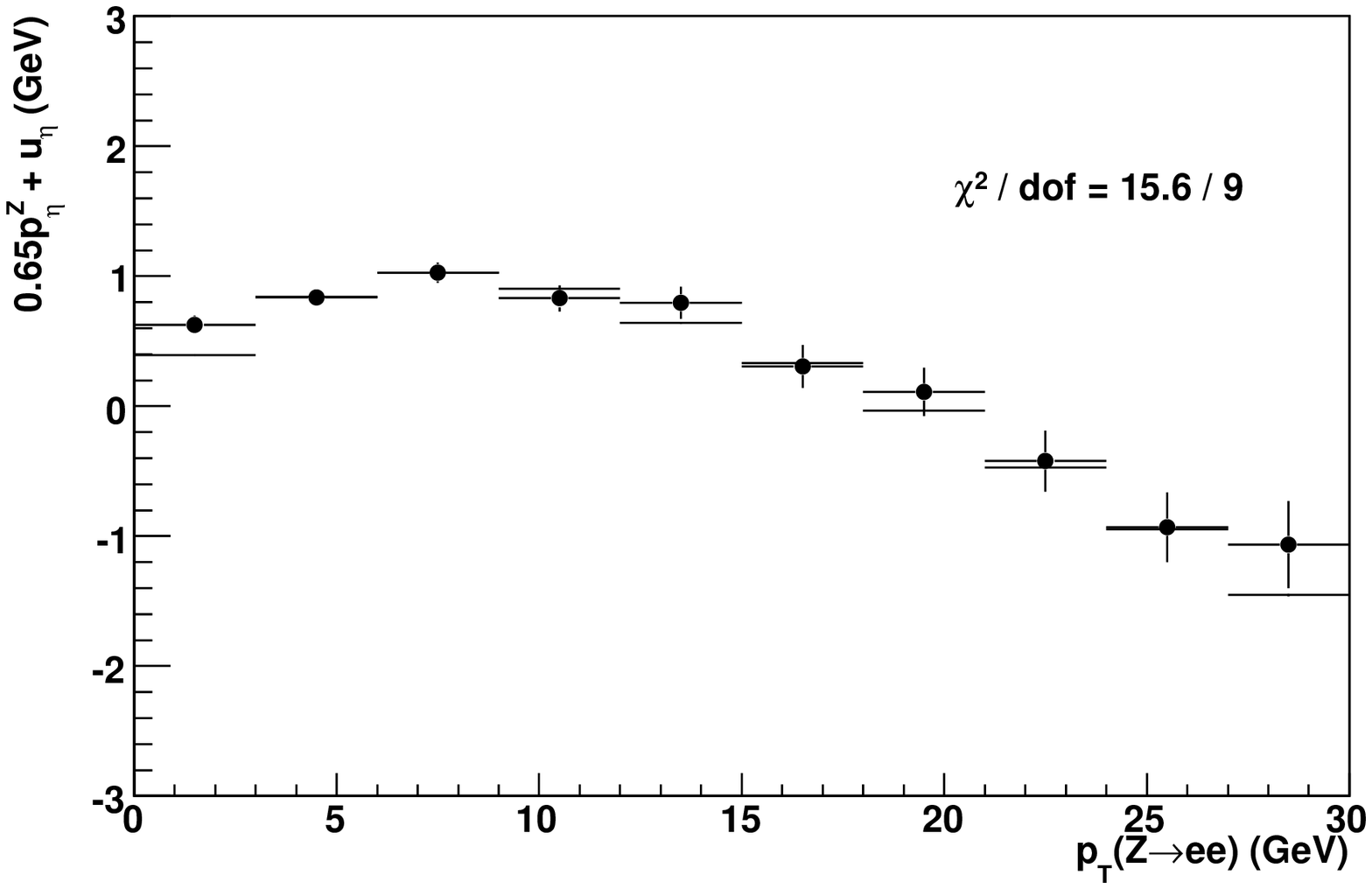}
\caption{Distribution of $0.65 p^{\ell\ell}_{\eta} + u_{\eta}$ for $Z$-boson decays to muons (top) and electrons (bottom) as a function of $Z$-boson $p_T$ in simulated (lines) and experimental (circles) data. The detector response parameters are 
obtained by minimizing the combined $\chi^2$ of these distributions. }
\label{fig:recoilscalefit}
\end{center}
\end{figure}

The parameters in Eq.~(\ref{eq:recoilresponse}) are determined using the balance between the 
recoil and dilepton momenta, $p_\eta^{\ell\ell}+u_\eta$, which is well defined when the boson 
is produced at rest.  In the case of perfect response this sum would be zero; in practice the 
calorimeter response to the recoil is about 65\% for the relevant $p_T$ range in this analysis.  
Figure~\ref{fig:recoilscalefit} shows $0.65 p_\eta^{\ell\ell}+u_\eta$ for the following best-fit 
values of $a$, $b$, and $c$:

\begin{equation}
\begin{split}
a &= 0.645 \pm 0.002\\
b &= 8.2 \pm 2.2\\
c &= 5.1 \pm 0.6~{\rm GeV}^{-1}, 
\end{split}
\end{equation}

\noindent
where the central values are obtained from minimizing the combined $\chi^2$ for electron and muon 
distributions, and the uncertainties are statistical.  These values are used to model the recoil 
response in simulated $W$ and $Z$ boson events. 

\subsubsection{Recoil resolution}
\label{sec:recoilresolution}
\begin{figure*}
\begin{center}
\includegraphics*[width=8.5cm]{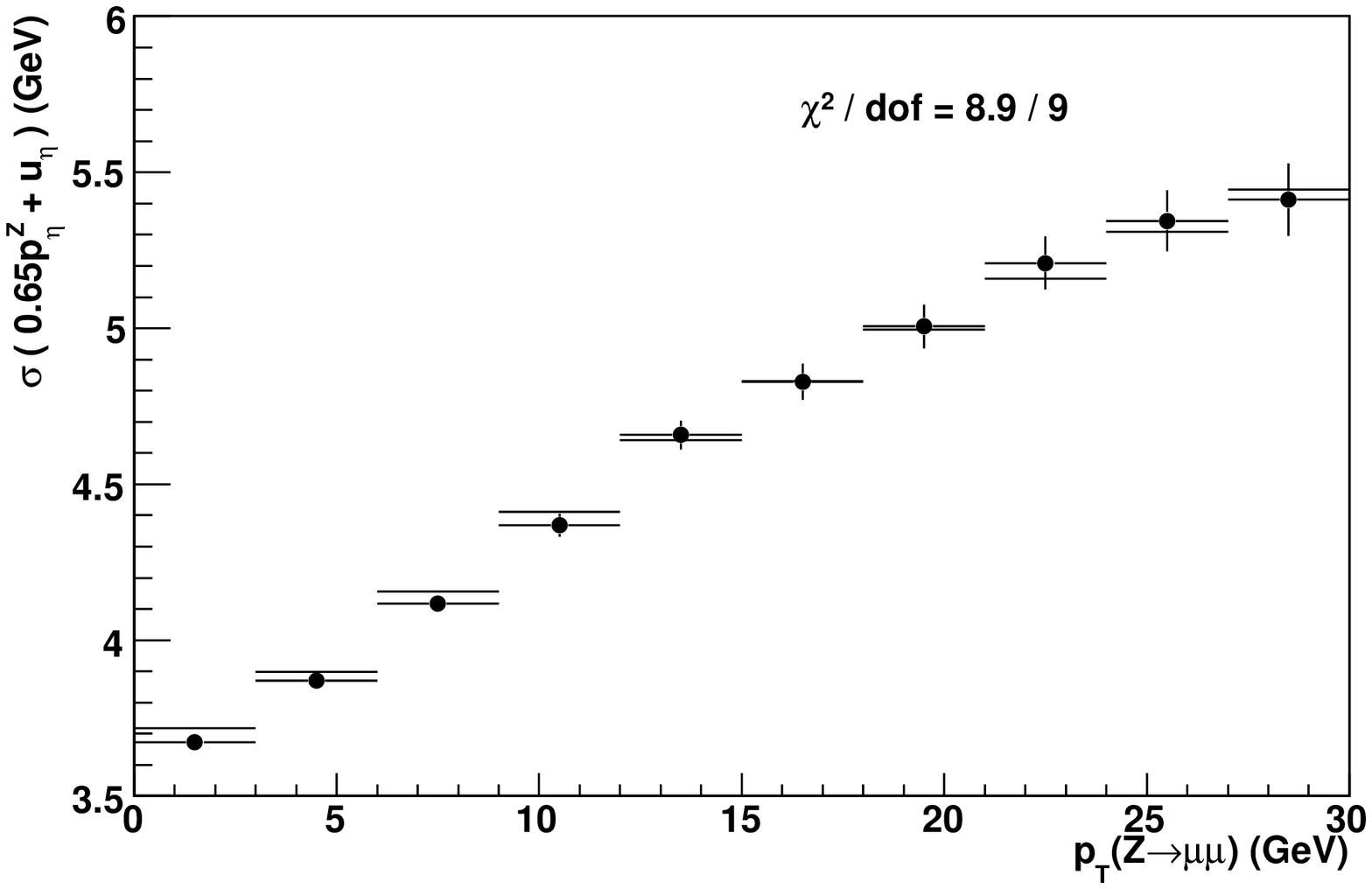}
\includegraphics*[width=8.5cm]{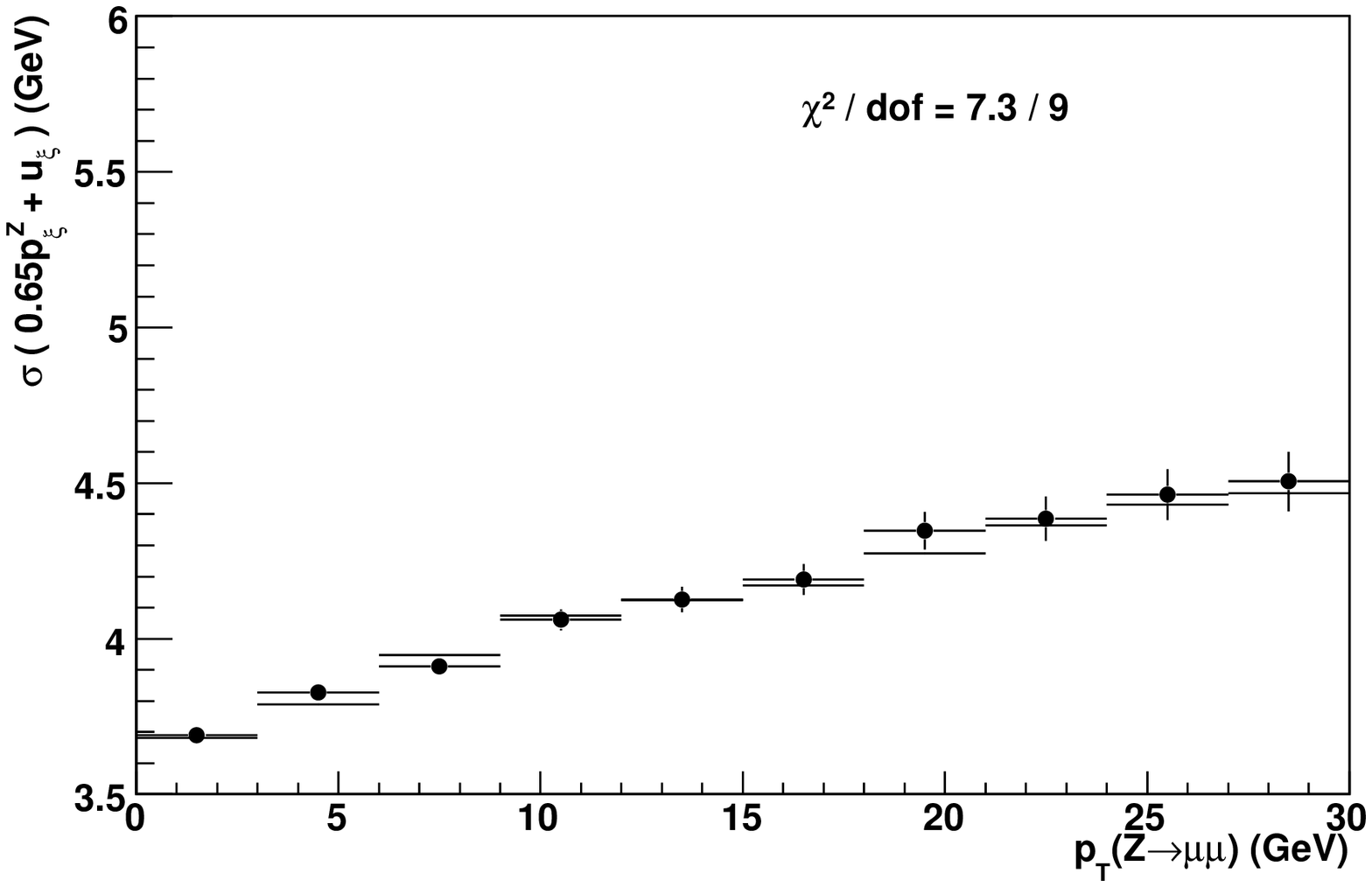}
\includegraphics*[width=8.5cm]{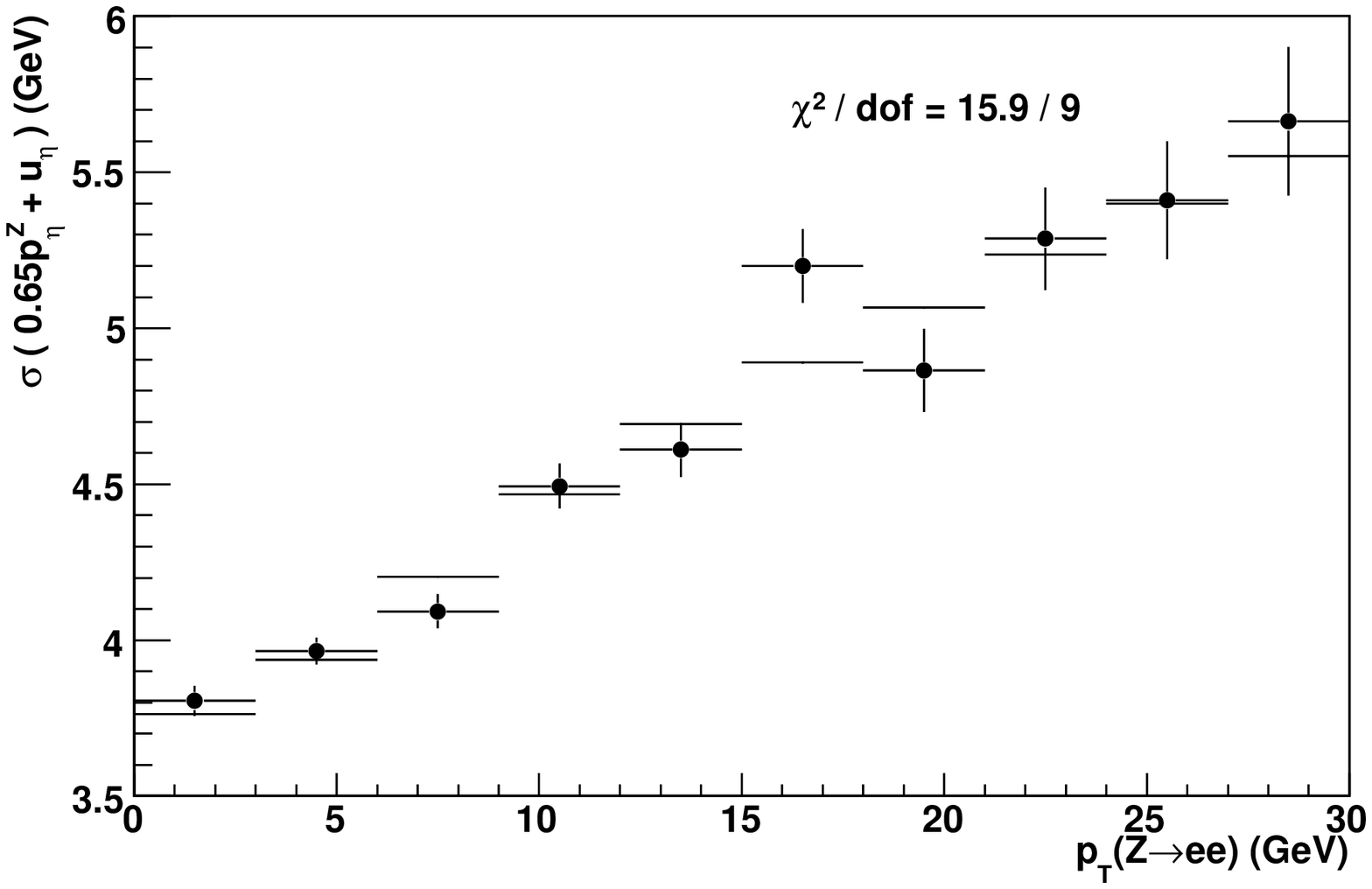}
\includegraphics*[width=8.5cm]{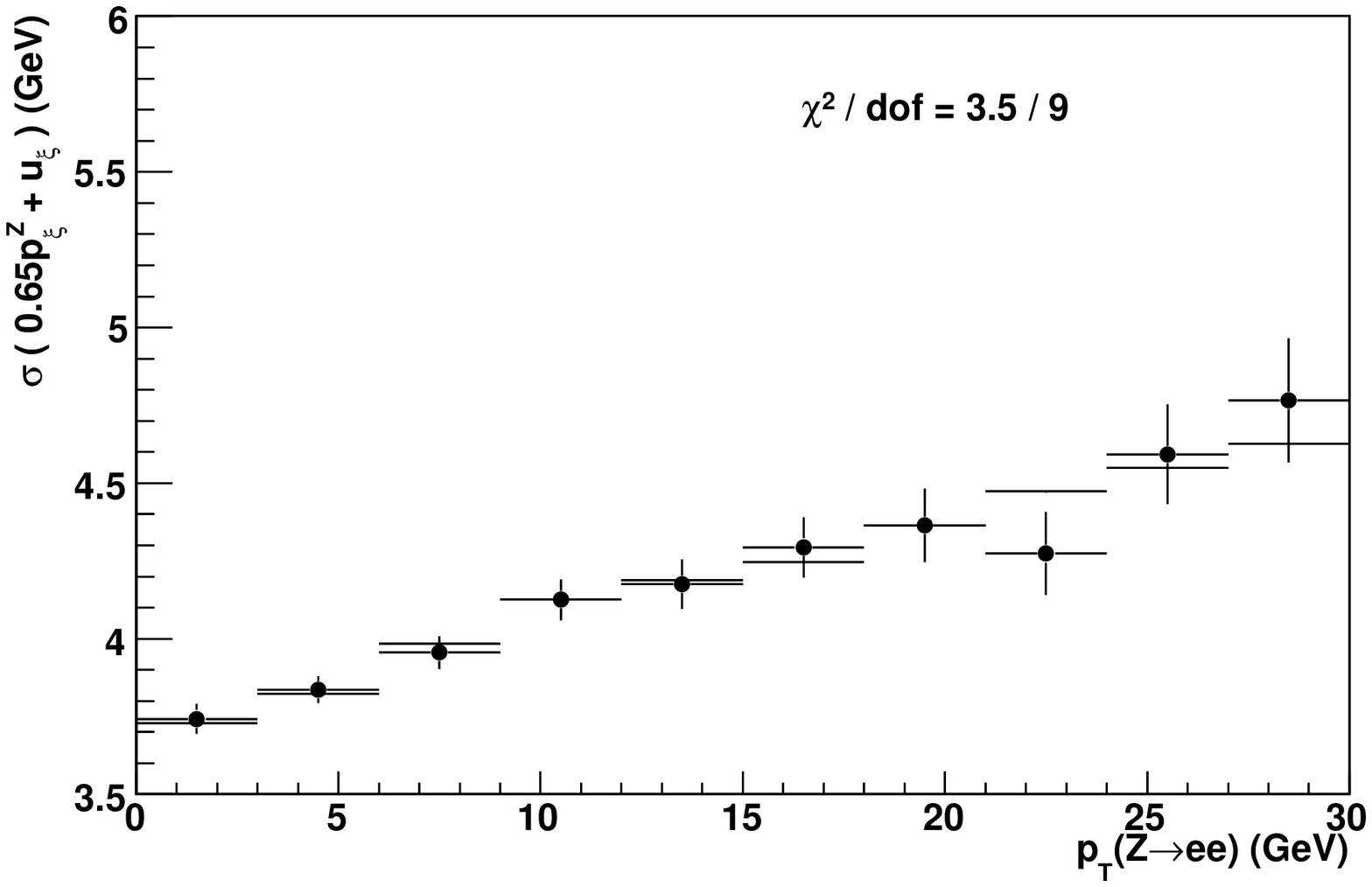}
\caption{Resolution on $0.65 p^{\ell\ell}_{\eta} + u_{\eta}$ (left) and
$0.65 p^{\ell\ell}_{\xi} + u_{\xi}$ (right) in simulated (lines) 
and data (circles) $Z$-boson decays to muons (top) and electrons (bottom).  The sum of 
the $\chi^2$ values in the $\xi$ or $\eta$ direction is minimized in the fit 
for the jet angular resolution parameters or the recoil resolution parameters
 ($N_{W,Z}$ and $s_\mathrm{had}$), respectively.  }
\label{fig:recoilres}
\end{center}
\end{figure*}

We parametrize the resolution on the recoil magnitude and direction using 
$Z\to \ell\ell$ data.  The dominant effect is the sampling resolution
\begin{equation}
\sigma(u_T) = s_\mathrm{had}\sqrt{u_T^\mathrm{true}},
\end{equation}

\noindent
where $s_\mathrm{had}$ is the calorimeter sampling constant.  The rms 
resolution on the sum $0.65 p_\eta^{\ell\ell}+u_\eta$ is used to fit for 
$s_\mathrm{had}$ in $Z\to\ell\ell$ data (Fig.~\ref{fig:recoilres}, top), giving 
\begin{equation}
s_\mathrm{had} = 0.820 \pm 0.009_{\rm stat} {\rm ~GeV}^{1/2}.
\end{equation}

\begin{figure}[!tp]
\begin{center}
\epsfysize = 6.cm
\includegraphics*[width=8.5cm]{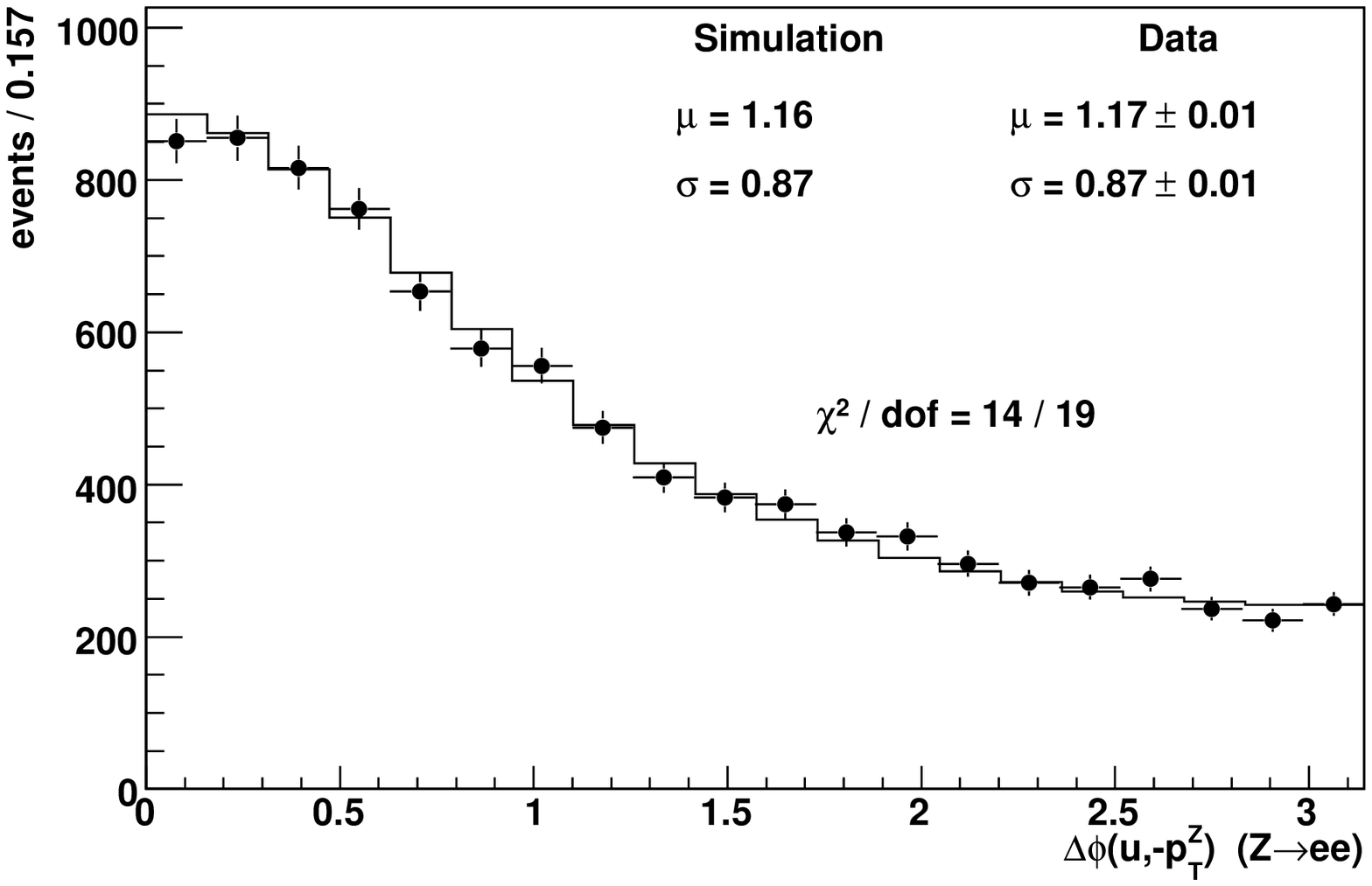}
\includegraphics*[width=8.5cm]{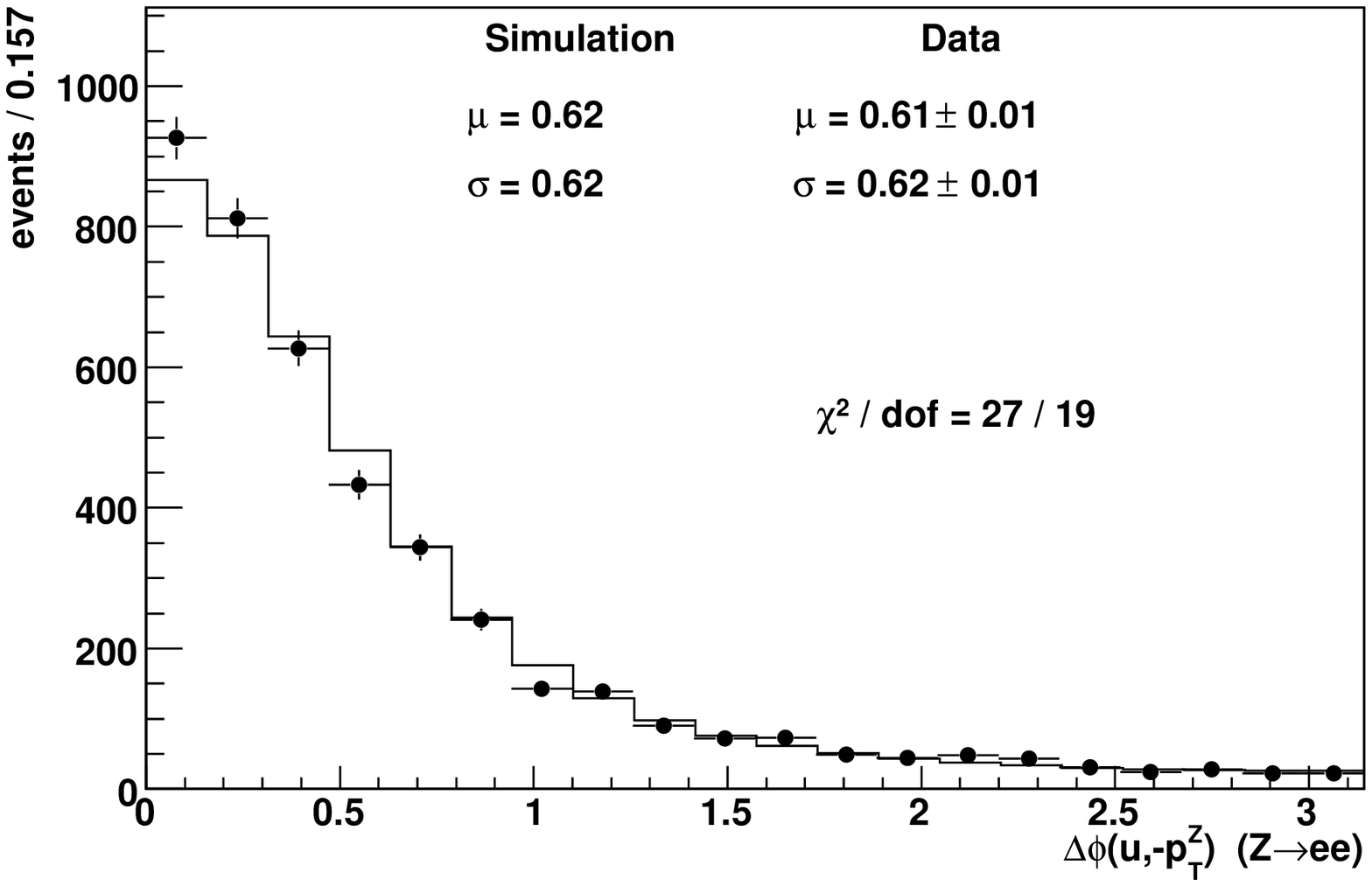}
\includegraphics*[width=8.5cm]{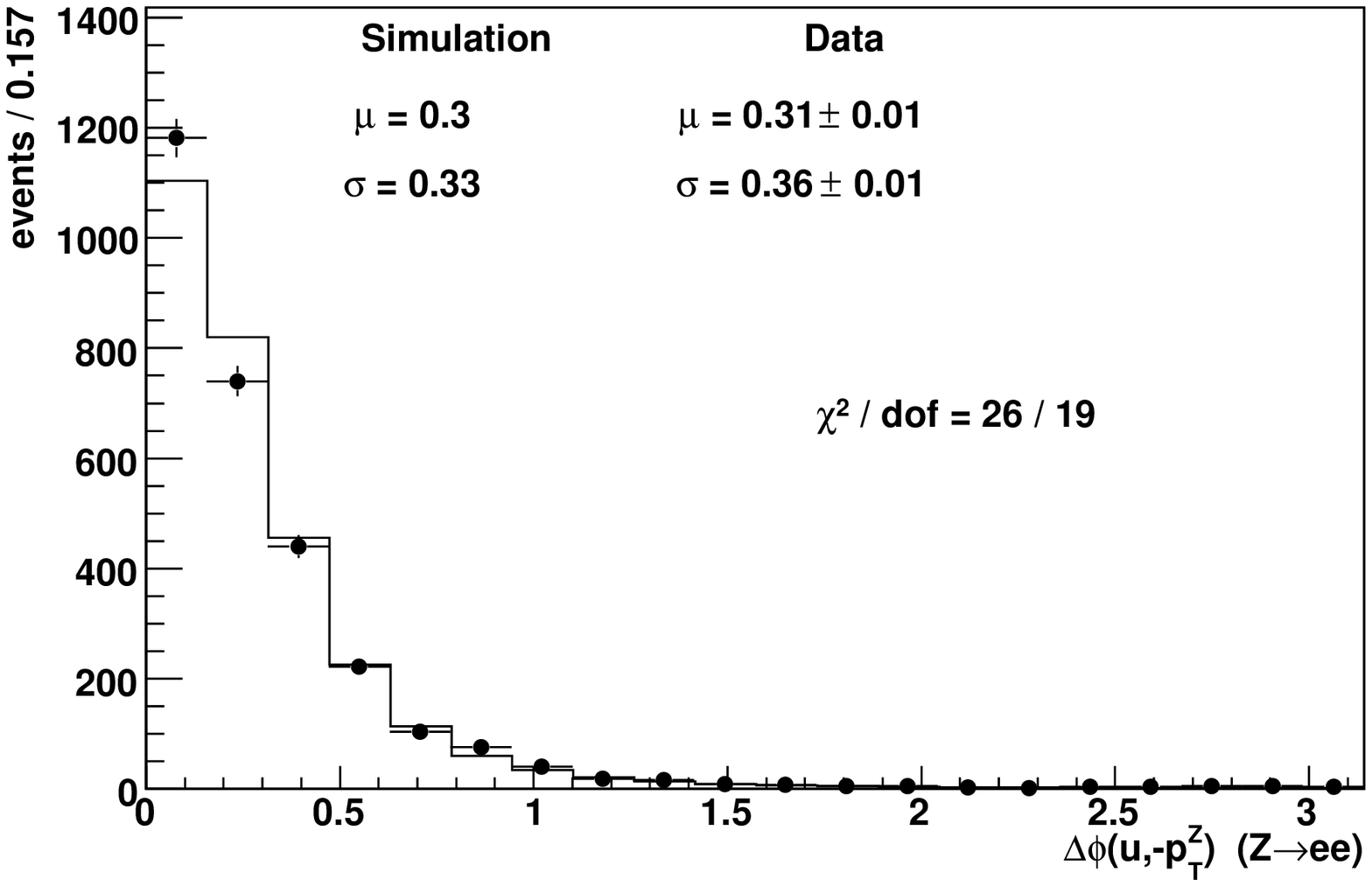}
\caption{Distributions of the angular separation between the
reconstructed recoil vector $\vec{u}_T$ and the dilepton
$-\vec{p}_T^{\ell \ell}$ vector in $Z \rightarrow ee$ decays, for simulation (lines) 
and data (circles) respectively. The distributions are shown for the
ranges $p_T^{\ell \ell} < 8$~GeV (top),  
$8 < p_T^{\ell \ell} < 15$~GeV (middle), 
and $p_T^{\ell \ell} > 15$~GeV (bottom).   }
\label{fig:zeudphipt}
\end{center}
\end{figure}

The angular resolution $\sigma(u_\phi)$ is modeled as a flat distribution with 
an rms parametrized as a continuous, piecewise-linear function separated into 
the ranges $0 < u_T^\mathrm{true} < 15$~GeV, $15<u_T^\mathrm{true}<30$~GeV and 
$u_T^\mathrm{true} > 30$~GeV. This monotonically-improving resolution
with increasing $u_T^\mathrm{true}$ is motivated by inspecting the
angular separation between $\vec{u}_T$ and $-\vec{p}_T^{\ell \ell}$ in
$Z \to \ell \ell$ decays. As illustrated in Fig.~\ref{fig:zeudphipt},
the distribution of this angular separation, which is sensitive to
$\sigma(u_\phi)$, narrows with increasing $p_T^{\ell \ell}$.  The
parameters defining the $\sigma(u_\phi)$ function are its 
values at $u^{\rm true}_T = 9.4$~GeV, 15~GeV, and 24.5~GeV, respectively, 
chosen so that the parameter uncertainties are uncorrelated.  We refer to 
these parameters as $\alpha$, $\beta$, and $\gamma$ respectively, such that  
\begin{align} 
      \sigma(u_\phi) - \alpha &~~ \propto  9.4 - u^{\rm true}_T/{\rm GeV} & 
   (u^{\rm true}_T < 15~{\rm GeV}) \nonumber \\
\sigma(u_\phi) &~~ =  \beta ~  & (u^{\rm true}_T = 15~ {\rm GeV}) \nonumber \\
	\sigma (u_\phi) - \gamma &~~ \propto   24.5 - u^{\rm true}_T/{\rm GeV} ~ & 
   (15 < u^{\rm true}_T < 30~{\rm GeV}) \nonumber \\
	\sigma (u_\phi) &~~ =  {\rm constant}~  & (u^{\rm true}_T > 30~{\rm GeV}),
\label{jetAngleSmearing}
         \end{align}

\noindent
where the unspecified coefficients are fixed by continuity.  The parameters 
are tuned on the rms resolution of $0.65 p_\xi^{\ell\ell}+u_\xi$ 
(Fig.~\ref{fig:recoilres}, bottom), since the $\xi$-projection is much more sensitive 
to recoil angular fluctuations than to energy fluctuations.  The fit using 
$Z\to\ell\ell$ data yields
\begin{eqnarray}
\alpha & = & 0.306 \pm 0.006_{\rm stat}, \nonumber \\
\beta & = & 0.190 \pm 0.005_{\rm stat}, \nonumber \\
\gamma & = & 0.144 \pm 0.004_{\rm stat} \; .  
\end{eqnarray}


\subsubsection{Spectator and additional $p\bar{p}$ interactions}
\label{sec:spectator}
The resolution on the measured recoil receives contributions from energy produced 
by spectator partons and additional interactions~\cite{CDF2}.  We propagate these 
effects as a function of $\sum E_T$, the scalar sum of transverse energies in the 
calorimeter towers, in our simulation.  For each simulated event, $\sum E_T$ is 
evaluated by adding two contributions sampled from distributions separately 
representing multiple interactions and spectator interactions accompanying the 
boson production.  Given the $\sum E_T$ in an event, a corresponding contribution 
to the recoil ($\Delta u$) is calculated. 

The $\sum E_T$ distributions are obtained from zero-bias and minimum-bias collision 
data for multiple interactions and spectator interactions, respectively.  The 
zero-bias data are weighted to have an instantaneous luminosity profile consistent 
with that of the $W$ and $Z$ boson data.  The minimum-bias distribution is scaled 
by a parameter $N_V$ to account for differences in spectator interactions in $W$ 
and $Z$ boson production relative to minimum-bias production.  The parameter is 
measured using a combined $\chi^2$ fit for $N_V$ and $s_\mathrm{had}$ 
(Sec.~\ref{sec:recoilresolution}) using the rms resolution on the sum 
$0.65 p_\eta^{\ell\ell}+u_\eta$.  The fit yields the result 
\begin{equation} 
N_V = 1.079 \pm 0.012_{\rm stat}.
\end{equation}

\noindent 
The net contribution to the measured recoil is calculated from the total $\sum E_T$ 
in a simulated event as 
\begin{equation}
\Delta u_{x,y} = A_{x,y}+B_{x,y}\sum E_T\oplus \sigma_{x,y}\left(\sum E_T\right),
\end{equation}

\noindent
where the parameters $A_{x,y} = (-11, 23)$~MeV and $(B_x,B_y)=(0.00083, -0.00087)$ 
are obtained from a linear fit to the mean $\Delta u$ in minimum-bias data, and 
the resolution parameters are determined from power-law fits to the resolution on 
$\Delta u$ in minimum-bias data, 
\begin{equation}
\sigma_{x,y} = 0.3852\left(\sum E_T\right)^{0.5452} \textrm{ GeV}.
\end{equation}

\subsection{Model tests}
\label{sec:recoilcheck}

\begin{figure}[!tp]
\begin{center}
\epsfysize = 6.cm
\includegraphics*[width=8.5cm]{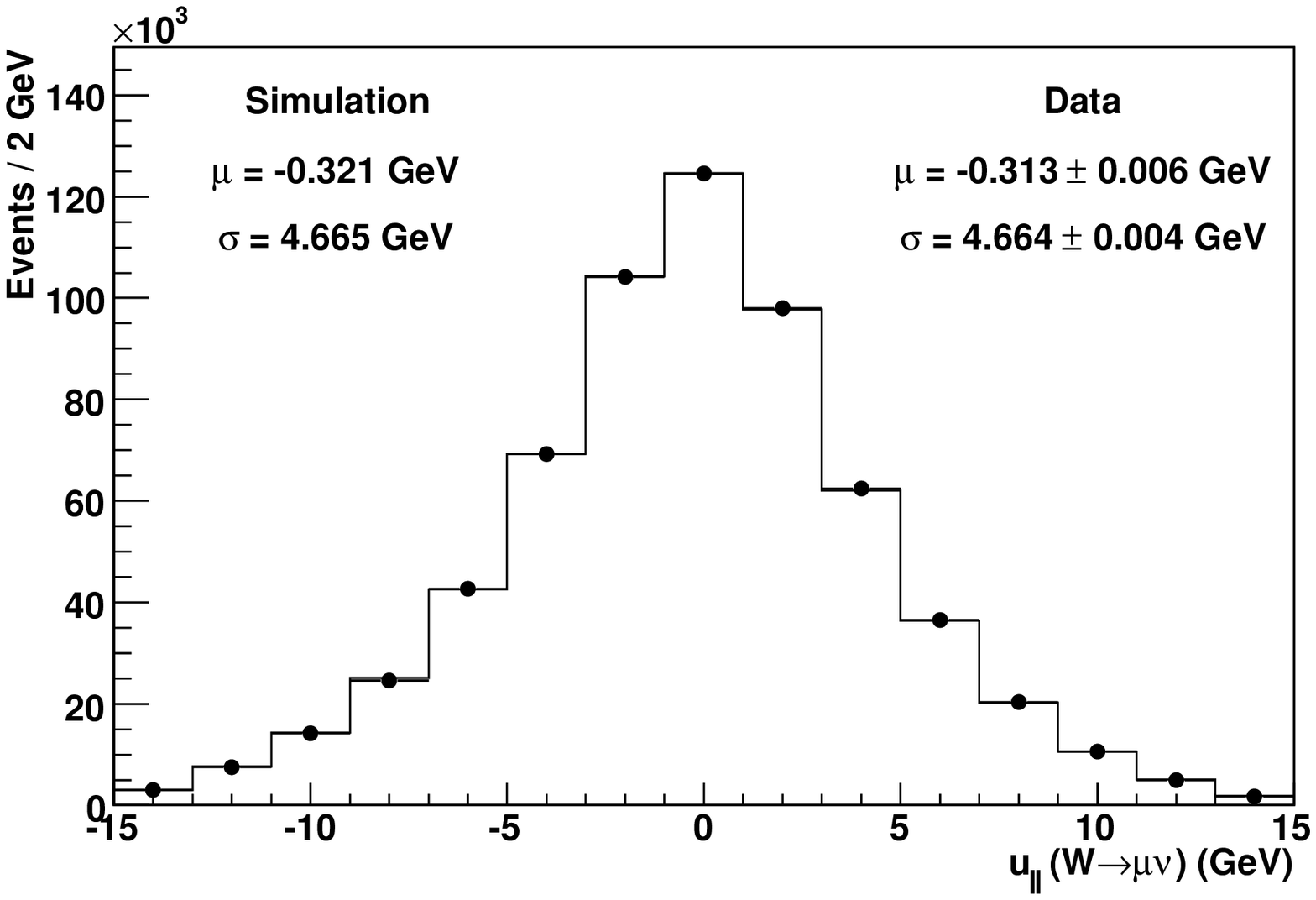}
\includegraphics*[width=8.5cm]{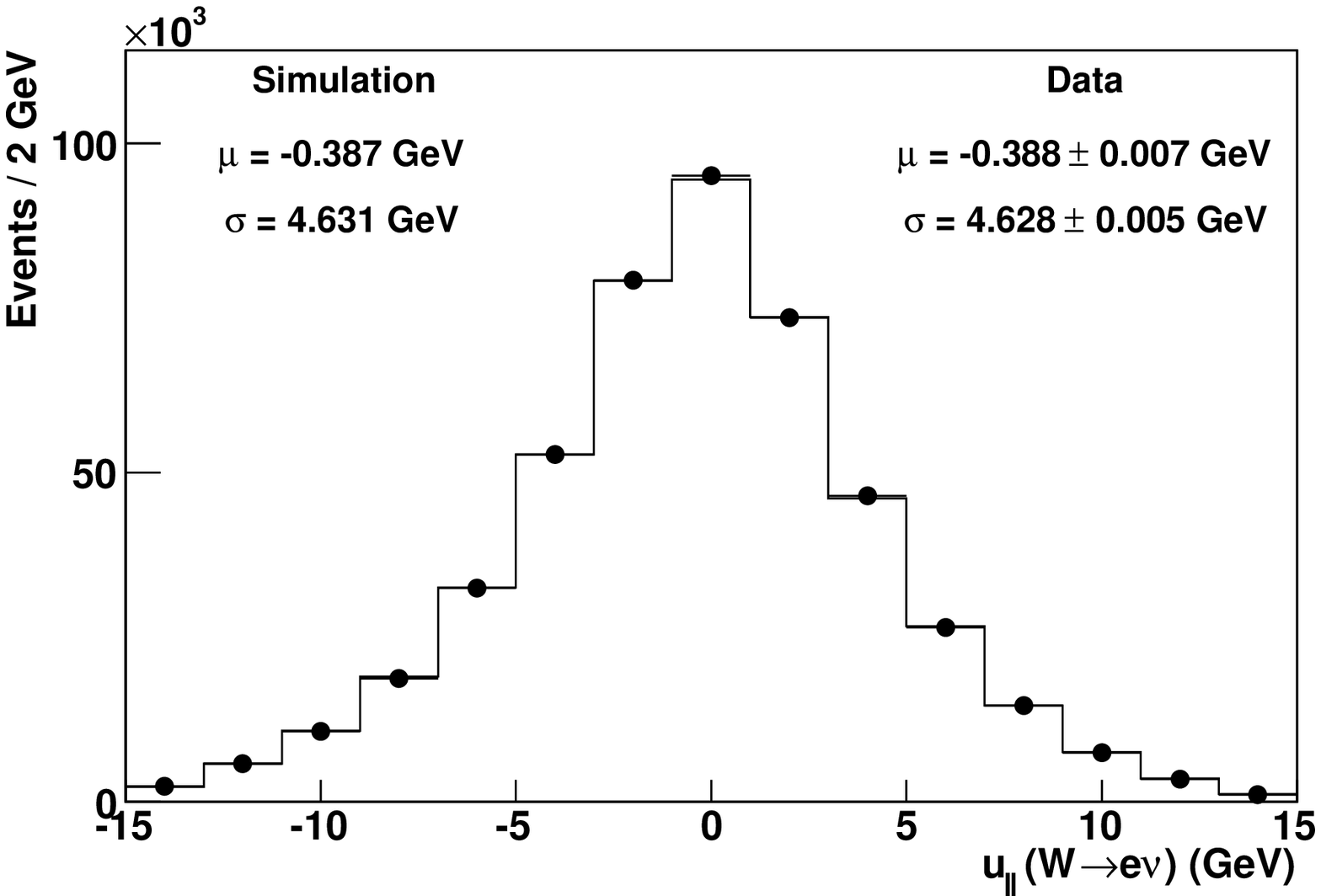}
\caption{Distributions of $u_{||}$ from simulation (histogram) and data (circles) 
 for $W$ boson decays to $\mu\nu$ (top) and 
$e\nu$ (bottom) final states.  The simulation uses parameters fit from 
$Z$ boson data, and the uncertainty on the simulation is 
due to the statistical uncertainty on these parameters.  The 
data mean ($\mu$) and rms spread ($\sigma$) are well-modeled by 
the simulation.}
\label{fig:upar}
\end{center}
\end{figure}

\begin{figure}[!htp]
\begin{center}
\epsfysize = 6.cm
\includegraphics*[width=8.5cm]{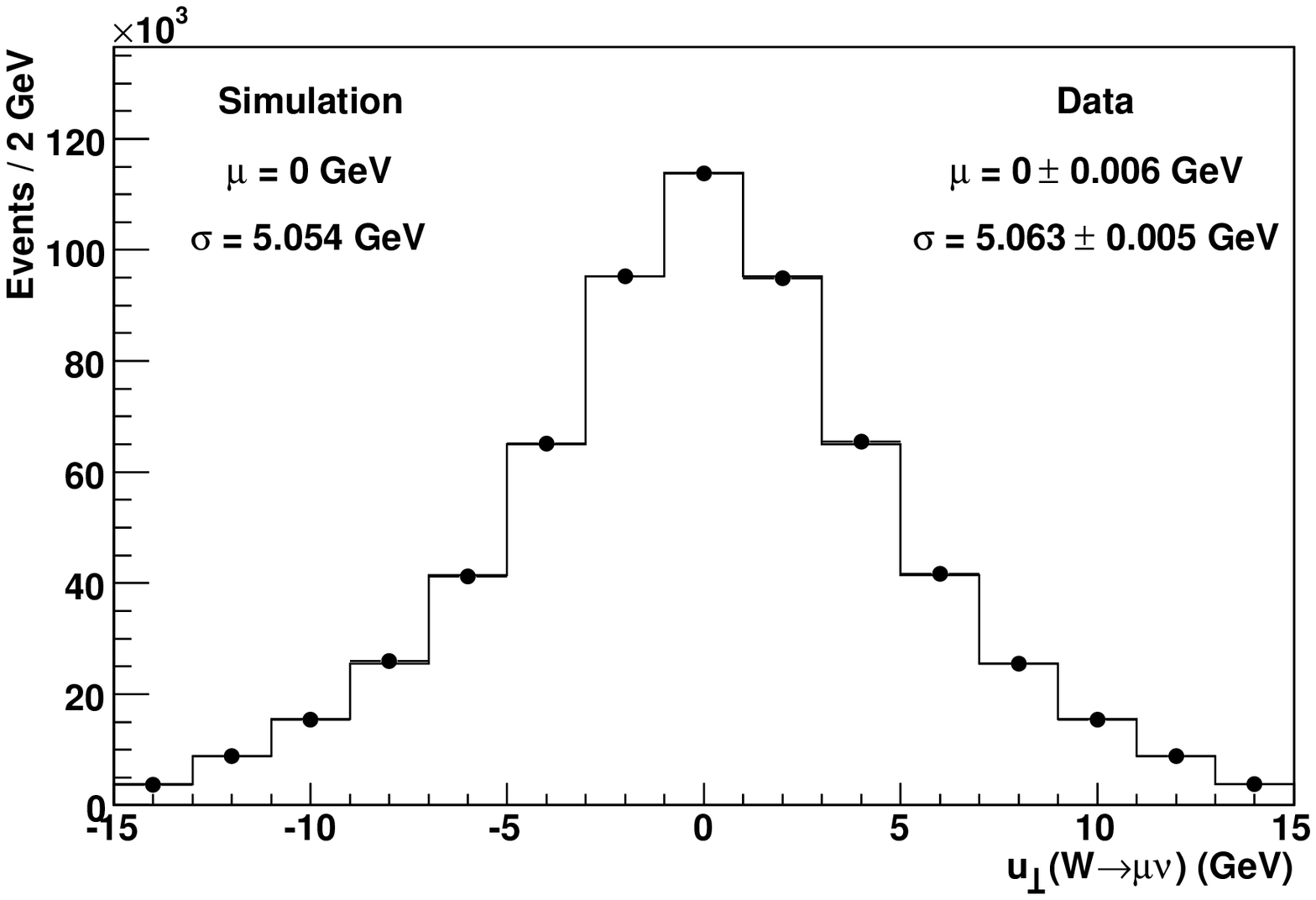}
\includegraphics*[width=8.5cm]{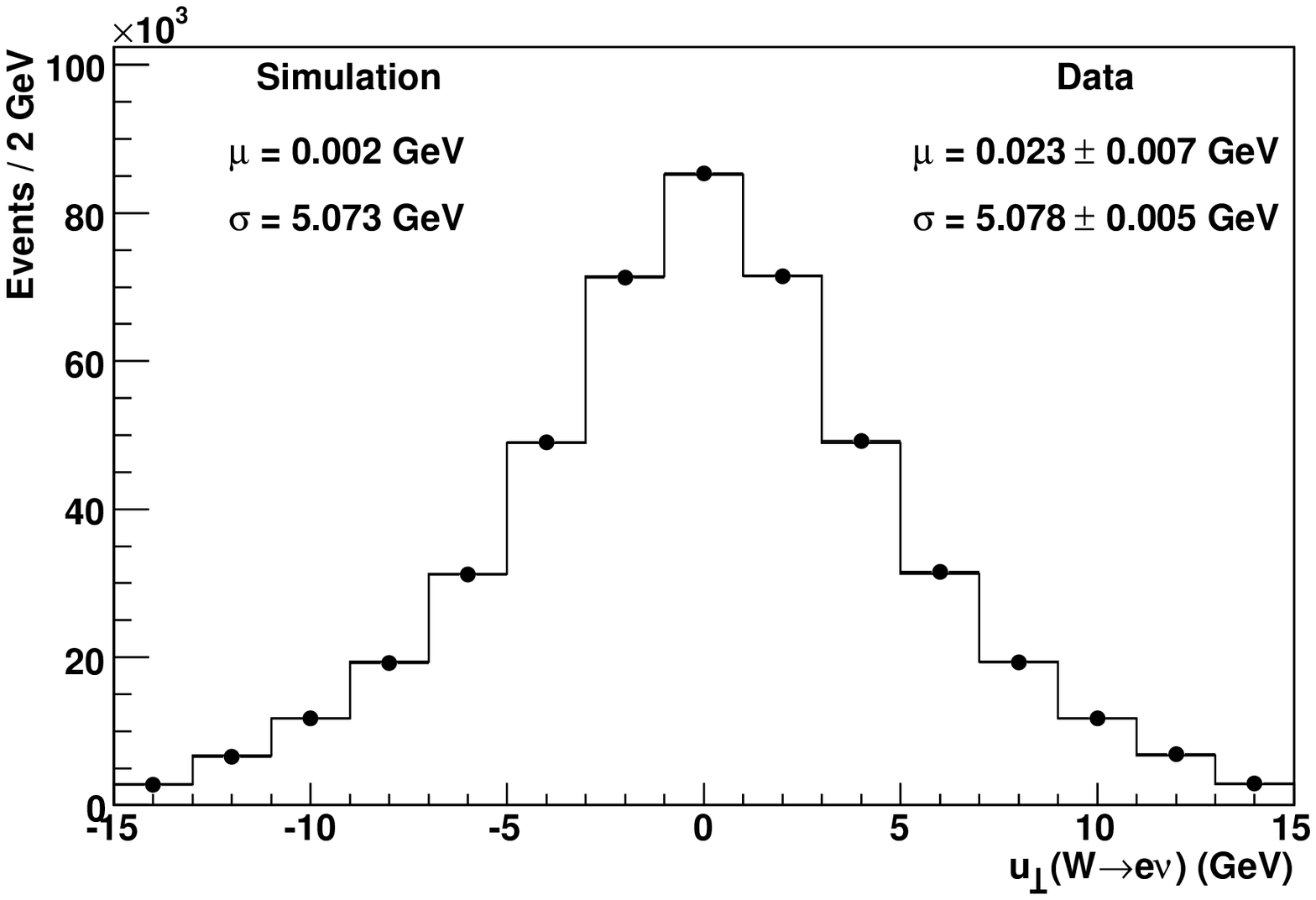}
\caption{Distributions of $u_{\perp}$ from simulation (histogram) and data (circles) 
 for $W$ boson decays to $\mu\nu$ (top) and 
$e\nu$ (bottom) final states.  The simulation uses parameters fit from 
$Z$ boson data, and the uncertainty on the simulation is 
due to the statistical uncertainty on these parameters.  The 
data mean ($\mu$) and rms spread ($\sigma$) are well-modeled by 
the simulation.}
\label{fig:uper}
\end{center}
\end{figure}

\begin{figure}[!htp]
\begin{center}
\epsfysize = 6.cm
\includegraphics*[width=8.5cm]{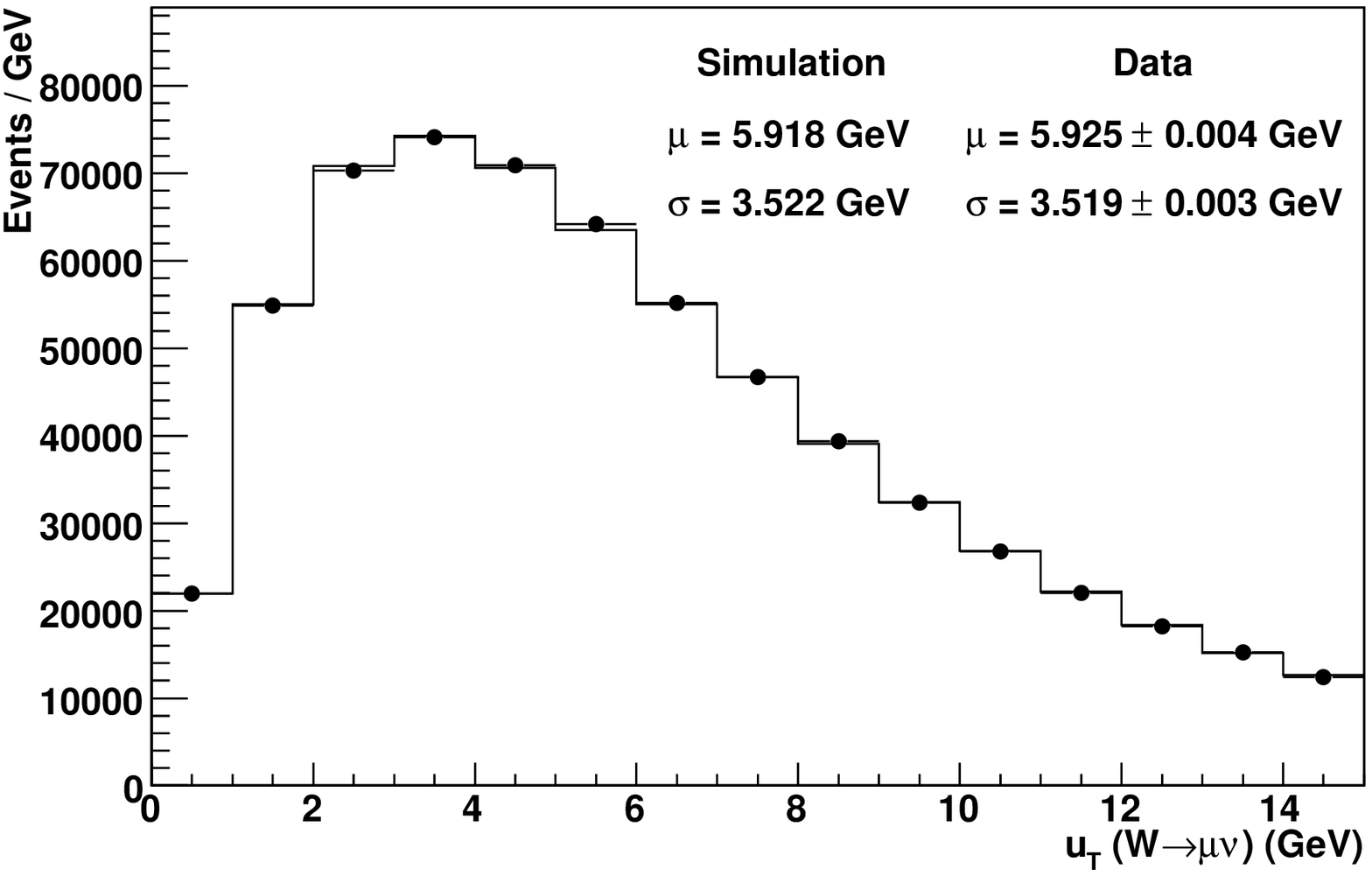}
\includegraphics*[width=8.5cm]{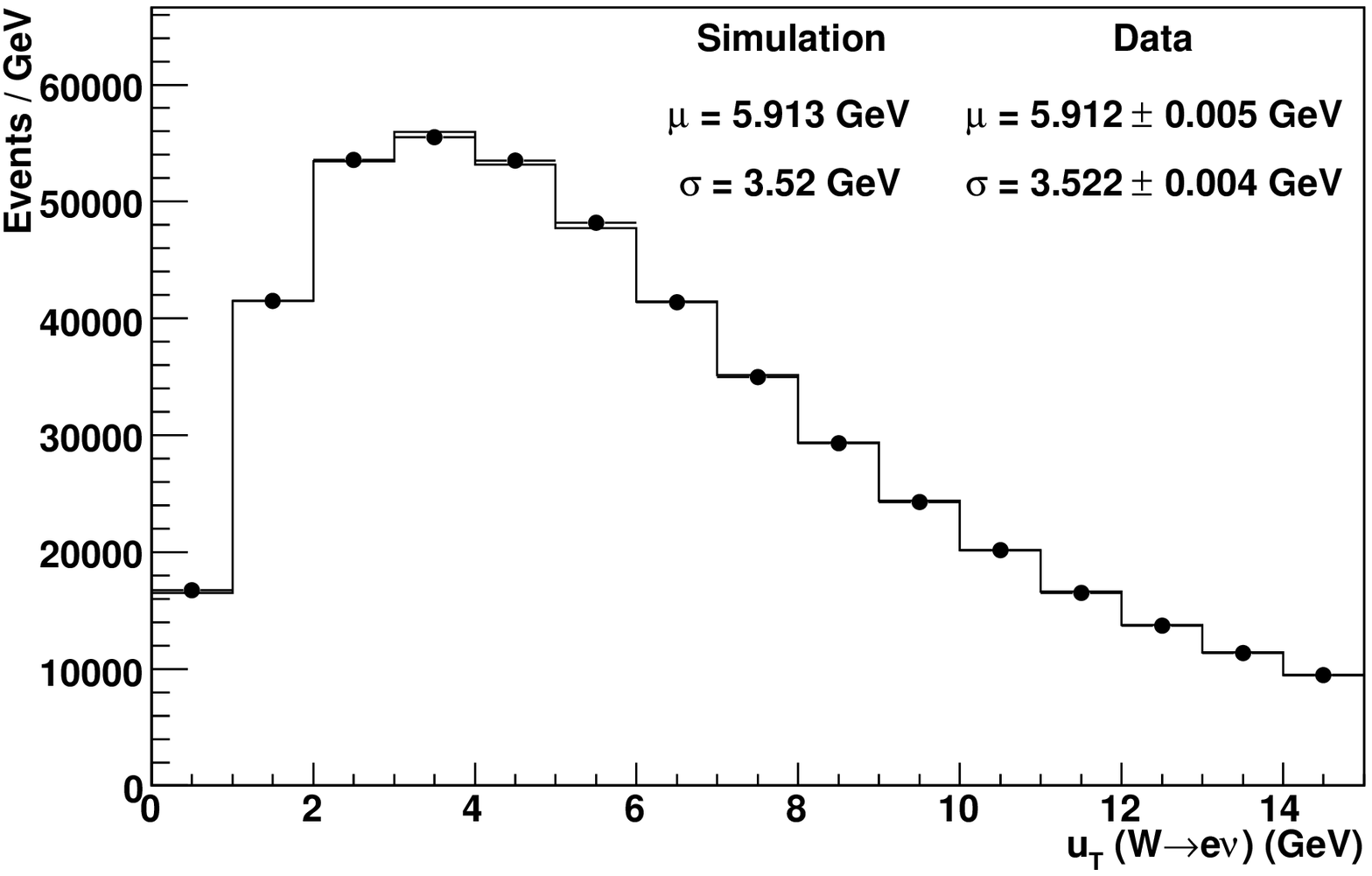}
\caption{Distributions of $u_T$ from simulation (histogram) and data (circles) 
 for $W$ boson decays to $\mu\nu$ (top) and 
$e\nu$ (bottom) final states.  The simulation uses parameters fit from 
$Z$ boson data, and the uncertainty on the simulation is 
due to the statistical uncertainty on these parameters.  The 
data mean ($\mu$) and rms spread ($\sigma$) are well-modeled by 
the simulation.}
\label{fig:ut}
\end{center}
\end{figure}

The recoil model, tuned from $Z$ boson events, is applied to simulated $W$ boson events. We validate the model by comparing the simulated $W$-boson recoil to the recoil measured in data. We utilize two projections of the recoil, along ($u_{||}$) and perpendicular to ($u_{\perp}$) the charged lepton momentum (Sec.~\ref{sec:definitions}), as well as the total recoil $u_T$. 
Comparing the $u_{||}$ distributions in data and in simulation (Fig.~\ref{fig:upar}) shows no evidence of a bias.  We also compare the $u_\perp$ distribution (Fig.~\ref{fig:uper}), which is dominantly affected by recoil resolution, in data and in simulation and find no evidence of a bias. The distribution of $u_T$ in both $W\to\mu\nu$ and $W\to e\nu$ data is well-modeled by our tuned simulation (Fig.~\ref{fig:ut}). Consistency checks with $Z$ bosons decaying to forward ($|\eta| > 1$) electrons show consistency of the relative central-to-plug calorimeter calibration~\cite{malik}.

We estimate the uncertainties on the $M_W$ fits from the recoil model by varying each parameter in the model by $\pm 1\sigma$ and assuming linear parameter-dependent variations of the $M_W$ estimate (Table~\ref{tab:massShifts}). The total uncertainty on $M_W$ due to the recoil model is 9 MeV, 8 MeV, and 11 MeV from the $m_T$, $p_T^\ell$, and $p_T^\nu$ fits, respectively. The uncertainties are entirely correlated for the electron and muon channels as the parameters are obtained from combined fits to $Z\to ee$ and $Z\to \mu\mu$ data.

\begin{table}[ht!]
\begin{center}
\caption{Signed shifts in the $M_W$ fits, in MeV, due to $1\sigma$ increases in recoil model parameters. }
\begin{ruledtabular}
\begin{tabular}{lccc}
Parameter  & $m_T(l \nu)$ & $p_T(l)$ & $p_T(\nu)$ \\
\hline
$a$ & $+2$ & $+5$ & $-1$ \\
$b$ &  $+4$  & $-2$ & $+1$ \\
$c$ & $+3$ & $+3$ & $-1$ \\
\hline
Response total & 5 & 6 & 2 \\
\hline
$s_{had}$ & $+1$ & $+2$ & $0$ \\
$N_V$ &  $+4$ &  $-2$ &  $-9$ \\
$\alpha$ &  $-4$ & $0$  & $-6$ \\
$\beta$ &  $0$ & $+3$ & $-1$ \\
$\gamma$ & $-2$  & $-3$ & $+1$ \\
\hline
Resolution total & 7 & 5 & 11 \\
\end{tabular}
\end{ruledtabular}
\label{tab:massShifts}
\end{center}
\end{table}


\section{Backgrounds}
\label{sec:background}

While the $W\to\mu\nu$ and $W\to e\nu$ event selections (Sec.~\ref{sec:wsample}) result in high-purity samples, several small sources of background persist and can affect the distributions used for $M_W$ fits. Both the $W\to\mu\nu$ and $W\to e\nu$ samples have backgrounds resulting from $Z/\gamma^*\to ll$, where one lepton is not detected; $W\to\tau\nu$ where the $\tau$ decay products include a reconstructed lepton; and multijet events where at least one jet is mis-reconstructed as a lepton. The $W\to\mu\nu$ sample also contains backgrounds from cosmic rays as well as from long lived hadrons decaying to $\mu\nu X$ final states.

\subsection{$W \rightarrow \mu\nu$ Backgrounds}
\label{sec:wmbd}
We model the $Z/\gamma^*\to \mu\mu$ background using events generated with {\sc pythia}~\cite{pythia} and simulated with a full {\sc geant}~\cite{GEANT}-based detector simulation. The $W\to\tau\nu$ background is estimated using the custom simulation and checked with the full {\sc geant} simulation. We use control regions in the data to model the multijet, cosmic ray, and hadronic decay-in-flight backgrounds.

As the full {\sc geant}-based CDF II detector simulation, or ``CDFSim'', models global detector inefficiencies, it is more suitable for estimating background normalizations than our custom fast simulation is. However, CDFSim does not model the detector response to underlying event energies as accurately as our fast simulation. Therefore, we tune the calorimeter energies simulated in CDFSim based on the tunings described in Sec.~\ref{sec:recoil}. The uncertainty on this tuning is propagated to the $M_W$ measurement as an uncertainty in the background normalization and shapes estimated for $Z/\gamma^*\to\mu\mu$ decays.

The $Z/\gamma^*\to\mu\mu$ background is determined from the ratio of $Z/\gamma^*\to\mu\mu$ to $W\to\mu\nu$ acceptances determined from CDFSim, multiplied by the corresponding ratio of cross sections times branching ratios. In the standard model, the ratio $\mathcal{R}\equiv\sigma\mathcal{B}(W\to\mu\nu)/\sigma\mathcal{B}(Z\to\mu\mu)$ has been calculated to be $10.69\pm 0.08$~\cite{wzprd}. In our estimation of the $Z\to\mu\mu$ background, we take an uncertainty of $\pm 0.13$, which includes an additional 1\% uncertainty due to the uncertainty on the ratio of $W$ and $Z$ boson acceptances. From this value of $\mathcal{R}$ and our measured acceptances, we estimate the $Z/\gamma^*\to\mu\mu$ background in the $W\to\mu\nu$ candidate sample to be $(7.35\pm 0.09)\%$. The background due to $Z$ boson decays constitutes a larger portion of the $W\to\mu\nu$ candidate sample than of the $W\to e\nu$ sample due to the limited coverage of the muon detection system in $\eta.$

To estimate the $W\to\tau\nu$ background, we incorporate the $W\to\tau\nu$ kinematic distributions, radiative corrections, predicted $\mathcal{B}(\tau\to\mu\nu\bar{\nu}\nu)$, and $\tau\to\mu\nu\bar{\nu}\nu$ decay spectrum, including $\tau$ polarization, into the custom simulation. Then, we estimate the ratio of events from $W\to\tau\nu\to\mu\nu\bar{\nu}\nu$ to $W\to\mu\nu$ to be 0.963\%, with negligible statistical uncertainty. We verify this prediction  using CDFSim, adopting the same approach used to predict the $Z\to\mu\mu$ background, and obtain a ratio of $(0.957\pm 0.009)\%$, where the uncertainty is due to limited Monte Carlo sample size. The ratio $W\to\tau\nu\to\mu\nu\bar{\nu}\nu/W\to\mu\nu$  predicted by the custom simulation, 0.963\%, normalized to the observed candidate sample including all backgrounds, corresponds to an estimate of $(0.880\pm 0.004)\%$ for the $W\to\tau\nu$ background contribution.

Multijet events where a jet mimics a muon track contribute background to the $W\to\mu\nu$ candidate sample. To estimate this background, we use an artificial neural network (NN)~\cite{jetnet} that differentiates prompt muons (from $W$ and $Z$ boson decay) and muon candidates that arise from jets. As input variables to the NN, we utilize the calorimeter energy and track momenta in an $\eta-\phi$ cone of size 0.4 surrounding the muon candidate. We then construct histograms of the NN discriminant for control samples of pure signal and pure background events. For our signal control sample, we select muons from $W\to\mu\nu$ generated with {\sc pythia} and simulated with CDFSim. For our background control sample, we select events from the data that satisfy the $W\to\mu\nu$ selection criteria except with $p_T^{\nu}<10$~GeV and $u_T<45$~GeV. We combine these spectra such that the summed spectrum matches the discriminant spectrum for muons from $W \rightarrow \mu \nu$ data. In this fitting process, the background fraction is allowed to vary as a free parameter and is extracted via $\chi^2$ minimization. Using this method~\cite{zeng}, we estimate the fraction of the $W\to\mu\nu$ candidate sample resulting from hadronic jets to be $(0.035\pm 0.025)\%$. 

Long lived hadrons, such as $K$ and $\pi$ mesons, decaying into muons before the hadrons reach the calorimeter can mimic $W\to\mu\nu$ events. This decay-in-flight (DIF) background enters our candidate sample when a low-momentum meson decaying to a muon results in the reconstruction of a single high-$p_T$ track with an abrupt change of curvature in it ({\it i.e.}, a kinked pattern). As described in Sec.~\ref{sec:muselection}, DIF events are reduced by imposing a cut on the number of times the hit residuals change sides along a COT track as well as imposing restrictions on track impact parameter and reconstruction quality. To estimate the residual DIF background, we fit the track $\chi^2/$dof distribution of $W\to\mu\nu$ candidates in the data to a sum of signal and DIF background templates. We use data events passing the $W\to\mu\nu$ selection criteria except that large track impact parameters ($2<d_0<5$~mm) are required for the DIF-enriched background template and $Z\to\mu\mu$ events for the signal template. After correcting for the presence of real $W\to\mu\nu$ events in the background template, we estimate the DIF background to be $(0.24 \pm 0.08)\%$ in the $W\to\mu\nu$ candidate sample.

High-energy muons from cosmic rays can mimic $W\rightarrow \mu\nu$ events when passing close to the beam line and reconstructed as a muon track on only one side of the COT.  The cosmic-ray identification algorithm searches for unreconstructed tracks and removes cosmic rays with approximately 99\% efficiency~\cite{cosmic}.  The residual cosmic-ray background is estimated using the 
reconstructed interaction time $t_0$ and transverse impact-parameter magnitude $|d_0|$ from the COT track 
fit.  We use the estimate for the cosmic-ray background from the smaller data set reported in Ref.~\cite{CDF2} and scale it by the ratio of run-time to integrated luminosity to obtain the cosmic-ray background fraction in the $W\to\mu\nu$ candidate sample of $(0.02\pm 0.02)\%$. 

\begin{table}[!ht]
\caption{Background fractions from various sources in the
$W\rightarrow \mu\nu$ data set, and the corresponding uncertainties
on the $m_T$, $p_T^\mu$, and $p_T^\nu$ fits for $M_W$. }

\begin{ruledtabular}
\begin{tabular}{lcccc}
  & Fraction of & \multicolumn{3}{c}{$\delta M_W$ (MeV)} \\
Source        & $W\rightarrow \mu\nu$ data (\%) & ~$m_T$ fit~
                  & ~$p_T^\mu$ fit~ & ~$p_T^\nu$ fit~ \\
\hline
$Z/\gamma^*\rightarrow \mu\mu$ & $7.35\pm 0.09$   & 2.2 & 4.0 & 5.4 \\
$W\rightarrow\tau\nu$          & $0.880 \pm 0.004$ & 0.2 & 0.2  & 0.2 \\
Hadronic jets                  & $0.035\pm 0.025$   & 0.5 & 0.7  & 1.0 \\
Decays in flight               & $0.24\pm 0.08$    & 0.9 & 3.1 & 3 \\
Cosmic rays                    & $0.02\pm 0.02$ & 0.5 & 1  & 0.7 \\
\hline
Total                          & 8.53 $\pm$ 0.12   & 3 &  5.2 & 5.7 \\
\end{tabular}
\end{ruledtabular}
\label{tbl:mubd}
\end{table}

The  $m_T$, $p_T^\mu$, and $p_T^\nu$  distributions are obtained from the {\sc geant}-based simulation for $W$ and $Z$ boson backgrounds, from identified cosmic ray events for the cosmic ray background, and from events in the $W\rightarrow \mu\nu$ sample with high-$\chi^2$ (isolation) muons for the decay-in-flight (hadronic jet) background. Including uncertainties on the shapes of the distributions, the total uncertainties on the background estimates result in uncertainties of 3, 5, and 6 MeV on $M_W$ from the $m_T$, $p_T^\ell$, and $p_T^\nu$ fits, respectively (Table~\ref{tbl:mubd}).

\subsection{$W \rightarrow e\nu$ Backgrounds}
\label{sec:webd}
We model the $Z/\gamma^*\to ee$ background using {\sc pythia}-generated events simulated with the {\sc geant}-based CDFSim. We follow the same procedure used to estimate the $Z/\gamma^*\to\mu\mu$ background (Sec.~\ref{sec:wmbd}), correcting the reconstructed energies in CDFSim, and using the theoretical prediction of $R=10.69\pm 0.08$. We estimate the $Z/\gamma^*\to ee$ background in the $W\to e\nu$ candidate sample to be $(0.139\pm 0.014)\%$.

We model the $W\to\tau\nu$ background using our fast simulation, as with the $W\to\mu\nu$ channel. We estimate the ratio of events from $W\to\tau\nu \to e \nu \bar{\nu} \nu$ to $W\to e \nu$ to  be 0.945\%, which is consistent with the CDFSim prediction of $(0.943 \pm 0.009_{\rm MC \; stat})$\%. This ratio yields a prediction of the $W\to\tau\nu$ background in the $W\to e\nu$ candidate sample of $(0.93\pm 0.01)\%$.

As in the $W\to\mu\nu$ sample, multijet events enter the $W\to e\nu$ sample when a hadronic jet is misreconstructed as an electron. To estimate this background, we fit distributions of electron identification variables from $W\to e\nu$ candidate data to a sum of simulated electrons and background shapes. For the background sample, we select jet-enriched data events by applying the $W\to e\nu$ selection criteria (Sec.~\ref{sec:wesample}), except that the $m_T$ requirement is removed, $u_T$ is required not to exceed 45~GeV, and $p_T^\nu$ is required not to exceed 10~GeV. The identification variables are the weighted track isolation and the output of an artificial neural network (NN). The weighted track variable is the sum of the $p_T$ of particles within a $\eta-\phi$ cone of 0.4 and $\delta z_0 = 5$~cm around the identified electron's track. The NN uses several kinematic variables used in $W\to e\nu$ selection, such as $E_{\rm had}/E_{\rm EM}$ and $\delta z$. As an alternative estimate, we fit the $p_T^\nu$ distribution of the $W\to e\nu$ candidate events to a sum of simulated $W\to e\nu$ events and jet-enriched events obtained using the NN. Using the results from all three fits, we obtain an estimate of the multijet background in the $W\to e\nu$ candidate sample of $(0.39\pm 0.14)\%$. 

\begin{table}
\caption{Background fractions from various sources in the
$W\rightarrow e\nu$ data set, and the corresponding uncertainties
on the $m_T$, $p_T^\mu$, and $p_T^\nu$ fits for $M_W$. }
\begin{center}
\begin{ruledtabular}
\begin{tabular}{lcccc}
  			  & Fraction of & \multicolumn{3}{c}{$\delta M_W$ (MeV)} \\
Source        & $W\rightarrow e\nu$ data (\%)& ~$m_T$ fit~& ~$p_T^e$ fit~ & ~$p_T^\nu$ fit~ \\
\hline
$Z/\gamma^*\to ee$ & $0.139\pm 0.014$ & 1.0  & 2.0  & 0.5 \\
$W\to\tau\nu$ &  $0.93\pm 0.01$ & 0.6  & 0.6  & 0.6  \\
Hadronic jets & $0.39\pm 0.14$ & 3.9 & 1.9 & 4.3\\
\hline
Total & $1.46\pm 0.14$ & 4.0 & 2.8 & 4.4 \\
\end{tabular}
\end{ruledtabular}
\end{center}
\label{tab:elebd}
\end{table}

The distributions for the $M_W$ fit variables are obtained from simulated events for $W$ and $Z$ boson backgrounds, and from events in the $W\rightarrow e\nu$ sample with low-NN electron candidates for the hadronic jet background.  We fit these distributions and include their shapes and relative normalizations in the $M_W$ template fits.  The uncertainties on the background estimates result in uncertainties of 4, 3, and 4 MeV on $M_W$ from the $m_T$, $p_T^e$, and $p_T^\nu$ fits, respectively (Table~\ref{tab:elebd}).


\section{\boldmath $W$-boson-mass fits}
\label{sec:fits}

The $W$ boson mass is extracted by performing fits to a sum of background and simulated signal templates of the $m_T$, $p_T^\ell$, and $p_T^\nu$ distributions. The fits minimize $-\ln {\cal L}$, where the likelihood $\cal L$ is given by

\begin{equation}
{\cal L} = \prod^N_{i=1}\frac{e^{-m_i}m_i^{n_i}}{n_i!},
\end{equation}
where the product is over $N$ bins in the fit region with $n_i$ entries (from data) and $m_i$ expected entries (from the template) in the $i$th bin. The template is normalized to the data in the fit region.  The likelihood is a function of $M_W$, where $M_W$ is defined by the relativistic Breit-Wigner mass distribution

\begin{equation}
\frac{d\sigma}{dm}\propto\frac{m^2}{(m^2-M_W^2)^2+m^4\Gamma_W^2/M_W^2},
\end{equation}
where $m$ is the invariant mass of the propagator. We assume the standard model $W$ boson width $\Gamma_W = 2\,094\pm2$~MeV. The uncertainty on $M_W$ resulting from $\delta\Gamma_W = 2$~MeV is negligible.  

\subsection{Fit Results}
\begin{figure}
\begin{center}
\epsfysize = 6.cm
\includegraphics*[width=8.5cm]{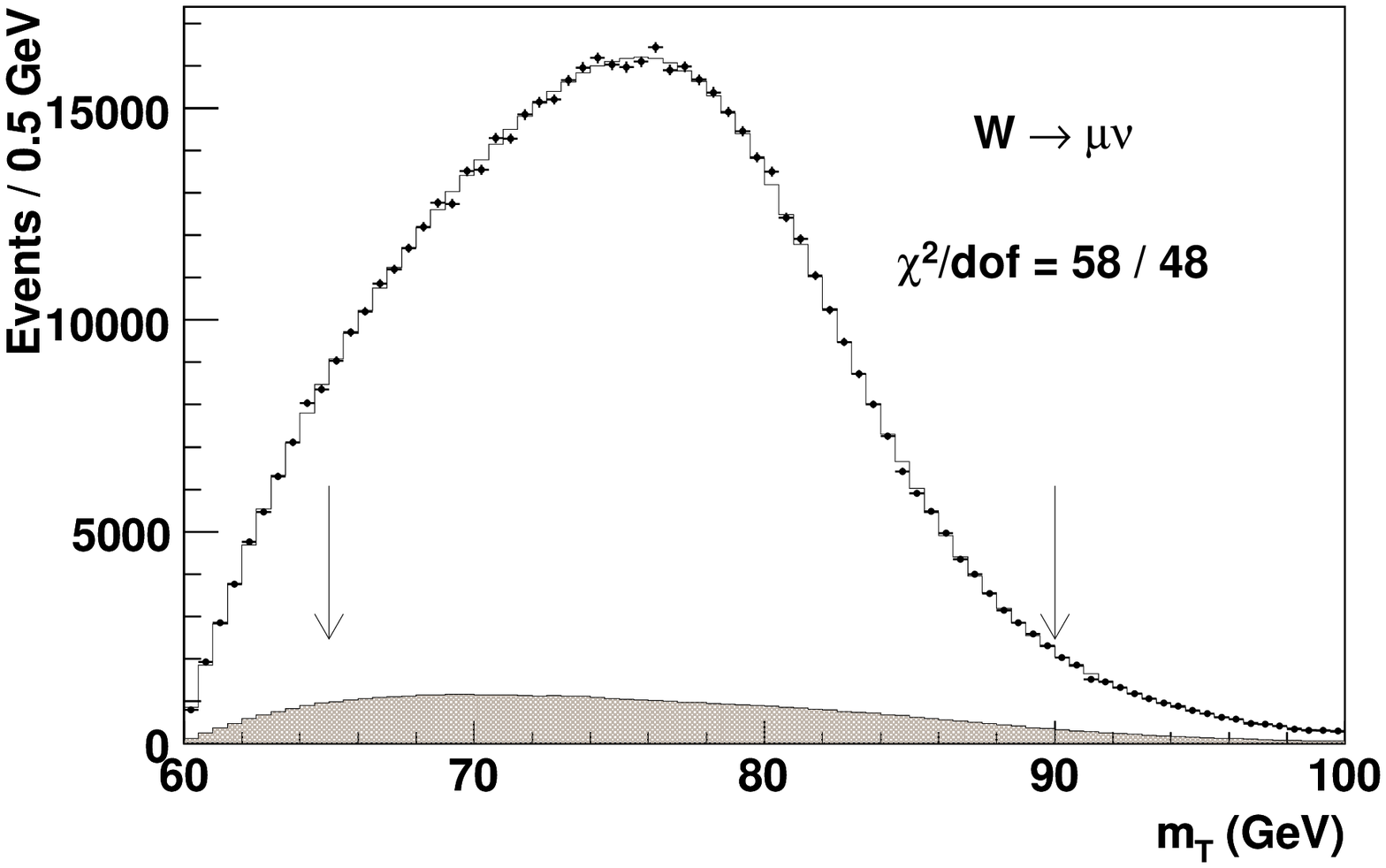}
\includegraphics*[width=8.5cm]{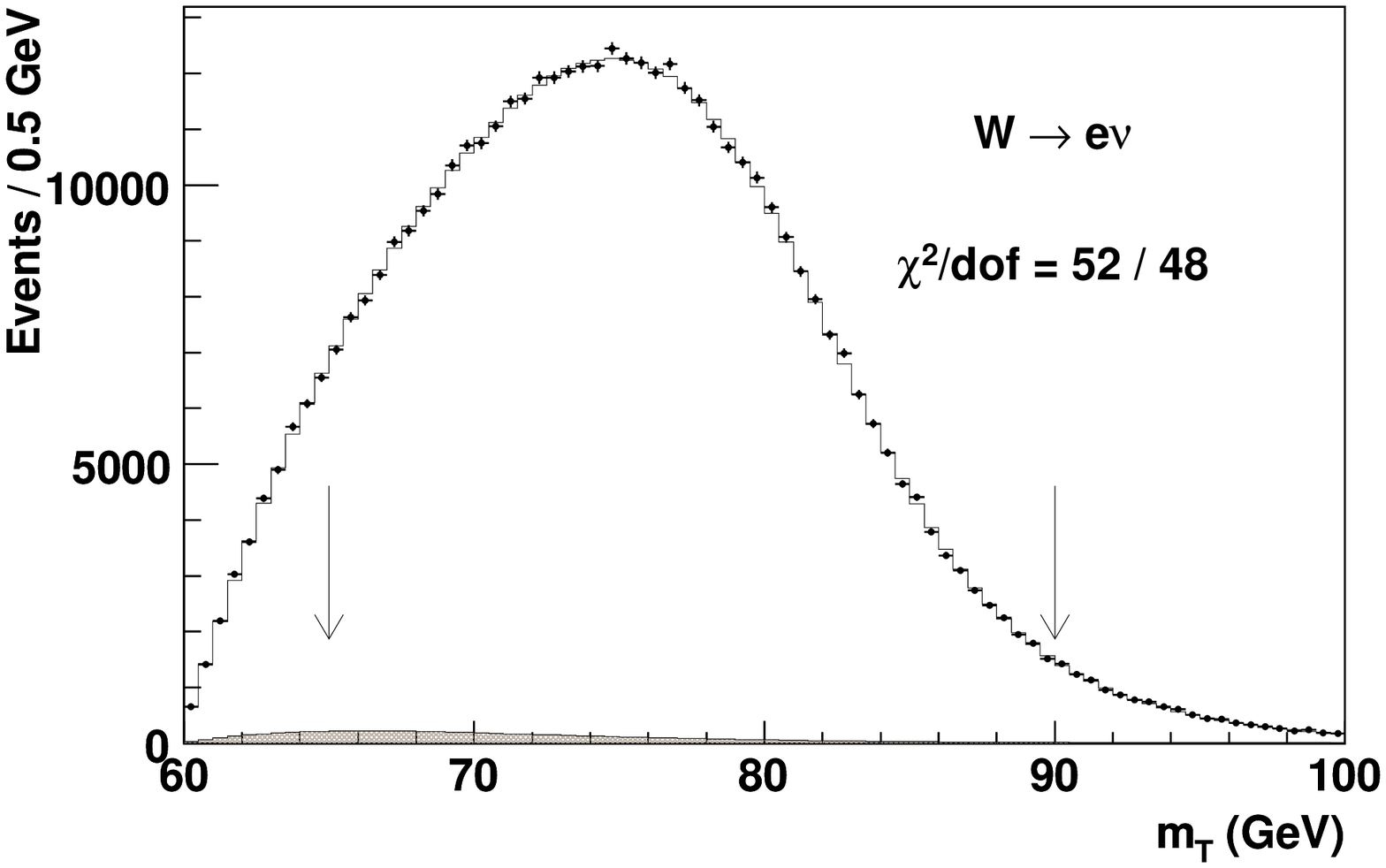}
\caption{Distributions of $m_T$ 
for $W$ boson decays to $\mu\nu$ (top) and $e\nu$ (bottom) final states in simulated (histogram) and experimental (points) data.  The 
simulation corresponds to the maximum-likelihood value of $M_W$ and includes backgrounds (shaded). The likelihood is computed using events 
between the two arrows.   }
\label{fig:mt}
\end{center}
\end{figure}

\begin{figure}
\begin{center}
\epsfysize = 6.cm
\includegraphics*[width=8.5cm]{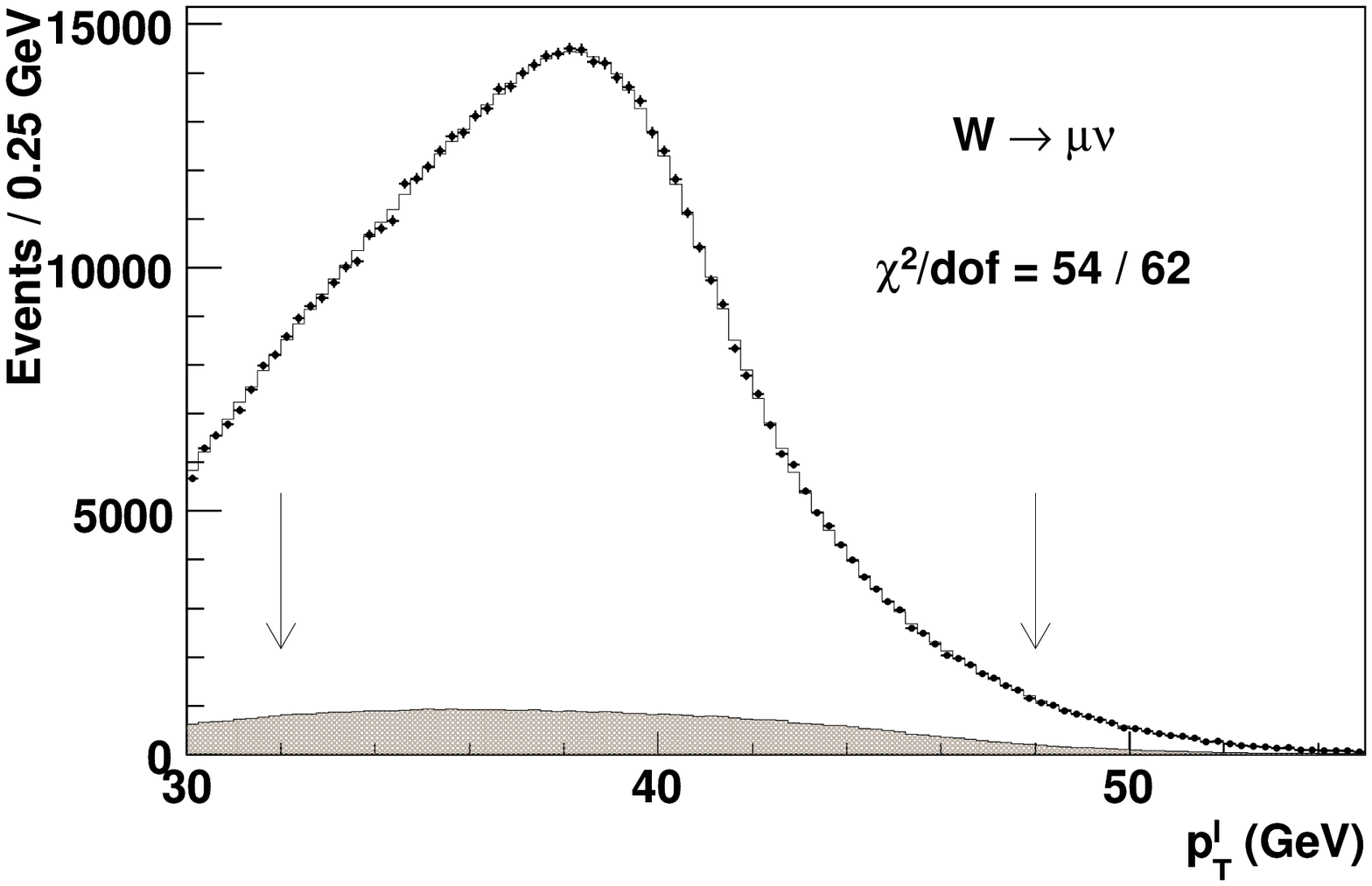}
\includegraphics*[width=8.5cm]{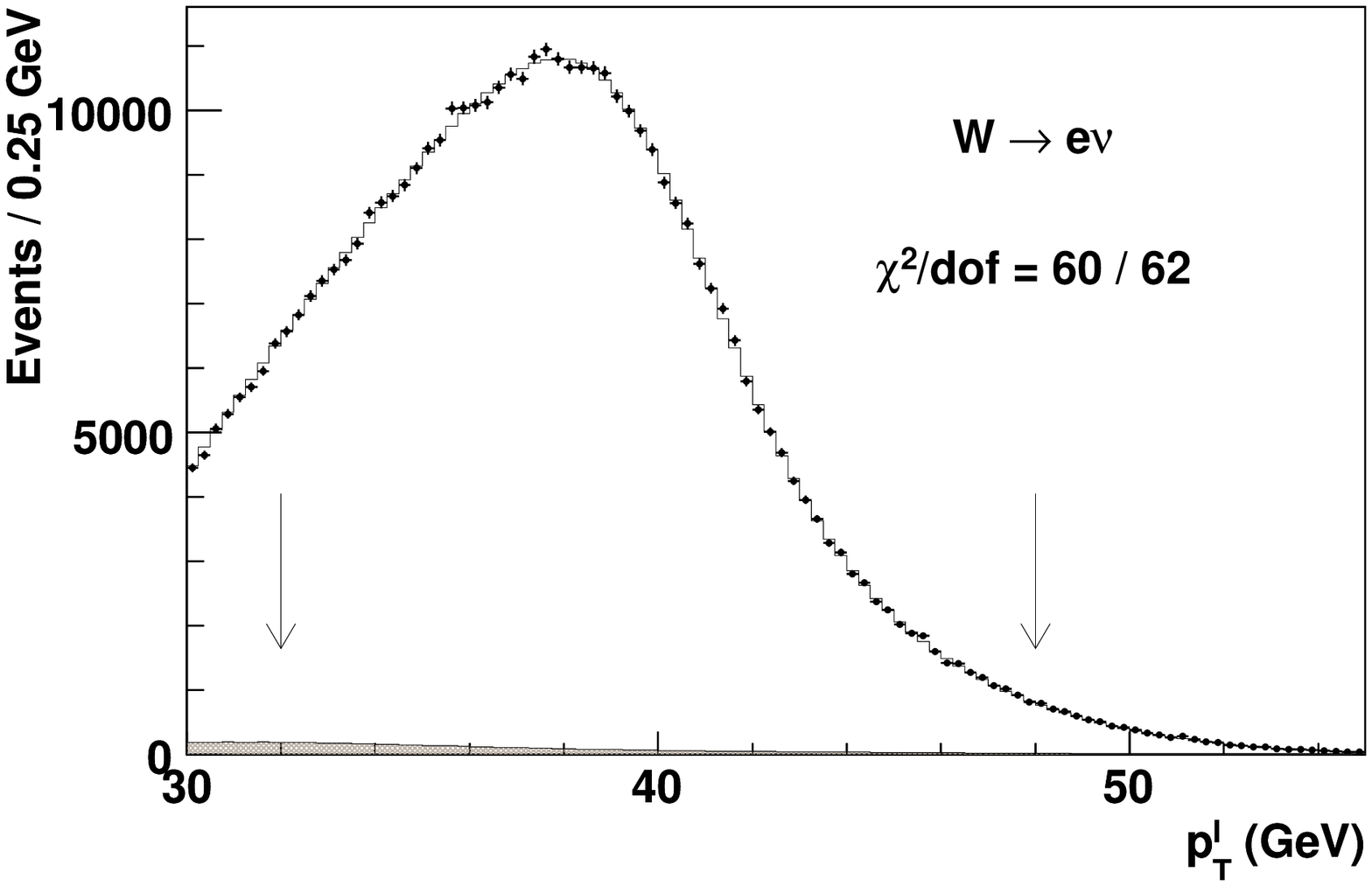}
\caption{Distributions of $p_T^\ell$ 
for $W$ boson decays to $\mu\nu$ (top) and $e\nu$ (bottom) final states in simulated (histogram) and experimental (points) data.   The 
simulation corresponds to the maximum-likelihood value of $M_W$ and includes backgrounds (shaded). The likelihood is computed using events 
between the two arrows.     }
\label{fig:pt}
\end{center}
\end{figure}

\begin{figure}[!tp]
\begin{center}
\epsfysize = 6.cm
\includegraphics*[width=8.5cm]{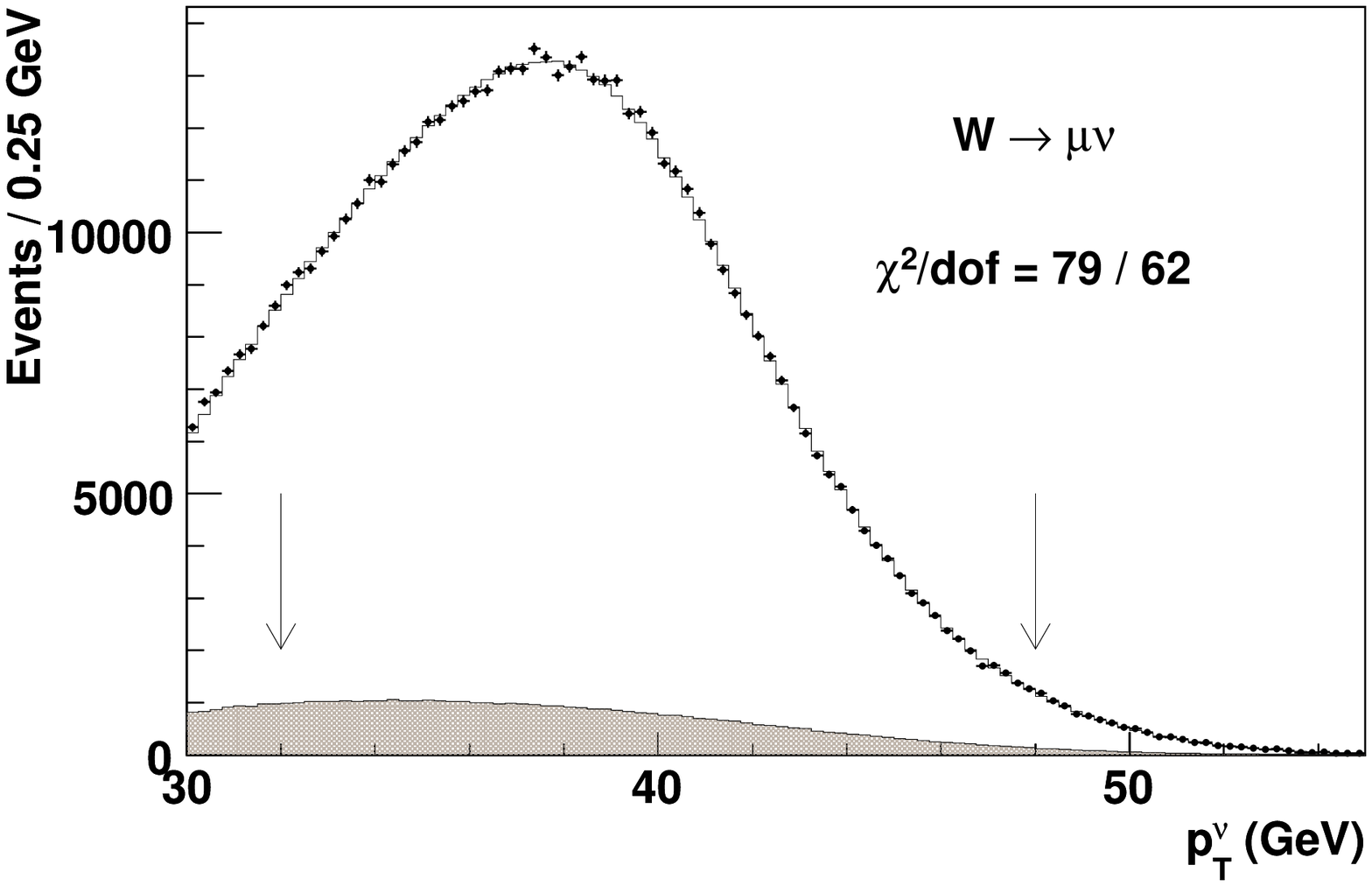}
\includegraphics*[width=8.5cm]{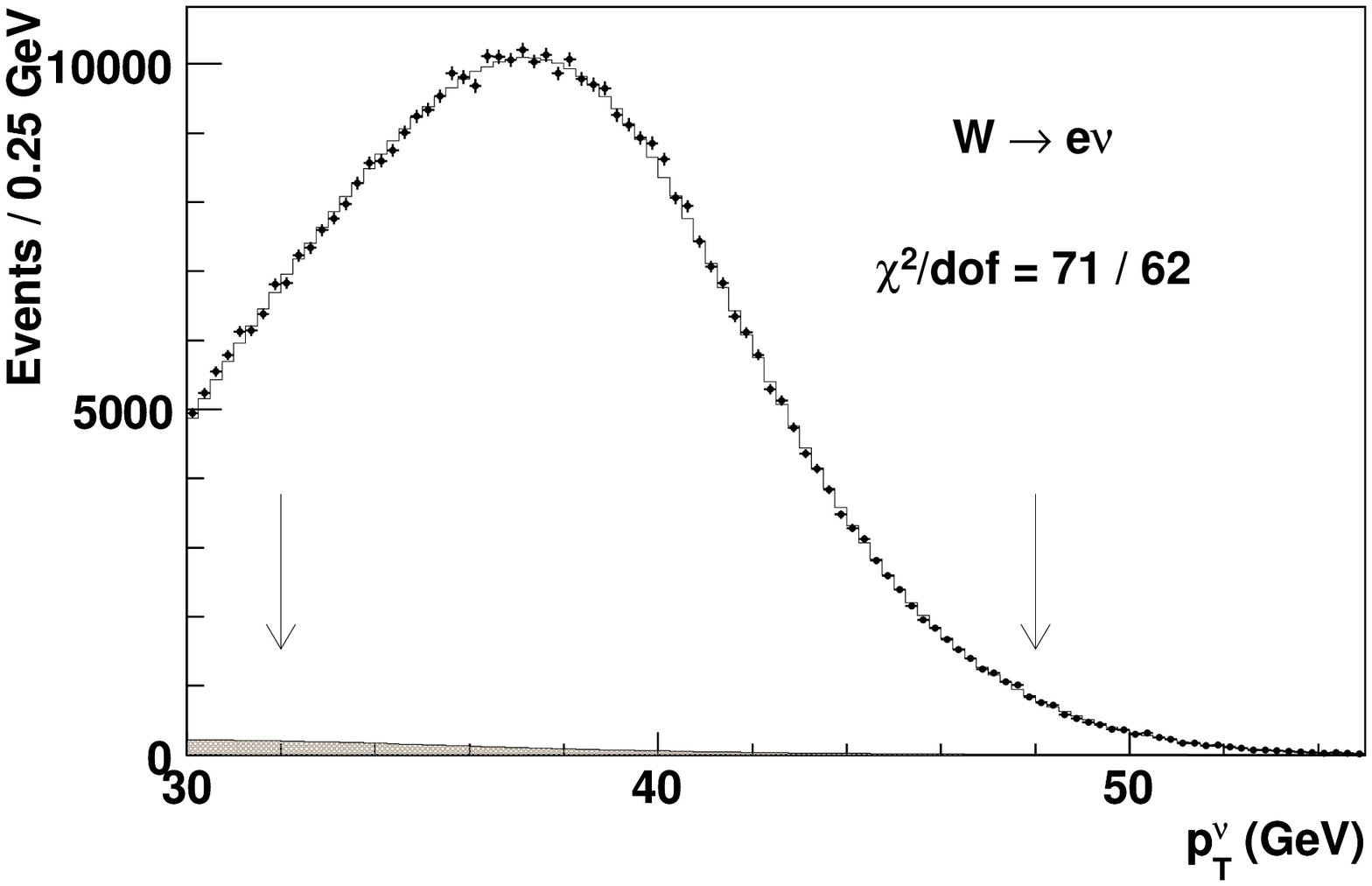}
\caption{Distributions of $p_T^\nu$ 
for $W$ boson decays to $\mu\nu$ (top) and $e\nu$ (bottom) final states in simulated (histogram) and experimental (points) data.  The 
simulation corresponds to the maximum-likelihood value of $M_W$ and includes backgrounds (shaded). The likelihood is computed using events 
between the two arrows.   }
\label{fig:met}
\end{center}
\end{figure}

\begin{figure}[!tp]
\begin{center}
\epsfysize = 6.cm
\includegraphics*[width=8.5cm]{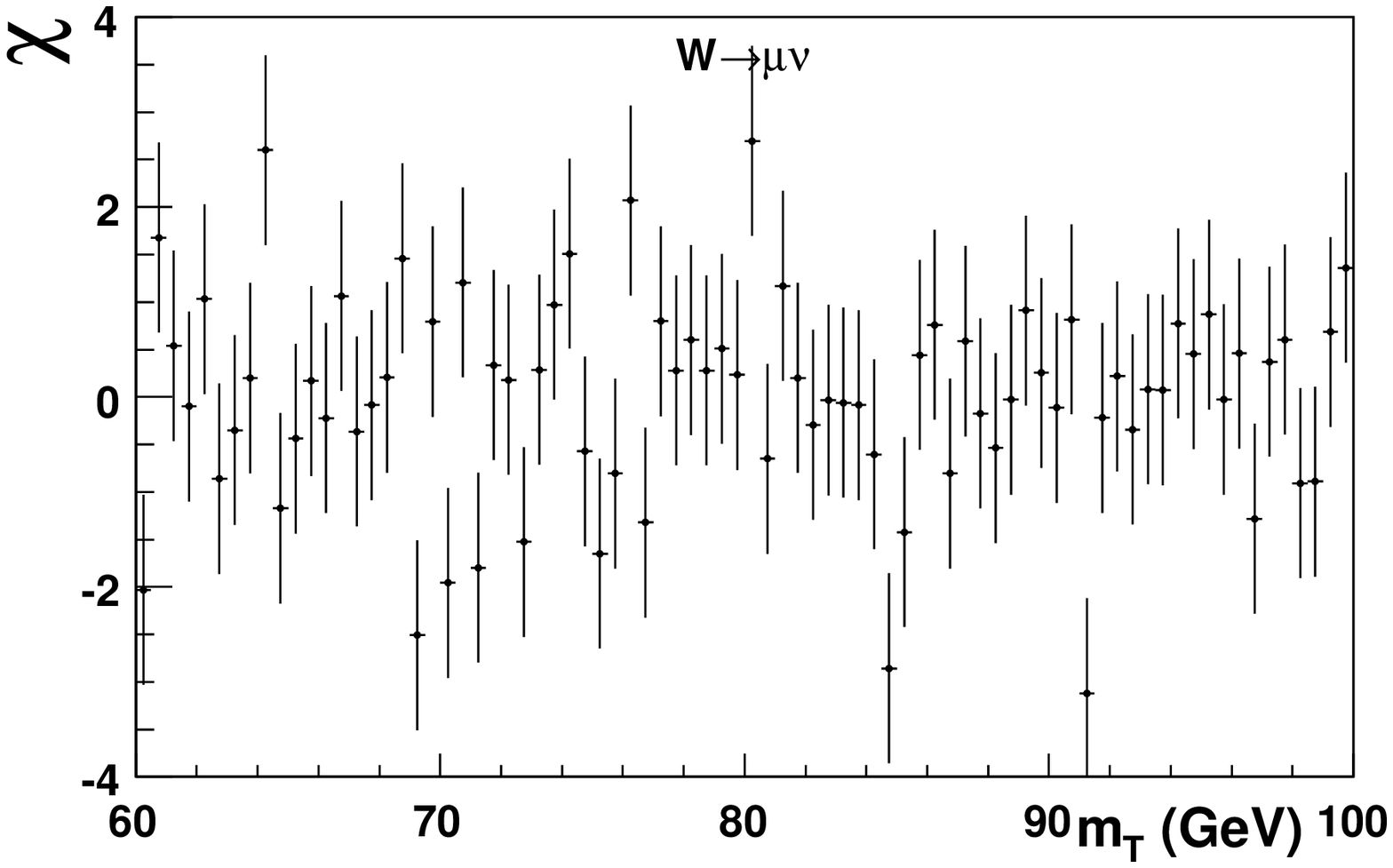}
\includegraphics*[width=8.5cm]{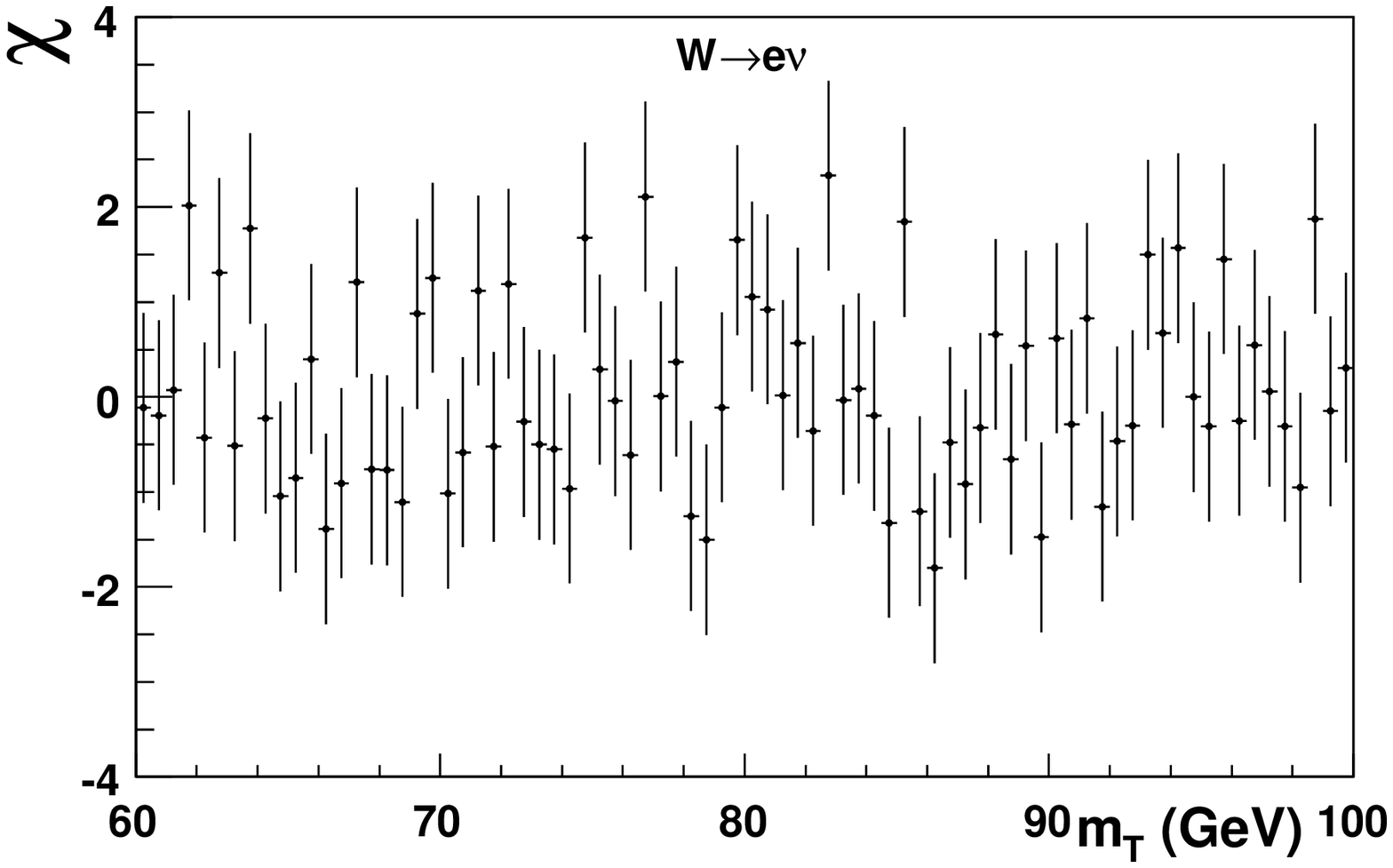}
\caption{Differences between the data and simulation, divided by the expected  
statistical uncertainty, for the $m_T$ distributions in the 
muon (top) and electron (bottom) channels.  }
\label{fig:signedchimt}
\end{center}
\end{figure}

\begin{figure}[!tp]
\begin{center}
\epsfysize = 6.cm
\includegraphics*[width=8.5cm]{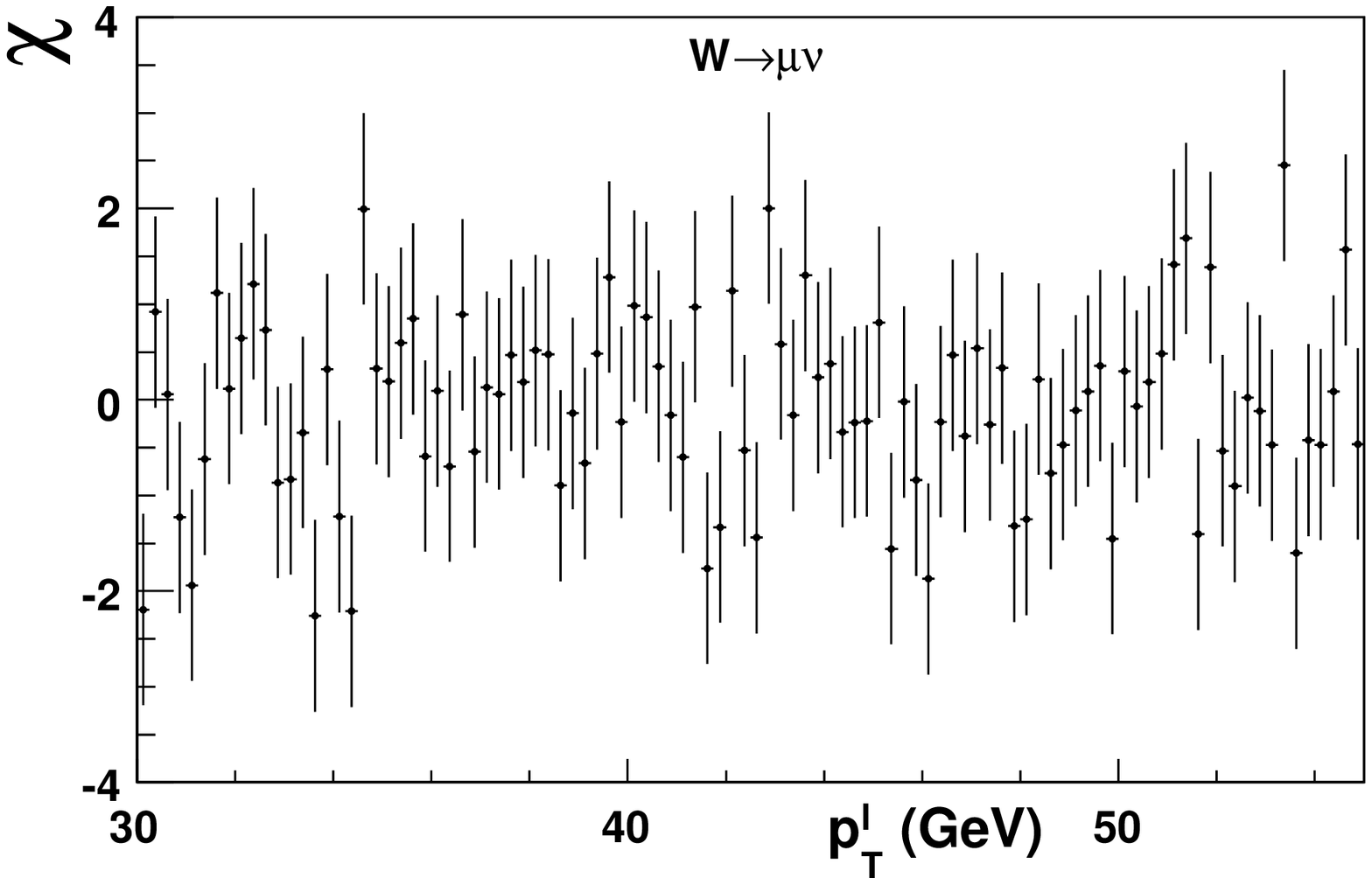}
\includegraphics*[width=8.5cm]{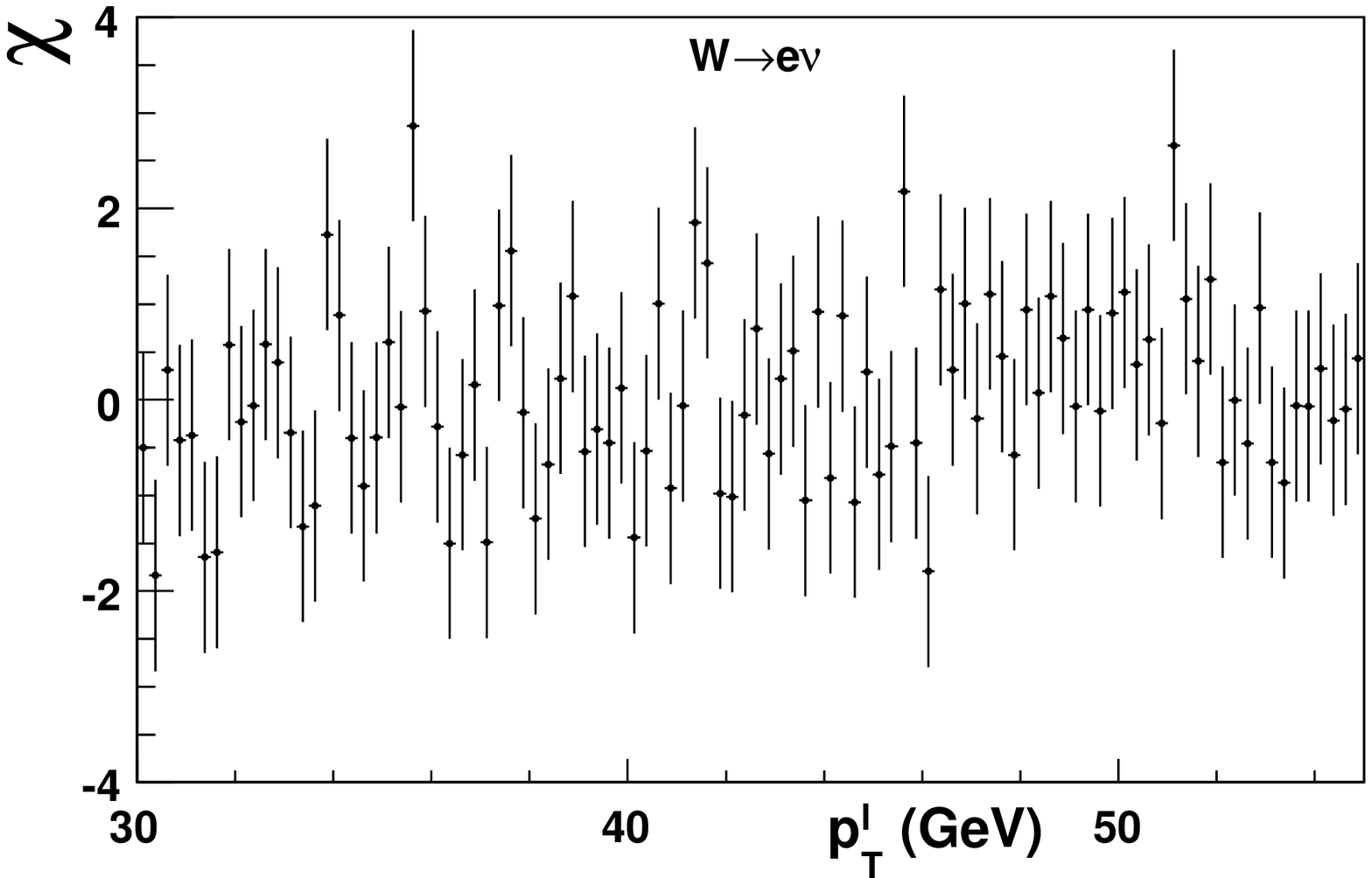}
\caption{Differences between the data and simulation, divided by the 
expected statistical uncertainty, for the $p_T$ distributions in the 
muon (top) and electron (bottom) channels.  }
\label{fig:signedchipt}
\end{center}
\end{figure}

\begin{figure}[!tp]
\begin{center}
\epsfysize = 6.cm
\includegraphics*[width=8.5cm]{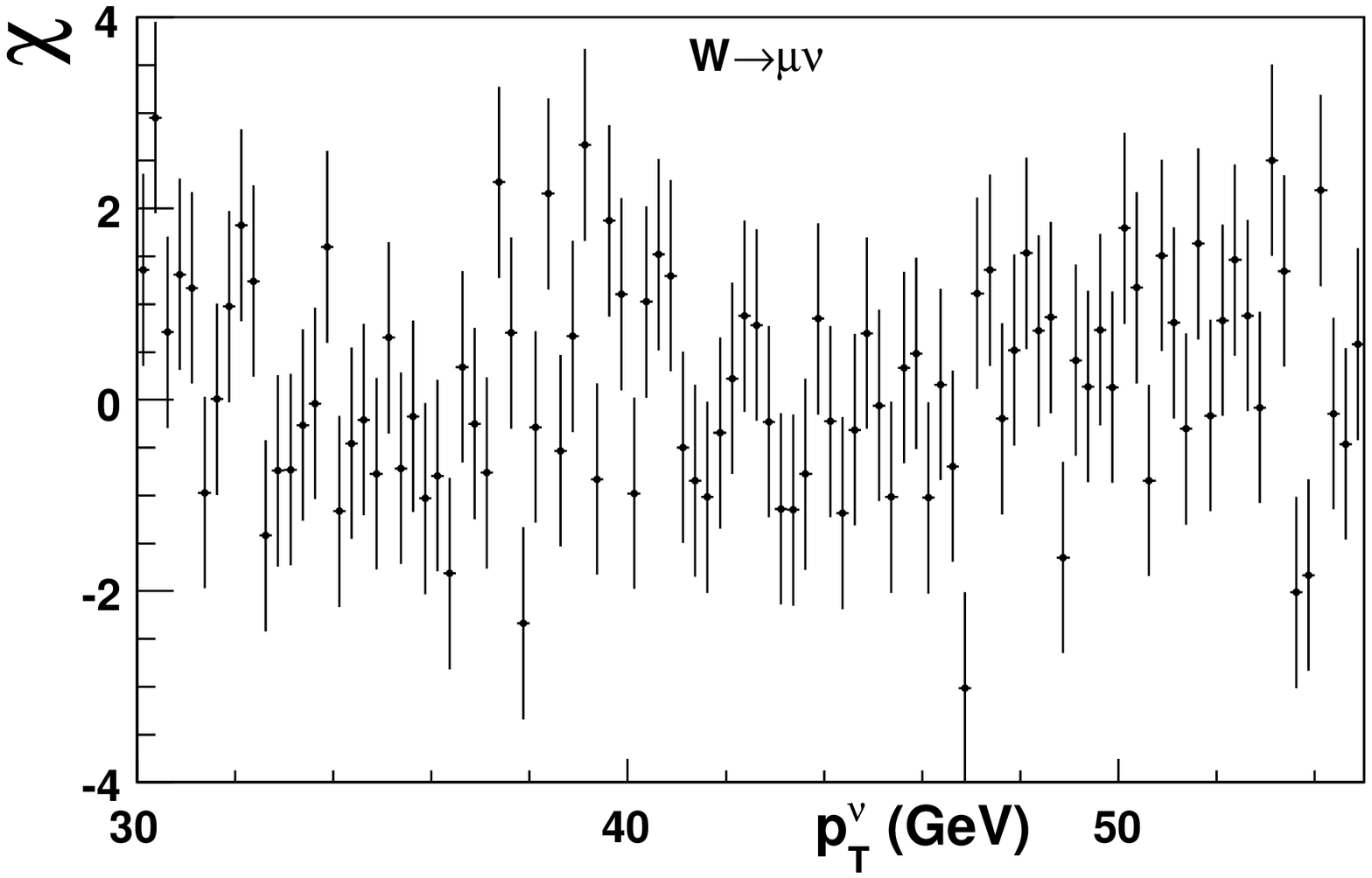}
\includegraphics*[width=8.5cm]{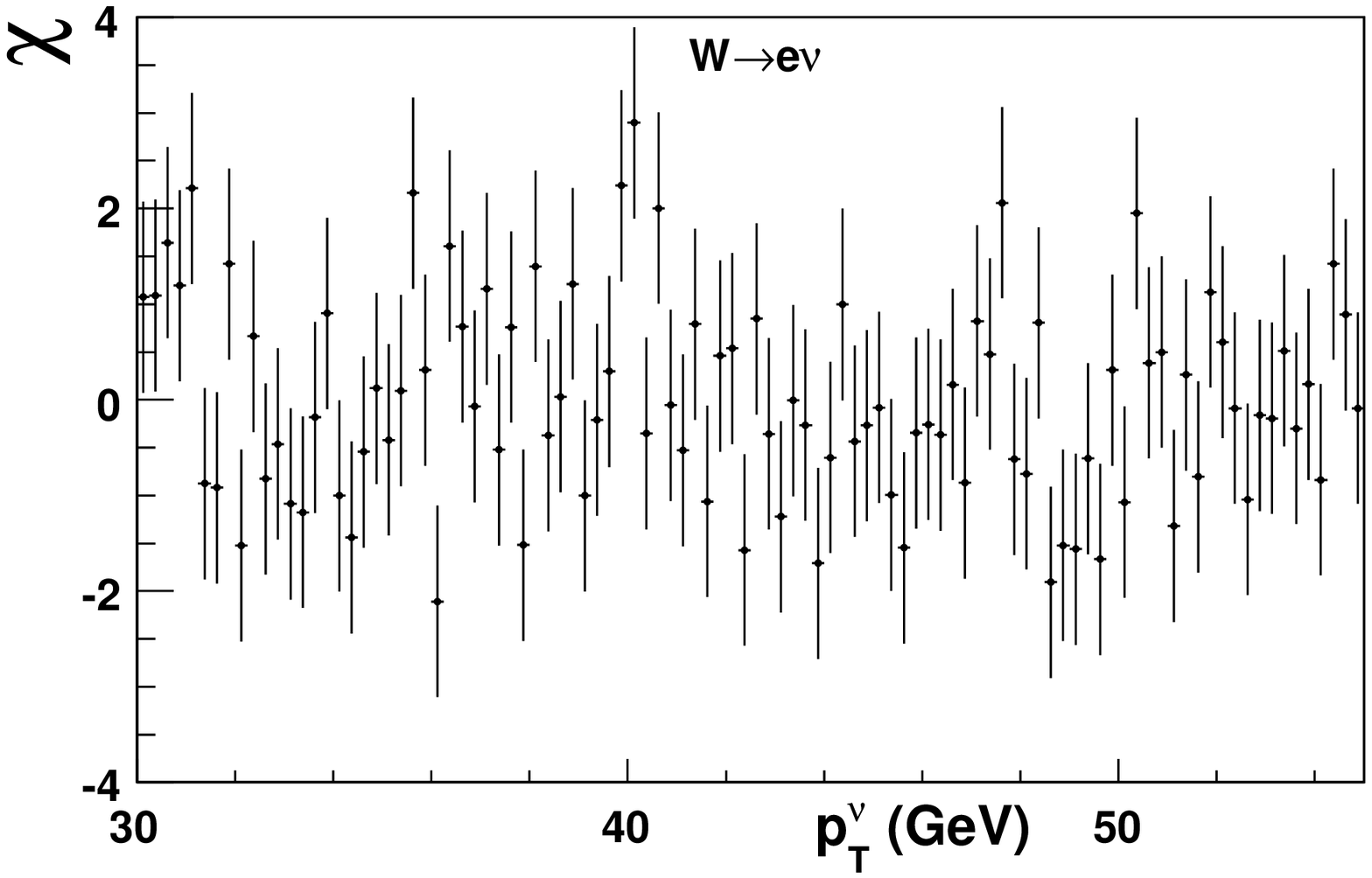}
\caption{Differences between the data and simulation, divided by the expected  
statistical uncertainty, for the $p_T^\nu$ distributions in the 
muon (top) and electron (bottom) channels. }
\label{fig:signedchimet}
\end{center}
\end{figure}

The $m_T$ fit is performed in the range $65 < m_T < 90$ GeV. Figure~\ref{fig:mt} shows the results of the $m_T$ fit for the $W\to\mu\nu$ and $W\to e\nu$ channels while a summary of the 68\% confidence uncertainty associated with the fit is shown in Table~\ref{tbl:mt}. The $p_T^\ell$ and $p_T^\nu$ fits are performed in the ranges $32 <p_T^\ell< 48$~GeV  and  $32 <p_T^\nu< 48$~GeV, respectively, and are shown in Figs.~\ref{fig:pt} and \ref{fig:met}, respectively. The uncertainties for the $p_T^\ell$ and $p_T^\nu$ fits are shown in Tables~\ref{tbl:pt} and \ref{tbl:met}, respectively. The differences between data and simulation for the three fits, divided by the statistical uncertainties on the predictions,  are shown in Figs.~\ref{fig:signedchimt}-\ref{fig:signedchimet} and the fit results are summarized in Table~\ref{tbl:fitsummary}.

\begin{table}[htbp]
\caption{Uncertainties on $M_W$ (in MeV) as resulting from transverse-mass fits in the $W\to \mu\nu$ and $W\to e\nu$ samples. The last column reports the portion of the uncertainty that is common in the $\mu\nu$ and $e\nu$ results.  }
\begin{ruledtabular}
\begin{center}
\begin{tabular}{lccc}
\multicolumn{4}{c}{$m_T$ fit uncertainties} \\
Source                   & $W\rightarrow \mu\nu$ & $W\rightarrow e\nu$ & Common \\
\hline
Lepton energy scale   & 7  &  10  &  5 \\
Lepton energy resolution        & 1   &  4   &  0  \\
Lepton efficiency	 & 0   &  0   &  0  \\
Lepton tower removal     & 2   &  3   &  2  \\
Recoil scale	         & 5   &  5   &  5 \\
Recoil resolution	 & 7   &  7   &  7  \\
Backgrounds 	         & 3   &  4   &  0  \\
PDFs                     & 10  &  10  &  10 \\
$W$ boson $p_T$          & 3   &  3   &  3  \\
Photon radiation         & 4  &  4  &  4 \\
\hline
Statistical              & 16   & 19  &  0  \\
\hline
Total                    & 23   & 26  &  15 \\

\end{tabular}
\label{tbl:mt}
\end{center}
\end{ruledtabular}
\end{table}

\begin{table}[htbp]
\begin{ruledtabular}

\begin{center}
\caption{Uncertainties on $M_W$ (in MeV) as resulting from charged-lepton transverse-momentum fits in the $W\to \mu\nu$ and $W\to e\nu$ samples. The last column reports the portion of the uncertainty that is common in the $\mu\nu$ and $e\nu$ results.  }
\begin{tabular}{lccc}

\multicolumn{4}{c}{$p_T^\ell$ fit uncertainties} \\
Source                   & $W\rightarrow \mu\nu$ & $W\rightarrow e\nu$ & Common \\
\hline
Lepton energy scale   			& 7  &  10  &  5 \\
Lepton energy resolution        	& 1   &  4   &  0  \\
Lepton efficiency				& 1   &  2   &  0  \\
Lepton tower removal     		& 0   &  0   &  0  \\
Recoil scale	         			& 6   &  6   &  6 \\
Recoil resolution	 			& 5   &  5   &  5  \\
Backgrounds 	         			& 5   &  3   &  0  \\
PDFs                     		& 9  &  9  &  9 \\
$W$ boson $p_T$          		& 9   &  9   &  9  \\
Photon radiation         		& 4  &  4  &  4 \\
\hline
Statistical              		& 18   & 21  &  0  \\
\hline
Total                    		& 25   & 28  &  16 \\

\end{tabular}
\label{tbl:pt}
\end{center}
\end{ruledtabular}
\end{table}

\begin{table}[htbp]
\begin{ruledtabular}

\begin{center}
\caption{Uncertainties on $M_W$ (in MeV) as resulting from neutrino-transverse-momentum fits in the $W\to \mu\nu$ and $W\to e\nu$ samples. The last column reports the portion of uncertainty that is common in the $\mu\nu$ and $e\nu$ results.  } 

\begin{tabular}{lccc}

\multicolumn{4}{c}{$p_T^\nu$ fit uncertainties} \\
Source                   & $W\rightarrow \mu\nu$ & $W\rightarrow e\nu$ & Correlation \\
\hline
Lepton energy scale   			& 7  &  10  &  5 \\
Lepton energy resolution        	& 1   &  7   &  0  \\
Lepton efficiency				& 2   &  3   &  0  \\
Lepton tower removal     		& 4   &  6   &  4  \\
Recoil scale	         			& 2   &  2   &  2 \\
Recoil resolution	 			& 11   &  11   &  11  \\
Backgrounds 	         			& 6   &  4   &  0  \\
PDFs                     		& 11  &  11  &  11 \\
$W$ boson $p_T$          		& 4   &  4   &  4  \\
Photon radiation         		& 4  &  4  &  4 \\
\hline
Statistical              		& 22   & 25  &  0  \\
\hline
Total                    		& 30   & 33  &  18 \\

\end{tabular}
\label{tbl:met}
\end{center}
\end{ruledtabular}
\end{table}


\begin{table}
\caption{Summary of fit results to the $m_T$, $p_T^\ell$, and $p_T^\nu$ distributions for the electron and muon decay channels.}
\begin{ruledtabular}
\begin{tabular}{lcc}
Distribution & $M_W$ (MeV) & $\chi^2$/dof\\
\hline
$W\to e\nu$ & & \\
\ \ $m_T$  	&  $80\,408 \pm 19$   & 52/48   \\
\ \ $p_T^\ell$   		&  $80\,393 \pm 21$   & 60/62   \\
\ \ $p_T^\nu$	  	&  $80\,431 \pm 25$   & 71/62   \\
$W\to\mu\nu$ & & \\
\ \ $m_T$  	&  $80\,379 \pm 16$   & 57/48   \\
\ \ $p_T^\ell$  		&  $80\,348 \pm 18$   & 58/62   \\
\ \ $p_T^\nu$    	&  $80\,406 \pm 22$   & 82/62   \\
\end{tabular}
\label{tbl:fitsummary}

\end{ruledtabular}
\end{table}

\begin{table}
\caption{Statistical correlations between the $m_T$, $p_T^\ell$, and $p_T^\nu$ fits 
in the muon and electron decay channels.}
\begin{ruledtabular}
\begin{tabular}{lcc}
Correlation      & $W\rightarrow \mu\nu$ (\%) & $W\rightarrow e\nu$ (\%) \\
\hline
$m_T-p_T^\ell$        & $67.2\pm 2.8$  & $70.9\pm 2.5$ \\
$m_T-p_T^\nu$       & $65.8\pm 2.8$  & $69.4\pm 2.6$ \\
$p_T^\ell-p_T^\nu$	 & $25.5\pm 4.7$  & $30.7\pm 4.5$ \\
\end{tabular}
\label{tab:fitcorr}
\end{ruledtabular}
\end{table}%

We utilize the best-linear-unbiased-estimator (BLUE)~\cite{BLUE} algorithm to combine individual fits. Each source of systematic uncertainty is assumed to be independent from all other sources of uncertainty within a given fit. We perform simulated experiments~\cite{zeng} to estimate the statistical correlation between fits to the $m_T$, $p_T^\ell$, and $p_T^\nu$ distributions (Table~\ref{tab:fitcorr}). 

Combining the $M_W$ fits to the $m_T$ distributions in muon and electron channels, we obtain

\begin{equation}
M_W = 80\,390\pm 20 \textrm{ MeV}.
\end{equation}
The $\chi^2$/dof for this combination is 1.2/1 and the probability that two measurements would have a $\chi^2/$dof at least as large is 28\%.

Combining fits to the $p_T$ distributions in both muon and electron channels yields

\begin{equation}
M_W = 80\,366\pm 22 \textrm{ MeV}.
\end{equation}
The $\chi^2$/dof for this combination is 2.3/1 with a 13\% probability for the two measurements to result in $\chi^2/$dof $\geq 2.3$.

The result of combining the muon and electron channel fits to the $p_T^\nu$ distributions is

\begin{equation}
M_W = 80\,416\pm 25 \textrm{ MeV},
\end{equation}
with a 49\% probability of obtaining a $\chi^2/$dof value at least as large as the observed 0.5/1.

The combination of all three fits in the muon channel yields

\begin{equation}
M_W = 80\,374\pm 22 \textrm{ MeV},
\end{equation}
with a $\chi^2$/dof of 4.0/2. Combining all three fits in the electron channel results in the value

\begin{equation}
M_W = 80\,406\pm 25 \textrm{ MeV},
\end{equation}
with a $\chi^2$/dof of 1.4/2.

We combine all six fits to the final result,

\begin{equation}
M_W = 80\,387\pm 19 \textrm{ MeV}.
\end{equation}
The relative weights, as calculated by the BLUE method~\cite{BLUE}, of the $m_T$, $p_T^\ell$, and $p_T^\nu$ fits in this combination are 53\%, 31\% and 16\%, respectively. The contribution of the muon (electron) channel in the final combination is 62\% (38\%). The $\chi^2/$dof of this combination is 6.6/5 with a 25\% probability of obtaining a $\chi^2/$dof at least as large. We evaluate the contribution from each source of systematic uncertainty in the combined measurement; these uncertainties are presented in Table~\ref{tbl:combinedsys}.

\begin{table}
\caption{Uncertainties in units of MeV on the final combined result on $M_W$. }
\begin{ruledtabular}
\begin{center}
\begin{tabular}{lc}

Source                   & Uncertainty\\
\hline
Lepton energy scale and resolution & 7 \\
Recoil energy scale and resolution & 6 \\
Lepton tower removal     		& 2  \\
Backgrounds 	         			& 3  \\
PDFs                     		& 10\\
$p_T(W)$  model         		& 5  \\
Photon radiation         		& 4 \\
\hline
Statistical              		& 12  \\
\hline
Total                    		& 19 \\

\end{tabular}
\label{tbl:combinedsys}
\end{center}
\end{ruledtabular}
\end{table}

\subsection{Consistency checks}

We test our results for unaccounted systematic biases by dividing the data into several subsamples and comparing the electron and muon $p_T^\ell$ fit results obtained from these subsamples (Table~\ref{tbl:crosschecks}). The uncertainty shown for $M_W$($\ell^+$)$-M_W$($\ell^-$) in the muon channel includes the systematic uncertainty on the mass fits in the $W^+\to\mu^+\nu$ and $W^-\to\mu^-\nu$ channels due to the COT alignment parameters $A$ and $C$ (Sec.~\ref{sec:alignment}), which contribute to this mass splitting. For the electron channel, we show the mass fit differences with and without applying an $E/p$-based calibration from the corresponding subsample. A residual dependence of the CEM energy scale on azimuth and time is observed. By suppressing this dependence through a calibration, the remaining variation of the electron channel mass fit is eliminated. 

\begin{table}
\caption{Charged-lepton $p_T$-fit mass shifts (in MeV) for subdivisions of our data. For the spatial and time dependence of the electron channel fit result, we show the dependence without (with) the corresponding cluster energy calibration using the subsample $E/p$ fit. The variation observed without cluster energy recalibration is eliminated upon recalibration, proving that the effect arises dominantly due to residual variation of the energy scale.}
\begin{ruledtabular}
\begin{tabular}{lcc}
Fit difference      & $W\rightarrow \mu\nu$  & $W\rightarrow e\nu$ \\
\hline
$M_W$($\ell^+$)$-M_W$($\ell^-$)                 & $ 71\pm 70$   & $-49 \pm 42$ \\
$M_W$($\phi_\ell > 0$)$ - M_W$($\phi_\ell < 0$) & $-54 \pm 36$     & $-117\pm 42 (-58\pm 45)$ \\ 
$M_W$(Aug 2006-Sep 2007) $-$ & & \\ 
$M_W$(Mar 2002-Aug 2006) & $116 \pm 36$ & $-266 \pm 43 (39\pm 45)$ \\
\end{tabular}
\label{tbl:crosschecks}
\end{ruledtabular}
\end{table}

\begin{figure*}
\begin{center}
\includegraphics [width=3.2in] {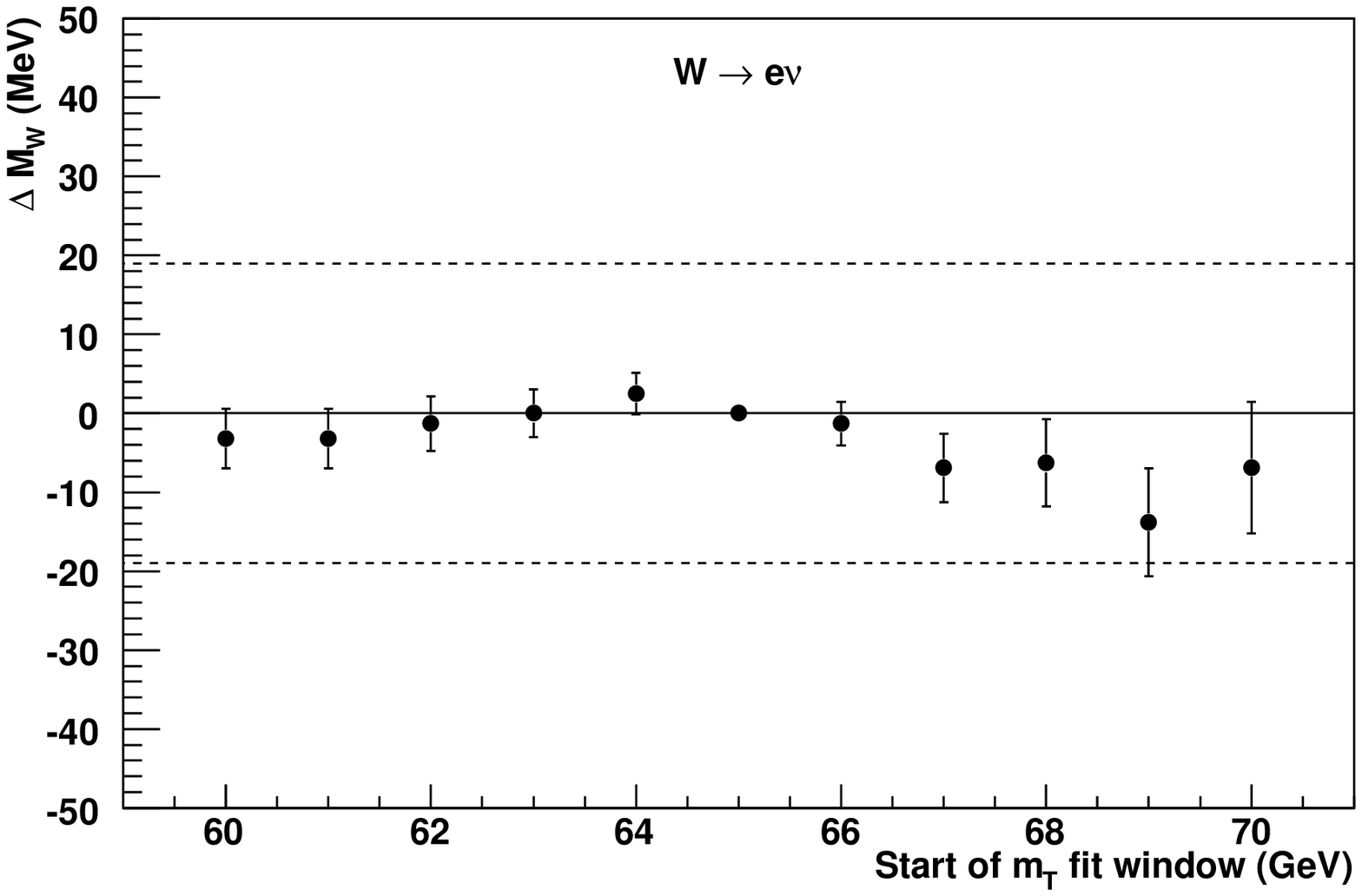}
\includegraphics [width=3.2in] {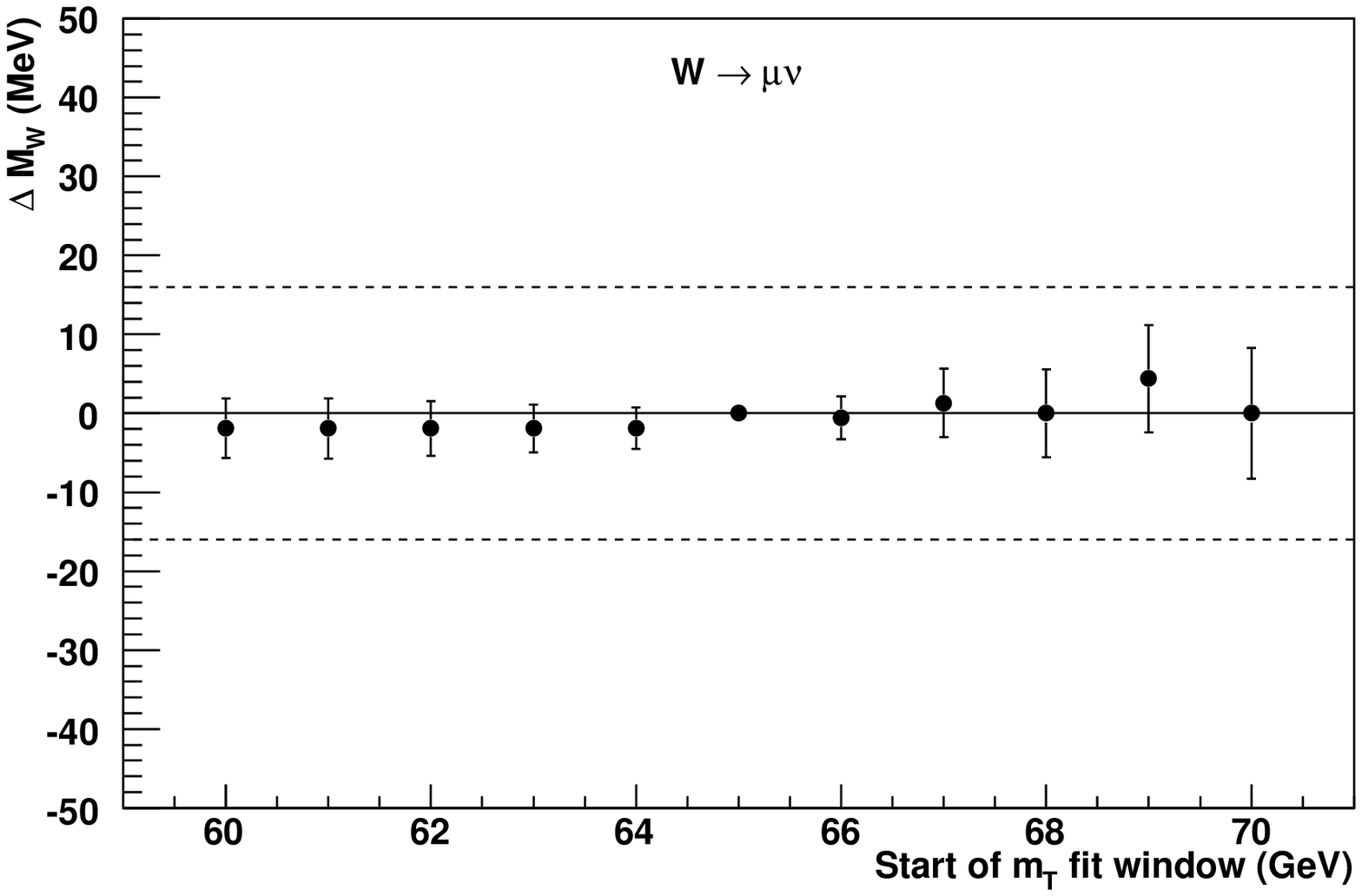}
\includegraphics [width=3.2in] {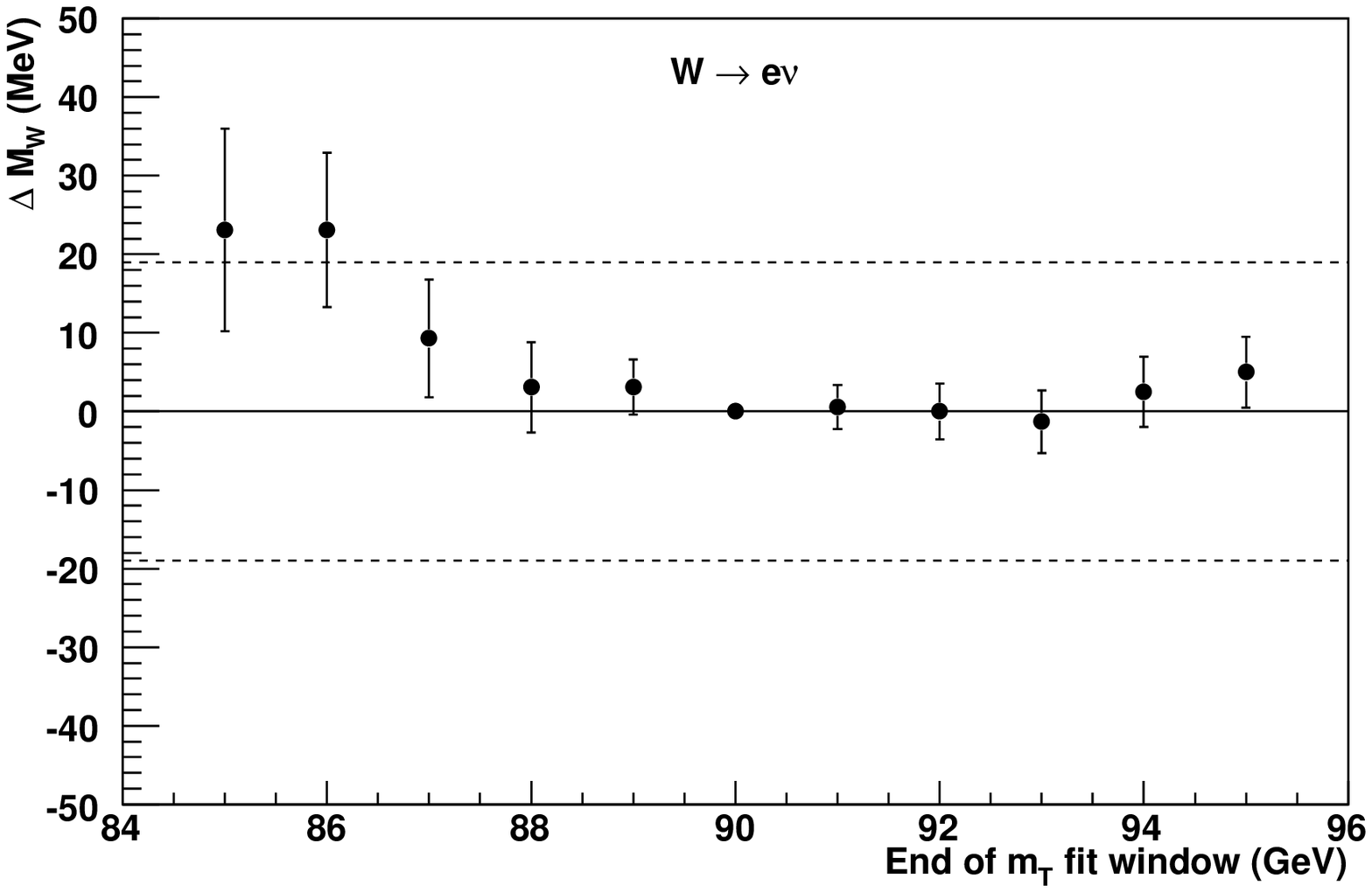}
\includegraphics [width=3.2in] {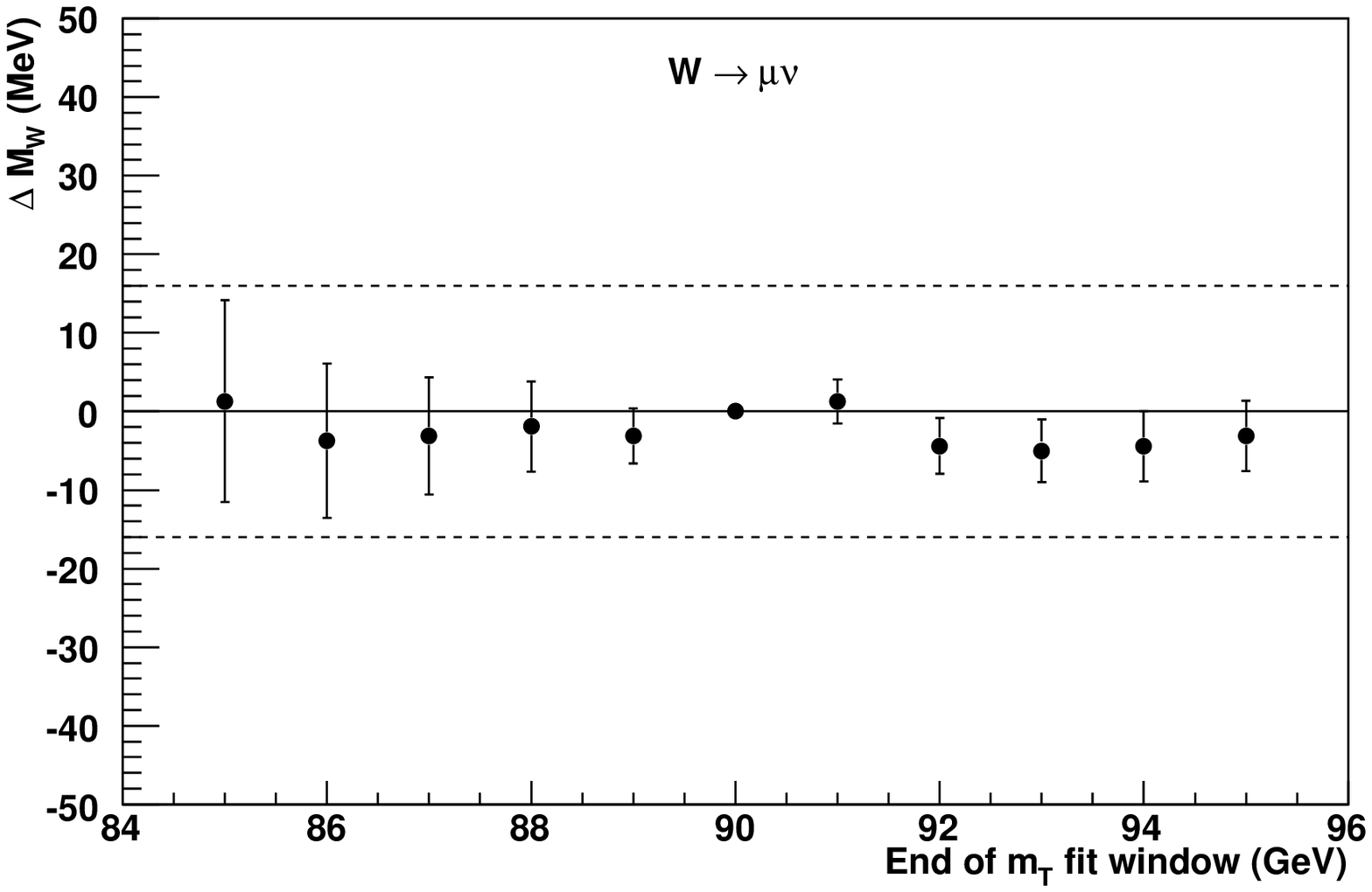}
\caption{Variations of the $M_W$ value determined from the transverse-mass fit as a function of the choice of the (top) lower and (bottom) upper edge of the fit range, for the electron (left) and muon (right) channels.  Uncertainty 
bars indicate expected variationwith respect to the default fit window, as computed using pseudoexperiments. The dashed lines
 indicate the statistical uncertainty from the default mass fit.}
\label{mass_vs_end_mt}
\end{center}
\end{figure*}

\begin{figure*}
\begin{center}
\includegraphics [width=3.2in] {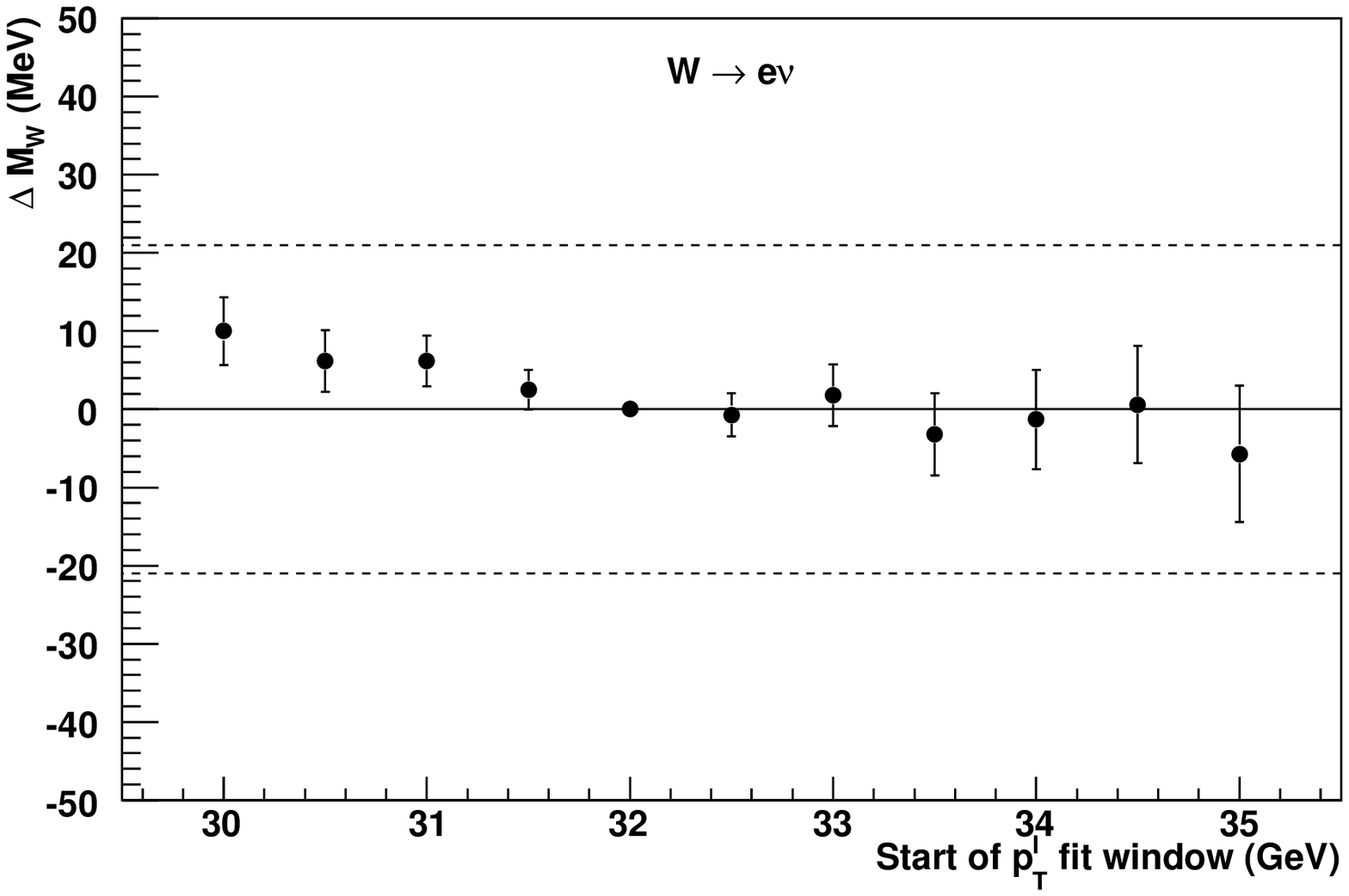}
\includegraphics [width=3.2in] {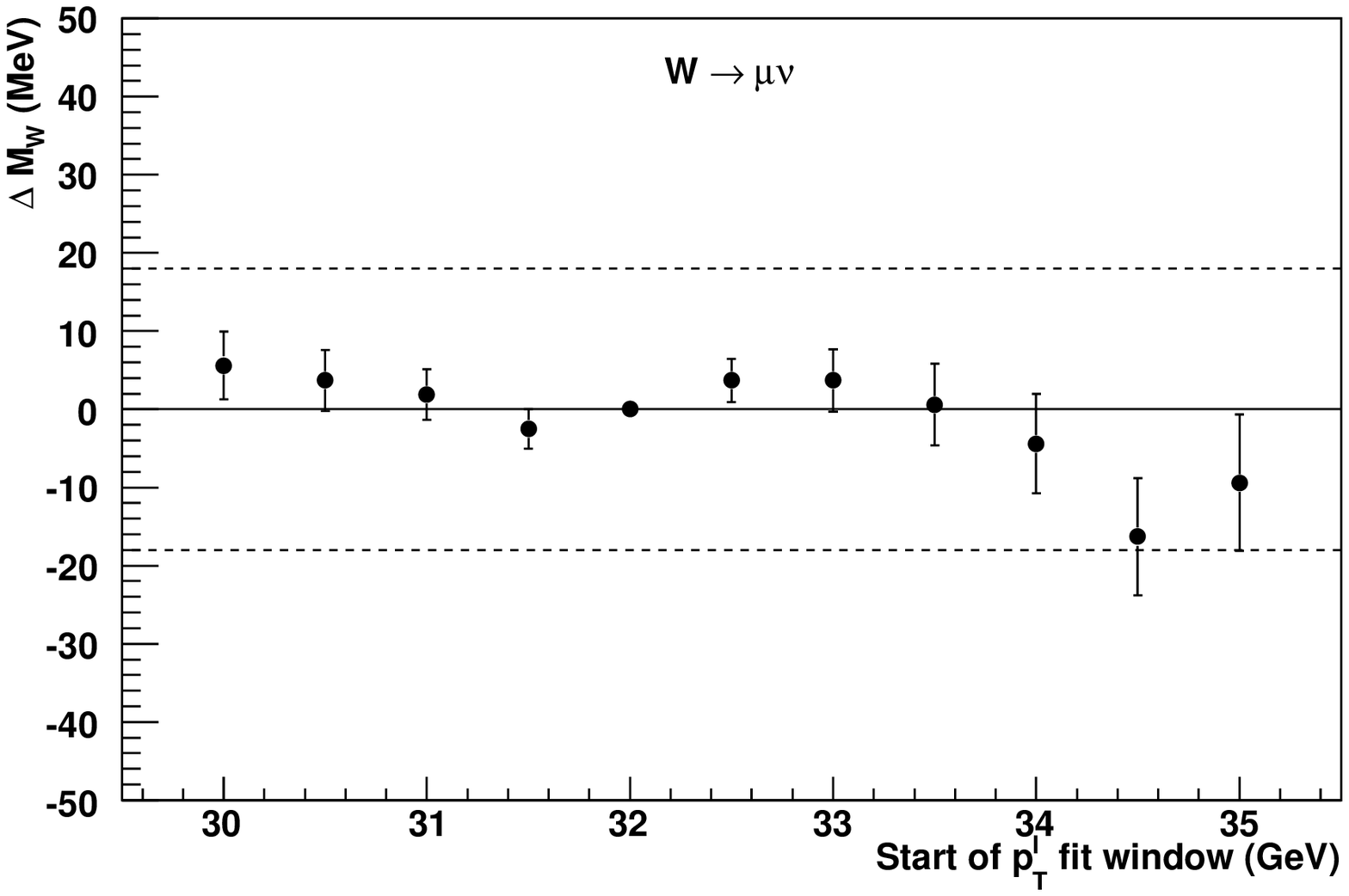}
\includegraphics [width=3.2in] {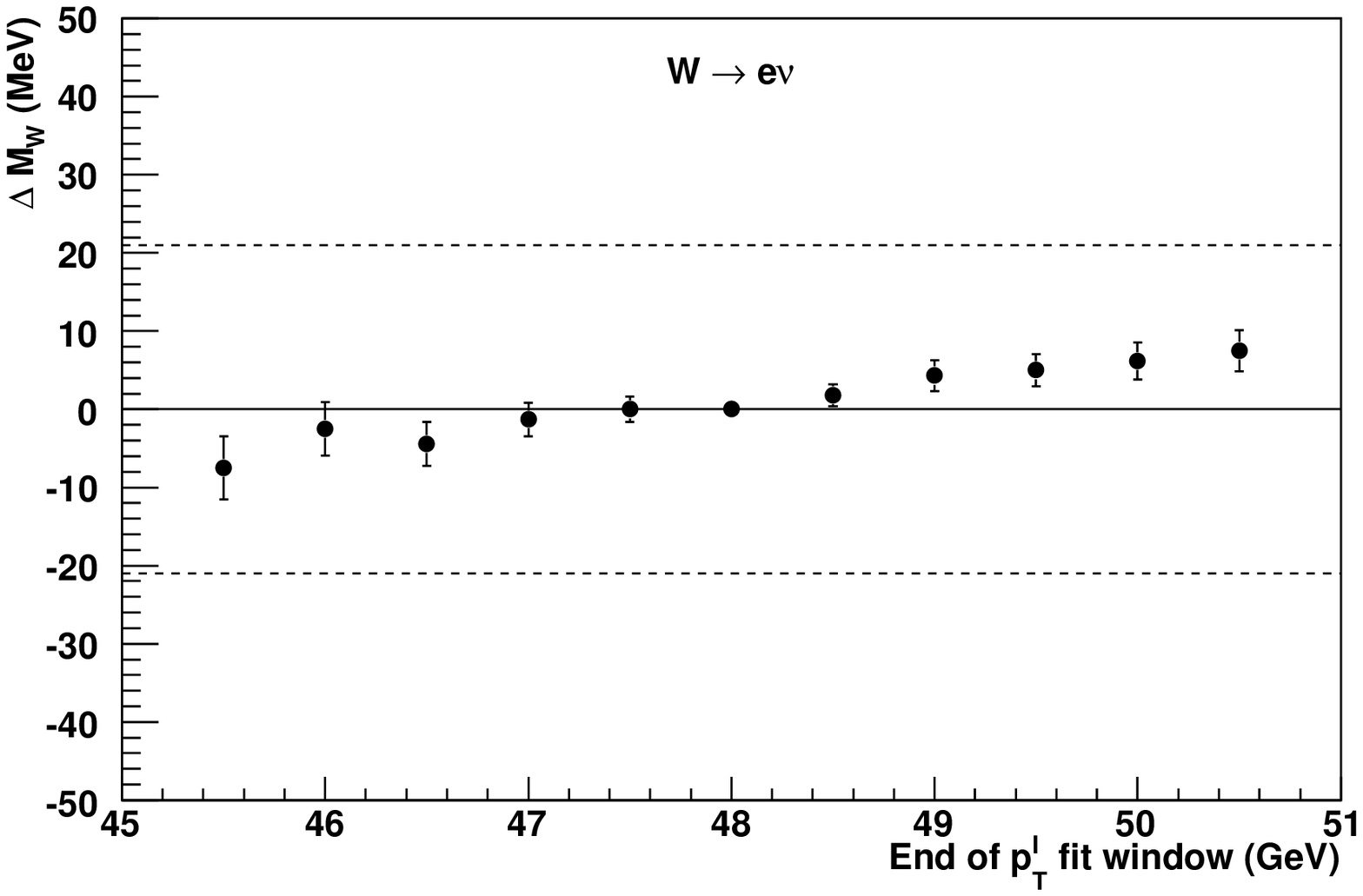}
\includegraphics [width=3.2in] {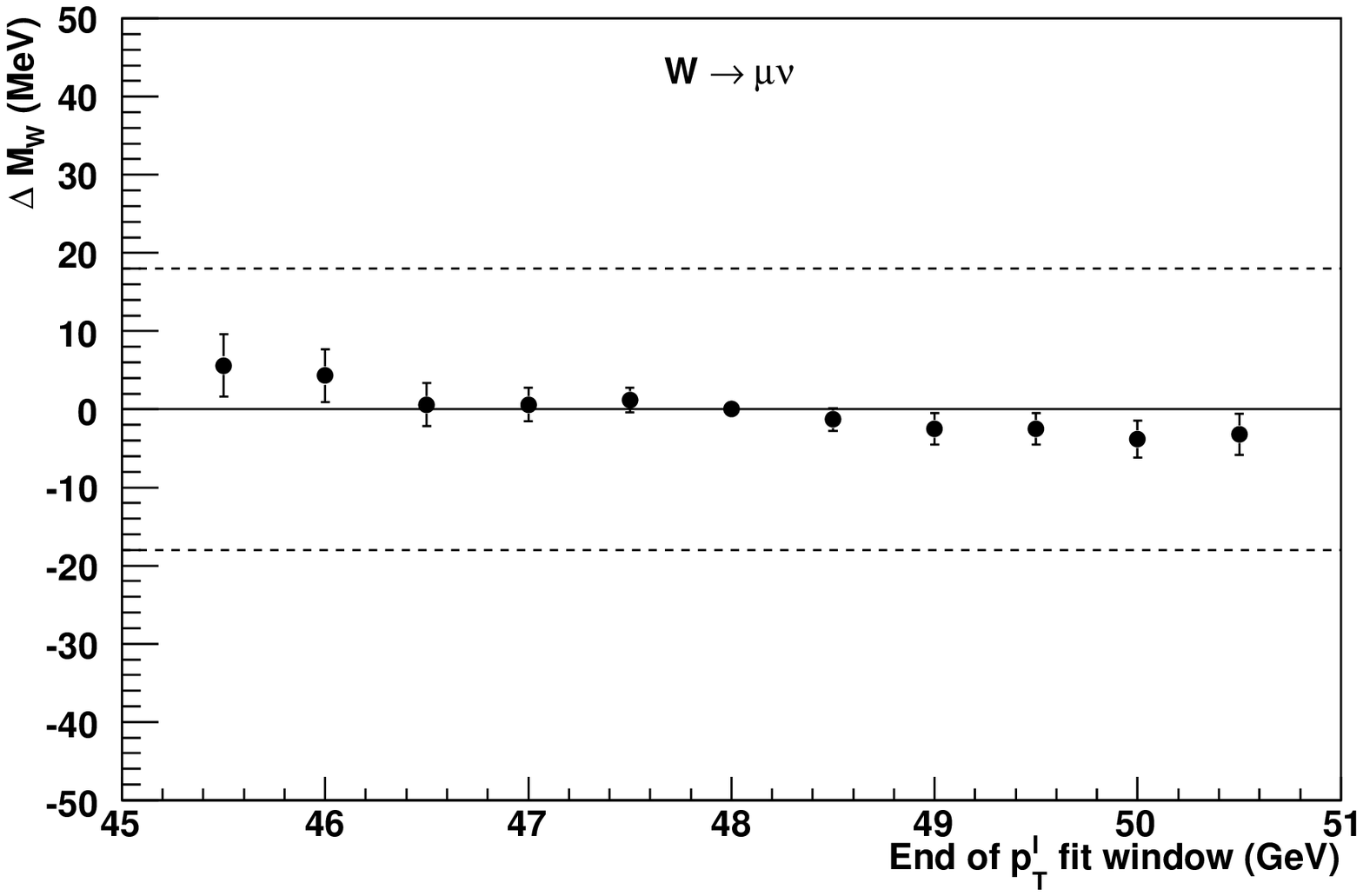}
\caption{Variations of the $M_W$ value determined from the charged-lepton transverse-momentum fit as a function of the choice of the (top) lower and (bottom) upper edge of the fit range, for the electron (left) and muon (right) channels.
 Uncertainty bars indicate expected variation with respect to the default fit window, as computed using pseudoexperiments.
The dashed lines indicate the statistical uncertainty from the default mass fit.}
\label{mass_vs_end_pt}
\end{center}
\end{figure*}

\begin{figure*}
\begin{center}
\includegraphics [width=3.2in] {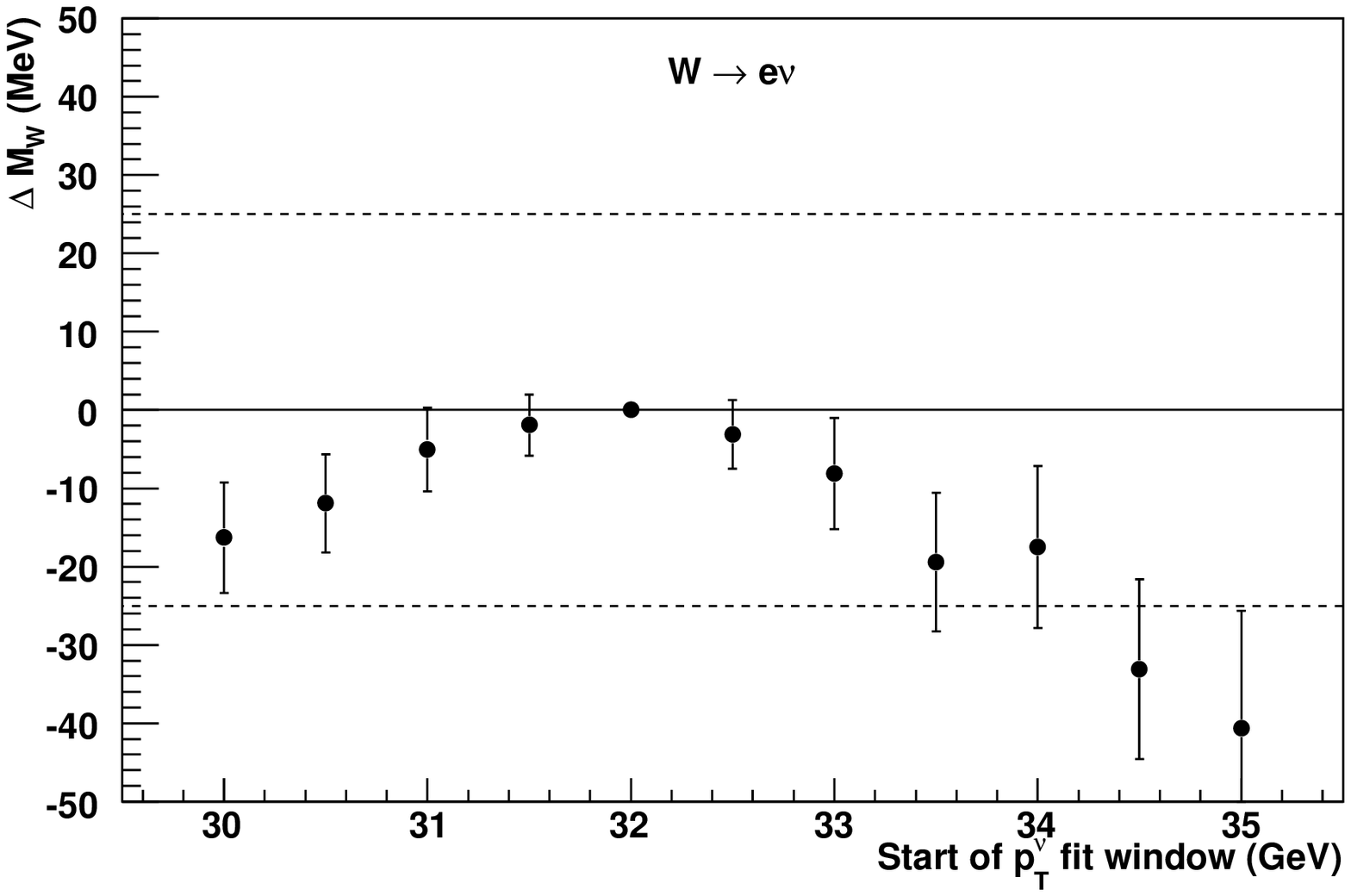}
\includegraphics [width=3.2in] {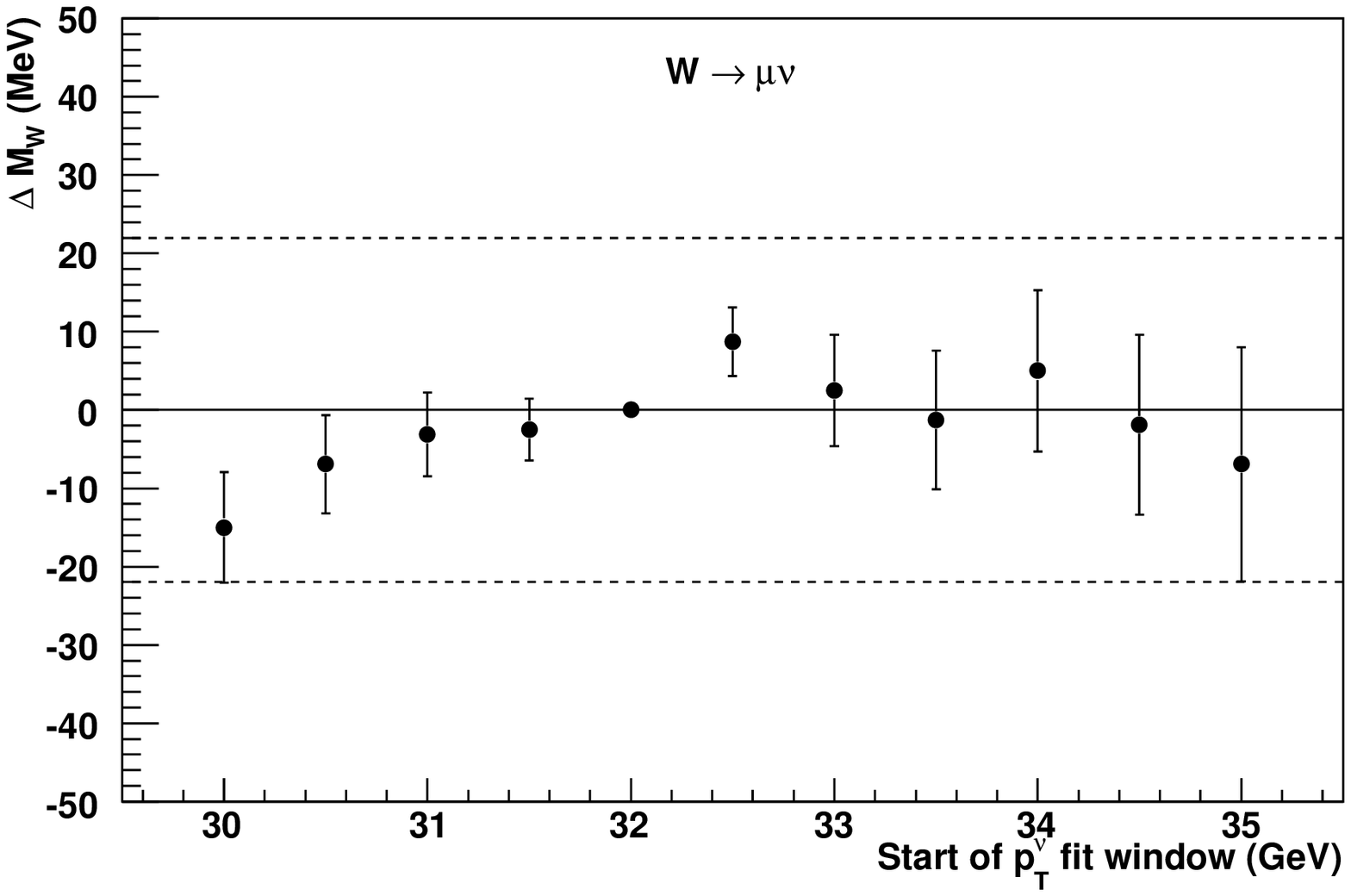}
\includegraphics [width=3.2in] {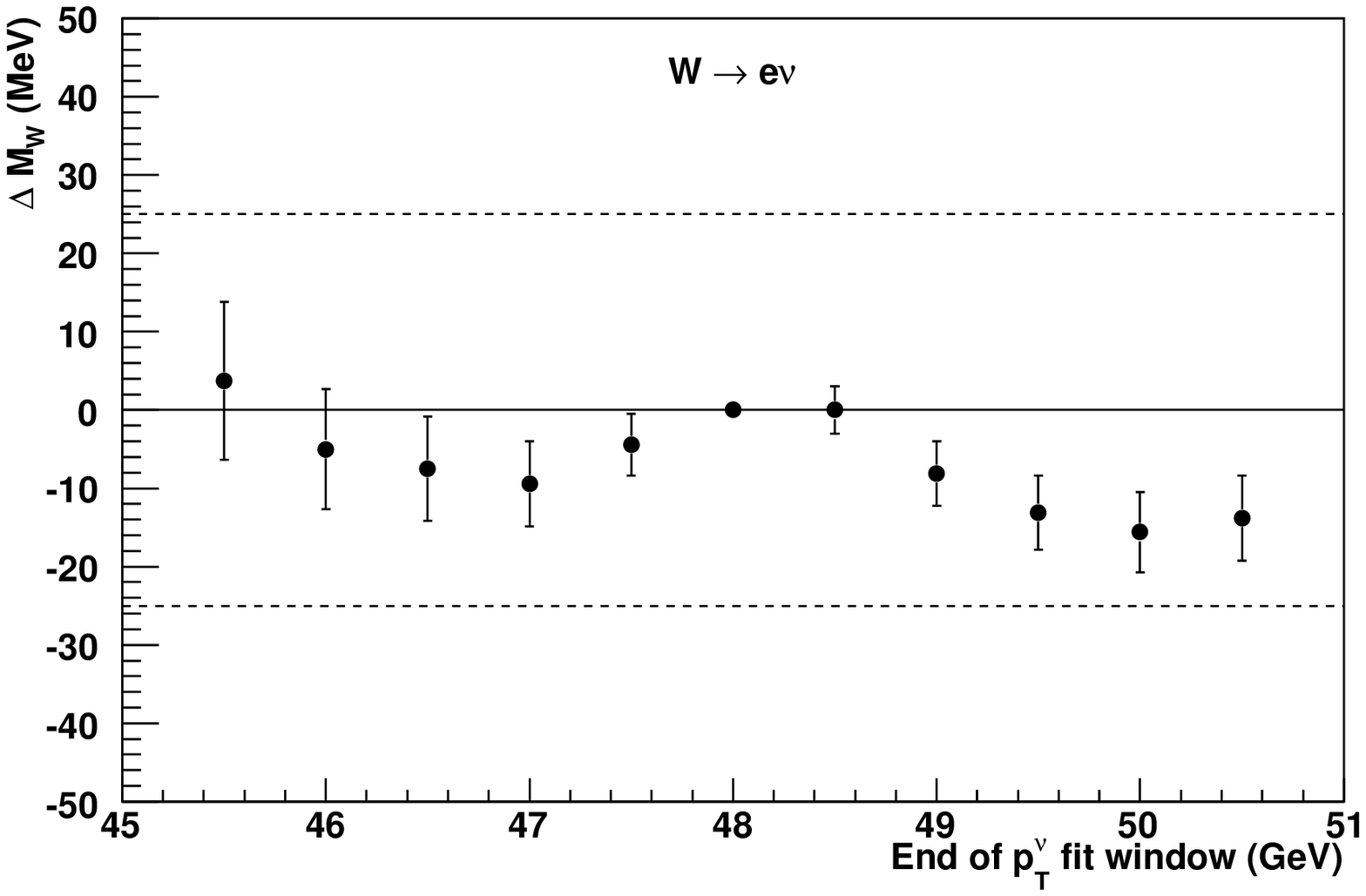}
\includegraphics [width=3.2in] {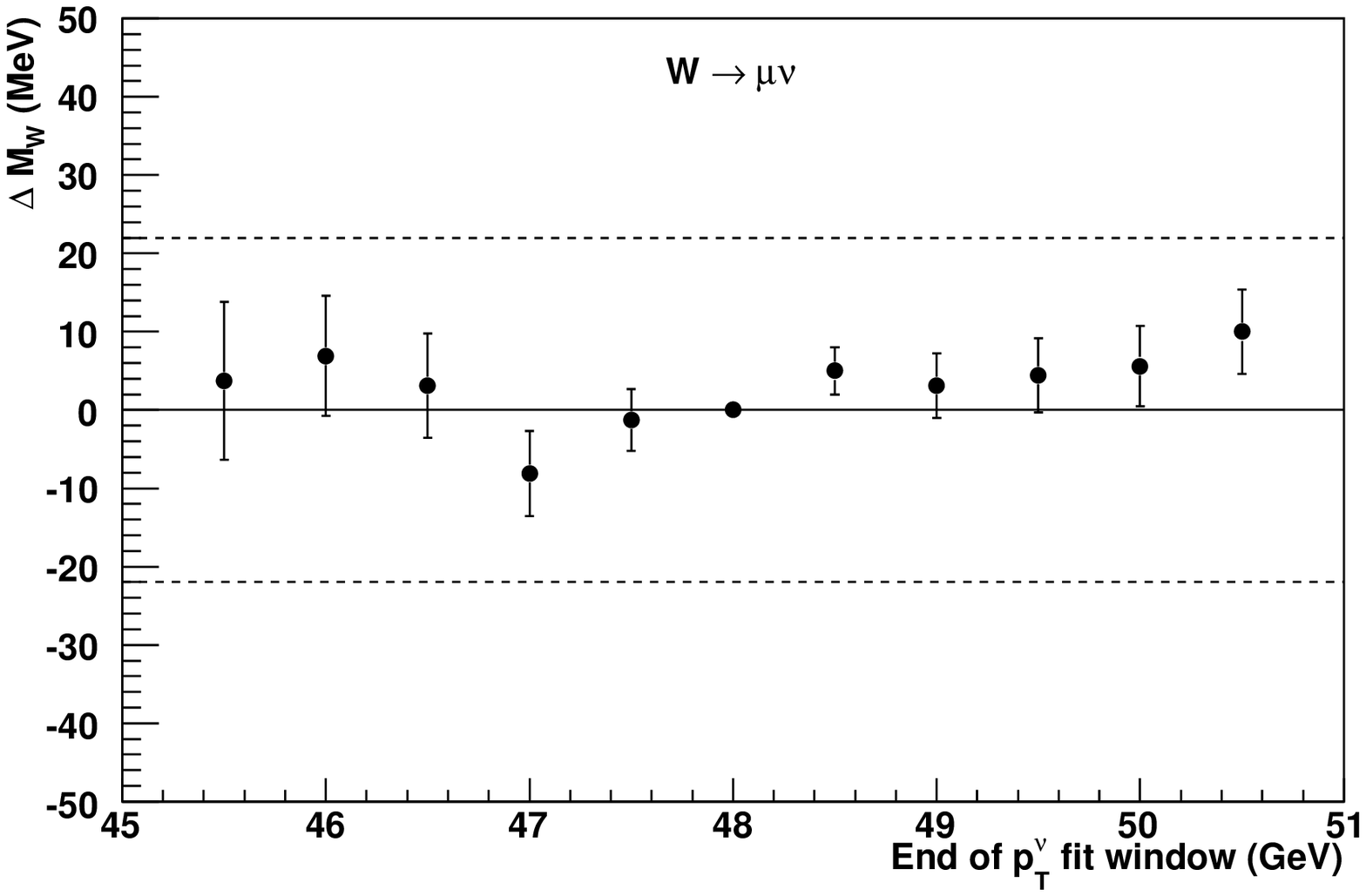}
\caption{Variations of the $M_W$ value determined from the neutrino-transverse-momentum fit as a function of the choice of the (top) lower and (bottom) upper edge of the fit range, for the electron (left) and muon (right) channels.
 Uncertainty bars indicate expected variation with respect to the default fit window, as computed using pseudoexperiments.
The dashed lines  indicate the statistical uncertainty from the default mass fit.}
\label{mass_vs_end_met}
\end{center}
\end{figure*}

The variations of the fitted mass values relative to the nominal results, as the fit regions are varied, are consistent with statistical fluctuations, as shown in Figs.~\ref{mass_vs_end_mt}-\ref{mass_vs_end_met}~\cite{zeng}. Furthermore, this consistency check is conservative, as the known systematic uncertainties are not included in displayed error bars. The systematic uncertainties that we consider (Tables~\ref{tbl:mt}-\ref{tbl:met}) would induce additional expected shifts between shift regions. The observed shifts in Figs.~\ref{mass_vs_end_mt}-\ref{mass_vs_end_met} are typically substantially smaller than these systematic unertainties.

\section{Summary}
\label{sec:summary}
We  measure the $W$-boson mass using a sample of proton-antiproton collision data corresponding to an integrated luminosity of  
$2.2$ fb$^{-1}$ collected by the CDF II detector at 
$\sqrt{s} = 1.96$ TeV.  We  use fits to $m_T$, $p_T$, and \met 
distributions of the $W\rightarrow \mu\nu$ and $W\rightarrow e\nu$ 
data samples to obtain

\begin{equation*}
M_W = 80387 \pm 12_{\rm stat} \pm 15_{\rm syst}~{\rm MeV} =  80387 \pm 19~{\rm MeV},
\end{equation*}

\noindent
which is the single most precise measurement of $M_W$ to date.  This 
 measurement subsumes the previous CDF measurement from a 200 pb$^{-1}$ subset of the present data~\cite{CDF2}. 
\par
Using the method described in Ref.~\cite{run2combo}, we obtain a combined Tevatron result of

\begin{equation*}
M_W = 80385 \pm 16~{\rm MeV.}
\end{equation*}

\noindent
which includes the most recent measurement of $M_W$ from D0~\cite{dzero5fbprl}. 

Assuming no correlation between the Tevatron and LEP 
measurements, we obtain a new world average of 

\begin{equation*}
\label{worldaverage}
M_W = 80385 \pm 15~{\rm MeV.}
\end{equation*}

Following the discovery of the Higgs boson at the LHC and the measurement of its mass~\cite{lhchiggs}, all of the SM parameters required to make a prediction of the $W$-boson mass are now known. Including the radiative corrections mentioned in Eq.~(\ref{eq:mwtheory}), the mass of the $W$ boson is predicted to be~\cite{gfitter}

\begin{equation*}
\label{worldaverage}
M_W = 80359 \pm 11~{\rm MeV.}
\end{equation*}

The comparison of this prediction with our measurement over-constrains the SM and provides a stringent test of the radiative corrections. The level of consistency between the prediction and the measurement places bounds on non-SM physics that can affect $M_W$ at tree-level or via loops.

We thank the Fermilab staff and the technical staffs of the
participating institutions for their vital contributions. This work
was supported by the U.S. Department of Energy and National Science
Foundation; the Italian Istituto Nazionale di Fisica Nucleare; the
Ministry of Education, Culture, Sports, Science and Technology of
Japan; the Natural Sciences and Engineering Research Council of
Canada; the National Science Council of the Republic of China; the
Swiss National Science Foundation; the A.P. Sloan Foundation; the
Bundesministerium f\"ur Bildung und Forschung, Germany; the Korean
World Class University Program, the National Research Foundation of
Korea; the Science and Technology Facilities Council and the Royal
Society, United Kingdom; the Russian Foundation for Basic Research;
the Ministerio de Ciencia e Innovaci\'{o}n, and Programa
Consolider-Ingenio 2010, Spain; the Slovak R\&D Agency; the Academy
of Finland; the Australian Research Council (ARC); and the EU community
Marie Curie Fellowship Contract No. 302103.


\begin{thebibliography}{99}
\bibitem{g2value}
Since the best-fit value of $g_2$ depends on the values of other parameters and PDFs used 
in the fitting model, we do not quote a measurement of $g_2$. As a consistency check, we compare 
our best-fit value of $g_{CDF} = (0.649 \pm 0.013_{\textrm stat})$~GeV with the global-fit 
value of $g_2 = 0.68\pm 0.01$, and we find consistency considering that different PDFs yield 
an uncertainty on $g_2$ of $\mathcal{O}(0.02)$.
\end{thebibliography}
\end{document}